\documentclass[preprint,sor4-1,longbibliography]{revtex4-1}
\usepackage{mathtools}
\usepackage[table]{xcolor}
\definecolor{lightgray}{gray}{0.9}
\usepackage{graphics}
\usepackage{graphicx}

\begin{document}

\title{Single Polymer Dynamics for Molecular Rheology}

\author{Charles M. Schroeder}
\email{cms@illinois.edu}
\affiliation{Department of Chemical and Biomolecular Engineering, University of Illinois at Urbana-Champaign, 600 S. Mathews Avenue, Urbana, IL 61801, USA}

\date{\today}

\begin{abstract}
Single polymer dynamics offers a powerful approach to study molecular-level interactions and dynamic microstructure in materials. Direct visualization of single chain dynamics has uncovered new ideas regarding the rheology and non-equilibrium dynamics of macromolecules, including the importance of molecular individualism, dynamic heterogeneity, and molecular sub-populations that govern macroscale behavior. In recent years, the field of single polymer dynamics has been extended to increasingly complex materials, including architecturally complex polymers such as combs, bottlebrushes, and ring polymers and entangled solutions of long chain polymers in flow. Single molecule visualization, complemented by modeling and simulation techniques such as Brownian dynamics and Monte Carlo methods, allow for unparalleled access to the molecular-scale dynamics of polymeric materials. In this review, recent progress in the field of single polymer dynamics is examined by highlighting major developments and new physics to emerge from these techniques. The molecular properties of DNA as a model polymer are examined, including the role of flexibility, excluded volume interactions, and hydrodynamic interactions in governing behavior. Recent developments in studying polymer dynamics in time-dependent flows, new chemistries and new molecular topologies, and the role of intermolecular interactions in concentrated solutions are considered. Moreover, cutting-edge methods in simulation techniques are further reviewed as an ideal complementary method to single polymer experiments. Future work aimed at extending the field of single polymer dynamics to new materials promises to uncover original and unexpected information regarding the flow dynamics of polymeric systems. 
\end{abstract}

\maketitle

\section{Introduction}
A grand challenge in the field of rheology is to understand how the emergent or macroscopic properties of complex materials arise from microscopic interactions. To this end, a tremendous amount of research has been focused on the development of molecular-level constitutive equations and kinetic theory for polymeric liquids \cite{Larsonbook,Larsonbook2,Doiedwards,Bird1987,Bird1987b,McLeish1998}. In tandem, bulk-level experimental methods including mechanical rheometry and optical rheometry have been used to study the macroscopic response of polymer solutions and melts in flow \cite{Bird1987,Fullerbook,Bairdbook}. Moreover, a substantial amount of our current understanding of macromolecular behavior has been provided by spectroscopic methods such as nuclear magnetic resonance (NMR) \cite{Koenigbook} and neutron spin echo spectroscopy \cite{Richter2005}, where the latter technique has been particularly useful in studying densely entangled and glassy systems. Together, these approaches have provided tremendous insight into the properties of polymer solutions and melts. 

An alternative and particularly powerful method for studying non-equilibrium properties of polymers involves direct observation of single molecules. In this review, we define the term molecular rheology as the use of experimental and computational molecular-based methods to directly observe and study the dynamics of polymers under equilibrium and non-equilibrium conditions. In the context of experimental molecular rheology, these methods generally rely on single molecule fluorescence microscopy (SMFM) to observe the dynamics of fluorescently labeled polymers such as DNA in flow. Single polymer dynamics offers the ability to directly observe the molecular conformations of polymer chains near equilibrium or in strong flows \cite{Shaqfeh2005,Mai2016}, thereby providing a window into viewing non-equilibrium chain conformations that ultimately give rise to bulk-level properties such as stress and viscosity. 

The modern field of single polymer dynamics began in earnest over 20 years ago with the development of new methods for directly visualizing single DNA molecules \cite{Chu1991,ChuKron1990}. In the mid-1990's, advances in fluorescence imaging and low-light level detection were complemented by new methods for precise fabrication of microfluidic devices \cite{Quake2000} and optical tweezing for manipulating single DNA \cite{Ashkin1986}. Together, these efforts allowed for pioneering experiments of single DNA dynamics in well defined flows and conditions, including DNA relaxation from high stretch \cite{Perkins1994a} and direct observation of the tube-like motion of DNA in a concentrated polymer solution \cite{Perkins1994b}. Fortuitously, around the same time, the development of cutting-edge methods in optical microscopy and DNA manipulation were complemented by advances in theoretical modeling of polymer elasticity \cite{Marko1995} and coarse-grained polymer models for computer simulations \cite{Ottingerbook,Doyle1997}. The confluence of these advances in related fields, including both computation, theory, and experiment, acted in synergy to usher in unprecedented and fundamentally new methods in molecular-scale analysis of polymer dynamics. 

Over the last few years, researchers have used these techniques to uncover fundamentally new information regarding polymer chain dynamics in non-equilibrium conditions, including the importance of distributions in polymer conformations, heterogeneous chain dynamics at the single polymer level, and molecular individualism \cite{deGennes1997}. Today, the vibrant field of single polymer dynamics continues to advance into new territory, extending the direct observation of polymer chain dynamics to architecturally complex polymers, densely entangled solutions, and chemically heterogeneous polymers. In this review article, the current state of the field of single polymer dynamics is explored with a particular emphasis on the physical properties of model polymers and new directions in the field. For discussions on related topics, I refer the reader to recent reviews on microfluidic and nanofluidic devices for performing single polymer studies \cite{Mai2012,Rems2016}, DNA dynamics under confinement \cite{Dai2016}, and the electrophoretic motion of DNA \cite{Dorfman2010}. In this article, I focus specially on the hydrodynamics and non-equilibrium behavior of single polymers in dilute, semi-dilute, and entangled solutions.  

The review article is organized as follows. In Section \ref{DNAprops}, the physical properties of DNA as a model polymer are discussed, including issues surrounding flexibility, persistence length, and monomer aspect ratio. The role of excluded volume (EV) interactions and solvent quality are considered in the context of an effective excluded volume exponent, the theta temperature, and the chain interaction parameter for DNA. The effect of EV interactions on the elasticity of double stranded and single stranded DNA is further considered. In Section \ref{dilutepre2007}, the dynamics of single DNA molecules in dilute solutions is reviewed with a focus on early progress through 2007, with work including dynamics of single DNA chains in shear flow, extensional flows, and linear mixed flows. The role of intramolecular hydrodynamic interactions (HI) is discussed in detail, including the development of computational methods to accurately model HI and the emergence of polymer conformation hysteresis. In Section \ref{diluterecent}, recent progress in the study of dilute solution single polymer dynamics for linear chains is reviewed since 2007, including single polymer observation of single stranded DNA, dynamics in time-dependent oscillatory flows, and the development of non-equilibrium work relations for polymer dynamics. The framework of successive fine graining for modeling polymer chain dynamics is further discussed. In Section \ref{nondilute}, single chain dynamics in semi-dilute unentangled and entangled solutions is discussed, along with observation of elastic instabilities and shear banding in DNA solutions. In Section \ref{branched}, recent work in extending single polymer dynamics to polymers with complex molecular architectures such as combs and bottlebrush polymers is discussed. Finally, the review article concludes in Section \ref{conclusion} with an evaluation of recent progress in the field and perspectives for future work.

\section{DNA as a model polymer: physical properties}
\label{DNAprops}
Double stranded DNA has served as a model polymer for single molecule experiments for many years. Although DNA is an interesting and biologically relevant polymer, it was selected for single polymer dynamics because of several inherent properties. First, DNA is water-soluble and can be studied in aqueous buffered solutions, which are generally compatible with microfluidic devices fabricated from poly(dimethylsiloxane) (PDMS) using standard methods in soft lithography. Moreover, a wide variety of fluorescent dyes has been developed for imaging applications in molecular and cellular biology, and these experiments are commonly carried out in aqueous solutions. From this view, single polymer studies of DNA have benefited from these efforts in developing bright and photostable fluorescent dyes and by further leveraging strategies for minimizing photobleaching in aqueous solutions. Second, DNA is a biological polymer and can be prepared as a perfectly monodisperse polymer samples using polymerase chain reaction (PCR) \cite{Mai2015} or by extracting and purifying genomic DNA from viruses or microorganisms \cite{Laib2006}. Monodisperse polymer samples greatly simplify data analysis and interpretation of physical properties. Third, DNA can be routinely prepared with extremely high molecular weights, thereby resulting in polymer contour lengths $L$ larger than the diffraction limit of visible light ($\approx$ 300 nm). For example, bacteriophage lambda DNA (48,502 bp) is a common DNA molecule used for single polymer dynamics with a natural (unlabeled) crystallographic contour length $L$ = 16.3 $\mu$m. Such large contour lengths enable the direct observation of polymer chain conformation dynamics in flow using diffraction-limited fluorescence imaging. Moreover, $\lambda$-DNA is commercially available, which circumvents the need for individual polymer physics laboratories to prepare DNA in-house using biochemical or molecular biology techniques. Fourth, the physical properties of DNA are fairly well understood, which enables complementary quantitative modeling and simulation of experimental data. Finally, DNA can be often prepared in linear or ring topologies because many forms of genomic DNA occur in naturally circular form \cite{Laib2006}. The ability to prepare DNA as a linear macromolecule or as a ring polymer allows for precise investigation of the effect of topology on dynamics \cite{Robertson2007}.

\subsection{Persistence length, effective width, and monomer aspect ratio}
DNA is a negatively charged semi-flexible polymer with a fairly large persistence length compared to most synthetic polymers \cite{Marko1995}. In this review, I use the term semi-flexible to denote a polymer that is described by the wormlike chain or Kratky-Porod model \cite{Kratky1949}. As noted below, the elasticity of wormlike chains is qualitatively different compared to flexible polymers, especially in the low force regime. Natural B-form DNA (0.34 nm/bp, 10.5 bp/helix turn) has a persistence length of $l_p \approx 50$ nm in moderate ionic strength conditions consisting of at least 10 mM monovalent salt \cite{Hagerman1998}. At low ionic strengths ($<$ 1 mM monovalent salt), $l_p$ can increase to over 200 nm due to the unusually high linear charge density of DNA \cite{Marko1995,Dobrynin2006}. Nevertheless, most single polymer dynamics experiments are performed under reasonable ionic strength conditions with Debye lengths $l_D \approx$ 1-2 nm, conditions under which DNA essentially behaves as a neutral polymer under moderate or high ionic strength conditions. The generally accepted value of the persistence length for unlabeled DNA is $l_p$ = 53 nm \cite{Bustamante1994}. A further consideration is the effect of the fluorescent dye on the persistence length $l_p$ of DNA. A broad class of nucleic acid dyes such as the cyanine dimer or TOTO family of dyes (TOTO-1, YOYO-1) are known to intercalate along the backbone of DNA, which is thought to change local structure by slightly unwinding the DNA double helix. The precise effect of intercalating dyes on the persistence length of DNA has been widely debated, with some recent atomic force microscopy (AFM) experiments suggesting that the action of YOYO-1 does not appreciably change the persistence length of DNA upon labeling \cite{Kundukad2014}. For the purposes of this review, the persistence length is taken as $l_p$ = 53 nm and the Kuhn length $b=2l_p$ = 106 nm, such that $\lambda$-DNA contains approximately $N = L/b \approx$ 154 Kuhn steps.

The effective width $w$ of DNA can be envisioned as arising from electrostatic and steric interactions along the DNA backbone, and these interactions play a role in determining the static properties of DNA such as the radius of gyration ($R_g$) and excluded volume (EV) interactions. It should be emphasized that the effective width $w$ is different than the hydrodynamic diameter $d$ of DNA, the latter of which is generally smaller than $w$ (such that $d\approx$ 2 nm) and is important in modeling hydrodynamic friction, chain dynamics, and hydrodynamic interactions (HI), as discussed below. In any event, the rise of the double helix can be roughly estimated by calculating bond sizes to yield an approximate width of 2 nm. However, DNA is a charged polymer, and $w$ is also dependent on the ionic strength of the solution \cite{Tree2013}. Calculations by Stigter that considered the second virial coefficient of stiff charged rods predict an effective width $w\approx$ 4-5 nm under conditions of $\approx$ 150 mM monovalent salt \cite{Stigter1977}. 

Monomer aspect ratio $b/w$ provides a quantitative measure of monomer stiffness or anisotropy. Using the Kuhn length $b$ = 106 nm and an effective width $w$ = 4 nm, DNA has a monomer aspect ratio $b/w \approx$ 25 under moderate salt concentrations around 150 mM monovalent salt. On the other hand, most synthetic flexible polymers have much smaller monomer aspect ratios such that $b/w \approx \mathcal{O}(1)$. As a comparison, single stranded DNA (ssDNA) has a persistence length $l_p \approx 1.5$ nm under moderate salt conditions (150 mM Na$^+$) \cite{Chen2012} and a bare, non-electrostatic persistence length of 0.62 nm \cite{Saleh2009}. Limited experimental data exist on the effective width of ssDNA, but magnetic tweezing experiments on ssDNA elasticity suggest that $w$ is relatively independent of salt for ssDNA \cite{McIntosh2009}. A reasonable assumption is an effective width $w \approx$ 1.0 nm \cite{Tree2013} for ssDNA, which yields a monomer aspect ratio $b/w \approx$ 2-3 for ssDNA.

\subsection{Excluded volume interactions \& solvent quality}
In order to understand the non-equilibrium dynamics of single polymers such as double stranded DNA or single stranded DNA, it is important to consider key physical phenomena such as excluded volume (EV) interactions. Blob theories and scaling arguments are useful in revealing the underlying physics of polymer solutions and melts \cite{deGennesbook,Rubinsteinbook}. For the ease of analysis, we often think about polymer chain behavior in the limits of a given property such as solvent quality (or temperature) and/or polymer concentration. For example, the average end-to-end distance $R$ for a long flexible polymer chain scales as $R \sim N^{0.5}$ in a theta solvent and $R \sim N^{0.59}$ in an athermal solvent, where $N$ is the number of Kuhn steps. A theta solvent is defined such that two-body interactions between monomers are negligible, which occurs when the attractive interactions between monomer and solvent exactly cancel repulsive interactions between monomer-monomer pairs \cite{Rubinsteinbook}. In a theta solvent, a polymer chain exhibits ideal chain conformations such that $R \sim N^{0.5}$. Athermal solvents generally refer to the high-temperature limit, wherein monomer-monomer repulsions dominate and excluded volume interactions are governed by hard-core repulsions, such that the Mayer-$f$-function has a contribution only from hard-core repulsions in calculating the excluded volume $v$ \cite{Rubinsteinbook}. In reality, polymer chains often exist in good solvent conditions, which occurs in the transition region between a theta solvent and an athermal solvent. For many years, there was major confusion surrounding the description of DNA in aqueous solution due to the complex influence of solvent quality, chain flexibility, and polymer molecular weight on the static and dynamic scaling properties. In the following sections, we review recent progress in elucidating these phenomena for DNA as it pertains to single polymer dynamics.

\subsubsection{Theta temperature $T_{\theta}$, chain interaction parameter $z$, and hydrodynamic radius $R_H$} 
\begin{figure}[t]
\includegraphics[height=6.5cm]{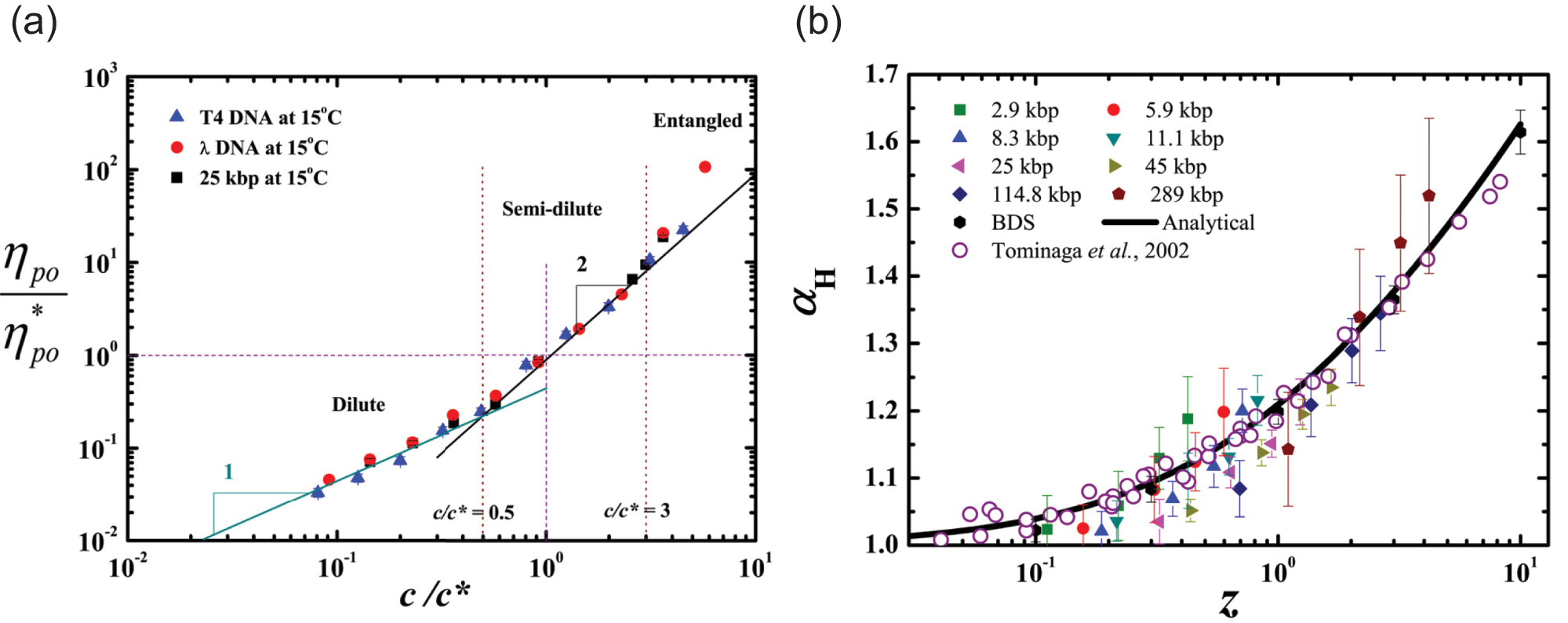}
  \caption{\label{fig:fig1} Bulk rheology and universal scaling functions in DNA solutions. (a) Normalized zero shear viscosity $\eta_{p,0}$ as a function of scaled concentration $c/c^*$ for three different molecular weight DNA samples at $T=T_{\theta}$. The zero shear viscosity $\eta_{p,0}^*$ is the value of $\eta_{p,0}$ at $c=c^*$. (b) Universal swelling ratio $\alpha_H=R_H / R_H^{\theta}$ for the hydrodynamic radius $R_H$ for DNA solutions and synthetic polymer solutions. Filled color symbols represent the swelling ratio $\alpha_H$ for DNA samples of variable molecular weight, BDS shows results from Brownian dynamics simulations, and data from Tominaga et al. (2002) \cite{Tominaga2002} show results for the swelling ratio $\alpha_H$ for synthetic polymers (polystyrene). The analytical curve is the function $\alpha_H = (1+az+bz^2+cz^3)^{m/2}$, where $a$ = 9.528, $b$ = 19.48, $c$ = 14.92, and $m$ = 0.0995 as determined from BD simulations. Reproduced with permission from Ref. \cite{Pan2014}.} 
\end{figure}

Double stranded DNA is a complex polymer to model due to the influence of chain flexibility, molecular weight, and solvent quality in determining scaling properties. As discussed above, blob theories are instructive in revealing the underlying physics of polymer chains \cite{Rubinsteinbook}, but many blob theories are derived by considering either the effects of polymer concentration or temperature, but not necessarily both in the same scaling relation. In 2012, Prakash and coworkers pointed out the need to consider the double cross-over behavior for static and dynamical scaling properties of polymer solutions in the semi-dilute solution regime, wherein polymer properties are given by power-law scaling relations as a function of scaled concentration and solvent quality \cite{Jain2012}. The chain interaction parameter $z$ effectively captures the influence of both temperature $T$ and polymer molecular weight $M$ on the behavior of polymer solutions in the region between theta solvents to athermal solvents:
\begin{equation} \label{eq:chaininteract} z = k \left( 1-\frac{T_{\theta}}{T} \right) \sqrt{M} \end{equation}
where $k$ is a numerical prefactor that depends on chemistry. In theory, the chain interaction parameter $z$ can be extremely useful in modeling DNA solutions, but in order to make this relation quantitative and practical, the prefactor $k$ and the theta temperature $T_{\theta}$ need to be determined. In 2014, Prakash and coworkers performed bulk rheological experiments and light scattering experiments on a series of linear DNA molecules ranging in size from 2.9 kbp to 289 kbp \cite{Pan2014}. Static light scattering experiments were used to determine the theta temperature $T_{\theta}$ of DNA in aqueous solutions containing monovalent salt, and it was found that $T_{\theta}$ = 14.7 $\pm$ 0.5$^o$ C. At the theta temperature $T=T_{\theta}$, the second virial coefficient $A_2$ is zero. Interestingly, these authors further showed that the second virial coefficient $A_2$ is a universal function of the chain interaction parameter $z$ in the good solvent regime when suitably normalized \cite{Pan2014,Rubinsteinbook}. Moreover, Prakash and coworkers showed that the polymer contribution to the zero shear viscosity $\eta_{p,0}$ obeys the expected power-law scaling with polymer concentration in the semi-dilute unentangled regime such that $\eta_{p,0} \sim (c/c^{\ast})^2$ at $T=T_{\theta}$, where $c^{\ast}$ is the overlap concentration (Figure \ref{fig:fig1}a).

Dynamic light scattering (DLS) experiments were further used to determine the hydrodynamic radii $R_H$ for these monodisperse DNA polymers in dilute solution ($c/c^{\ast}$ = 0.1) \cite{Pan2014}. First, the authors determined that the hydrodynamic radius $R_H$ for DNA was independent of salt concentrations $c_s >$ 10 mM monovalent salt, which ensures that charges along the DNA backbone are effectively screened and that putative polyelectrolyte effects are absent for these conditions. Second, the authors found the expected power law scaling of the hydrodynamic radius at $T=T_{\theta}$ such that $R_H^{\theta} \sim M^{0.5}$. At this point, the significance of the chain interaction parameter $z$ should be noted; any equilibrium property for a polymer-solvent system can be given as a universal value when plotted using the same value of $z$ in the crossover region between theta and athermal solvents \cite{Pan2014}. Therefore, determination of the value $k$ is essential. In order to determine the value of $k$ for DNA, Prakash and coworkers measured the swelling ratio $\alpha_H = R_H / R_H^{\theta}$ from DLS experiments. It is known that the swelling ratio can be expressed in an expansion such that $\alpha_H = (1+az+bz^2+cz^3)^{m/2}$, where $a,b,c,m$ are constants \cite{Kumar2003}. In brief, the value of $k$ was determined by quantitatively matching the swelling ratio $\alpha_H$ between experiments and BD simulations that give rise to the same degree of swelling in good solvents \cite{Pan2014}. Remarkably, the swelling ratio for DNA was found to collapse onto a universal master curve when plotted as a function of the chain interaction parameter $z$ across a wide range of DNA molecular weights (Figure \ref{fig:fig1}b). The numerical value of $k$ was determined to be $k$ = 0.0047 $\pm$ 0.0003 (g/mol)$^{-1/2}$, thereby enabling the chain interaction parameter $z$ to be determined as a function of molecular weight and temperature over a wide range of parameters relevant for single polymer experiments \cite{Pan2014}. These results further speak to the universal scaling behavior of DNA as a model polymer relative to synthetic polymers.

\subsubsection{Radius of gyration $R_g$ and overlap concentration $c^{\ast}$} 
The overlap concentration $c^{\ast} = M/ \left[ (4\pi / 3) R_g^3 N_A \right]$ is a useful characteristic concentration scale for semi-dilute polymer solutions, where $N_A$ is Avogadro's number and $R_g$ is the radius of gyration. A scaled polymer concentration of $c/c^*=1$ corresponds to a bulk solution concentration of polymer that is equivalent to the concentration of monomer within a polymer coil of size $R_g$. In order to calculate $c^*$ for an arbitrary temperature and molecular weight DNA, the value of $R_g$ must be determined for a particular size DNA and solution temperature. Prakash and coworkers determined the radius of gyration $R_g$ for DNA over a wide range of $M$ and $T$ using the expression $R_g = R_g^{\theta}\alpha_g(z)$, where $R_g^{\theta}=L/\sqrt{6N}$ is the radius of gyration under theta conditions and $\alpha_g(z)$ is the swelling ratio for the radius of gyration as a function of the chain interaction parameter $z$. With knowledge of $z$, the radius of gyration swelling ratio $\alpha_g(z) = (1+a'z+b'z^2+c'z^3)^{m'/2}$ can be determined at a given $M$ and $T$, where the constants $a',b',c'$, and $m'$ have been determined using BD simulations using a delta function potential for the excluded volume interactions \cite{Kumar2003}. Therefore, with knowledge of the solvent quality $z$, the radius of gyration $R_g$ and hence the overlap concentration $c^*$ can be determined at a given temperature and molecular weight for DNA \cite{Pan2014}. Clearly, systematic experiments on the bulk rheology and static and dynamic properties of DNA, combined with complementary BD simulations, have enabled an extremely useful quantitative understanding of the physical properties of DNA that can be leveraged for single polymer dynamics. 

\begin{figure}[t]
\includegraphics[height=8cm]{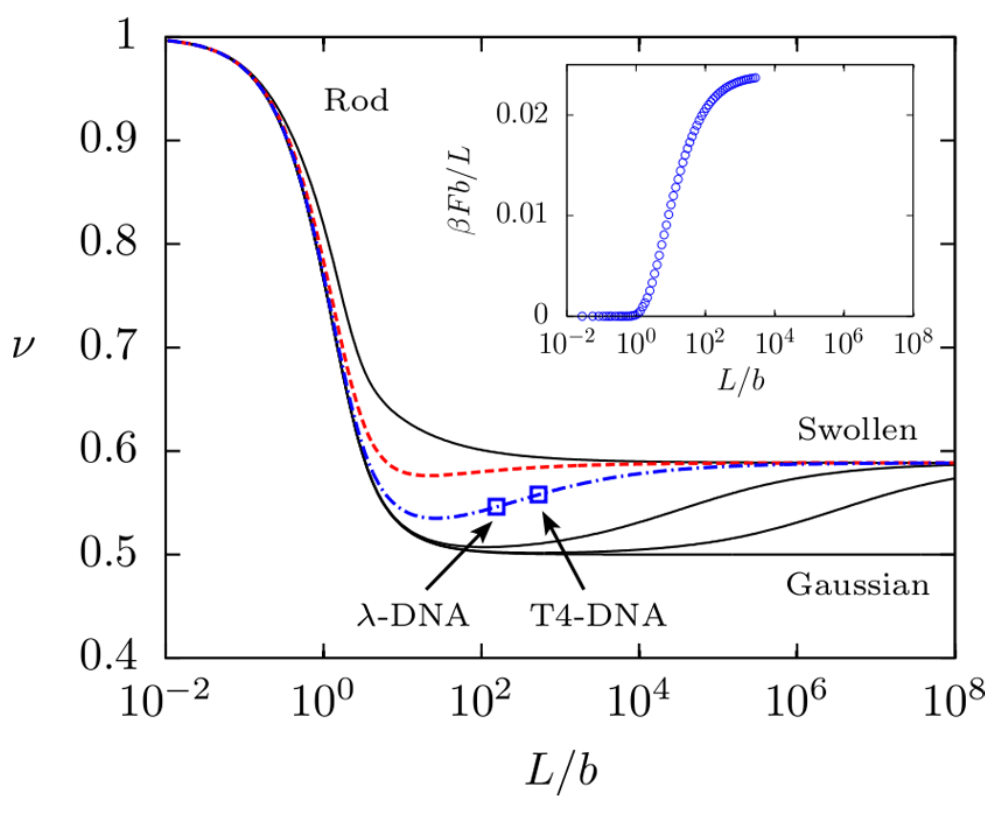}
  \caption{\label{fig:fig2} Apparent excluded volume exponent $\nu$ for a discrete wormlike chain model (DWLC) calculated by renormalization group theory of Chen and Noolandi \cite{Chen1992} reported in the recent work of Dorfman and coworkers \cite{Tree2013}. The apparent EV exponent was determined for static chain properties, corresponding to the RMS end-to-end distance $R_E$ such that $R_E \sim L^{\nu}$, where $\nu \equiv d \ln R_E / d \ln L$. Here, $\nu$ = 1 corresponds to rodlike behavior, $\nu$ = 0.5 to a Gaussian chain, and $\nu$ = 0.588 to a swollen chain. Results are shown for five different values of of the monomer aspect ratio $b/w$ (from top to bottom): 1.0, 4.5 (corresponding to ssDNA), 25 (corresponding to DNA), 316, 3160, and 0 (no excluded volume). Inset: PERM results for the excess free energy per Kuhn length due to EV interactions in a dilute solution of DNA. Reproduced with permission from Ref. \cite{Tree2013}.}
\end{figure}

\subsubsection{Excluded volume exponent $\nu$ and static chain properties}
The average size of a polymer chain can be determined using static measurements such as light scattering (thereby leading to $R_g$) or by dynamic measurements based on diffusion (thereby leading to $R_H$) \cite{deGennesbook,Sunthar2006}. It has long been known that these two different measures of average polymer size exhibit different power-law scalings in the limit of large molecular weights \cite{deGennesbook}, such that $R_g \sim N^{0.59}$ and $R_H \sim N^{0.57}$ under good solvent conditions. First, we consider the power-law scaling behavior of the static properties of DNA chains associated with the root-mean-square end-to-end distance:
\begin{equation} R_E \equiv \left< R^2 \right>^{1/2} = \left< (\mathbf{r}_{N} - \mathbf{r}_1)^2 \right>^{1/2} \end{equation}
where the brackets $\left< \cdot \right>$ represent an average over an ensemble and the vectors $\mathbf{r}_{N}$ and $\mathbf{r}_{1}$ represent the ends of a polymer chain. Because we are considering static properties, the power-law scaling behavior for $R_E$ should be equivalent to the scaling behavior of $R_g$. In 2010, Clisby performed high precision calculations of static chain properties to show that the ratio $R_E^2/R_g^2$ exhibits a universal value $R_E^2/R_g^2 \approx 6.254$ in the limit of large molecular weight for self-avoiding chains, together with a Flory excluded volume exponent of $\nu=0.587597(7)$ \cite{Clisby2010}.

In 2013, Dorfman and coworkers investigated the static and dynamic equilibrium properties of DNA and ssDNA using a Monte Carlo modeling approach based on the pruned-enriched Rosenbluth method (PERM) \cite{Tree2013}. In this work, the authors used a discrete wormlike chain model (DWLC), which is a coarse-grained model for polymers that incorporates a series of inextensible bonds of length $a$ linked with a bending potential. Excluded volume interactions are included with a hard-core repulsive potential, which essentially amounts to athermal solvent conditions and does not consider the effect of solvent quality. Hydrodynamic interactions were further included using an Oseen tensor and a bead hydrodynamic radius $d$. Taken together, the model parameters for the DWLC include the Kuhn step size $b$, an effective width $w$, a hydrodynamic diameter $d$, and a bond length $a$, where it was assumed that $d=a$ \cite{Tree2013}. Using this parametrized model for DNA, PERM calculations were used to investigate the power-law scaling behavior for the average end-to-end distance $R$ for a set of parameters correspond to DNA over a limited range of molecular weights. 

In order to expand the range of parameters and molecular weights under investigation, Dorfman and coworkers further used renormalization group (RG) theory of Chen and Noolandi \cite{Chen1992}. In this way, these authors determined an apparent excluded volume exponent $\nu$, where $\nu \equiv d \ln R_E / d \ln L$, as a function of polymer molecular weight $N = L/b$ and the monomer aspect ratio $b/w$. Interestingly, these results provided a quantitative description of chain flexibility on the equilibrium structural properties of DNA (Figure \ref{fig:fig2}). The results clearly show that the apparent excluded volume exponent $\nu$ is a sensitive function of the number of Kuhn steps $N$ and the monomer aspect ratio $b/w$. For double stranded DNA with $b/w \approx 25$, the RG theory predicts a value of $\nu \approx 0.546$ for $\lambda$-DNA. Clearly, the properties of $\lambda$-DNA (and most common molecular weight DNA molecules used for single molecule studies) appear to lie in the transition regime between an ideal Gaussian chain (without dominant EV interactions) and a fully swollen flexible chain (with dominant EV interactions). However, it should be emphasized that these results only considered hard-core repulsive interactions between monomers in the context of excluded volume interactions, which corresponds to the limit of athermal solvents in the long-chain limit. Therefore, these calculations do not consider the effect of arbitrary solvent quality on the static or dynamic properties of DNA or ssDNA in the crossover regime. In other words, these results show that the intermediate value of the apparent excluded volume exponent $\nu$ is dictated by the flexibility (or semi-flexibility) of the DNA polymer chain, but approaching the limit of a theta solvent would only serve to further decrease the value of $\nu$ for DNA. 

\subsubsection{Effective excluded volume exponent $\nu_{eff}$ and dynamic chain properties}
In addition to static measures of average polymer size such as $R_E$ and $R_g$, the hydrodynamic radius $R_H$ is an additional measure of coil dimensions that can be determined by dynamic light scattering experiments \cite{deGennesbook,Douglas1985,Sunthar2006,Clisby2016}. As previously discussed, $R_H$ and $R_g$ exhibit different power-law scaling relations in the limit of high molecular weight, essentially because these different physical quantities represent distinct averages (over different moments) of fluctuating variables such as local distances between monomers along a polymer chain. A full discussion of this phenomenon is beyond the scope of this review article, though we provide a brief overview here in order to interpret single molecule diffusion experiments. In brief, Sunthar and Prakash \cite{Sunthar2006} used an Edwards continuum model \cite{Edwards1965} to show that differences in the crossovers between the swelling of the hydrodynamic radius $\alpha_H$ and radius of gyration $\alpha_g$ arise due to dynamic correlations to the diffusivity. These differences are important because dynamic correlations are ignored when determining hydrodynamic radius using the Kirkwood expression for hydrodynamic radius \cite{Sunthar2006,Clisby2016}:
\begin{equation} \frac{1}{R_H^*} = \frac{1}{N} \sum_{i \ne j} \left< \frac{1}{r_{ij}} \right> \end{equation}
where $r_{ij}$ is the distance between two monomers $i$ and $j$. In brief, the Kirkwood value of the hydrodynamic radius $R_H^*$ can be used to calculate a short-time approximation to the chain diffusivity, whereas the Stokes-Einstein relation provides the long-time value of the chain diffusivity:
\begin{equation} D = \frac{k_BT}{6 \pi \eta R_H} \end{equation}
where $R_H$ is the hydrodynamic radius determined from long-time diffusion measurements. The Stokes-Einstein value of the chain diffusivity is commonly determined from mean-squared values of the polymer chain center-of-mass from single molecule experiments, as discussed below. In any event, Sunthar and Prakash showed that the swelling ratio determined for the Kirkwood expression for the hydrodynamic radius is nearly identical to the swelling ratio for the radius of gyration, which suggests that both of these pertain to static measures of chain size \cite{Sunthar2006}. These results are consist with RG theory results from Douglas and Freed for chains with EV interactions \cite{Douglas1985}. However, the long-time diffusivity (calculated using $R_H$) needs consideration of dynamic correlations to explain the power-law scaling and crossover behavior. In any event, the ratio $R_g/R_H^*$ is known to exhibit a universal value for self-avoiding walks in the limit of long chains, such that $R_g/R_H^* = 1.5803940(45)$ as determined by recent high-precision Monte Carlo simulations by Clisby and D{\"u}nweg \cite{Clisby2016}. 
 
Single molecule experiments have been used to directly measure chain diffusivity by fluorescently labeling DNA molecules and tracking their diffusive motion over time at thermal equilibrium. In these experiments, the mean-squared displacement of the polymer center-of-mass is determined as a function of time for an ensemble of molecules, and these data are used to extract a diffusion coefficient. Based on the discussion above, these experiments yield the long-time diffusion coefficient $D$ from the Stokes-Einstein relation, thereby yielding the hydrodynamic radius $R_H$, which is distinct from the hydrodynamic radius $R_H^*$ calculated from the Kirkwood approximation. In 1996, the diffusion coefficients $D$ for a series of variable molecular weight linear DNA molecules were determined using fluorescence microscopy \cite{Smith1996}, and it was found that the apparent excluded volume exponent $\nu_{app} = 0.61 \pm 0.016$, where $D \sim L^{-\nu_{app}}$. However, uncertainly in the actual molecular weights of these DNA molecules was later found to increase uncertainty in these results \cite{Robertson2006}, which unfortunately added to the confusion surrounding the equilibrium properties of DNA. In 2006, these single molecule diffusion experiments were repeated on linear DNA, and it was found that $\nu_{app} = 0.571 \pm 0.014$ \cite{Robertson2006}. 

Recently, Prakash and coworkers applied the concepts of dynamical scaling the cross-over region between theta and athermal solvents \cite{Jain2012} to bulk rheological experiments on semi-dilute unentangled DNA solutions \cite{Pan2014}. In particular, it is known that the polymer contribution to the zero-shear viscosity $\eta_{p,0}$ should depend on both solvent quality and polymer concentration in the cross-over region in semi-dilute solutions, such that $\eta_{p,0}/\eta_s = f(z,c/c^*)$ \cite{Jain2012}. Using scaling arguments, it can be shown that $\eta_{p,0}/\eta_s \sim (c/c^*)^{(1/3 \nu_{eff}(z) - 1)}$, where $\nu_{eff}$ is an effective excluded volume exponent that depends on solvent quality in the cross-over region \cite{Jain2012}. Bulk rheological experiments on the zero-shear viscosity of DNA solutions in the semi-dilute regime revealed $\nu_{eff} = 0.52$ $\pm$ 0.01 for 25 kbp DNA, $\lambda$-DNA (48.5 kbp), and for T4 DNA (169 kbp) (all there experiments were conducted at $z=0.7$, which corresponds to different $T$ for each $M$), whereas quantitative BD simulations yielded $\nu_{eff} = 0.54$  $\pm$ 0.02 under the same conditions \cite{Pan2014}. Using the framework of dynamical scaling the cross-over regime, prior single molecule experimental data on T4 DNA relaxation in semi-dilute solutions can be reinterpreted. In particular, the longest relaxation time scales with concentration as: $\tau_1 \sim (c/c^*)^{(2-3 \nu_{eff}(z))/(3 \nu_{eff}(z) -1)}$ in semi-dilute unentangled solutions. Prior single polymer experiments on T4 DNA at $T=22$ $^oC$ reveal that $\nu_{eff}$ = 0.56. Moreover, recent single molecule experiments on the longest relaxation time of DNA in semi-dilute unentangled solutions at $T=22$ $^oC$ revealed $\nu_{eff} = 0.56$ \cite{Hsiao2017}, which is in good agreement with the T4 data and within the error bounds of the BD simulations at $z\approx 1$. As an aside, it is possible to define an alternative longest relaxation time based on zero-shear viscosity $\tau_{\eta}$ from bulk rheological experiments that can be directly compared to single molecule relaxation times \cite{Ottingerbook,Tsortos2011,Hsiao2017}. In sum, these experiments and simulations show that physical properties such as longest relaxation time and zero-shear viscosity obey power laws in the cross-over regions between theta and athermal solvents. Taken together, this work has elucidated the role of the solvent quality and concentration on the equilibrium properties of DNA.

\subsubsection{Thermal blobs}
The thermal blob size $\xi_T$ is the length scale at which the monomer interaction energy is comparable to thermal energy $k_BT$. On length scales smaller than $\xi_T$, excluded volume interactions are weaker than thermal energy and the chains follow ideal statistics on these length scales. The thermal blob size is given by $\xi_T \equiv c b^4/v(T)$, where $c$ is a numerical constant of order unity and $v(T)$ is the excluded volume which is a function of temperature \cite{Rubinsteinbook}. First, we consider the asymptotic limit of an athermal solvent, where only hard-core repulsions contribute to monomer interactions and the properties are independent of temperature. In an athermal solvent, the excluded volume for polymers with anisotropic monomers is given by $v_{a}=b^2w$, such that the excluded volume is much larger than the monomer occupied volume $v_o=bw^2$ because $b \gg w$ \cite{Rubinsteinbook}. In any case, the concept of a thermal blob generally refers to scaling arguments and is not generally taken as a quantitative property. For this reason, estimates of the thermal blob size can vary widely depending on how the prefactor $c$ is considered \cite{Latinwo2011,Tree2013}. Nevertheless, it is instructive to consider a specific definition of the thermal blob size as it pertains to DNA. Consider the case where the thermal blob size (length of sub-chain) is defined where the excess free energy from excluded volume interactions equals the thermal energy $k_BT$. In the context of this definition, the thermal blob size (length of chain) was determined using PERM simulations and was found to equal 16.8 kbp for DNA, which equates to $c \approx$ 0.1 in the scaling equation for $\xi_T$. On the other hand, if the prefactor is assumed to be of order unity $c \approx$ 1, then the thermal blob size corresponds to 166 kbp, which illustrates that care is required in defining such properties. Nevertheless, a thermal blob length of 16.8 kbp implies that most common DNA molecules used in single polymer experiments such as $\lambda$-DNA (48.5 kbp) are comprised of at least 3 thermal blobs in the limit of a very good solvent.

Moving away from the asymptotic limit of athermal solvents, thermal blobs have also been considered in the cross-over regime of intermediate solvent quality \cite{Jain2012}. It can be shown that the thermal blob size in the cross-over region is estimated by:
\begin{equation} \xi_T \sim bN^{1/2} z^{-1} \end{equation} 
where $z$ is the chain interaction parameter given by Eq. (\ref{eq:chaininteract}). Moreover, the temperature dependence of the excluded volume is given by: 
\begin{equation} v \approx \left( 1-\frac{T_{\theta}}{T} \right) v_a \end{equation}
where $v_a$ is the excluded volume in an athermal solvent \cite{Rubinsteinbook}. Using this approach, Prakash and coworkers estimated the molecular weight contained in a thermal blob $M_{blob}$ as a function of temperature (to within a numerical prefactor) for DNA in the vicinity of the theta temperature \cite{Pan2014}. Moreover, these results suggest that the hydrodynamic radius $R_H$ for DNA scales with a molecular weight power-law given by ideal chain statistics for molecular weights $M<M_{blob}$ (such that $R_H \sim M^{0.5}$), followed by the expected molecular weight power-law scaling for self-avoiding chains for $M>M_{blob}$ (such that $R_H \sim M^{0.59}$) \cite{Pan2014}. 

Finally, it should be noted that an alternative theoretical framework can be used to model semi-flexible polymers, and it has been shown that the properties of semi-flexible polymers differ from truly flexible chains in semi-dilute solutions \cite{Schaefer1980}. This theoretical model focused on marginal solutions and is applicable to polymers with large aspect ratio Kuhn monomers under conditions of intermediate solvent quality. Marginal solutions are qualitatively different than theta solutions because the dominant interactions are given by mean-field two-body interactions, whereas the dominant interactions in theta solvents are three-body because the second virial coefficient is exactly zero in a theta solvent. It is thought that the third virial coefficient for polymers with large aspect ratio Kuhn segments is extremely small, and two-body interactions are stronger than three-body interactions in these so-called marginal solutions, even under conditions in which the concentration blob size $\xi_c$ is smaller than the thermal blob $\xi_T$ \cite{Schaefer1980}. Although these concepts are intriguing, they are admittedly beyond the scope of the present review article and could represent an interesting direction of future analysis in the context of DNA.

\subsection{Elasticity of DNA and ssDNA: effect of excluded volume}
\label{elasticity}
DNA is well described by the wormlike chain model, wherein DNA conformations can be modeled by a space curve $r_i(s)$ with a contour length $L$ \cite{Kratky1949,Marko1995}. Marko and Siggia considered the entropic elasticity of a wormlike chain, such that the elastic restoring force arises from conformational entropy without consideration of enthalpic interactions such as excluded volume or solvent quality. Using this approach, Marko and Siggia developed an approximate interpolation equation for the force as a function of extension for DNA:
\begin{equation} \frac{fl_p}{k_BT} = \frac{x}{L} + \frac{1}{4(1-x/L)^2} - \frac{1}{4} \end{equation}
where $f$ is the applied force, $x$ is the end-to-end extension of a DNA molecule, and $k_BT$ is thermal energy. The Marko-Siggia formula generally provides an excellent fit to single molecule elasticity data describing the force-extension of DNA over a wide range of extensions \cite{Smith1992}, including the low force regime and up to fractional extensions $x/L \approx$ 0.97, whereupon the stretching force is large enough to disrupt base pairing and base stacking ($\approx$ 300 $k_BT/l_p$). The development of a simple analytic expression for the entropic elasticity of DNA enabled the direct simulation of DNA dynamics in flow using coarse-grained bead-spring models and Brownian dynamics simulations \cite{Larson1997,Hur2000a}. It is obvious from Eq. (1) that the low force elasticity of DNA is {\em linear}, as given by the WLC interpolation formula. In fact, the low force linearity is consistent with flexible polymers or freely-jointed polymers described by Gaussian coil statistics, such that a Gaussian chain yields a linear entropic restoring force in the end-to-end extension \cite{Larsonbook,Rubinsteinbook}. In other words, an ideal chain described by random walk statistics or theta-solvent conditions yields a low-force linear elasticity:
\begin{equation} \frac{fl_p}{k_BT} \approx \frac{x}{L} \end{equation}
Several years ago, however, Pincus considered the effect of excluded volume (EV) interactions on the low force elasticity of flexible polymers \cite{Pincus1976}, and his analytical results showed that real polymer chains with EV interactions exhibit a {\em non-linear} low force elasticity:
\begin{equation} \frac{fl_p}{k_BT} \approx \left( \frac{x}{L} \right)^{3/2} \end{equation}
where an EV exponent of $\nu = 3/5$ corresponding to good solvents has been assumed in the derivation \cite{Rubinsteinbook}. The key idea in the Pincus analysis is that the applied force generates a tensile screening length, known as a tensile blob or a Pincus blob $\xi_P$:
\begin{equation} \xi_P \approx \frac{k_BT}{f} \end{equation}
such that long-range EV interactions are screened for distances greater than $\xi_P$. In other words, a polymer chain under tension will break up into a series of tensile blobs of size $\xi_P$; within each tensile blob, chain conformation is described by a self-avoiding walk in good solvent conditions such that $\xi_P = bg^{3/5}$, where $g$ is the number of monomers in a tensile blob. For length scales larger than $\xi_P$, chains are extended into an array of tensile blobs. Using this framework, Pincus blobs form for applied forces greater than $f \approx k_BT/R_F$, where the Pincus blob size is on the order of the coil dimension $R_F$. For applied tensions $f>k_BT/\xi_T$, Pincus blobs are smaller than thermal blobs such that $\xi_P<\xi_T$, and excluded volume interactions are screened at all length scales.

\begin{figure}[t]
\includegraphics[height=7cm]{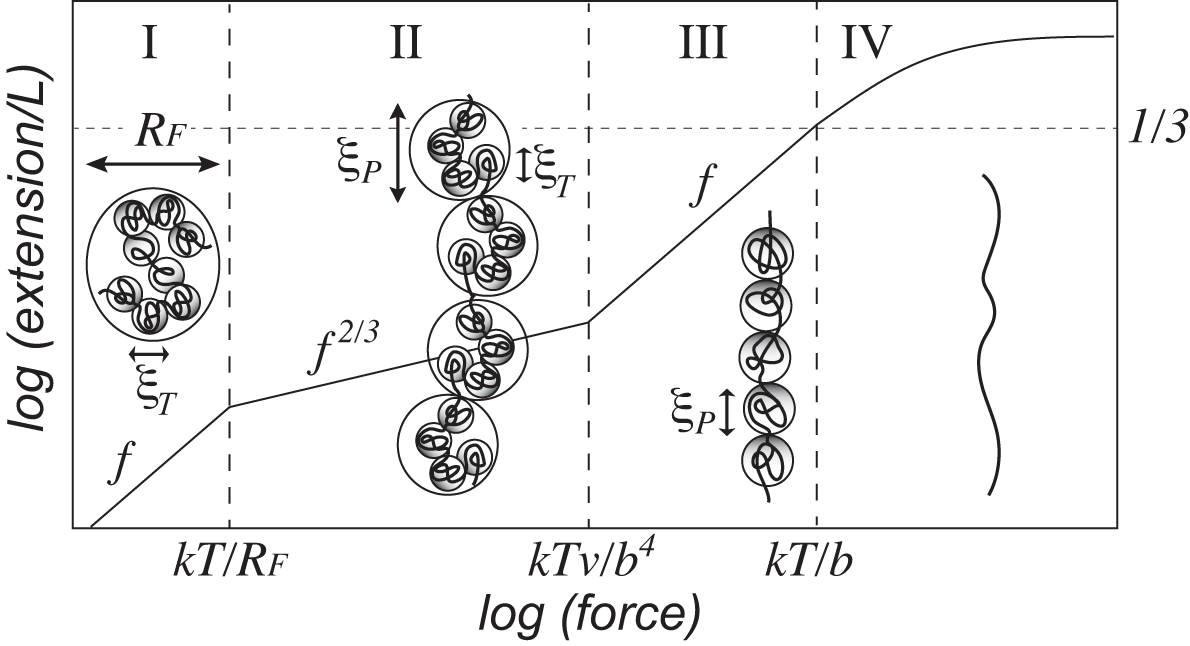}
  \caption{\label{fig:fig3}Force-extension behavior of a real polymer chain in a good solvent. Regime I corresponds to a weakly perturbed polymer coil with a Flory radius $R_F$ and a thermal blob size $\xi_T$. Regime II corresponds to a non-linear low force region due to a self-avoiding polymer chain with a Pincus blob size $\xi_P$. Regime III corresponds to a moderately stretched polymer with a linear force relation. At the transition between Regime III and IV (occurring at a fraction extension $x/L \approx 1/3$), the finite extensibility of the polymer begins to play a role. The details of the non-linear elasticity of Regime IV depend on the polymer chemistry \cite{McIntosh2009,Dittmore2011,Radhakrishnan2012}.}  
\end{figure}

Although Pincus predicted a non-linear low force elasticity for flexible polymers in 1976, this phenomenon was not experimentally observed until 2009 by the Saleh group, who used magnetic tweezing experiments to measure the force-extension of single stranded DNA (ssDNA) under denaturing conditions \cite{Saleh2009}. These experiments were the first to demonstrate a clear scaling of the applied force $f \sim \left( x/L \right)^{3/2}$ over a wide range of forces and salt concentrations for ssDNA. Interestingly, at high salt concentrations (3 M NaCl), the elasticity of ssDNA was well fit by the Marko-Siggia force relation, which describes ideal chain statistics in the low force regime in the absence of monomer-monomer interactions, thereby suggesting a transition to theta-solvent conditions under high salt. These experiments also revealed a `bare' non-electrostatic persistence length of ssDNA of $l_p \approx$ 0.62 nm. These single molecule elasticity experiments on ssDNA were followed by equivalent single chain `pulling' experiments on poly(ethylene glycol) (PEG), which is a synthetic flexible polymer, and the results again showed a non-linear low force elasticity $f \sim \left( x/ L \right)^{3/2}$ which is clearly different than the entropic force response of an ideal chain $f \sim x / L$ \cite{Dittmore2011}. Moreover, the initial experiments on ssDNA elasticity were complemented with a detailed scaling analysis \cite{McIntosh2009}. Recently, the elasticity of ssDNA was further elucidated by developing a new model incorporating intrapolymer electrostatic repulsion, which generates a salt-dependent internal tension \cite{Jacobson2017}. This model was inspired by a mean-field approach, and it was shown that the internal tension can be related to the linear charge density along a charged polymer backbone. This work shows that mesoscopic polymer conformation emerges from microscopic structure.

A schematic of the elasticity of a single polymer chain as a function of applied force is shown in Figure \ref{fig:fig3}. In Regime I, a polymer chain is weakly perturbed from equilibrium at very low forces ($f<k_BT/R_F$), where the Flory radius of the chain is $R_F=L^{3/5}l_p^{1/5}w^{1/5}$ \cite{Rubinsteinbook}. Here, the thermal blob size $\xi_T \approx b^4/v$ is the length scale at which the monomer interaction energy is comparable to thermal energy $k_BT$. Upon increasing the applied force into Regime II, the chain begins to extend with a {\em non-linear} low force elasticity wherein $x/L \sim f ^{2/3}$ due to excluded volume interactions. At higher forces, the self-avoidance effects weaken and the behavior transitions to a linear elasticity region, where the polymer acts as an ideal chain. The transition between Regimes II and III occurs when the thermal blob size is on the order of the Pincus blob size $\xi_T=\xi_P$. Regime IV corresponds to high forces and large fractional extensions $x/L \ge 1/3$, wherein the finite extensibility of the chain plays a role in the elasticity. 

The physical properties of polymer chains with different chemistries dictate the dominance and transition between the different regions in the force-extension curve in Figure \ref{fig:fig3}. Interestingly, the linear elastic behavior in Regime III was not observed for ssDNA \cite{Saleh2009}, but it was observed for PEG \cite{Dittmore2011}. ssDNA has isotropic or nearly spherical monomers such that the excluded volume $v \approx b^3$ and the monomer aspect ratio $b/w$ is near unity, which means that the thermal blob size is essentially equal to the Kuhn step size ($\xi_T=b$). This effectively shrinks Regime III to a vanishingly small size, and the non-linear low force elastic region in Regime II dominates the low-force elastic behavior. On the other hand, PEG has slightly more anisotropic monomers compared to ssDNA, with a PEG monomer aspect ratio $b/w \approx 5$ \cite{Dittmore2011}, which results in a non-negligible Regime III for slightly less flexible polymers. 

How does this picture change for double stranded DNA? As previously discussed, DNA has a large persistence length and a correspondingly large monomer aspect ratio $b/w \approx 25$. For polymers such as DNA, Regime III extends further into the low force region such that the non-linear elastic response in Regime II is essentially non-existent. In other words, due to the large monomer aspect ratio for DNA, the entire low-force elasticity regime is essentially linear. How do these concepts relate to linear and non-linear rheology for DNA? First, consider the limit of large forces which corresponds to non-linear rheological experiments. For any polymer subjected to high stretching forces $f > k_BT/\xi_T$, Pincus blobs are smaller than thermal blobs, which suggests that EV interactions are screened at all scales and the elastic force relation is linear (Figure \ref{fig:fig3}). However, DNA has a large Kuhn step size, and certainly no Pincus blobs will form when the Kuhn step size is larger than the Pincus blob size such that $b>\xi_P$. A rough estimate for the force at which the Pincus blob size equals the DNA Kuhn step size $b$ is $f \approx$ 0.04 pN, which is a relatively small force. Therefore, over the practical force range for most non-linear rheological experiments, $b > \xi_P$ for DNA, which necessarily precludes the formation of tensile blobs. In the limit of low forces (corresponding to linear rheological experiments), Pincus blobs only form between applied tensions $k_BT/R_F < f < k_bTw /b^2$ (assuming an athermal solvent), which is a vanishingly small window for most reasonable size DNA molecules using in single molecule experiments due to the large values of $b/w$ and $b$. Clearly, the difference in physical properties between DNA and flexible polymers result in major differences in elasticity, which impacts the emergent rheological properties of these materials \cite{Latinwo2011}. 

The notion that solvent quality can impact the elastic (force-extension) properties of polymer chains was generally known in the polymer physics community, but quantitative chain elasticity models capturing this effect were largely absent from the rheological community until recently. For example, the Warner force relation and the Pade approximant to the inverse Langevin function \cite{Larsonbook,Cohen1991} both neglect the role of EV and the non-linear low-force elasticity for flexible polymers in a good solvent. In 2012, Underhill and coworkers postulated a new elastic (spring force) relation for flexible polymers that accounts for solvent quality ranging from theta solvents to good solvents \cite{Radhakrishnan2012}. This force extension relation smoothly interpolates between the non-linear low-force elasticity for flexible polymers in a good solvent (Regime II) and the linear (Regime III) and ultimately non-linear finite extensibility regions (Regime IV) of the force-extension diagram, as shown in Figure \ref{fig:fig3}. Interestingly, this force extension relation also captures a scale-dependent Regime III, such that polymers with larger aspect ratio monomers $b/w$ tend to exhibit a larger Regime III, consistent with experimental data on single molecule pulling experiments on PEG \cite{Dittmore2011}. The main advantage of this analytic form of the force-extension relation is the ability to use it in coarse-grained bead-spring Brownian dynamics simulations for flexible polymers, where the force-extension relation is used to represent the elasticity of chain sub-segments in a coarse-grained model. This force-extension relation was subsequently used to study the impact of solvent quality on the coil-stretch transition and conformation hysteresis for flexible polymers (without HI) in a good solvent \cite{Radhakrishnan2012b}. 

Recently, the concept of EV-influenced elasticity was further extended to wormlike chains by Li, Schroeder, and Dorfman \cite{Li2015}. Here, a new analytic interpolation formula (called the EV-WLC model) was developed for wormlike chains such as DNA in the presence of excluded volume interactions. Again, this model was found to smoothly interpolate between the relevant regions of the force-extension diagram for wormlike chains; however, the parameters in the EV-WLC interpolation formula were determined by rigorous calculations using Monte Carlo/PERM simulations, rather than phenomenological estimation. In 2016, Saadat and Khomami developed a new force-extension relation for semi-flexible chains by incorporating bending potential \cite{Saadat2016}. This force-extension model accurately describes correlations along the backbone of the chain, segmental length, and the elastic behavior of semi-flexible macromolecules in the limit of 1 Kuhn step per spring. 

Finally, it should be noted that the elastic behavior in the high-force region (Regime IV) qualitatively differs between ssDNA and either freely-jointed chains or wormlike chains.  In the limit of high forces, the chain stretch scales as $x \sim (1-f^{-\alpha})$, where $\alpha$ = 0.5 for wormlike chains and $\alpha$ = 1.0 for freely-jointed chains \cite{Shaqfeh2004}. Moreover, it is known that there is a cross-over between freely-jointed and semi-flexible force-extension behavior at large polymer chain extensions \cite{Dobrynin2010}. However, single molecule pulling experiments on ssDNA showed a qualitatively different response, such that $x \sim \ln f$ in the high-force regime \cite{Saleh2009}. It was found that this unusual behavior arises due to polyelectrolyte or charge effects for flexible polymers in the high-force regime, supported by scaling theory \cite{Toan2012} and simulations of the stretching of flexible polyelectrolytes under high force \cite{Stevens2012,Saleh2015}.

\section{Dilute solution single chain dynamics: Pre-2007}
\label{dilutepre2007}
It has long been appreciated that DNA can serve as a model polymer for understanding the rheology and dynamics of long chain macromolecules in solution due to monodispersity and molecular weight selectivity \cite{Zimm1965,Pecora1990,Pecora1991}. In the early to mid-1990's, advances in fluorescence microscopy and optical imaging allowed for the direct observation of single DNA molecules at equilibrium and in flow. A major focus of the early single polymer experiments was to directly probe molecular conformations under non-equilibrium conditions in the ultra-dilute concentration regime ($\approx 10^{-5}c^{\ast}$), conditions under which bulk experimental methods based on mechanical rheometry or bulk optical properties such as birefringence cannot detect an appreciable signal due to the extremely dilute nature of the polymer solutions. To this end, single molecule methods enabled a truly new and unique window into viewing non-equilibrium molecular conformations in uniform flow \cite{Perkins1995}, shear flow \cite{Smith1999}, and extensional flow \cite{Perkins1997,Smith1998}. This molecular-scale information on polymer chain dynamics can be further used to inform the development of molecular-based constitutive models for relating polymer microstructure to bulk stress and viscosity \cite{Bird1987b,Doiedwards} and can be directly compared to coarse-grained Brownian dynamics simulations of polymer chains in flow \cite{Larsonbook}. Taken together, these efforts fundamentally enhanced our understanding of polymer physics under non-equilibrium conditions. 

Single DNA dynamics was reviewed in a comprehensive article in 2005 \cite{Shaqfeh2005}, and the main content of this prior review article is not considered here. For the purposes of this review, I will explore a few interesting and perhaps under-appreciated topics in dilute solution single polymer dynamics that could inspire future investigation. To begin, it is useful to consider a few of the major results together with intriguing and outstanding questions in the area of dilute solution single polymer dynamics. 

\subsubsection{Relaxation and stretching in uniform flow}
The relaxation of single DNA molecules from high stretch was first observed using single molecule techniques 1994 by Chu and coworkers \cite{Perkins1994a}. Interestingly, these results showed that the longest relaxation time $\tau$ scaled with contour length $L$ as $\tau \sim L^{1.65 \pm 0.10}$, which suggests that the apparent excluded volume exponent $\nu_{app} \approx$ 0.55 for DNA. These results agree with our modern understanding of excluded volume interactions and the effect of flexibility in the DNA backbone, as discussed in the prior section. However, subsequent experiments on the diffusion of single DNA molecules yielded an excluded volume exponent of $\nu_{app}$ = 0.611 $\pm$ 0.016 \cite{Smith1996}, which appeared to be in conflict with the longest polymer relaxation time data until these single polymer diffusion experiments were repeated in 2006 with more uniform molecular weight samples, which yielded a value of $\nu_{app}$ = 0.571 $\pm$ 0.014. Nevertheless, the disparity in these results lead to confusion in the field regarding the influence or role of EV in DNA for many years, though this has now been largely resolved, as discussed above. Following relaxation experiments, Chu and coworkers performed the first single polymer experiments on the stretching dynamics of a tethered chain in a uniform flow \cite{Perkins1995}, a planar extensional flow \cite{Perkins1997,Smith1998}, and a simple shear flow \cite{Smith1999}. In 1997, Larson showed that the experimental data on uniform flow stretching could be quantitatively described by a simple dumbbell model using the non-linear wormlike chain (Marko-Siggia) force relation and a conformation-dependent hydrodynamic drag \cite{Larson1997}. These ideas were some of the first to analyze single polymer dynamics results in a quantitative manner by considering the role of intramolecular hydrodynamic interactions. 

\subsubsection{Dynamics in extensional flow: molecular individualism}
Several interesting phenomena were observed from single polymer dynamics experiments in an extensional flow \cite{Perkins1997,Smith1998}. First, the coil-stretch transition was observed to be extremely sharp when considering only the subset of molecules that reached a steady extension at any value of the dimensionless flow strength called the Weissenberg number $Wi = \dot{\epsilon} \tau$, where $\dot{\epsilon}$ is the strain rate. In fact, the sharpness of coil-stretch transition was striking compared to prior bulk rheological measurements based on flow birefringence, which typically average over a large number of molecules that may (or may not) have reached a steady-state extension. Moreover, it was observed that single polymer chains generally adopt a rich set molecular conformations during transient stretching in extensional flow such as dumbbell, hairpin, and kink shapes \cite{Smith1998}. 

From these experiments emerged the notion of `molecular individualism', wherein a single polymer molecule may adopt a series of different stretching pathways given the same initial conditions and the same dynamic stretching experiment \cite{deGennes1997}. These concepts began to show the true value of single polymer experiments in revealing unexpected and rich sets of molecular sub-populations, and further, how these molecular populations served to influence bulk properties such as stress and viscosity. In tandem, major progress was being made in the development of coarse-grained Brownian dynamics (BD) simulations of polymer chains \cite{Ottingerbook,Doyle1997}, which provide a direct complement to single polymer experiments. Larson and coworkers performed BD simulations on DNA stretching in extensional flow using a multi-bead-spring polymer model using the Marko-Siggia force relation, albeit in the absence of hydrodynamic interactions and excluded volume interactions \cite{Larson1999}. Nevertheless, these simulations provided good agreement with single molecule experiments, including the emergence of different molecular conformations such as folds, kinks, and dumbbells. These simulations were useful in revealing the origin of the heterogeneity in molecular conformations, which essentially arise from variability in the initial polymer conformation and emerge as a balance between the conformational macromolecular diffusion and the imposed flow \cite{Larson1999}. A related question is the impact of molecular individualism on bulk rheological properties of polymer solutions. On the one hand, it is clear that one needs to be extremely careful in implementing methods such as pre-averaging (for modeling hydrodynamic interactions) or making {\em a priori} assumptions regarding an underlying probability distribution for molecular properties. Molecular individualism necessarily broadens distributions in molecular behavior. Moreover, it is theoretically possible that a non-majority molecular conformation may dominate a bulk rheological property such as stress; future experiments that aim to couple bulk rheological measurements with single molecule observations can address these compelling questions.

\subsubsection{Dynamics in shear flow and linear mixed flows}
\begin{figure}[t]
\includegraphics[height=6cm]{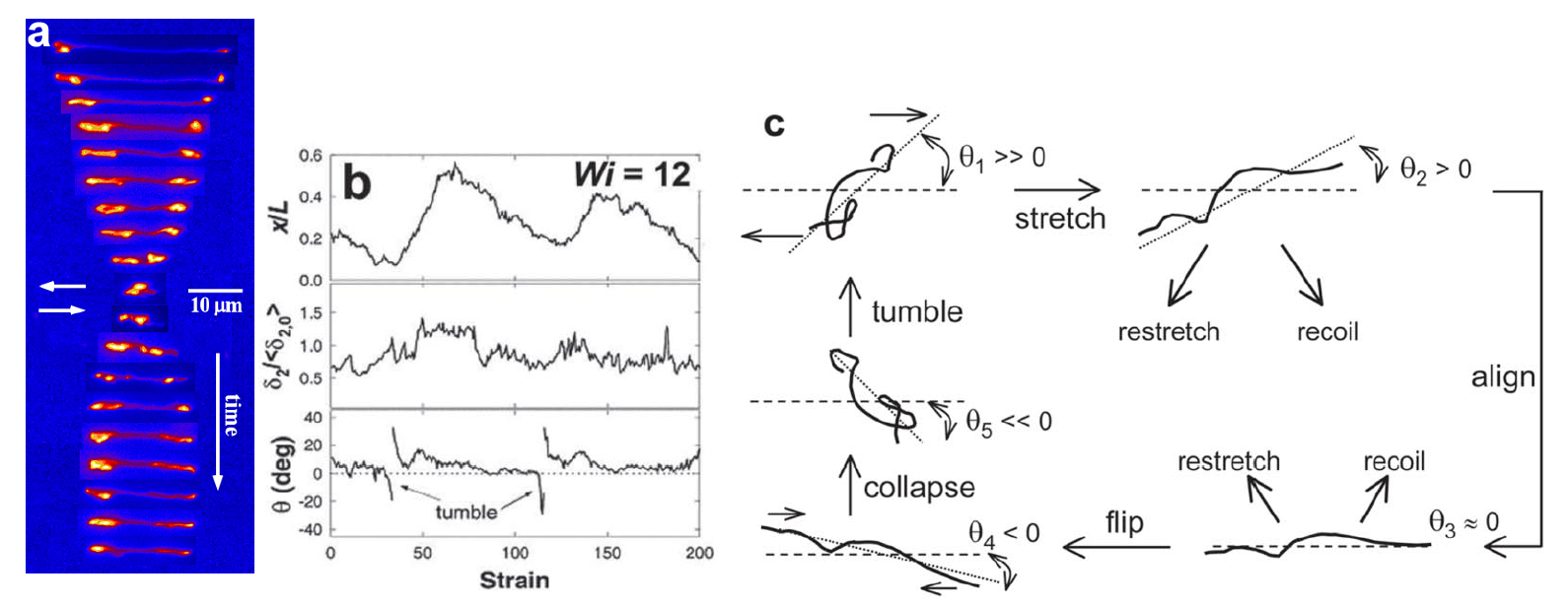}
  \caption{\label{fig:fig4} Dynamics of a single DNA polymer in steady shear flow. (a) Time sequence of images of fluorescent DNA ($L$ = 84 $\mu$m) in shear flow illustrating end-over-end tumbling motion. Orientation of shearing motion is shown by arrows, dimensionless flow strength $Wi$ = 109, and time between images is $\approx$ 10 $s$. (b) Trajectories of fractional polymer extension $x$/$L$, gradient direction polymer thickness $\delta_2$, and polymer orientation angle $\theta$ for lambda DNA at $Wi$ = 12. A discontinuous transition in orientation angle $\theta$ from negative to positive values is a signature of a polymer tumbling event. Accumulated fluid strain $\gamma = t_{obs} \dot{\gamma}$, where observation time is $t_{obs}$. (c) Descriptive cycle of periodic polymer motion in unbounded shear flow. Reproduced with permission from Ref. \cite{Schroeder2005}.}  
\end{figure}

Experiments in extensional flow were followed by single polymer dynamics in steady shear flow \cite{Smith1999}, which provided direct experimental evidence for the relatively weaker stretching dynamics of polymers in a simple shear flow due to the influence of vorticity. Unlike single chain dynamics in extensional flow, polymers do not exhibit a steady-state extension in steady shear flow; rather, single polymer chains undergo repeated end-over-end tumbling events in shear flow \cite{Smith1999}. Direct imaging in the flow-gradient plane of shear flow allowed for polymer `thickness' in the gradient direction to be measured and interpreted in the context of shear viscosity in dilute polymer solutions \cite{Teixeira2005}; these experiments were complemented by BD simulations with HI and EV \cite{Schroeder2005b}. Interestingly, the power spectral density of polymer chain extension fluctuations in steady shear suggest that the end-over-end tumbling events are aperiodic, with no preferred frequency for the tumbling cycle \cite{Smith1999}. Similar conclusions were drawn from single polymer experiments for tether chains in shear flow \cite{Doyle2000}. However, a characteristic periodic motion for single polymers in shear flow (untethered shear and tethered shear flow) was found by considering the coupling between the advection of the polymer chain flow direction and diffusion in the shear gradient direction (Figure \ref{fig:fig4}) \cite{Schroeder2005}. In other words, polymer chain motion appears to be non-periodic when only considering the chain stretch in the flow direction, but quantities such as polymer orientation angle which rely on coupled chain dynamics between the gradient direction and flow direction reveal a characteristic periodic motion in flow. Here, it was found that the power spectral density of polymer orientation angle $\theta$ exhibited a clear peak at a characteristic frequency, and scaling relations were further developed to describe the physics of the characteristic tumbling frequency and cyclic polymer motion in shear flow \cite{Schroeder2005}. 

Experimental results were complemented by Brownian dynamics simulations of single polymers in steady shear flow \cite{Hur2000a}, again considering freely-draining chains and no excluded volume interactions. Here, it was found that BD simulation results are in qualitative agreement with single molecule experiments in terms of capturing average extension as a function of $Wi$. The dynamics of single polymers was also investigated in the start-up of shear flow, which included bulk rheological measurements and Brownian dynamics simulations for solutions of $\lambda$-DNA between $10^{-5}c^{\ast}$ and 6.0 c$^{\ast}$ \cite{Babcock2000,Hur2001}. Interestingly, these results showed that the overshoot in the transient shear viscosity occurs at earlier strains compared to the overshoot in the polymer fractional extension. These results were key in revealing that the overshoot in transient shear viscosity occurs when a polymer chain begins to rotate toward the stagnation line, whereupon the `thickness' of the polymer in the gradient direction is maximum. This rotational motion leads to polymer stretching event, thereby showing that (on average) polymer chain extension reaches a maximum after the overshoot in the bulk solution viscosity. Moreover, results from Brownian dynamics simulations essentially showed that plots of average fractional polymer extension, and separately plots of shear viscosity, collapse onto similar curves for both ultra-dilute $10^{-5}c^{\ast}$ and semi-dilute solutions 6.0 c$^{\ast}$ when time is scaled by the longest relaxation time in either dilute or semi-dilute solutions \cite{Hur2001}. These results suggest that dynamics in the semi-dilute regime are qualitatively similar to polymer chain dynamics in steady and transient shear flow in the dilute regime. These results were interpreted by concluding that the increased polymer concentration merely serves to increase the background solution viscosity in semi-dilute solutions in shear flow rather than altering chain dynamics. Interestingly, however, recent experiments and BD simulations have shown that this is not the case for polymer dynamics in extensional flow \cite{Hsiao2017,Sasmal2017}, where chain dynamics change significantly in semi-dilute unentangled solutions, as discussed below. 

The dynamics of single polymers were further studied in linear mixed flows, wherein the degrees of rotational and extensional character are varied between purely extensional flow and purely rotational flow \cite{Babcock2003,Hur2002}. Interestingly, the coil-stretch transition was observed to be sharp for flows with dominant extensional character, and the steady-state polymer extension was observed to collapse onto a universal curve for extension-dominated flows when plotted against a rescaled $Wi^{eff} = Wi \sqrt{\alpha}$, where $\alpha$ is the flow-type parameter. Here, the eigenvalue of the velocity gradient tensor sets the scale of the effective strain rate along the extensional axis \cite{Fuller1980}. In brief, the $Wi^{eff}$ is rescaled by the magnitude of the extensional rate along the extensional axis eigenvector, which yields a universal curve for steady polymer extension for extension-dominated flows. These dynamics were further corroborated by an analytical model based on the finitely extensible Rouse model \cite{Dua2003}. Moreover, it was observed that polymer chains align along the extensional eigenvector in extension-dominant mixed flows, while occasionally experiencing a Brownian `kick' that knocks the polymer chain along the compressional axis, followed by a compression and re-extension event \cite{Babcock2003,Hur2002}. In an additional study, single polymer experiments and BD simulations were used to probe the conformational dynamics of DNA in linear mixed flows with dominant rotational character \cite{Lee2007}. Single polymer experiments were performed using a microfluidic four-roll mill \cite{Lee2007b}, which allows for generation of flows with arbitrary flow type between pure shear $\alpha=0$ and purely rotational flow $\alpha =$ -1.0 for varying $Wi$. In rotational dominated flows, it was observed that the polymer trajectory essentially follows an ellipsoid and the tumbling motion is `vanelike' and approaches simple chain rotation without extension changes as $\alpha$ approaches -1.0.

\subsubsection{Hydrodynamic interactions (HI)}
It has long been appreciated that intramolecular hydrodynamic interactions (HI) affect the dynamic properties of polymer chains \cite{Kirkwood1948,Kirkwood1949,Zimm1956}. Over the last few decades, advances in computational power have allowed for the simulation of long-chain polymers with intramolecular HI using coarse-grained Brownian dynamics methods. The Kirkwood-Riseman diffusion tensor was developed to incorporate the effects of HI, but the tensor was not generally amenable to simulation because it does not remain positive definite for all chain conformations. In 1969, Rotne and Prager developed a diffusion tensor for incorporating HI in coarse-grained simulations of polymers that remains positive semi-definite for all chain conformations \cite{Rotne1969}. The Rotne-Prager diffusion tensor allows for proper inclusion of HI, though simulation method based on this tensor are relatively slow due to requirement to decompose the diffusion tensor at each time step in order to calculate the Brownian force. In 1986, Fixman developed an approximate method for determining the Brownian forces with reduced computational cost \cite{Fixman1985}, however, due to limited computational power and the need for expensive computations, this method was not implemented until 2000 by Graham, de Pablo and coworkers \cite{Jendrejack2000}. 

Brownian dynamics simulations allows for the direct calculation of polymer chain dynamics with dominant HI. On the other hand, several approximate methods were developed to apply polymer kinetic theory to model the effect of HI on polymer chain dynamics \cite{Bird1987b}, including the preaveraging approximation \cite{Zimm1956}, the consistent averaging approximation \cite{Ottinger1985,Ottinger1987}, and the Gaussian approximation \cite{Ottinger1989}. The latter incorporates the effects of fluctuating HI (rather than constant HI at a pre-chosen chain conformation) and is perhaps the most realistic of the models \cite{Zylka1989}. Nevertheless, solution of the diffusion equation (Fokker-Planck equation) with HI is extremely difficult, and the approximate methods merely offer an approximate answer to the problem. In the case of dominant hydrodynamic interactions giving rise to polymer conformation hysteresis, this resulted in major confusion in the field for many years, as discussed below. 

Brownian dynamics simulation, complemented by single polymer dynamics, offers the ability to directly probe the influence of hydrodynamic interactions without resulting to approximate analytic methods. In 1978, Ermak and McCammon developed a Brownian dynamics simulation algorithm that yields the particle position Langevin equations in the presence of HI \cite{Ermak1978}, and this algorithm set the stage for nearly all BD simulations of polymer chains with intramolecular HI in subsequent years \cite{Ottingerbook}. Using the BD simulation approach, the dynamics of single polymers in shear flow \cite{Cascales1992,Lyulin1999,Kroger2000} and extensional flow \cite{Agarwal1998,Cifre1999,Agarwal2000,Cifre2001,Neelov2002} were extensively studied. These BD simulations were some of the first to consider the impact of HI on the dynamics of single polymer molecules, though the specific case of DNA polymers and comparison to single polymer experiments were soon to follow. 

The dynamics of single DNA molecules in the presence of intramolecular HI in dilute solution flows was specifically considered using Brownian dynamics simulations around 2000. Using the Chebyshev polynomial approximation for Fixman's method \cite{Fixman1985,Jendrejack2000}, the dynamics of single DNA molecules in shear flow and extensional flow were simulated with fluctuating HI and EV interactions in 2002 by Jendrejack, Graham, and de Pablo \cite{Jendrejack2002}. These BD simulations yielded results that were in quantitative agreement with single molecule diffusion experiments and steady shear and extension flow experiments. However, the simulation generally requires selection of 5 model parameters, including polymer contour length $L$, discretization level (number of springs $N_s$), Kuhn step size $b_K$, HI parameter (hydrodynamic bead radius) $a$, and EV parameter $v$ \cite{Jendrejack2002}. Polymer contour length and Kuhn step size for DNA are fairly well defined, but one concern is that some degree of parameter matching was required in order to choose $a$ and $v$ for a given level of discretization. An alternative Brownian dynamics simulation with HI and EV was subsequently reported by Hsieh and Larson which involved a systematic method for choosing the HI parameter $a$ (or $h^{\ast}$) by matching the longest relaxation time or diffusion coefficient and the hydrodynamic drag on the polymer at full extension \cite{Hsieh2003}. This method again yielded good agreement for DNA dynamics compared to single polymer experiments. Moreover, this work made use of an efficient second order semi-implicit predictor-corrector scheme for the time stepping algorithm \cite{Somasi2002}, which enables relatively low computational expense for systems with large numbers of beads. Nevertheless, all of these methods require selection of model parameters that generally depend on the level of discretization. The method of successful fine graining allows for parameter-free modeling of polymers with HI and EV, as discussed below.

In the context of DNA, a key question arises: {\em are hydrodynamic interactions important or necessary for modeling the dynamics of DNA?} The answer is absolutely yes.  Near equilibrium, we have seen that the power-law scaling of DNA diffusion constants with molecular weight are consistent with a non-draining coil \cite{Smith1996,Robertson2006}. PERM simulations show that the theoretical Zimm limit for full non-draining coils is reached only in the limit of very high molecular weight DNA ($L \approx 10^2$-$10^3$ $\mu$m), however, DNA polymer coils are nearly fully non-draining for DNA molecules of size $\lambda$-DNA \cite{Tree2013}. For these reasons, HI is clearly dominant at equilibrium for DNA molecules of size at least $\approx$40-50 kbp, which corresponds to polymers of size $\lambda$-DNA. Moving away from equilibrium and considering the role of HI on flow dynamics, it should be noted that DNA has a fairly large persistence length, which suggests that the increase in hydrodynamic drag (so called conformation-dependent drag) might be fairly minor for DNA compared to flexible polymers. Indeed, slender body theory predicts that the increase in drag between the coiled state to extended state is only a factor of $\approx$ 1.7 for $\lambda$-DNA \cite{Batchelor1970,Larson1997}. For these reasons, the role of HI was difficult to answer and it was a major question in the field for many years. Nevertheless, this question has been suitably addressed using the simulations and experiments discussed in this article. The bottom line is the following. In the absence of HI, simulations can only provide {\em qualitative} agreement with dynamic experiments. Upon including HI (and EV) in molecular simulations, one can achieve {\em quantitative} agreement with experiments, though it is preferred and desirable to use methods such as successive fine graining (SFG) for BD simulations to systematically choose model parameters that do not depend on the level of discretization of the model. Given the complexity of the problem, many questions surrounding the role of HI on DNA dynamics in non-dilute solutions are still being answered. For example, the role of HI in semi-dilute solutions of DNA was only recently addressed in a quantitative fashion in 2017, as discussed below \cite{Hsiao2017,Sasmal2017}. In any event, it is clear that simulations play an indispensable role in revealing the underlying physics of polymer solution dynamics, in addition to complementing single polymer dynamics experiments.

\subsubsection{Conformation hysteresis}
In the limit of high molecular weight polymer chains, intramolecular HI gives rise to conformation-dependent drag, wherein the hydrodynamic drag in the stretched state of polymer is significantly larger than the drag in the coiled state \cite{Larsonbook,Larson1997}. In particular, the hydrodynamic drag on a polymer coil scales with the equilibrium radius-of-gyration such that $\zeta^{coil} \sim N^{\nu}$, and the drag on the polymer in the extended state is given by slender body theory and scales as $\zeta^{stretch} \sim N / \ln(L/d)$, where $N$ is the number of Kuhn steps, $L$ is the contour length, and $d$ is the hydrodynamic width of a monomer. In effect, the frictional `grip' on the polymer chain by the surrounding medium increases as the chain stretches. It was long ago predicted that this effect can give rise polymer conformation hysteresis in extensional flow \cite{deGennes1974,Tanner1975,Hinch1977}, though the early predictions led to a vigorous debate in the field for many years \cite{Leal1984,Fan1985,Magda1988,Wiest1989,Carrington1996}. Several challenges existed which complicated a clear answer to the question of polymer conformation hysteresis, including difficulties in analytical solution to polymer kinetic theories incorporating finite extensibility and hydrodynamic interactions, lack of suitable computational power to simulate the dynamics of long-chain polymers with HI, and lack of the ability to directly observe the dynamics of single polymer chains in flow. Polymer conformation hysteresis has been reviewed elsewhere \cite{Shaqfeh2005}, so here I only focus on the key aspects and recent considerations of the problem. 

In 2003, polymer conformation hysteresis was directly observed using single molecule techniques \cite{Schroeder2003}. The dynamics of large DNA molecules ($L$ = 1.3 mm) were observed in extensional flow; in brief, single DNA molecules were initially prepared in either the coiled state or the extended state, followed by observing the polymer chain extension at a single value of the dimensionless flow strength $Wi$ in extensional flow. In the vicinity of the coil-stretch transition, initially stretched polymers remained stretched (and initially coiled chains remained coiled) for over 12 units of strain. Single molecule experiments were complemented by Brownian dynamics simulations with intramolecular HI and EV \cite{Schroeder2003,Schroeder2004}, and good agreement was obtained between both methods. These single molecule experiments required several advances in order to obtain the results confirming hysteresis. First, the experiment required handling of extremely large DNA polymers in excess of 1 mm in contour length, which was required to achieve a large value of the drag ratio $\zeta^{stretch}/ \zeta^{coil}$ to induce conformation hysteresis. For $\lambda$-DNA ($L$=21 $\mu$m for stained DNA), the ratio $\zeta^{stretch}/ \zeta^{coil}\approx$ 1.6, which is fairly small, but for bacterial genomic DNA ($L$=1.3 mm), the ratio increases to $\zeta^{stretch}/ \zeta^{coil} \approx$ 5 \cite{Schroeder2003}. Second, the observation of single polymers for large strains in extensional flow required the use of feedback control over the stagnation point, which effectively traps objects for long times in flow \cite{Schroeder2003}. This method has been further optimized and automated in the development of the Stokes trap, which is a new method to manipulate multiple arbitrary particles or molecules using only fluid flow \cite{Shenoy2016}. Moreover, these experiments showcased the power of combining single polymer experiments with Brownian dynamics simulations to probe physical phenomena in polymer solutions. The initial experiments on DNA were followed by BD simulations of polystyrene in dilute solution, which again predicted the emergence of hysteresis at a critical molecular weight \cite{Hsieh2005}. It was further shown that conformation hysteresis can be induced by non-linear extensional flows and that hysteresis can be formally understood by considering Kramers theory and the process of activated transitions over an energy barrier separating coiled and extended states \cite{Beck2007}. Moreover, the role of solvent quality on the coil-stretch transition was also studied using BD simulations \cite{Somani2010}. 

In 2007, the first evidence of polymer conformation hysteresis using bulk rheology was observed in filament stretching rheometry (FiSER) experiments by Sridhar, Prakash, Prabhakar and coworkers \cite{Sridhar2007}. These experimental observations showed the existence of history-dependent stretching and so-called glassy dynamics near the coil-stretch transition in dilute solution extensional flows. Recently, additional evidence of polymer conformation hysteresis was shown in semi-dilute solutions of synthetic polymers \cite{Prabhakar2017}. Strikingly, the transient extensional viscosity in semi-dilute solutions of flexible synthetic polymers (polystyrene) was observed to reach vastly different values (2 orders of magnitude apart) depending on whether the sample was pre-treated to a large extension rate or a low extension rate before making the measurement. These results show that the hysteresis window in the coil-stretch transition widens as the polymer concentration is increased into the semi-dilute regime, with the window reaching a maximum at $c/c^* = 1.0$. Experimental results are in agreement with BD simulations for semi-dilute solutions and an analytical theory describing the effect of concentration on hydrodynamics in semi-dilute solutions \cite{Prabhakar2016}. Taken together, these are the first bulk rheological experiments to corroborate the phenomenon of hysteresis in extensional flow. 

In recent years, some attention has been drawn to the underlying nature of the coil-stretch transition \cite{Celani2006,Steinberg2008}. By making an analogy to a thermodynamic process (at true thermal equilibrium), a phase transition may be classified as a first-order or second-order transition \cite{Reifbook}. A first-order transition (such as liquid vaporization) is generally described by a discontinuity in the order parameter across a transition, whereas a second-order transition is described by a continuous order parameter across the transition. In many respects, the classification is of academic interest as it pertains to the coil-stretch transition, but a practical aspect is the emergence of a potentially hysteretic stress-strain curve in polymer processing, which would have major implications for polymer rheology. In any event, hysteresis is a classic signature of a first-order transition. A recent paper has claimed that the transition is second-order due to the evidence of `critical slowing down' in polymer dynamics near the coil-stretch transition \cite{Steinberg2008}. However, these results only consider DNA polymers up to the size of T4 genomic DNA (169 kbp), which is known to be much smaller than the critical molecular weight $MW_{crit}$ required to observe polymer hysteresis. Moreover, an increase in chain fluctuations is expected as the polymer molecular weight approaches $MW_{crit}$, as the effective non-equilibrium free energy in flow will broaden and exhibit a deep and wide energy well in the vicinity of the coil-stretch transition for polymers near $MW_{crit}$ \cite{deGennes1974,Schroeder2003}. In fact, these general observations were made in the original work on hysteresis \cite{Schroeder2003}, which showed that DNA polymers of size $L \approx$ 575 $\mu$m were sluggish to recoil to the collapsed state when initially prepared in the stretched state near $Wi \approx$ 0.5. Taken together, essentially all evidence points toward a first-order-like transition for polymers in extensional flow. 

\subsubsection{Dynamics of DNA polymers in post arrays}
Significant attention has been given to the stretching dynamics of single DNA molecules in microfabricated or nanofabricated arrays of posts. In the context of electrophoretically driven motion, the dynamics of single DNA in post arrays was studied many years ago by Austin and coworkers \cite{Volkmuth1992}. Moreover, the dynamics of a single DNA molecule with an isolated (insulating) post has been studied in detail by Randall and Doyle \cite{Randall2004,Randall2005,Randall2006}. Single DNA-post collision events were systematically studied as a function of the initial transverse offset $b/R_g$ of the DNA from the post, where $b$ is distance between the DNA center-of-mass and the post center, and as a function of the post size. Hooking and roll-off events were observed, and taken together, this information in important in designing microfluidic-based separation tools for genomic DNA. In addition to single post collision events, the dynamics of electrophoretically driven DNA in large scale post arrays was studied experimentally using fluorescence microscopy \cite{Volkmuth1992,Volkmuth1994}. These experiments were followed by a detailed study of the dynamics of single DNA molecules in an array of posts driven by hydrodynamics instead of electrophoretics \cite{Teclemariam2007}. In particular, BD simulations were combined with single molecule experiments to reveal the dynamic behavior of DNA in post arrays driven by hydrodynamics, which differs fundamentally from the case of electrophoretic motion, because the disturbance velocity decay as $r^{-1}$ in the case of hydrodynamics. These experiments and simulations provided key insight into how polymer stretching is induced by single and multiple collisions and interaction events with stationary posts. The field of electrophoretic DNA manipulation continues to be vibrant today, as reviewed in an extensive article in 2010 \cite{Dorfman2010}, with recent work showing that nanofabricated `funnels' can be used to enhance the translocation and localization of genomic DNA in small devices \cite{Zhou2017}. 

\section{Dilute solution linear polymer dynamics: 2007 to present}
\label{diluterecent}
Major advances in the field of single polymer dynamics have occurred over the last decade. In terms of experiments, single polymer dynamics have been extended in fundamentally new directions, including direct observation of flexible polymer chains based on single stranded DNA and direct observation of topologically complex polymer architectures such as rings, combs, and bottlebrushes. In terms of modeling and simulations, major strides have been made in the treatment of hydrodynamic interactions and in the extension of non-equilibrium work relations (such as the Jarzynski equality) to polymer dynamics. In this section, I focus on new directions in single molecule studies of linear polymers, followed by consideration of recent work on semi-dilute solutions, concentrated solutions, and non-linear chain topologies in subsequent sections. 

\subsubsection{Single stranded DNA: a flexible polymer}
For many years, double stranded DNA has served as a model polymer for single molecule studies of non-equilibrium polymer dynamics. As discussed in previous sections, the concept of universality in polymer solutions indeed applies to double stranded DNA, at least in the limit of high molecular weight chains. Indeed, the equilibrium properties of DNA molecules in the size range of 50-150 kbp (corresponding to $N \approx$ 160-480 Kuhn steps) are well described by dynamical scaling relations in terms of diffusion constants and longest relaxation times. However, recent work has shown that for most practical molecular weight DNA molecules, physical properties lie in the transition region between purely theta solvent and athermal solvents \cite{Pan2014}; moreover, in the asymptotic limit of good solvents where EV interactions are governed only by hard-core repulsive potentials, the apparent EV exponent was found to be $\nu$ = 0.546 due to the large monomer aspect ratio $b/w$ for DNA \cite{Tree2013}. Moreover, it was found that the diffusion coefficient approaches the full non-draining Zimm value only in the limit of extremely large molecular weight DNA molecules ($L \approx$ 1 mm) \cite{Tree2013}. These predictions are consistent with the results from the hysteresis experiments on long DNA molecules ($L \approx$ 1.3 mm) in that extremely long DNA polymers are required to achieve full non-draining behavior. Nevertheless, despite these differences, DNA can be described by concepts of universality in polymer physics within the context of dynamical scaling. 

However, DNA differs fundamentally from truly flexible polymer chains in several respects. First, flexible polymers such as ssDNA and PEG exhibit a low-force non-linear elasticity in the limit of low forces due to monomer-monomer excluded volume interactions. As discussed in Section \ref{elasticity} above, the non-linear low force elastic regime is essentially absent for most reasonable size DNA molecules used in single polymer experiments. Moreover, the extensibility (ratio of contour length to equilibrium size) of DNA is substantially less than that of flexible polymer chains. For example, a double stranded DNA molecule with a contour length $L$ = 20 $\mu$m contains only $N \approx$ 190 Kuhn steps, whereas a flexible polymer with Kuhn step size $b_K$ = 2 nm would contain $N \approx 10^4$ Kuhn steps for the same contour length $L$. In effect, this results in enhanced roles of conformation-dependent hydrodynamic drag and hydrodynamic interactions in flexible polymer chains, thereby resulting in non-linear phenomena such as hysteresis at much lower molecular weights and contour lengths for flexible polymers compared to semi-flexible DNA \cite{Schroeder2003,Hsieh2005,Latinwo2011}. In addition, most practical polymer processing applications are focused on synthetic flexible polymers. Taken together, there is a strong motivation to perform direct single molecule imaging experiments on truly flexible polymers. 

\begin{figure}[t]
\includegraphics[height=8cm]{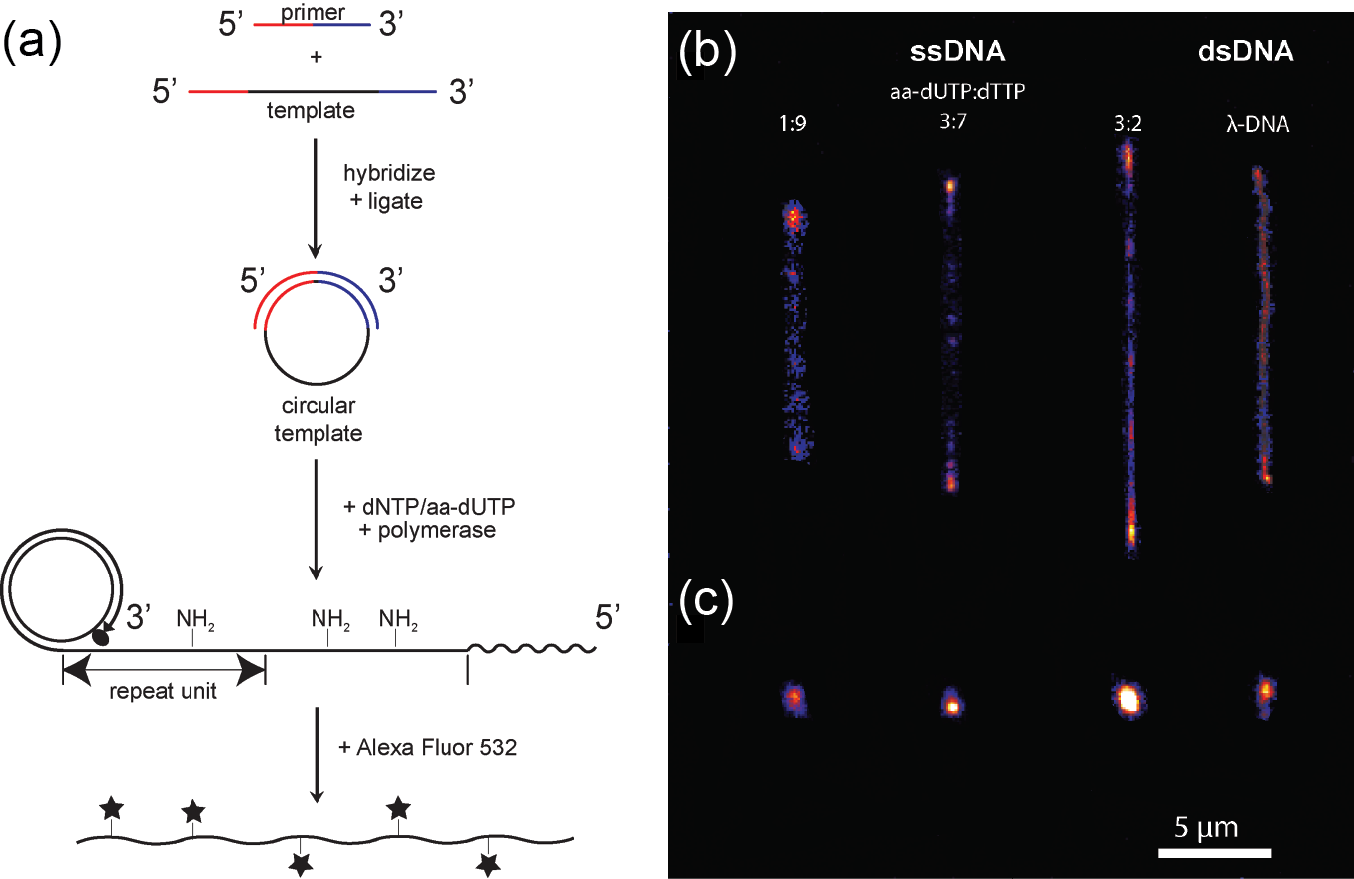}
  \caption{\label{fig:fig5} Single polymer studies of single stranded DNA (ssDNA), a flexible polymer chain. (a) Schematic of rolling circle replication (RCR) scheme for ssDNA synthesis. DNA polymerase is represented by an oval shape and stars represent fluorescent dye molecules. (b) Direct visualization of fluorescently-labelled ssDNA molecules using fluorescence microscopy. Single molecules of ssDNA and double stranded $\lambda$-DNA are shown for (b) stretched and (c) coiled configurations. Reproduced with permission from Ref. \cite{Brockman2011}.}  
\end{figure}

In 2011, Schroeder and coworkers performed the first set of single polymer imaging experiments on single stranded DNA (ssDNA), a flexible polyelectrolyte (Figure \ref{fig:fig5}) \cite{Brockman2011}. Using a biochemical synthesis reaction known as rolling circle replication (RCR), long chain polymers of ssDNA were synthesized with designer sequences to contain only 2 or 3 out of the 4 possible nucleobases (A, C, T, G), thereby preventing base pairing interactions. In this way, long flexible polymers consisting of ssDNA can be synthesized and studied in the absence of enthalpic interactions such as hairpin formation due to base pairing. Moreover, the RCR reaction was performed using chemically modified nucleotides (containing a primary amine group), which serve as reactive chemical sites for fluorescent labeling along the polymer backbone. Following synthesis, purification, and characterization, fluorescently-labeled ssDNA molecules were directly imaged stretching in an extensional flow in a cross-slot microfluidic device using epifluorescence microscopy. Images of single polymer chains stretching in fluid flow were obtained and show uniformly labeled polymer backbones with relatively low dye loadings ($\approx$ 1 dye per 100 bases). However, polymer products generated by RCR are fairly polydisperse, which complicates single molecule data acquisition and analysis. Future efforts focused on decreasing the polydispersity of the RCR method and/or in more stringent purification methods to selectively isolate relatively monodisperse fractions of ssDNA would enable more quantitative studies of ssDNA polymer dynamics in flow. Interestingly, the elasticity of single ssDNA molecules containing the same repeated nucleotide sequence in the experiments of Schroeder and coworkers were determined using single molecule magnetic tweezing experiments by Saleh and coworkers \cite{McIntosh2014}. Here, it was found that short repeats of the purine base adenine (A) result in some degree of base stacking interactions, which appears in the force-extension relation. Clearly, enthalpic interactions affect the elasticity of real polymer chains in many ways that are not captured by simple models that only capture entropic elasticity such as the Warner relation or the inverse Langevin chain \cite{Larsonbook}. Finally, it should be noted that the elastic force relations of other biological and synthetic flexible polymers have been studied using single molecule force microscopy, which has been reviewed elsewhere \cite{Marciel2013}. 

\subsubsection{Dynamics in time-dependent, oscillatory flows}
The vast majority of single polymer dynamics experiments has focused on the steady-state or transient response of polymers when exposed to a step strain rate in shear or extensional flow \cite{Shaqfeh2005}. However, there is a need to understand single polymer dynamics in more complicated, time-dependent transient flows such as oscillatory shear flow or oscillatory extensional flow. In recent years, there has been a renewed interest in the dynamics of soft materials in large amplitude oscillatory shear flow (LAOS) \cite{Ewoldt2010,Ewoldt2013,Rogers2012a,Rogers2012b}, and a clear understanding of the molecular conformations in oscillatory shear could reveal how macroscopic stress emerges. Moreover, flow through porous media practically involves repeated exposure to time-dependent extensional flows (contraction-expansion events) within a packed bed. However, despite the continued interest and relevance of time-dependent or oscillatory flows, the dynamics of single polymers in these complex flows was not directly studied until recently. 

\begin{figure}[t]
\includegraphics[height=7cm]{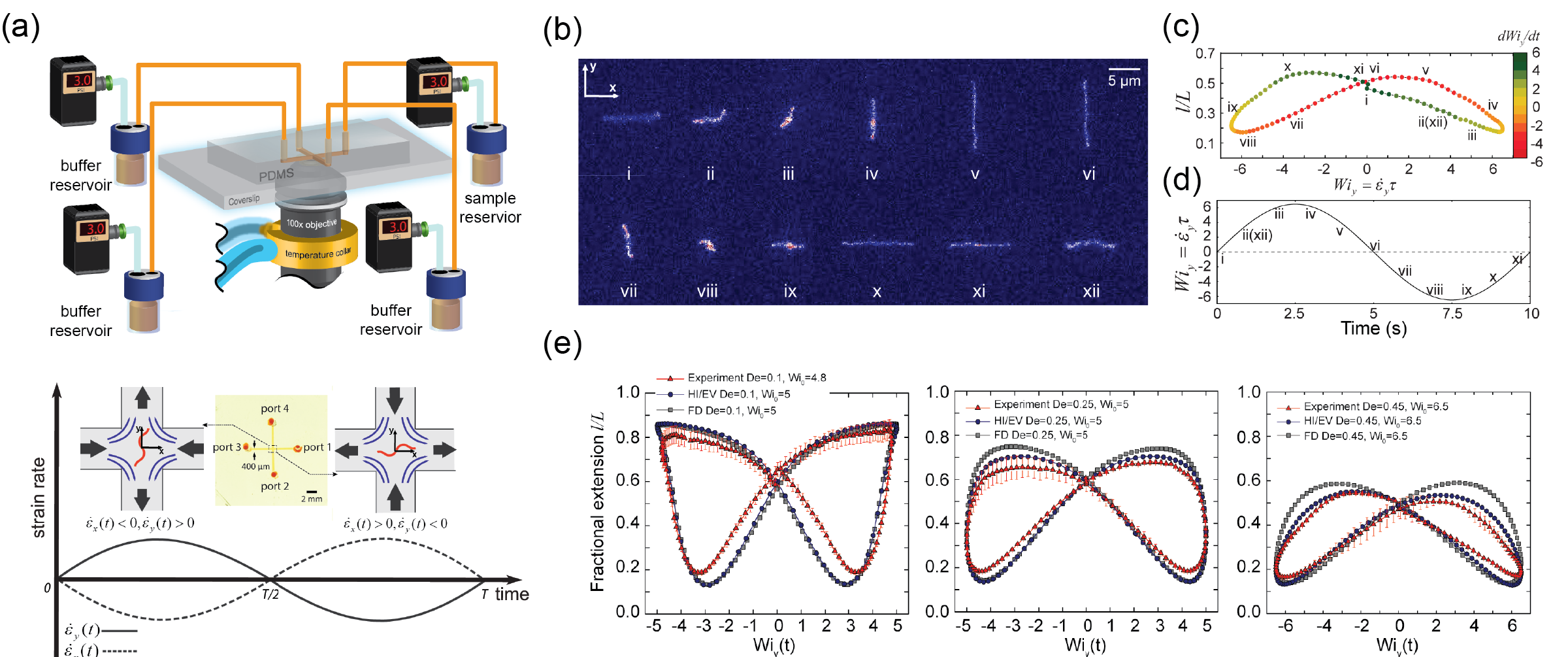}
  \caption{\label{fig:fig6} Single polymer dynamics in large amplitude oscillatory extension (LAOE). (a) Schematic of the experimental setup for the LAOE experiment. Sinusoidal strain rate input function for a full cycle is also shown, with an inset schematic showing the oscillatory extensional flow profile in the microfluidic cross-slot during the first half ($0<t<T/2$) and second half period ($T/2<t<T$) of the cycle. (b) Single polymer LAOE at $Wi_0=6.5$ and $De=0.45$. A series of single polymer snapshots is used to characterize polymer motion during one sinusoidal strain rate input cycle. The time between each snapshot is 1 s. (c) Experimental single polymer Lissajous plot showing the average polymer fractional extension $l/L$ as a function of $Wi_y(t)$ at $Wi_0=6.5$ and $De=0.45$. The color scale denotes the rate of change of the strain rate input function $dWi_y/dt$. (d) Experimental strain rate input function with period $T=10 $ s. (e) Single polymer Lissajous curves from experiments, BD simulations with hydrodynamic interactions (HI) and excluded volume (EV) interactions, and free-draining (FD) BD simulations at (left) $De=0.1$ and $Wi_0=5$, (center) $De=0.25$ and $Wi_0=5$, and (right) $De=0.45$ and $Wi_0=6.5$. Reproduced with permission from Refs. \cite{Zhou2016a,Zhou2016b}.}  
\end{figure}

In 2016, Schroeder and coworkers published two articles on the direct observation of single polymer dynamics in large amplitude oscillatory extensional flow (LAOE) \cite{Zhou2016a,Zhou2016b}. In general, it is quite challenging to study the dynamics of single polymers in controlled oscillatory extensional flows, mainly because fluid elements separate exponentially in time in extensional flow, which complicates the observation of long-time dynamic behavior in these flows. To address this problem, two-dimensional LAOE flow was generated using a feedback-controlled stagnation point device known as the Stokes trap \cite{Shenoy2016}, which results in an oscillatory planar extensional flow with alternating principal axes of extension and compression. For these experiments, the Stokes trap is implemented using a four-channel cross-slot microfludic device (Figure \ref{fig:fig6}), wherein opposing channels serve as alternating directions of compression/extension. The LAOE experiment is performed by alternating the positive pressures in the top/bottom and left/right channels in a sinusoidal manner. During the first half of the sinusoidal strain rate input ($0<t<T/2$), fluid is pumped into the flow cell from ports $p1$ and $p3$ such that the $y$-axis is the extensional axis and the $x$-axis is the compressional axis. During the second half of the cycle ($T/2<t<T$), fluid is pumped into the flow cell from ports $p2$ and $p4$ such that the $y$-axis is the compressional axis and the $x$-axis is the extensional axis. During this process, single DNA molecules are trapped near the stagnation point using a feedback controller \cite{Shenoy2016} that applies small pressures to the opposing ports. The feedback control pressures $\delta$ are negligible compared to the primary pressure $P$ used to generate the oscillatory extensional flow, such that $\delta \ll P$ for the majority of the cycle. Using this approach, a sinusoidal oscillatory planar extensional flow is applied in the cross-slot device, and the local fluid velocity in the vicinity of the stagnation point is described by:
     \begin{equation} \textbf{v}=(v_x, v_y, v_z)=(\dot{\epsilon}_x(t)x, \dot{\epsilon}_y(t)y, 0) = \left(-\dot{\epsilon}_0 \sin \left(\frac{2\pi}{T}t \right)x,\dot{\epsilon}_0 \sin \left( \frac{2\pi}{T}t \right)y, 0 \right)
     \end{equation}
     where $\dot{\epsilon}_x(t)$ and $\dot{\epsilon}_y(t)$ are the time-dependent fluid strain rates in the $x$ and $y$ directions, $x$ and $y$ are distances measured from the stagnation point, $\dot{\epsilon}_0$ is the maximum strain rate amplitude, and $T$ is the cycle time. 

The results from single molecule LAOE experiments show that polymers experience periodic cycles of compression, reorientation, and extension, and dynamics are generally governed by a dimensionless flow strength known as the Weissenberg number $Wi_0=\dot{\epsilon}_0\tau$, where $\dot{\epsilon}_0$ is the maximum strain rate applied during a sinusoidal cycle and $\tau$ is the longest polymer relaxation time, and a dimensionless frequency is given by the Deborah number $De = \tau/T$. Several single polymer Lissajous curves (polymer extension-strain rate) were determined as a function of $Wi$ and $De$, and qualitatively different shapes and signatures of single molecule Lissajous curves were observed across the two-dimensional Pipkin space defined by $Wi$ and $De$. A series of Lissajous curves is shown in Figure \ref{fig:fig6}, where the flow strength is maintained at $Wi_0 \approx 5$ and cycle frequency is changed from $De=0.1$ to $De=0.45$. It was observed that single polymer Lissajous curves open up from an arch shape at $De=0.45$ and $Wi_0=6.5$, to a bow tie shape at $De=0.25$ and $Wi_0=5$, and finally to a butterfly shape at $De=0.1$ and $Wi_0=5$. As the cycle time becomes longer (smaller $De$), the maximum in the polymer chain extension occurs at the same time as the maximum in strain rate $Wi$, which results in the sharp cusps in the butterfly shapes. As the frequency increases (larger $De$), the polymer relaxation time becomes longer than the cycle time, and the polymer chain cannot dynamically respond before the next phase of the periodic cycle. The single polymer experiments were complemented by Brownian dynamics simulations with and without intramolecular hydrodynamic interactions (HI) and excluded volume (EV) interactions, and the BD simulations accurately capture the dynamics of single polymers in LAOE over a wide range of control parameters. It was also found that the average unsteady extension in LAOE (polymer extension versus $Wi_0$) for different $De$ values collapsed onto a master curve when plotted versus an effective $Wi_{eff} = Wi_0 / (k \, De + 1)$, where $k$ is a numerical constant. Physically, the effective $Wi_{eff}$ can be motivated by considering the amount of fluid strain applied in a half-cycle, which is $\epsilon_{T/2} = Wi_0 / \pi De$ \cite{Zhou2016b}. 

In addition to LAOE, the dynamics of single polymers in large amplitude oscillatory shear flow (LAOS) was studied by Brownian dynamics simulation in 2009 \cite{Thomas2009}. Interestingly, it was found that single chain dynamics can essentially be described as experiencing a steady shear flow for dimensionless frequencies $De < De_{T}$, where $De_T =  \tau f_T/2$ is a critical Deborah number defined based on the characteristic tumbling frequency $f_T$ of single polymers in simple shear flow \cite{Schroeder2005}. For the case of $De < De_T$, single polymer chains undergo periodic tumbling events in each phase of the half cycle. However, for higher forcing frequencies $De > De_T$, chain flipping events are observed to dominate the chain dynamics, and the average polymer extension asymptotes toward the equilibrium value as $De$ is further increased. These simulations provide an intriguing link to prior work on the characteristic tumbling frequency of polymers in steady shear flow which was determined using a combination of single molecule experiments and BD simulations \cite{Schroeder2005,Schroeder2005b}. The dynamics of single polymers in LAOS has not yet been studied experimentally, though this would provide key insights into connecting microscale dynamics and bulk rheological phenomena in LAOS such as strain softening and strain hardening.

\subsubsection{Non-equilibrium work relations for polymer dynamics}
In 2013, it was shown that equilibrium properties such as polymer chain elasticity can be determined from far-from-equilibrium information such as polymer stretching dynamics in flow \cite{Latinwo2013}. This demonstration represents a fundamentally new direction in the field for determining near-equilibrium properties and thermodynamic quantities such as free energy, and it was made possible by applying recent methods in non-equilibrium statistical mechanics to the field of single polymer dynamics. In particular, the Jarzynski equality allows for determination of a free energy change $\Delta F$ between two states of a system by sampling the work distribution for transitioning the system from state 1 to state 2 \cite{Jarzynski1997}:
\begin{equation} 
e^{-\beta \Delta F} = \left< e^{-\beta w} \right> = \int p(w) e^{-\beta w} \, \text{d}w 
\end{equation}
where $w$ is the work done on a system connecting states 1 and 2, $\beta = 1/k_BT$ is the inverse Boltzmann temperature, and $p(w)$ is the probability distribution associated with the work. Using this framework, repeated measurements of the work performed on a molecular system upon transitioning between states 1 and 2 enables determination of a free energy change. The Jarzynski equality is intrinsically amenable to single molecule experiments, assuming that the work $w$ can be determined for a process. In early experiments, optical tweezers were used to transition a single RNA strand between two states described by a specified molecular extension $x$, thereby enabling determination of the free energy of an RNA hairpin, wherein work is simply defined as force applied over a distance \cite{Liphardt2002}. However, calculation of work done by a flowing fluid in stretching a polymer between two states of molecular extension is fundamentally different due to dissipation, and therefore required a different (and careful) definition of work for polymer dynamics \cite{Latinwo2013,Speck2008}. 

\begin{figure}[t]
\includegraphics[height=9cm]{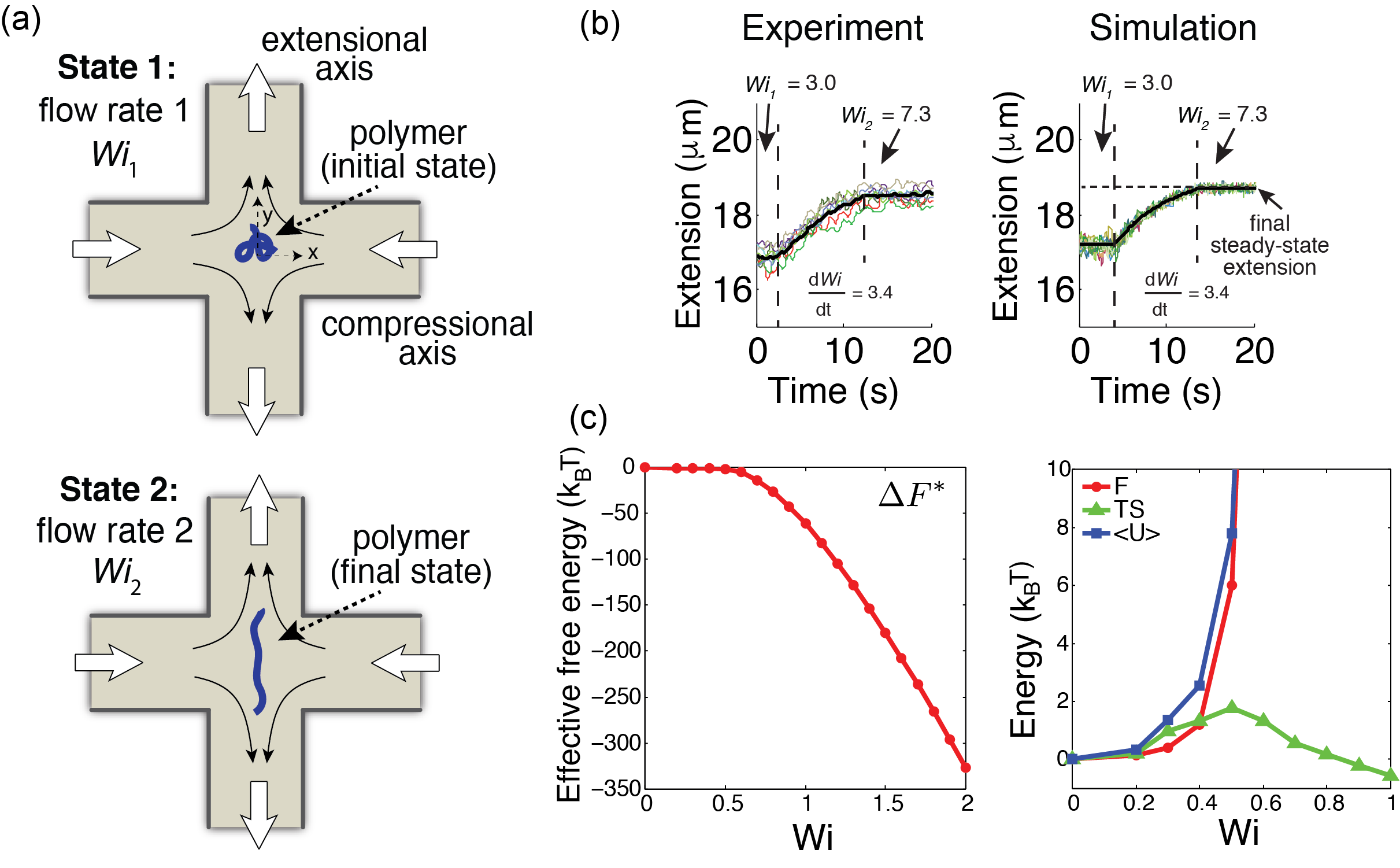}
  \caption{\label{fig:fig7} Non-equilibrium work relations for single polymer dynamics. (a) Schematic of experimental protocol for transitioning single polymers between state 1 (at $Wi_1$) and state 2 (at $Wi_2$) in extensional flow. (b) Molecular stretching trajectories for single DNA chains in extensional flow upon transitioning between $Wi_1$ = 3.0 and $Wi_2$ = 7.3 at a dimensionless transition rate of $dWi/dt$ = 3.4. Trajectories from single polymer experiments are shown together with results from Brownian dynamics simulations. (c) Determination of a non-equilibrium effective free energy $\Delta F^*$ in extensional flow and a Helmholtz free energy $\Delta F$, an average potential energy $\Delta \left< U \right>$, and a non-equilibrium entropy $\Delta S$. Reproduced with permission from Ref. \cite{Latinwo2014b}.}  
\end{figure}

Using this approach, Schroeder and coworkers published a series of papers showing that polymer chain elasticity (force-extension relations) and relaxation times can be determined for single polymers by sampling non-equilibrium properties such as stretching dynamics in flow (Figure \ref{fig:fig7}) \cite{Latinwo2013,Latinwo2014a,Latinwo2014b}. First, the stored elastic energy (or alternatively, the elastic force as a function of extension) was determined by systematically stepping single polymers between defined states of molecular extension in extensional flow \cite{Latinwo2014a}. Using Brownian dynamics simulations of free-draining bead-spring polymer chains, the entropic elasticity (force-extension) was determined for wormlike chains described by the Marko-Siggia relation and for polymer chains described by the inverse Langevin force relation. The general framework was also applied to prior single polymer experimental data on large concatemers of 7-$\lambda$ DNA, which further validates the results. In related work, the method was extended to multi-bead-spring chains with hydrodynamic interactions (HI) and for polymer dynamics in shear flow which contains vorticity \cite{Latinwo2013}. In all cases, this approach also allows for determination of the housekeeping power, which is the rate of work required to maintain polymer extension at an average constant extension in a given flow field. Finally, it was shown that the external control parameter could be taken to be flow strength ($Wi$), rather than polymer extension, which is generally more applicable to experiments \cite{Latinwo2014b}. Here, the polymer is transitioned between non-equilibrium steady states (at constant $Wi$), which required a new definition of the work done through the process. Nevertheless, and quite remarkably, this method allowed for determination of an effective non-equilibrium free energy in flow, a Helmholtz free energy, and a non-equilibrium entropy for single polymers in flow \cite{Latinwo2014b}. Finally, the general framework of non-equilibrium work relations was further applied to polymer dynamics using an analytical path integral approach by Cherayil and coworkers, which further extends the applicability of the method \cite{Ghosal2016a,Ghosal2016b}. 

From a broad perspective, this work is intriguing and significant from several different perspectives. Traditionally, the modeling of non-equilibrium polymer dynamics follows a fairly prescribed approach, wherein model parameters based on chemistry or solution conditions are chosen to match equilibrium materials properties, followed by the determination of non-equilibrium information such as steady or transient stress in flow. However, the application of non-equilibrium work relations to single polymer dynamics essentially reverses the flow of information and allows for equilibrium properties to be determined by sampling non-equilibrium dynamics in flow. Moreover, this method prescribes a relatively straightforward approach for determining non-equilibrium energies (such as a flow energy or an effective Helmholtz free energy) under highly non-equilibrium conditions. These concepts may be useful in designing optimized processing methods for polymers with consideration of flow energies, for example in designing a process to minimize dissipation or heat (as lost work), or to maximize stored elastic energy, while further minimizing energy input.

\subsubsection{Dynamics of polymer globules}
The dynamics of polymer globules has studied intensively over the last decade. A key motivation for this work is to understand the dynamics of von Willebrand factor (vWF), which is a large multimeric glycoprotein found in blood plasma and plays a key role in blood clotting. It is known that shear flow induces unfolding and subsequent adhesion of vWF \cite{Schneider2007}. In 2006, the dynamics of single polymer globules in shear flow was studied using a coarse-grained bead-spring model with the stiff-spring approximation, which essentially amounts to a bead-rod model \cite{Katz2006}. Poor-solvent conditions were simulated by introducing an attractive potential between beads using a Lennard-Jones potential. It was found that below a critical shear rate $\dot{\gamma}^*$, the polymer chain remains collapsed, but for shear rates $\dot{\gamma}>\dot{\gamma}^*$, single polymer chains undergo repeated collapsed/unfolding cycles. Importantly, hydrodynamic interactions were necessary to capture the proper dynamics during the collapse transitions \cite{Katz2006,Katz2008}. Simulations of polymer globules were later extended to extensional flow \cite{Sing2010,Sing2010b} and linear mixed flows \cite{Sing2011}. In related work, the role of internal friction in collapsed globules was examined \cite{Einert2011}, followed by further studies on the effects of hydrodynamic-induced lift forces on tethered polymers in shear flow \cite{Sing2011b,Sing2011c}. Simulations were further modified to incorporate the effects of local chemistry by modeling `stickers' or regions of the polymer chain that result in adhesion. In 2011, Sing and Alexander-Katz incorporated Monte Carlo-based self-association stickers into a BD simulation, which was used to study coil-globule transitions and the transition between Rouse chain dynamics and self-association dynamics \cite{Sing2011d}. This approach was further extended to study the dynamics of self-associating polymer chains in shear flow \cite{Sing2011e}. In 2014, Larson and coworkers developed a systematic method for coarse-graining bead-rod chains with attractive monomer interactions and variable bending stiffness, which serves as  suitable model for semi-flexible biopolymers such as cellulose \cite{Kong2014}. Interestingly, this work revealed an intriguing range of collapsed polymer structures including tori, helices, and and folded bundles for different ratios of the bead diameter to the persistence length. Recently, the role of shear flow on the dynamic formation of globules in shear flow was studied using bead-spring BD simulations by Underhill and coworkers \cite{Radhakrishnan2015}. 

\subsubsection{Successive fine graining}
In the early 2000's, Prakash and coworkers embarked on a systematic study of the influence of excluded volume (EV) interactions on the dynamics of polymers \cite{Prakash1999,Prakash2001,Prakash2002a,Prakash2002b}. EV interactions between beads within a dilute solution of Hookean dumbbells \cite{Prakash1999} and Rouse chains was modeled using a narrow Gaussian potential \cite{Prakash2001}. Using this approach, the linear viscoelastic properties \cite{Prakash2001} and the dynamics under shear flow for Rouse chain were modeled in the presence of EV \cite{Prakash2002a}. In continuing work, the coupling and additive influence of HI an EV in the dynamics of Hookean dumbbells was considered using a regularized Oseen-Burgers tensor for the HI \cite{Prakash2002b}. These publications were followed by the work of Graham and coworkers for simulating DNA dynamics by implementing the Chebyshev approximation for Fixman's method \cite{Jendrejack2000}, and soon after by the work of Larson and coworkers \cite{Hsieh2003} and Shaqfeh and coworkers \cite{Schroeder2004} for related coarse-grained multi-bead spring models with HI and EV. However, despite the impressive advances reported in these publications, parameter selection and the dependence of model parameters on the level of coarse graining continued to be an issue and arguably amounted to finding the best fit of parameters to match experimental data.

To address this issue, Prakash and coworkers developed a new method called successive fine graining that provided a systematic method for choosing model parameters \cite{Prakash2004,Prakash2005}. Importantly, in the limit of high molecular weight polymer chains, the predictions from Brownian dynamics simulations become independent of model parameters \cite{Prakash2005}. In essence, the method relies on representing the polymer chain as a long but finite macromolecule using a coarse-grained bead-spring model. In a series of simulations, the polymer chain is successively fine-grained by increasing the number of beads. Extrapolating the results obtained using BD simulations to the large $N$ limit enables determination of equilibrium or non-equilibrium properties that are essentially independent of the model parameters. The method was first developed for Hookean chains with HI interactions \cite{Kroger2000}, where the general methodology of extrapolation was developed, followed by extending the approach to finitely extensible polymers in extensional flow \cite{Prakash2005}. The method was validated by comparison to dilute solution experiments in terms of the elongational viscosity of a polystyrene solution using filament stretching rheometry \cite{Prakash2004b}. Remarkably, the method of SFG was further shown to generate quantitatively accurate predictions of DNA stretching dynamics in semi-dilute unentangled solutions in extensional flow \cite{Sasmal2017}.

\subsubsection{Computationally efficient BD algorithms \& conformational averaging of HI}
Several recent efforts have been directed at improving the computational efficiency of Brownian dynamics simulations of polymer chains, even in the limit of dilute solutions. In 2014, Saadat and Khomami developed new computationally efficient algorithms for incorporating hydrodynamic interactions and excluded volume interactions in BD simulations, with a particular emphasis on comparing Krylov subspace and Chebyshev based techniques \cite{Saadat2014}. This work was followed by a simulation-based study of the extensional rheology of high molecular weight polystyrene in dilute solutions \cite{Saadat2015b}. Here, Saadat and Khomami developed a hi-fidelity Brownian dynamics approach to simulation high molecular weight polymers using the Krylov framework and a semi-implicit predictor-corrector scheme. Moreover, Moghani and Khomami further developed computationally efficient BD simulations for high molecular weight chains with HI and EV using bead-rod (instead of bead-spring) models \cite{Moghani2017}. In recent work, these concepts and models have been extended to polyelectrolyte chains \cite{Moghani2016}.

In 2017, a new method for simulating the dynamics of single polymer chains with HI in dilute solutions was developed \cite{Sing2017}. The conformational averaging (CA) method essentially treats the intramolecular HI as a mean-field, which is an extremely efficient approximate method for determining hydrodynamic interactions. An iterative scheme is used to establish self-consistency between a hydrodynamic matrix that is averaged over simulation and the hydrodynamic matrix used to run the simulation. Results from this method were compared to standard BD simulations and polymer theory, which show that this method quantitatively captures both equilibrium and steady-state dynamics in extensional flow after only a few iterations. The use of an averaged hydrodynamic matrix allows the computationally expensive Brownian noise calculation to be performed infrequently, so that it is no longer the bottleneck of the simulation calculations. The method has only been applied to dilute solution systems (linear chains and ring polymers), though it can be extended to provide an extremely efficient method for studying semi-dilute polymer solutions in flow.  

\section{Semi-dilute unentangled and entangled dynamics}
\label{nondilute}
Recent work in the field of single polymer dynamics has pushed into semi-dilute unentangled and entangled solutions. In these experiments, the general approach is to fluorescently label a small amount of tracer or probe molecules in a background of unlabeled polymers. In this way, the influence of intermolecular interactions on the dynamics of a single chain can be directly observed in non-dilute polymer solutions. In tandem with single polymer experiments, significant advances were made in the modeling and simulation of semi-dilute unentangled polymer solutions, which is extremely challenging due to many-body effects and the role of HI and EV. These recent single molecule experiments and simulations have begun to probe the precise roles of HI in the semi-dilute regime in order to quantitatively understand DNA dynamics in flow \cite{Hsiao2017,Sasmal2017}. As discussed below, the effect of HI is important and necessary for understanding the dynamics of semi-dilute polymer solutions.

\subsubsection{Semi-dilute unentangled solution dynamics}
\begin{figure}[t]
\includegraphics[height=9.5cm]{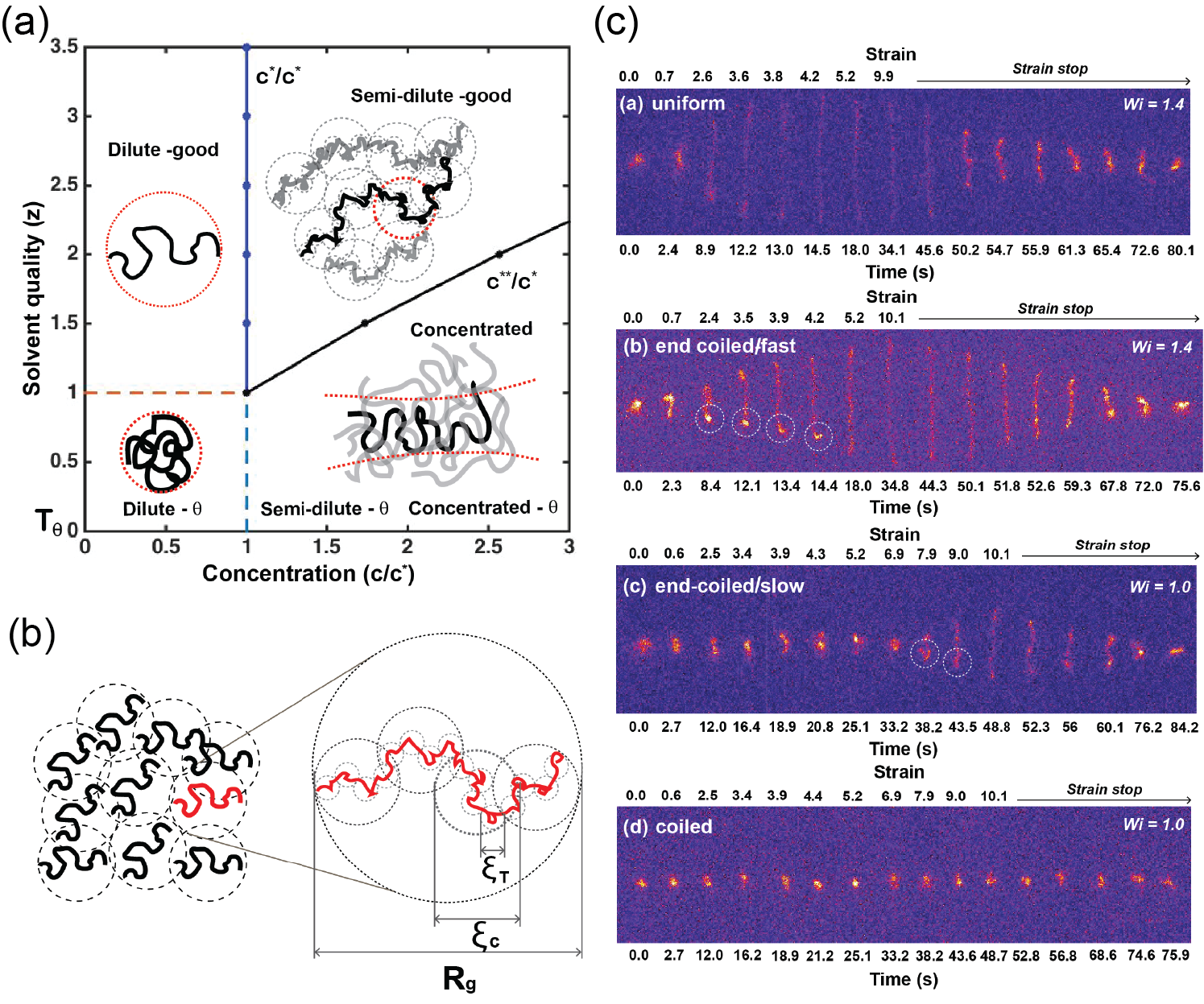}
  \caption{\label{fig:fig8} Single polymer dynamics in semi-dilute solutions. (a) Phase diagram for polymer solutions as a function of scaled polymer concentration $c / c^{\ast}$ and chain interaction parameter $z$. For display purposes, monomer size is chosen as $b$ = 1 and an excluded volume exponent was chosen as $\nu = 0.56$, based on experimental results \cite{Hsiao2017}. (b) Semi-dilute polymer solutions in the context of the blob picture, showing a schematic of a polymer solution near-equilibrium with interpenetrating polymer coils and a zoomed-in view of single polymer chain in good solvent conditions. Characteristic length scales are the thermal blob size $\xi_T$, the concentration blob size $\xi_c$, and the radius of gyration $R_g$. (c) Characteristic snapshots of stretching dynamics for single polymers in different molecular conformations in semi-dilute solutions in an extensional flow. Single molecule images show single chains adopting a uniform stretching conformation, an end-coiled fast stretching and an end-coiled slow stretching, and a coiled conformation near $Wi$ = 1.0. Circles drawn in the end-fast and end-coiled slow conformations denote putative hooking events with unlabeled polymers in the background solution. Reproduced with permission from Refs. \cite{Hsiao2017,Sasmal2017}.}  
\end{figure}

The dynamics of polymer chains in semi-dilute unentangled solutions is a particularly challenging problem in the field. Semi-dilute unentangled polymer solutions are characterized by a concentration $c$ that is larger than the overlap concentration $c^*$ but less than the entanglement concentration $c_e$, such that $c^* < c < c_e$. Semi-dilute solutions often exhibit large fluctuations in concentration, which precludes straightforward treatment using a mean-field approach. Moreover, the role of intra- and intermolecular HI may be significant, further complicating modeling and simulations due to many-body interactions. The near equilibrium properties of semi-dilute polymer solutions are governed by an interplay between polymer concentration and solvent quality (Figure \ref{fig:fig8}) \cite{Jain2012,Rubinsteinbook}. Recently, it was appreciated that polymer behavior can be described using a dynamic double crossover in scaling properties with concentration and solvent quality in the semi-dilute regime \cite{Jain2012}. In this framework, two parameters are commonly used to describe the equilibrium properties of semi-dilute solutions. First, the critical overlap concentration $c^{\ast} \approx M / N_A R_g^3$ is used as a characteristic polymer concentration in semi-dilute solutions, where $M$ is polymer molecular weight, $N_A$ is Avogadro's number, and $R_g$ is the radius of gyration. Using the overlap concentration, a scaled polymer concentration of $c/c^{\ast}=1$ corresponds to a bulk solution concentration of polymer that is roughly equivalent to the concentration of monomer within a polymer coil of size $R_g$. In addition, solvent quality can be characterized by the chain interaction parameter $z$, which is a function of polymer molecular weight $M$ and temperature $T$ relative to the theta temperature $T_{\theta}$. 

In 2006, Smith and coworkers developed a series of well-defined, monodisperse DNA constructs for single polymer dynamics \cite{Laib2006}. In particular, methods to prepare linear or circular double stranded DNA ranging in molecular weight between 2.7 kbp and 289 kbp were reported, and it was shown that the DNA constructs could be propagated in bacteria and prepared using standard methods in bacterial cell culture and DNA purification. Importantly, the authors made these constructs publicly available, which has enabled many other researchers in the field to study DNA dynamics using these materials. Using these DNA constructs, Prakash and coworkers determined the zero-shear viscosity $\eta_{p,0}$ of semi-dilute linear DNA solutions in 2014 \cite{Pan2014}, with results showing that DNA polymer solutions generally obey universal scaling relations \cite{Pan2014b}. Moreover, the theta temperature for DNA in aqueous solutions was determined to be $T$=14.7 $^o$C by static light scattering, and the solvent quality and radius of gyration was determined using dynamic light scattering and rigorous quantitative matching to BD simulations \cite{Pan2014}. Using these results, the overlap concentration for $\lambda$-DNA was found to be $c^* = $ 44 $\mu$g/mL, and the chain interaction parameter $z \approx$ 1.0 for $T =$ 22 $^o$C These advances greatly improved our understanding of the fundamental physical properties of DNA.

\begin{figure}[t]
\includegraphics[height=14cm]{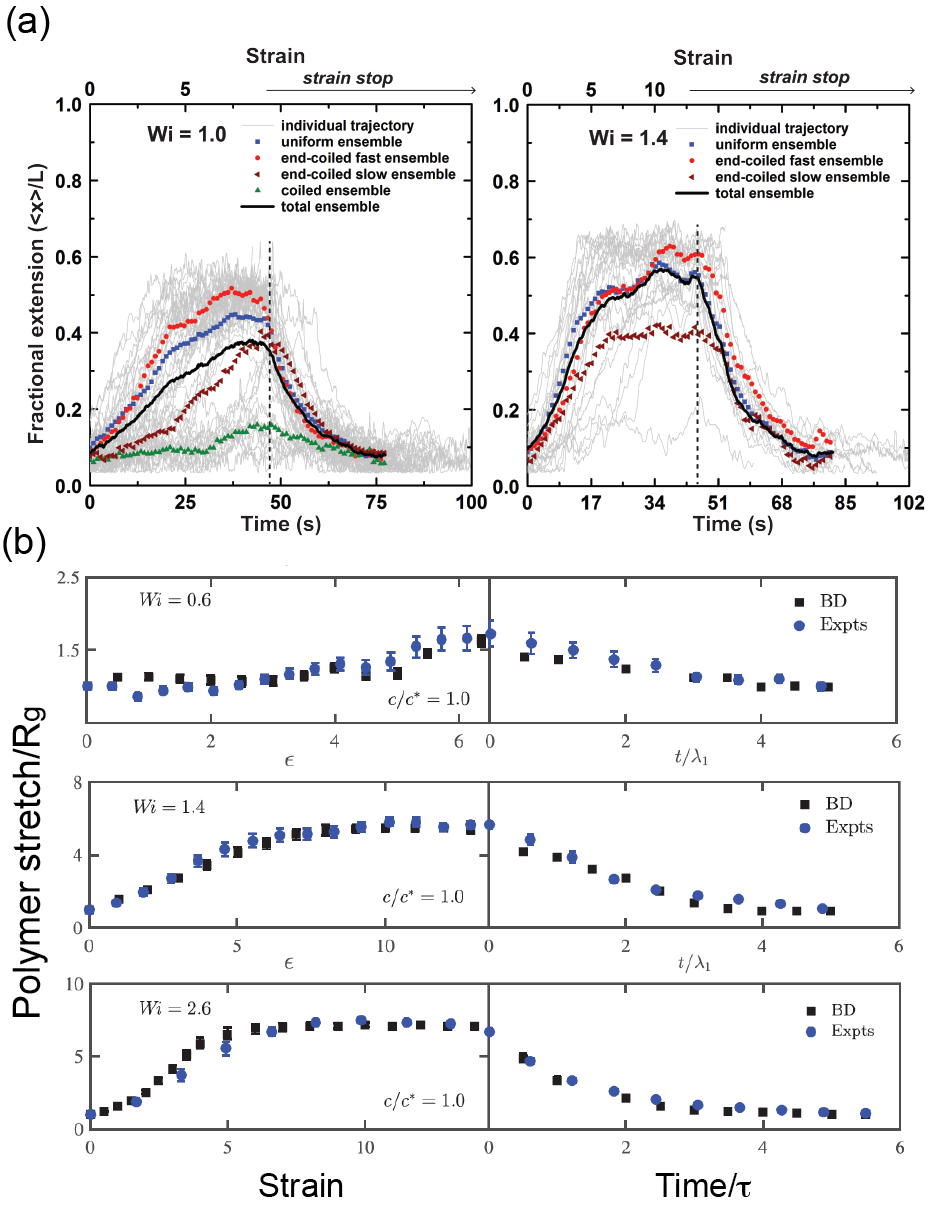}
  \caption{\label{fig:fig9} Molecular individualism in polymer stretching in semi-dilute solutions. Transient fractional extension of polymers in 1 $c^{\ast}$ solutions is shown as a function of molecular conformation. (a) Transient stretching dynamics in 1 $c^{\ast}$ solutions at $Wi$ = 1.0 and $Wi$ = 1.4, with results plotted as a function of polymer conformation in terms of the ensemble average stretch. The dotted lines indicate where the step-strain rate is stopped. (b) Comparison of single polymer stretching dynamics between BD simulations and single polymer experiments. BD simulations with HI and EV quantitatively match experimental data. Reproduced with permission from Refs. \cite{Hsiao2017,Sasmal2017}.}  
\end{figure}

In 2017, the dynamics of single DNA in semi-dilute unentangled solutions in extensional flow was studied using a combination of single molecule experiments \cite{Hsiao2017} and BD simulations \cite{Sasmal2017} (Figures \ref{fig:fig8} and \ref{fig:fig9}). Single molecule experiments were used to investigate the dynamics of semi-dilute solutions of $\lambda$-phage DNA in planar extensional flow, including polymer relaxation from high stretch, transient stretching dynamics in step-strain experiments, and steady-state stretching in flow. In terms of polymer relaxation, results showed a power-law scaling of the longest polymer relaxation time $\tau \sim (c/c^{\ast})^{0.48}$ in semi-dilute solutions. These results enabled determination of an effective excluded volume exponent $\nu$ $\approx$ 0.56, which is in good agreement with recent bulk rheological experiments on DNA \cite{Pan2014}, prior single molecule relaxation experiments in the semi-dilute regime \cite{Liu2009}, and PERM simulations for DNA \cite{Tree2013}. 

The non-equilibrium stretching dynamics of single polymers in semi-dilute solutions was further studied, including transient (1 $c^{\ast}$) and steady-state (0.2 $c^{\ast}$ and 1 $c^{\ast}$) stretching dynamics using an automated microfluidic device for extensional flow studies (Stokes trap \cite{Shenoy2016}). A decrease in transient polymer stretch in semi-dilute solutions at moderate $Wi$ compared was observed relative to dilute solutions (at the same $Wi$), and a milder coil-to-stretch transition for semi-dilute polymer solutions was observed at 0.2 $c^{\ast}$ and 1 $c^{\ast}$ compared to dilute solutions. Interestingly, a unique set of molecular conformations during the transient stretching process for single polymers in semi-dilute solutions is observed, which suggests transient stretching pathways for polymer chains in semi-dilute solutions are qualitatively different compared to dilute solutions due to intermolecular interactions. In particular, evidence of flow-induced entanglements was directly provided by single polymer conformations with an `end-hooked' conformation that appeared to stretch much faster than the average ensemble, which could be indicative of polymer chain hooking to an unlabeled polymer in the background solution in flow (see circles drawn in the end-coiled fast stretching single molecule snapshots in Figure \ref{fig:fig8}(c)). Interestingly, the stretching dynamics of DNA in semi-dilute solutions is qualitatively different in extensional flow compared to shear flow. In particular, it was earlier observed that the steady average fractional extension in semi-dilute solutions in shear flow collapsed onto the dilute solution values when plotted as a function of $Wi$, where $Wi$ is defined using the longest relaxation time in either dilute or semi-dilute solutions \cite{Babcock2000}. However, the results from DNA stretching in extensional flow did not show the same level of universality \cite{Hsiao2017}. 

In terms of new efforts in modeling semi-dilute polymer solutions, several mesoscopic techniques have been developed in recent years to study non-equilibrium dynamics \cite{Stoltz2006, Huang2010,Jain2012b,Saadat2015}. Prakash and coworkers developed an optimized BD algorithm for semi-dilute polymer solutions in the presence of HI and EV \cite{Jain2012b}. This algorithm was used in conjunction with the method of successive fine graining (SFG) to provide parameter-free predictions of the dynamics of DNA in semi-dilute solution in extensional flow \cite{Sasmal2017}, thereby providing a direct complement to single molecule experiments \cite{Hsiao2017}. Remarkably, BD simulation results provided quantitative agreement to single polymer experiments using the SFG method. Taken together, these results show that HI is important and necessary to quantitatively capture the dynamic behavior of DNA in semi-dilute solutions. Interestingly, an analytical model has been developed that incorporates the effects of polymer self-concentration and conformation-dependent drag in the semi-dilute regime \cite{Prabhakar2016}, and it was recently shown that this model captures hysteresis in the coil-stretch transition as observed in semi-dilute solutions using filament stretching rheometry \cite{Prabhakar2017}.

\subsubsection{Semi-dilute entangled solution dynamics}
Semi-dilute entangled solutions are generally defined to lie in a concentration regime $c_e < c < c^{**}$, where $c^{**}$ is the polymer concentration at which the concentration blob size $\xi_c$ equals the thermal blob size $\xi$. Recent single molecule experiments on longest polymer relaxation times \cite{Hsiao2017} and bulk rheological experiments on intrinsic viscosity \cite{Pan2014,Pan2014b} show that the critical entanglement concentration for $\lambda$-DNA occurs at $c_e \approx 4$ $c^*$, which is consistent with the range of onset of the entangled regime for different polymer chemistries \cite{Rubinsteinbook}. In 1994, Chu and coworkers used single polymer dynamics to directly observe the tube-like motion of single fluorescently labeled DNA molecules in a background of unlabeled entangled DNA \cite{Perkins1994b}. In these experiments, one end of a concatemer of $\lambda$-DNA ranging in contour length from 16 to 100 $\mu$m was linked to a micron-sized bead, and an optical trap was used to pull the bead through the solution of entangled $\lambda$-DNA at a concentration of 0.6 mg/mL ($c \approx 4$ $c_e$). In these experiments, the polymer chain was observed to relax along its stretched contour, which provides evidence for polymer reptation in entangled solutions. Some degree of concern was expressed that the motion of the large micron-sized bead through the polymer solution might disrupt the local entanglement network, thereby resulting in modified polymer relaxation behavior. However, the relaxation of the local network (equilibration time of a thermally diffusing chain in a tube) was much faster than the reptation time of the long polymer chain. In 1995, the diffusion of single DNA molecules ranging in size from 4.3 to 23 kbp was observed in background solutions of concentrated $\lambda$-DNA at 0.6 mg/mL \cite{Smith1995}. Results showed that the center-of-mass diffusion coefficient scaled with molecular weight as $D_{rep} \sim L^{-1.8}$, which is close to the predictions of reptation theory such that $D_{rep} \sim L^{-2}$ in a theta solvent. Upon including the effects of tube length fluctuations in reptation theory, it is found that $D_{rep} \sim L^{-2.3}$, as observed in experiments by Lodge that reconciled these predictions with experiments \cite{Lodge1999}. Additional single molecule diffusion experiments on entangled linear DNA in backgrounds of linear DNA solutions revealed that the length and concentration dependence of the center-of-mass diffusion coefficient scales as $D \sim L^{-2}c^{-1.75}$ in the limit of increasing solution concentrations in the vicinity of $c \approx$ 6 $c^*$ for 45 kbp DNA \cite{Robertson2007}.

\begin{figure}[t]
\includegraphics[height=9cm]{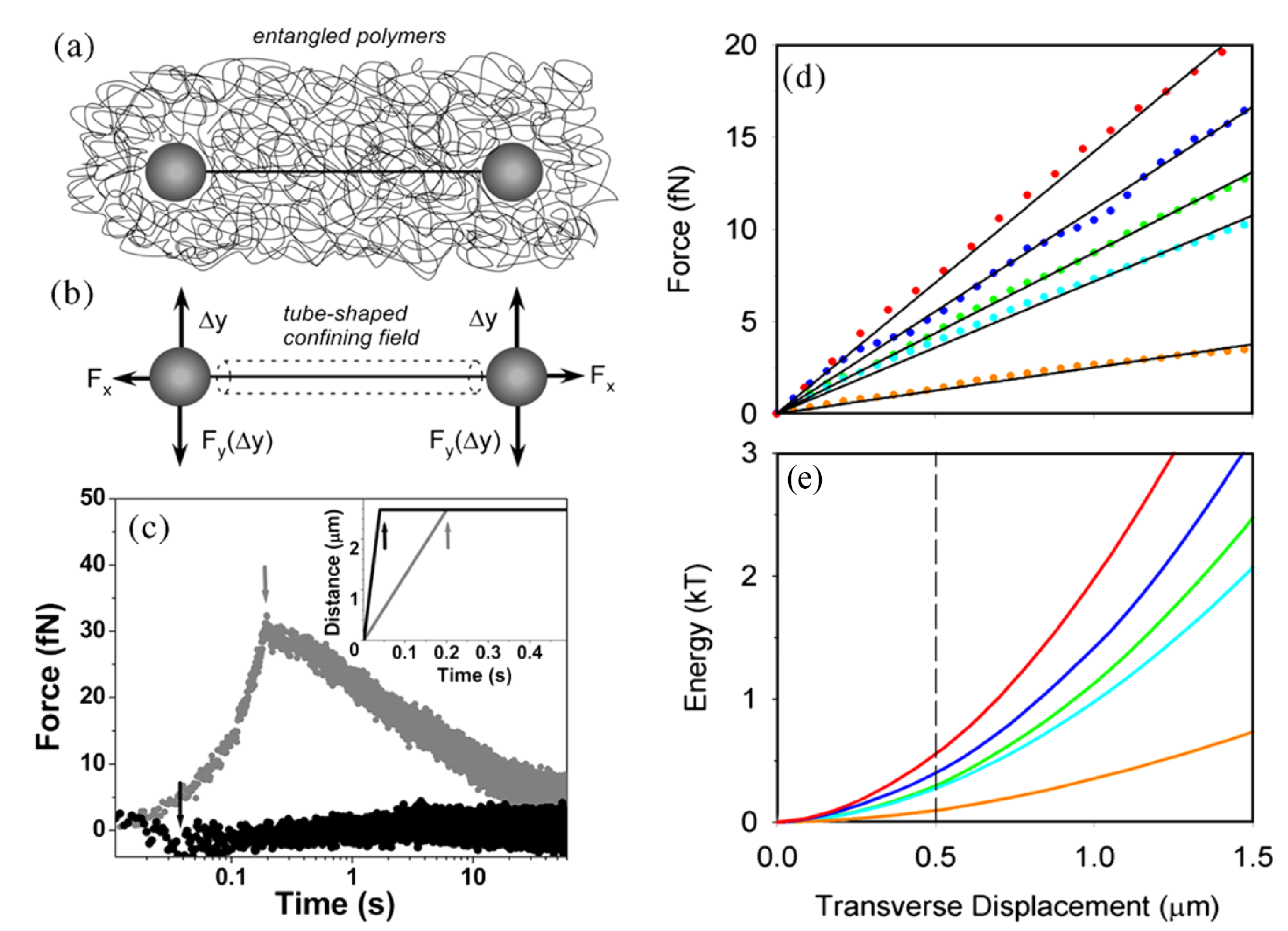}
  \caption{\label{fig:fig20} Direct measurement of the intermolecular confining forces for single polymers in entangled solutions. (a) Schematic of single molecule experiment. A single DNA molecule is held stretched between two optically trapped beads in an entangled solution of DNA. (b) The confining force per unit length was measured ($F_x$ and $F_y$) in response to an imposed displacement $x$ or $y$. (c) Average force induced by a displacement $y$ at 13 $\mu$m/s (gray) versus a displacement $x$ at 65 $\mu$m/s (black). Arrows mark the maximum displacements. The inset graph shows the displacement profiles. (d) $F_y$ versus $y$ at rates of 65 $\mu$m/s (red), 25 $\mu$m/s (blue), 13 $\mu$m/s (green), 0.52 $\mu$m/s (cyan), and 0.10 $\mu$m/s (orange). (e) Confining potential per unit length $U_y$ determined by integration of force data in plot (d). Reproduced with permission from Ref. \cite{Robertson2007b}.}  
\end{figure}

The bulk viscosity of DNA solutions was considered many years ago by Zimm using purified genomic DNA from bacteriophage T2 and T7 \cite{Zimm1956}. In the late 1990s, the linear viscoelastic properties of concentrated DNA solutions was studied by Wirtz and coworkers using calf thymus DNA (polydisperse with average molecular size of 13 kbp \cite{Mason1998}. These results showed that for DNA concentrations greater than the entanglement concentration $c_e \approx$ 2 mg/mL (for calf thymus DNA), a plateau modulus was observed in the storage modulus such that $G'_p \approx$ 6.1 dyn/cm$^2$ at $c=c_e$. These experiments were followed by bulk rheological experiments on the non-linear viscoelasticity of entangled DNA solutions in shear flow \cite{Jary1999}, with results showing a plateau in shear stress over a decade in shear rate for concentrated solutions of T4 DNA. 

In 2007, Robertson and Smith directly measured the intermolecular forces experienced by a single polymer chain in entangled DNA solutions using optical tweezers (Figure \ref{fig:fig20}) \cite{Robertson2007b}. In this experiment, a single DNA chain (25.3 kbp) was linked to two micron-sized beads, and a dual optical trap was used to confine both beads and to induce transverse displacement of the DNA-bead tether through the entangled polymer solution (1 mg/mL solution of 115 kbp linear DNA, such that $c \approx$ 40 $c^*$ for 115 kbp DNA). These results enabled estimation of the tube radius of 0.8 $\mu$m, which was close to the value predicted from Doi-Edwards theory \cite{Doiedwards} and from simulations from Larson and coworkers \cite{Zhou2006}. 

The dynamics of single DNA molecules in entangled solutions in shear flow was investigated in 2007 \cite{Teixeira2007}. Here, the dynamics of single fluorescently labeled $\lambda$-DNA was observed in background solutions of unlabeled $\lambda$-DNA at concentrations ranging between 0.65 mg/mL (16 $c^*$) and 2.2 mg/mL (55 $c^*$). Relaxation following a rapid shear deformation suggested that polymer relaxation followed two distinct timescales, including a fast retraction time and longer reptation time. Limited single polymer data on dynamics in steady shear and transient (start-up) shear suggested that polymer chains exhibit a large degree of molecular individualism in entangled solutions. This work also presented a relatively complete bulk rheological characterization of entangled $\lambda$-DNA solutions, including linear viscoelastic data and non-linear shear rheology. 

\subsubsection{Elastic instabilities in semi-dilute DNA solutions}
It has long been known that elastic polymer solutions can give rise to instabilities and secondary flows \cite{Bird1987,Pakdel1996,Shaqfeh1996}. The onset of secondary flows in DNA solutions has been studied using a wide array of microfluidic geometries \cite{Rems2016}. In particular, the small length scales in microfluidic devices and associated viscous-dominated flow conditions allows for flow phenomena to be studied in the limit of low Reynolds number $Re \ll 1$ and high $Wi$, thereby allowing access to the highly elastic regime defined by the elasticity number $El \equiv Wi/Re$ \cite{Rodd2005b}. In 2008, elastic secondary flows of semi-dilute DNA solutions were studied in microfluidic devices containing abrupt 90$^o$ microbends \cite{Gulati2008}. Although not strictly a single polymer visualization experiment, particle tracking velocimetry (PTV) can be applied to DNA solutions in flow, thereby revealing the onset of secondary flows and instabilities due to elasticity. These experiments revealed that a vortex flow developed in the upstream corner of the right-angle bend and tended to grow in size with increasing $Wi$. In related work, the flow of semi-dilute unentangled $\lambda$-DNA solutions ($0.5 <c/c^* < 3$) and lightly entangled $\lambda$-DNA solutions $c =$ 10 $c^*$ was studied in a gradual microfluidic contraction flow with combined on-chip pressure measurements \cite{Gulati2015}. Here, it was observed that large, stable vortices form about the centerline and upstream of the channel entrance. 

Direct visualization of single DNA conformation and stretching, combined with flow visualization measurements, were performed on a semi-dilute unentangled and entangled solution of DNA in a 4:1 planar micro-contraction flow \cite{Hemminger2010}. These experiments showed the ability to image single DNA polymers in non-canonical flow fields other than simple shear or extension. Recently, this approach has been used to study the necking and pinch-off dynamics of liquid droplets containing semi-dilute polymer solutions of polyacrylamide near the overlap concentration \cite{Sachdev2016}. Single fluorescently labeled DNA molecules were suspended in the semi-dilute polymer droplets, thereby enabling visualization of a DNA `tracer' polymer in this flow geometry. It was found that individual polymer molecules suddenly stretch from a coiled conformation at the onset of necking. The extensional flow inside the neck is strong enough to deform and stretch polymer chains; however, the distribution of polymer conformations was found to be quite broad, but the distribution remains stationary in time during the necking process. In addition, this approach was extended to visualize the dynamics of single DNA molecules in a microfluidic-based porous media flow \cite{Kawale2017}. A common feature in these experiments appears to be broad and heterogeneous distribution of polymer chain conformations and stretching dynamics in flow, features that can only be revealed using single molecule imaging. 

\subsubsection{Shear banding in entangled DNA solutions}
In 2008, Wang and coworkers studied the phenomena of wall slip and shear banding in concentrated DNA solutions \cite{Boukany2008,Boukany2009,Boukany2009b}. In a first study, calf thymus DNA (polydisperse, average molecular size of $\approx$75 kbp) was prepared at three different concentrations in an aqueous salt buffer corresponding to three levels of entanglement: $Z$ = 24, 60, and 156 entanglements per chain \cite{Boukany2009}. Here, it was found that only high levels of entanglement ($Z$ = 60, 156) resulted in shear banded profiles. In a follow up study, solutions of calf thymus DNA were prepared at extremely high concentrations in an aqueous salt buffer at 10 mg/mL, corresponding to a concentration of approximately 161 $c^*$ \cite{Boukany2008}. Although not technically a single molecule experiment, flow visualization was performed by dissolving a small amount of tracer particles in the DNA solutions, followed by imaging the flow profiles in Couette flow or a cone and plate geometry. It was observed that at very low shear rates $\dot{\gamma} <$ 0.1 s$^{-1}$, the flow profile was linear, however, wall slip was observed for 0.1 $s^{-1} < \dot{\gamma} <$ 40 s$^{-1}$. Upon increasing the shear rate further, a shear banded profile was observed for $\dot{\gamma} >$ 40 s$^{-1}$ after a few hundred strain units. 

Additional experiments by Wang and coworkers group in 2010 explored the effect of a gradual ramp up in shear rate, rather than an abrupt step change in shear rate \cite{Boukany2010}. Interestingly, it was observed that a graduate start-up of the shear rate does not result in a shear banded profile. These results indicate that shear banding may originate from the sudden fast deformation in startup shear. In other words, the authors concluded that when forced yielding is avoided, so is shear banding. These results further suggest that shear banding is not a unique response, that is, different pre-conditioning may result in vastly different steady responses. In 2010, the first direct imaging of wall slip was obtained using single molecule techniques in entangled DNA solutions \cite{Boukany2010b}. Here, confocal fluorescence microscopy and rheometry were used to capture molecular single DNA images in the nonlinear response regime of entangled DNA solutions. Conformations of DNA molecules were imaged in shear slow to correlate with the magnitude of wall slip. It was found that interfacial chain disentanglement results in wall slip beyond the stress overshoot, and disentanglement generally produces tumbling motion of individual DNA in the entangled solutions under shear. More recently, the phenomenon of wall slip was studied by single molecule imaging near adsorbing and non-adsorbing boundaries \cite{Hemminger2017}. Finally, recent methods in optical coherence tomography and velocimetry measurements have further confirmed the existence of shear banded flow profiles in highly entangled DNA solutions \cite{Malm2017}. 

In terms of computational modeling of shear banding, there has been significant effort by many different research groups directed at this problem. Here, we focus only on a few molecular-based computational methods that have been used to investigate shear banding. In 2015, Mohagheghi and Khomami used dissipative particle dynamics (DPD) simulations to uncover the molecular processes leading to shear banding in highly entangled polymer melts \cite{Mohagheghi2015}. The mechanism is complicated, but it essentially the stress overshoot in shear flow drives locally inhomogeneous chain deformation and spatially inhomogeneous chain disentanglement. In turn, the localized jump in the entanglement density along the velocity gradient direction results in a considerable jump in normal stress and viscosity, which ultimately leads to shear banding. This work was followed by two companion articles by the same authors that investigated flow-microstructure coupling in entangled polymer melts, which ultimately gives rise to shear banding \cite{Mohagheghi2016a,Mohagheghi2016b}. Finally, these authors further elucidated a set of molecular-based criteria for shear banding in 2016 \cite{Mohagheghi2016c}. 

\section{Architecturally complex DNA: combs, bottlebrushes, rings}
\label{branched}
The dynamics of architecturally complex polymers is an extremely important problem in the field of rheology. A main goal is to identify how molecular branching and non-linear molecular topologies affect non-equilibrium dynamics and relaxation processes in entangled solutions and melts \cite{Ruymbeke2014}. In an entangled solution of combs or comb polymer melts, branch points are known to substantially slow down the overall relaxation processes within the material. Branching results in a spectrum of relaxation times that can be attributed to molecular topology, including the branch segments, branch points, and the motion of the long chain backbone \cite{Daniels2001}. In concentrated and entangled solutions, these complex dynamics can be exceedingly complicated to discern using bulk techniques. 

Single molecule experiments enable the direct imaging of these processes, thereby allowing for a molecular-scale understanding of bulk rheological behavior. Early work by Archer and coworkers focused on the synthesis and direct single molecule imaging of star-shaped DNA generated by hybridizing short oligonucleotides to form a small star-branched junction, onto which long DNA strands of $\lambda$-DNA were hybridized and ligated \cite{Heuer2003,Saha2006,Saha2004} This method was also extended to create pom-pom polymers by connecting two stars with a $\lambda$-DNA crossbar. Conformational dynamics were mainly studied under the influence of electric fields, and electrophoretic mobility of star DNA polymers was measured in solutions of polyacrylamide \cite{Heuer2003,Saha2006}, polyethylene oxide \cite{Saha2004}, and agarose and polyacrylamide gels \cite{Heuer2003}. This early work represents a few examples of only a small number of studies on single branched polymers. 

In the following section, I summarize recent efforts in the field of single polymer dynamics to understand the role of molecular topology on non-equilibrium flow phenomena. Here, the focus is on branching and ring polymers and not on intramolecular topological interactions such as knots, though there has been a recent interest in single molecule studies of knot dynamics \cite{Grosberg2007,Tang2011,Renner2014,Renner2015,Reifenberger2015}. For a lengthy discussion of knot dynamics as probed by single molecule techniques, I refer the reader to a recent review published elsewhere \cite{Mai2016}.

\subsubsection{Comb-shaped DNA}
\begin{figure}[t]
\includegraphics[height=7cm]{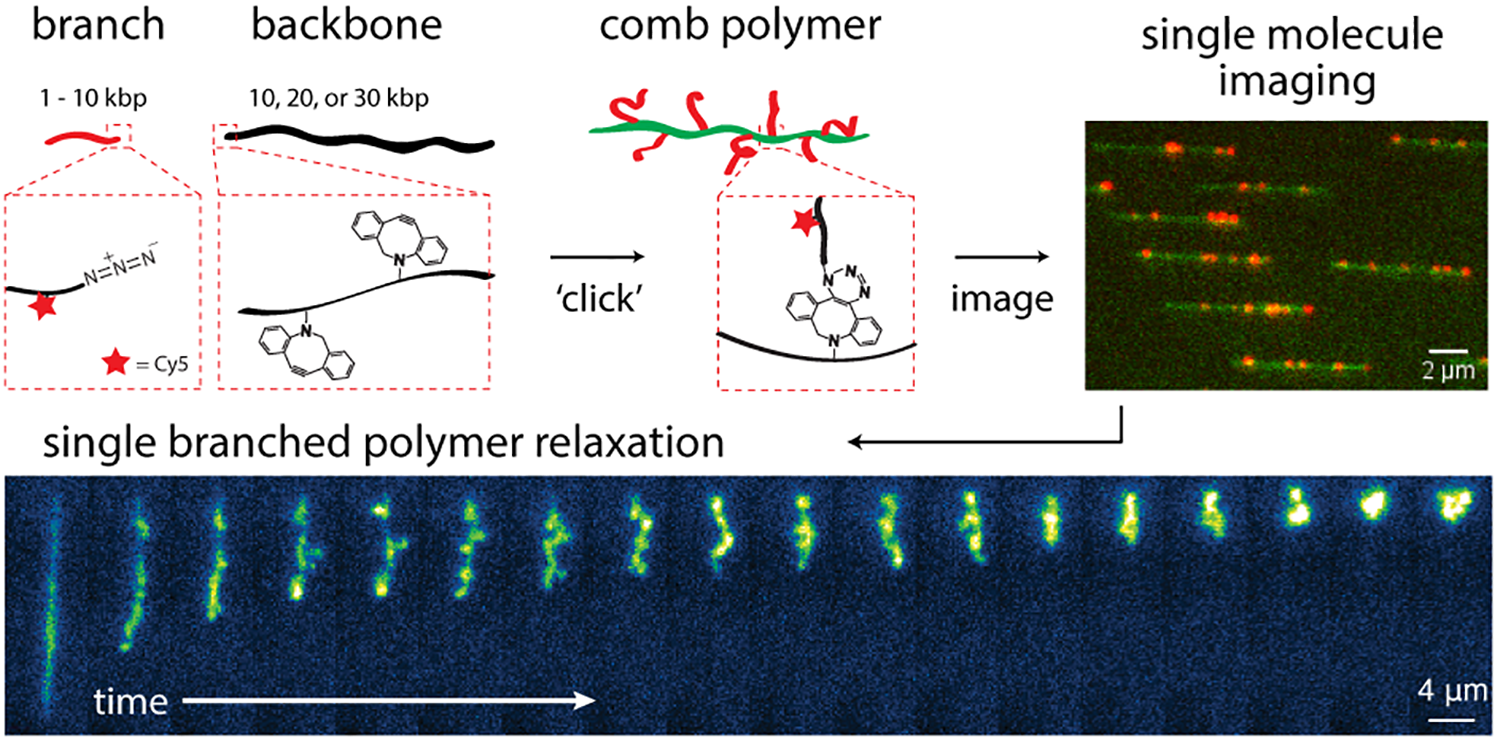}
  \caption{\label{fig:fig21} Synthesis and single molecule imaging of comb-shaped DNA polymers. A hybrid enzymatic-synthetic approach was used to synthesize long branched DNA for single molecule studies of comb polymers. Two-color imaging reveals simultaneous dynamics of branches and backbones. This approach was used to study the relaxation dynamics of single DNA-based combs as a function of comb architecture. Reproduced with permission from Ref. \cite{Mai2015}.}  
\end{figure}

Schroeder and coworkers recently developed new strategies to synthesize DNA-based polymers with comb-shaped architectures that are suitable single molecule imaging (Figure \ref{fig:fig21}) \cite{Marciel2015,Mai2015}. Here, a hybrid enzymatic-synthetic approach was used to synthesize long branched DNA for single molecule studies of comb polymers \cite{Mai2015}. Using PCR for the synthesis of the branches and backbones in separate reactions, precise control was maintained over branch length (1-10 kbp) and backbone length (10-30 kbp). However, a graft-onto reaction scheme was used to covalently link branches to DNA backbones, thereby resulting in average control over the degree of branching by tuning reaction stoichiometry. Overall, this method was made possible by the inclusion of chemically modified PCR primers (containing terminal azide moieties) and non-natural nucleotides (containing dibenzocyclooctyne or DBCO for copper-free click chemistry) during PCR. Moreover, side branches of DNA were synthesized to contain internal fluorescent dyes, thereby enabling simultaneous two-color imaging of branches (red) and backbones (green) during flow dynamics. 

Using this approach, the conformational relaxation of surface-tethered comb DNA was studied using single molecule imaging \cite{Mai2015}. In this work, DNA combs consisted of 30 kbp backbones and 10 kbp branches with an average branching density of 2-10 branches per backbone. At early times in the relaxation process, the backbones showed a rapid elastic recoil characterized by a decrease in backbone extension with respect to time. At intermediate times, branched polymers exhibited mixed relaxation dynamics of branches and backbones, such that branches explored various conformational breathing modes while the backbone relaxed. At long times, the conformational relaxation of the backbone dominated the process, and these timescales were quantified by tracking single polymer extension during the relaxation process. The longest relaxation time was found to increase with an increasing number of branches. Interestingly, the role of branch position was also studied, and a strong dependence on the location of the branch relative to the surface tether was found. Branches far from the tether slowed relaxation, whereas branches near the tether resulted in faster overall relaxation processes compared to a linear polymer. It was postulated that branches near the tether point may accelerate the relaxation process by inducing cooperative hydrodynamic flows. Taken together, these results clearly show that relaxation processes depend on molecular topology.

\subsubsection{Bottlebrush polymers}
The general methods developed by Schroeder and coworkers for synthesizing comb-shaped polymers based on double stranded DNA \cite{Marciel2015} were recently extended to synthesize bottlebrush polymers based on ssDNA \cite{Berezney2017}. The bottlebrush polymers consist of a ssDNA main chain backbone with poly(ethylene glycol) (PEG) side chains. First, ssDNA was synthesized using rolling circle replication (RCR) following a similar approach used for single molecule studies of linear unbranched ssDNA \cite{Brockman2011}. The RCR reaction is performed with a fraction of DBCO-modifed dUTPs, which replace thymine in the main chain, thereby serving as grafting points for side branches using a copper-free click chemistry reaction described above. PEG side chains (10 kDa) are relatively monodisperse (PDI=1.04-1.06). Grafting density was controlled in an average sense by tuning the ratio of DBCO-dUTP to natural dTTP nucleotides in the reaction, with a 1:4 ratio generating a sparsely-grafted comb polymer (one side chain per 35 bases) and a 4:1 ratio generating a bottlebrush polymer (one side chain per 8.75 bases). Finally, the ssDNA is terminally labeled with thiol and biotin moieties, enabling polymer immobilization on to a glass surface and a streptavidin-coated magnetic bead. Using this approach, a magnetic tweezing experimental setup was used to directly measure the elasticity (force extension relation) of single bottlebrush polymers. It was found that chain stiffening due to side branches was only significant on long length scales, with the main chain retaining flexibility on short length scales. From these experiments, an estimate of the internal tension generated by side-chain repulsion was determined. Taken together, these experiments represent the first measurements of the elasticity of bottlebrush polymers.

\subsubsection{Ring polymers}
Ring polymers represent a fascinating topic in polymer physics. Due to the constraint of ring closure, it is thought that ring polymers exhibit qualitatively different dynamics in dilute solutions, concentrated solutions, and melts compared to linear chains \cite{Kapnistos2008,Pasquino2013,Yan2016}. In recent years, there has been a renewed interest in the community in experimentally studying ring polymers using bulk rheology and molecular modeling techniques. However, it can be extremely difficult to prepare pure solutions and melts of ring polymers due to challenges in synthesis and purification to ensure high degrees of purity \cite{Pasquino2013}. Single molecule methods based on DNA offer an alternative approach to prepare highly pure solutions of ring polymers and to future probe their dynamics using single molecule techniques. For example, plasmids and many types of genomic DNA are naturally propagated in circular form in bacteria. Moreover, biochemical treatment based on endonucleases can specifically digest linear chains while leaving circular DNA chains intact. Using this general approach, DNA presents an advantageous polymeric system to study ring polymer dynamics using bulk rheology, microrheology, and single molecule imaging methods. 

In recent years, single molecule imaging has been used to study the center-of-mass diffusion of ring DNA and linear DNA in dilute solutions \cite{Robertson2006} and in concentrated solutions \cite{Robertson2007,Robertson2007c}. Importantly, these experiments began to probe the effect of polymer chain topology on the long-time diffusion dynamics of single chains. In dilute solution, it was observed that ring polymers generally follow a power-law scaling of the diffusion constant with molecular weight similar to linear DNA, where $D \sim L^{-0.589}$ \cite{Robertson2006}. In entangled solutions, a series of single molecule experiments were performed by varying the tracer chain topology (linear or circular) in a background solution of entangled, unlabeled polymer (linear or circular) \cite{Robertson2007,Robertson2007c}. The general trends were that the diffusion of circular chains in a circular background (C-C) was observed to be the fastest, with the complete trends as $D_{C-C} \ge D_{L-C} \gg D_{L-L} \gg D_{C-L}$. The slowest combination consisted of trace circular chains in a background of entangled linear polymers, with the slow dynamics attributed to circular chains becoming `hooked' on the surrounding linear chains, thereby inducing a local constraint that requires time for the linear chain unthreading event and associated constraint release of the linear chains. On the other hand, trace linear chains in a background of concentrated circular chains exhibit a relatively fast diffusive motion due to threading effects of linear chains through the matrix of ring polymers. These experiments were followed by a systematic study of polymer chain center-of-mass diffusion in blends of linear and circular DNA molecules \cite{Chapman2012}. Moreover, Habuchi and coworkers performed additional single molecule imaging experiments on ring polymer diffusion in entangled solutions \cite{Serag2014,Abadi2015}, further probing molecular relaxation processes in ring/linear polymer entangled solutions. For a more lengthy discussion of single molecule diffusion experiments, I refer the reader to a recent review article on the topic \cite{Mai2016}.

Moving beyond near-equilibrium polymer diffusion, recent work has focused on the dynamics of circular DNA in dilute solution extensional flow \cite{Li2015b,Hsiao2016}. First, the longest relaxation time $\tau$ of single polymers was measured as a function of molecular weight for 25, 45, and 114.8 kbp circular DNA \cite{Li2015b}. It was found that ring polymers relax faster than linear chains of the same molecular weight, which can be understood due to the differences in the mode structure. Ring boundary conditions do not permit the lowest mode that exists in the linear chain Rouse motion, and instead the lowest mode has half the wavelength $\lambda$ in the case of ring polymers such that $\lambda_{1,linear} = $ 2 $\lambda_{1,ring}$ \cite{Hsiao2016}. Therefore, the mode relaxes more quickly, in principle by a factor of 4 for the free-draining polymer case because $\tau_1 \sim 1/\lambda_1^2$ \cite{Doiedwards}. In particular, it was found that $\tau_{linear}/\tau_{ring} \approx $ 2 from single molecule experiments \cite{Li2015b} and $\tau_{linear}/\tau_{ring} \approx $ 4.0 from free-draining BD simulations \cite{Hsiao2016}. As HI and EV are included in the BD simulations, it was observed that $\tau_{linear}/\tau_{ring} \approx $ 1.1 \cite{Hsiao2016}, which is consistent with prior work using lattice Boltzmann simulations \cite{Hegde2011}. 

The power-law scaling of the longest relaxation time $\tau$ as a function of molecular weight was also considered for both ring and linear polymers. Single molecule experiments revealed that the longest relaxation time of ring polymers scaled as $\tau_{ring} \sim L^{1.58 \pm 0.06}$ over the range of 25, 45, and 114.8 kbp \cite{Li2015b,Hsiao2016}. Moreover, complementary Brownian dynamics simulations showed that $\tau_{linear} \sim L^{2.02 \pm 0.15}$ and $\tau_{ring} \sim L^{1.97 \pm 0.02}$ for the free-draining case, whereas $\tau_{linear} \sim L^{1.53 \pm 0.05}$ and $\tau_{ring} \sim L^{1.56 \pm 0.04}$ for the HI case. In other words, BD simulations suggest that linear and ring polymers exhibit similar power-law scalings with molecular weight for the free-draining and HI-only case. However, BD simulations with HI and EV showed that $\tau_{linear} \sim L^{1.93 \pm 0.09}$ and $\tau_{ring} \sim L^{1.65 \pm 0.04}$, which suggests that the inclusion of excluded volume interactions fundamentally changes the nature of chain relaxation for ring polymers. However, it should be noted that excluded volume interactions were included using a Lennard-Jones potential, and deviations from the expected relation of $\tau \sim N^{1.8}$ for the case of linear polymers could arise due to the remaining attractive portion of the L-J pair potential or due to errors in the high-$N$ data points that require long time averages \cite{Hsiao2016}. In any event, results from single molecule experiments on ring DNA relaxation are consistent with results from BD simulations with HI and EV to within the error.

Interestingly, ring polymers show a coil-stretch transition in extensional flow, though the onset of ring polymer stretch required a higher critical flow strength compared to linear polymers ($Wi_{crit,ring} \approx 1.25$ $Wi_{crit,linear}$). BD simulations reproduced the shift in the coil-stretch transition for ring polymers, but only in the presence of HI \cite{Hsiao2016}. A more detailed analysis suggests a strong influence of intramolecular HI for circular polymers, such that parallel strands within the ring polymer exert secondary backflows, thereby inducing an open `loop' ring conformation in extensional flow. Finally, transient dynamics of ring polymers revealed substantially less molecular individualism compared to linear polymers in extensional flow. Only two primary stretching pathways were identified for rings: continuous elongation and hindered stretching \cite{Li2015b}. The reduced degree of molecular individualism of circular DNA relative to linear DNA is consistent with the notion that circular molecules have fewer degrees of freedom due to intramolecular chain connectivity. Finally, BD simulations were successful in modeling the transient dynamic behavior of rings, including the hindered stretching conformation in a small sub-set of chains \cite{Hsiao2016}.

\section{Future directions \& perspectives}
\label{conclusion}
Single polymer dynamics has fundamentally changed our understanding of molecular rheology and the non-equilibrium dynamics of macromolecules in flow. This review article highlights several examples of new insights into the physical behavior of polymer solutions illuminated by single molecule techniques. The importance of dynamic heterogeneity and distributions in molecular conformation under non-equilibrium conditions have emerged from single molecule methods. In dilute solutions, single (identical) polymer molecules undergo a variety of conformational stretching pathways in strong flows, such as dumbbell, folded, and kink chain conformations in dilute solution extensional flows \cite{Smith1998}. Recent work in observing single polymer dynamics in semi-dilute unentangled solutions has revealed the influence of flow-induced entanglements in non-equilibrium polymer stretching dynamics \cite{Hsiao2017}. In particular, classifications of polymer solutions as being in the semi-dilute unentangled regime are based on equilibrium properties such as the polymer radius of gyration $R_g$. However, under non-equilibrium conditions such as strong fluid flows, a nominally unentangled solution at equilibrium may exhibit evidence of flow-induced entanglements or chain-chain interactions. Despite the intriguing and potentially important nature of these interactions, it can be challenging to experimentally determine their existence. Indeed, recent single molecule studies of DNA stretching in semi-dilute solutions in extensional flow suggest that flow-induced entanglements may govern a sub-population of transient chain stretching dynamics \cite{Hsiao2017}. These molecular-scale observations address similar issues that were examined in bulk rheological studies of polymer solutions in capillary breakup extensional rheometry experiments \cite{Clasen2006}. Moreover, ongoing single molecule experiments in the author's lab are examining the role of local solution properties (e.g. locally entangled  and locally unentangled behavior) in apparently well-mixed polymer solutions near the critical entanglement concentration $c_e$. From this perspective, single molecule experiments appear to provide an ideal method for probing dynamic behavior at the transition between physical regimes such as polymer concentration or molecular weight, where the latter property is related to associated effects of intramolecular HI in dilute and semi-dilute solutions and polymer conformational hysteresis \cite{Schroeder2004,Prabhakar2016,Prabhakar2017}. Numerous additional examples of dynamic heterogeneity of polymer chain dynamics in flow can be cited, ranging from polymer stretching in porous media \cite{Kawale2017} and chain collisions with single microfabricated posts \cite{Randall2006}. Taken together, these results showcase the importance of distributions in molecular behavior and molecular sub-populations in determining solution properties. 

In addition to revealing the importance of dynamic heterogeneity in polymer dynamics, single molecule methods are being used to directly observe the dynamics of topologically complex polymers. In recent work, the dynamics of comb polymers were observed at the single molecule level for the first time \cite{Mai2016}, with results showing that polymer chain topology (branch density, branch molecular weight, and position of branch points) directly determines polymer relaxation times following cessation of flow. These experiments are currently being extended to non-dilute solutions, which will be essential in comparing to molecular constitutive equations for comb polymer architectures that have so far been compared only to bulk rheological experiments \cite{Lentzakis2014}. Moreover, recent single molecule experiments on bottlebrush polymers have revealed the importance of an internal scale-dependent tension that impacts chain elasticity \cite{Berezney2017}, which fundamentally changes the force-extension behavior away from linear unbranched polymers. This work follows single molecule studies probing the role of excluded volume interactions on generating a non-linear low-force elasticity for linear polymers \cite{Saleh2009}, which subsequently inspired the development of several new force-extension relations for polymer chains that depend on solvent quality \cite{Radhakrishnan2012,Li2015}. To this end, single molecule experiments have directly informed on the elasticity of single polymers, information that can be used in coarse-grained simulations of polymer stretching in flow. In the realm of ring polymers, single molecule studies have revealed an intriguing and previously unexpected `ring-opening' chain conformation in dilute solution extensional flows that can be attributed to intramolecular HI \cite{Li2015b,Hsiao2016}. 

Despite recent progress, however, single molecule studies have only scratched the surface in addressing the broad range of polymer chemistries, topologies, solution concentrations, and non-equilibrium processing conditions for complex materials. Indeed, much work remains to be performed, and the coming years promise to yield exciting new forays in to the dynamics of increasingly complex polymeric systems using single molecule techniques. Even in the realm of dilute solution dynamics, several questions remain unanswered. For example, the modal structure of single polymers is not yet fully resolved from an experimental perspective. Early single molecule studies on partially stretched DNA showed that the motion of the DNA polymer chain backbone could be decomposed into a set of normal modes \cite{Quake1997}, however, these results suggest that hydrodynamic interactions do not play an appreciable role in extended chain dynamics for DNA molecules of size $\lambda$-DNA (48.5 kbp). Nevertheless, for increasingly flexible polymer chains with dominant intramolecular HI, we expect non-linear coupling interactions to invalidate the linearized approximations for ideal polymer chains \cite{deGennesbook}. Repeating the experiments on flexible polymer chains such as single stranded DNA \cite{Brockman2011} may yield different findings. 

The field of molecular rheology would benefit from efforts to combine measurements of bulk stress and high-resolution molecular scale imaging. For example, simultaneous measurement of stress and viscosity, coupled with direct imaging of single polymer chain dynamics, would yield invaluable information regarding how molecular-scale interactions give rise to macroscopic material properties. Indeed, recent work has begun to combine shear rheometry with direct single molecule imaging, for example by mounting a shear rheometer with a transparent lower surface onto an inverted fluorescence microscope \cite{Boukany2010}, thereby enabling simultaneous measurements of stress with non-equilibrium polymer conformations in flow. Moreover, increasingly creative experimental setups are enabling for direct imaging of single polymer dynamics in more complex flow fields, such as polymer chain dynamics spooling around rotating nanowires, as recently reported by Leslie and coworkers in 2017 \cite{Shendruk2017}. In other cases, single molecule techniques have inspired new methods in microrheology. In 2017, particle tracking in viscoelastic solutions was extended to extensional flow \cite{Hsiao2017b}, which enabled determination of extensional viscosity using microfluidics. This approach essentially amounts to passive non-linear microrheology, enabled by precise methods in particle trapping \cite{Shenoy2016}, which represents a new direction in the field of microrheology. 

Finally, single polymer dynamics has only been applied to a relatively small subspace of the range of possible chemistries and molecular topologies in soft materials. To a large degree, the vast parameter space of polymeric systems remains relatively unexplored by single molecule techniques. Recent work has attempted to move these highly powerful set of techniques beyond linear DNA polymers in dilute solutions, though much work remains. Indeed, the small number of recent studies that have explored non-linear polymer architectures or complex chemistries have revealed a wealth of molecular-scale information, which will only be increased by future investigations into new polymer and material systems. The next several years promise to yield exciting and new molecular-level insight into the non-equilibrium dynamics and rheology of polymer systems. Through these efforts, an improved understanding of bulk rheological phenomena will provide insights towards the molecular-scale design and processing of soft materials. 

\section{Acknowledgments}
I sincerely thank J. Ravi Prakash, Ronald Larson, and Charles Sing for critical reading of the manuscript and useful feedback. This work was funded by NSF CBET 1603925 for CMS.    

\bibliography{citation}

\begin{thebibliography}{246}%
\makeatletter
\providecommand \@ifxundefined [1]{%
 \@ifx{#1\undefined}
}%
\providecommand \@ifnum [1]{%
 \ifnum #1\expandafter \@firstoftwo
 \else \expandafter \@secondoftwo
 \fi
}%
\providecommand \@ifx [1]{%
 \ifx #1\expandafter \@firstoftwo
 \else \expandafter \@secondoftwo
 \fi
}%
\providecommand \natexlab [1]{#1}%
\providecommand \enquote  [1]{``#1''}%
\providecommand \bibnamefont  [1]{#1}%
\providecommand \bibfnamefont [1]{#1}%
\providecommand \citenamefont [1]{#1}%
\providecommand \href@noop [0]{\@secondoftwo}%
\providecommand \href [0]{\begingroup \@sanitize@url \@href}%
\providecommand \@href[1]{\@@startlink{#1}\@@href}%
\providecommand \@@href[1]{\endgroup#1\@@endlink}%
\providecommand \@sanitize@url [0]{\catcode `\\12\catcode `\$12\catcode
  `\&12\catcode `\#12\catcode `\^12\catcode `\_12\catcode `\%12\relax}%
\providecommand \@@startlink[1]{}%
\providecommand \@@endlink[0]{}%
\providecommand \url  [0]{\begingroup\@sanitize@url \@url }%
\providecommand \@url [1]{\endgroup\@href {#1}{\urlprefix }}%
\providecommand \urlprefix  [0]{URL }%
\providecommand \Eprint [0]{\href }%
\providecommand \doibase [0]{http://dx.doi.org/}%
\providecommand \selectlanguage [0]{\@gobble}%
\providecommand \bibinfo  [0]{\@secondoftwo}%
\providecommand \bibfield  [0]{\@secondoftwo}%
\providecommand \translation [1]{[#1]}%
\providecommand \BibitemOpen [0]{}%
\providecommand \bibitemStop [0]{}%
\providecommand \bibitemNoStop [0]{.\EOS\space}%
\providecommand \EOS [0]{\spacefactor3000\relax}%
\providecommand \BibitemShut  [1]{\csname bibitem#1\endcsname}%
\let\auto@bib@innerbib\@empty
%</preamble>
\bibitem [{\citenamefont {Larson}(1999)}]{Larsonbook}%
  \BibitemOpen
  \bibfield  {author} {\bibinfo {author} {\bibfnamefont {R.~G.}\ \bibnamefont
  {Larson}},\ }\href@noop {} {\emph {\bibinfo {title} {The Structure and
  Rheology of Complex Fluids}}}\ (\bibinfo  {publisher} {Oxford University
  Press},\ \bibinfo {year} {1999})\BibitemShut {NoStop}%
\bibitem [{\citenamefont {Larson}(1988)}]{Larsonbook2}%
  \BibitemOpen
  \bibfield  {author} {\bibinfo {author} {\bibfnamefont {R.~G.}\ \bibnamefont
  {Larson}},\ }\href@noop {} {\emph {\bibinfo {title} {Constitutive Equations
  for Polymer Melts}}}\ (\bibinfo  {publisher} {Butterworth-Heinemann},\
  \bibinfo {year} {1988})\BibitemShut {NoStop}%
\bibitem [{\citenamefont {Doi}\ and\ \citenamefont
  {Edwards}(1986)}]{Doiedwards}%
  \BibitemOpen
  \bibfield  {author} {\bibinfo {author} {\bibfnamefont {M.}~\bibnamefont
  {Doi}}\ and\ \bibinfo {author} {\bibfnamefont {S.~F.}\ \bibnamefont
  {Edwards}},\ }\href@noop {} {\emph {\bibinfo {title} {The Theory of Polymer
  Dynamics}}}\ (\bibinfo  {publisher} {Oxford University Press},\ \bibinfo
  {year} {1986})\BibitemShut {NoStop}%
\bibitem [{\citenamefont {Bird}\ \emph
  {et~al.}(1987{\natexlab{a}})\citenamefont {Bird}, \citenamefont {Armstrong},\
  and\ \citenamefont {Hassager}}]{Bird1987}%
  \BibitemOpen
  \bibfield  {author} {\bibinfo {author} {\bibfnamefont {R.B.}\ \bibnamefont
  {Bird}}, \bibinfo {author} {\bibfnamefont {R.C.}\ \bibnamefont {Armstrong}},
  \ and\ \bibinfo {author} {\bibfnamefont {O.}~\bibnamefont {Hassager}},\
  }\href@noop {} {\emph {\bibinfo {title} {Dynamics of Polymeric Liquids}}},\
  Vol.~\bibinfo {volume} {1}\ (\bibinfo  {publisher} {Wiley, New York, NY},\
  \bibinfo {year} {1987})\BibitemShut {NoStop}%
\bibitem [{\citenamefont {Bird}\ \emph
  {et~al.}(1987{\natexlab{b}})\citenamefont {Bird}, \citenamefont {Curtiss},
  \citenamefont {Armstrong},\ and\ \citenamefont {Hassager}}]{Bird1987b}%
  \BibitemOpen
  \bibfield  {author} {\bibinfo {author} {\bibfnamefont {R.B.}\ \bibnamefont
  {Bird}}, \bibinfo {author} {\bibfnamefont {C.~F.}\ \bibnamefont {Curtiss}},
  \bibinfo {author} {\bibfnamefont {R.C.}\ \bibnamefont {Armstrong}}, \ and\
  \bibinfo {author} {\bibfnamefont {O.}~\bibnamefont {Hassager}},\ }\href@noop
  {} {\emph {\bibinfo {title} {Dynamics of Polymeric Liquids}}},\ Vol.~\bibinfo
  {volume} {2}\ (\bibinfo  {publisher} {Wiley, New York, NY},\ \bibinfo {year}
  {1987})\BibitemShut {NoStop}%
\bibitem [{\citenamefont {McLeish}\ and\ \citenamefont
  {Larson}(1998)}]{McLeish1998}%
  \BibitemOpen
  \bibfield  {author} {\bibinfo {author} {\bibfnamefont {T.~C.~B.}\
  \bibnamefont {McLeish}}\ and\ \bibinfo {author} {\bibfnamefont {R.~G.}\
  \bibnamefont {Larson}},\ }\bibfield  {title} {\enquote {\bibinfo {title}
  {Molecular constitutive equations for a class of branched polymers: The
  pom-pom polymer},}\ }\href@noop {} {\bibfield  {journal} {\bibinfo  {journal}
  {Journal of Rheology}\ }\textbf {\bibinfo {volume} {42}},\ \bibinfo {pages}
  {81--110} (\bibinfo {year} {1998})}\BibitemShut {NoStop}%
\bibitem [{\citenamefont {Fuller}(1995)}]{Fullerbook}%
  \BibitemOpen
  \bibfield  {author} {\bibinfo {author} {\bibfnamefont {G.~G.}\ \bibnamefont
  {Fuller}},\ }\href@noop {} {\emph {\bibinfo {title} {Optical Rheometry of
  Complex Fluids}}}\ (\bibinfo  {publisher} {Oxford University Press},\
  \bibinfo {year} {1995})\BibitemShut {NoStop}%
\bibitem [{\citenamefont {Baird}\ and\ \citenamefont
  {Collias}(2014)}]{Bairdbook}%
  \BibitemOpen
  \bibfield  {author} {\bibinfo {author} {\bibfnamefont {D.~G.}\ \bibnamefont
  {Baird}}\ and\ \bibinfo {author} {\bibfnamefont {D.~I.}\ \bibnamefont
  {Collias}},\ }\href@noop {} {\emph {\bibinfo {title} {Polymer Processing:
  Principles and Design}}},\ \bibinfo {edition} {2nd}\ ed.\ (\bibinfo
  {publisher} {Wiley},\ \bibinfo {year} {2014})\BibitemShut {NoStop}%
\bibitem [{\citenamefont {Koenig}(1999)}]{Koenigbook}%
  \BibitemOpen
  \bibfield  {author} {\bibinfo {author} {\bibfnamefont {J.~L.}\ \bibnamefont
  {Koenig}},\ }\href@noop {} {\emph {\bibinfo {title} {Spectroscopy of
  Polymers}}},\ \bibinfo {edition} {2nd}\ ed.\ (\bibinfo  {publisher}
  {Elsevier},\ \bibinfo {year} {1999})\BibitemShut {NoStop}%
\bibitem [{\citenamefont {Richter}\ \emph {et~al.}(2005)\citenamefont
  {Richter}, \citenamefont {Monkenbusch}, \citenamefont {Arbe},\ and\
  \citenamefont {Colmenero}}]{Richter2005}%
  \BibitemOpen
  \bibfield  {author} {\bibinfo {author} {\bibfnamefont {D.}~\bibnamefont
  {Richter}}, \bibinfo {author} {\bibfnamefont {M.}~\bibnamefont
  {Monkenbusch}}, \bibinfo {author} {\bibfnamefont {A.}~\bibnamefont {Arbe}}, \
  and\ \bibinfo {author} {\bibfnamefont {J.}~\bibnamefont {Colmenero}},\
  }\enquote {\bibinfo {title} {Neutron spin echo in polymer systems},}\ \
  (\bibinfo  {publisher} {Springer Berlin Heidelberg},\ \bibinfo {year}
  {2005})\ Chap.\ \bibinfo {chapter} {Neutron Spin Echo in Polymer
  Systems}\BibitemShut {NoStop}%
\bibitem [{\citenamefont {Shaqfeh}(2005)}]{Shaqfeh2005}%
  \BibitemOpen
  \bibfield  {author} {\bibinfo {author} {\bibfnamefont {E.~S.~G.}\
  \bibnamefont {Shaqfeh}},\ }\bibfield  {title} {\enquote {\bibinfo {title}
  {The dynamics of single-molecule dna in flow},}\ }\href@noop {} {\bibfield
  {journal} {\bibinfo  {journal} {Journal of Non-Newtonian Fluid Mechanics}\
  }\textbf {\bibinfo {volume} {130}},\ \bibinfo {pages} {1--28} (\bibinfo
  {year} {2005})}\BibitemShut {NoStop}%
\bibitem [{\citenamefont {Mai}\ and\ \citenamefont
  {Schroeder}(2016)}]{Mai2016}%
  \BibitemOpen
  \bibfield  {author} {\bibinfo {author} {\bibfnamefont {D.~J.}\ \bibnamefont
  {Mai}}\ and\ \bibinfo {author} {\bibfnamefont {C.~M.}\ \bibnamefont
  {Schroeder}},\ }\bibfield  {title} {\enquote {\bibinfo {title} {Single
  polymer dynamics of topologically complex dna},}\ }\href@noop {} {\bibfield
  {journal} {\bibinfo  {journal} {Current Opinion in Colloid and Interface
  Science}\ }\textbf {\bibinfo {volume} {26}},\ \bibinfo {pages} {28--40}
  (\bibinfo {year} {2016})}\BibitemShut {NoStop}%
\bibitem [{\citenamefont {Chu}(1991)}]{Chu1991}%
  \BibitemOpen
  \bibfield  {author} {\bibinfo {author} {\bibfnamefont {S.}~\bibnamefont
  {Chu}},\ }\bibfield  {title} {\enquote {\bibinfo {title} {Laser manipulation
  of atoms and particles},}\ }\href@noop {} {\bibfield  {journal} {\bibinfo
  {journal} {Science}\ }\textbf {\bibinfo {volume} {253}},\ \bibinfo {pages}
  {861--866} (\bibinfo {year} {1991})}\BibitemShut {NoStop}%
\bibitem [{\citenamefont {Chu}\ and\ \citenamefont {Kron}(1990)}]{ChuKron1990}%
  \BibitemOpen
  \bibfield  {author} {\bibinfo {author} {\bibfnamefont {S.}~\bibnamefont
  {Chu}}\ and\ \bibinfo {author} {\bibfnamefont {S.}~\bibnamefont {Kron}},\
  }\bibfield  {title} {\enquote {\bibinfo {title} {Optical manipulation of
  dna},}\ }in\ \href@noop {} {\emph {\bibinfo {booktitle} {International
  Conference on Quantum Electronics, Technical Digest}}},\ Vol.~\bibinfo
  {volume} {8}\ (\bibinfo  {publisher} {Optical Society of America},\ \bibinfo
  {year} {1990})\ p.\ \bibinfo {pages} {202}\BibitemShut {NoStop}%
\bibitem [{\citenamefont {Quake}\ and\ \citenamefont
  {Scherer}(2000)}]{Quake2000}%
  \BibitemOpen
  \bibfield  {author} {\bibinfo {author} {\bibfnamefont {S.~R.}\ \bibnamefont
  {Quake}}\ and\ \bibinfo {author} {\bibfnamefont {A.}~\bibnamefont
  {Scherer}},\ }\bibfield  {title} {\enquote {\bibinfo {title} {From micro- to
  nanofabrication with soft materials},}\ }\href@noop {} {\bibfield  {journal}
  {\bibinfo  {journal} {Science}\ }\textbf {\bibinfo {volume} {24}},\ \bibinfo
  {pages} {1536--1540} (\bibinfo {year} {2000})}\BibitemShut {NoStop}%
\bibitem [{\citenamefont {Ashkin}\ \emph {et~al.}(1986)\citenamefont {Ashkin},
  \citenamefont {Dziedzic}, \citenamefont {Bjorkholm},\ and\ \citenamefont
  {Chu}}]{Ashkin1986}%
  \BibitemOpen
  \bibfield  {author} {\bibinfo {author} {\bibfnamefont {A.}~\bibnamefont
  {Ashkin}}, \bibinfo {author} {\bibfnamefont {J.}~\bibnamefont {Dziedzic}},
  \bibinfo {author} {\bibfnamefont {J.}~\bibnamefont {Bjorkholm}}, \ and\
  \bibinfo {author} {\bibfnamefont {S.}~\bibnamefont {Chu}},\ }\bibfield
  {title} {\enquote {\bibinfo {title} {Observation of a single-beam gradient
  force optical trap for dielectric particles},}\ }\href@noop {} {\bibfield
  {journal} {\bibinfo  {journal} {Optics Letters}\ }\textbf {\bibinfo {volume}
  {11}},\ \bibinfo {pages} {288--290} (\bibinfo {year} {1986})}\BibitemShut
  {NoStop}%
\bibitem [{\citenamefont {Perkins}\ \emph
  {et~al.}(1994{\natexlab{a}})\citenamefont {Perkins}, \citenamefont {Quake},
  \citenamefont {Smith},\ and\ \citenamefont {Chu}}]{Perkins1994a}%
  \BibitemOpen
  \bibfield  {author} {\bibinfo {author} {\bibfnamefont {T.~T.}\ \bibnamefont
  {Perkins}}, \bibinfo {author} {\bibfnamefont {S.~R.}\ \bibnamefont {Quake}},
  \bibinfo {author} {\bibfnamefont {D.~E.}\ \bibnamefont {Smith}}, \ and\
  \bibinfo {author} {\bibfnamefont {S.}~\bibnamefont {Chu}},\ }\bibfield
  {title} {\enquote {\bibinfo {title} {Relaxation of a single dna molecule
  observed by optical microscopy},}\ }\href@noop {} {\bibfield  {journal}
  {\bibinfo  {journal} {Science}\ }\textbf {\bibinfo {volume} {264}},\ \bibinfo
  {pages} {822--826} (\bibinfo {year} {1994}{\natexlab{a}})}\BibitemShut
  {NoStop}%
\bibitem [{\citenamefont {Perkins}\ \emph
  {et~al.}(1994{\natexlab{b}})\citenamefont {Perkins}, \citenamefont {Smith},\
  and\ \citenamefont {Chu}}]{Perkins1994b}%
  \BibitemOpen
  \bibfield  {author} {\bibinfo {author} {\bibfnamefont {T.~T.}\ \bibnamefont
  {Perkins}}, \bibinfo {author} {\bibfnamefont {D.~E.}\ \bibnamefont {Smith}},
  \ and\ \bibinfo {author} {\bibfnamefont {S.}~\bibnamefont {Chu}},\ }\bibfield
   {title} {\enquote {\bibinfo {title} {Direct observation of tube-like motion
  of a single polymer chain},}\ }\href@noop {} {\bibfield  {journal} {\bibinfo
  {journal} {Science}\ }\textbf {\bibinfo {volume} {264}},\ \bibinfo {pages}
  {819--822} (\bibinfo {year} {1994}{\natexlab{b}})}\BibitemShut {NoStop}%
\bibitem [{\citenamefont {Marko}\ and\ \citenamefont
  {Siggia}(1995)}]{Marko1995}%
  \BibitemOpen
  \bibfield  {author} {\bibinfo {author} {\bibfnamefont {J.~F.}\ \bibnamefont
  {Marko}}\ and\ \bibinfo {author} {\bibfnamefont {E.~D.}\ \bibnamefont
  {Siggia}},\ }\bibfield  {title} {\enquote {\bibinfo {title} {Stretching
  dna},}\ }\href@noop {} {\bibfield  {journal} {\bibinfo  {journal}
  {Macromolecules}\ }\textbf {\bibinfo {volume} {28}},\ \bibinfo {pages}
  {8759--8770} (\bibinfo {year} {1995})}\BibitemShut {NoStop}%
\bibitem [{\citenamefont {Ottinger}(1996)}]{Ottingerbook}%
  \BibitemOpen
  \bibfield  {author} {\bibinfo {author} {\bibfnamefont {H.~C.}\ \bibnamefont
  {Ottinger}},\ }\href@noop {} {\emph {\bibinfo {title} {Stochastic Processes
  in Polymeric Fluids}}}\ (\bibinfo  {publisher} {Springer},\ \bibinfo {year}
  {1996})\BibitemShut {NoStop}%
\bibitem [{\citenamefont {Doyle}\ \emph {et~al.}(1997)\citenamefont {Doyle},
  \citenamefont {Shaqfeh},\ and\ \citenamefont {Gast}}]{Doyle1997}%
  \BibitemOpen
  \bibfield  {author} {\bibinfo {author} {\bibfnamefont {P.~S.}\ \bibnamefont
  {Doyle}}, \bibinfo {author} {\bibfnamefont {E.~S.~G.}\ \bibnamefont
  {Shaqfeh}}, \ and\ \bibinfo {author} {\bibfnamefont {A.~P.}\ \bibnamefont
  {Gast}},\ }\bibfield  {title} {\enquote {\bibinfo {title} {Dynamic simulation
  of freely-draining flexible polymers in steady linear flows},}\ }\href@noop
  {} {\bibfield  {journal} {\bibinfo  {journal} {Journal of Fluid Mechanics}\
  }\textbf {\bibinfo {volume} {334}},\ \bibinfo {pages} {251--291} (\bibinfo
  {year} {1997})}\BibitemShut {NoStop}%
\bibitem [{\citenamefont {de~Gennes}(1997)}]{deGennes1997}%
  \BibitemOpen
  \bibfield  {author} {\bibinfo {author} {\bibfnamefont {P.~G.}\ \bibnamefont
  {de~Gennes}},\ }\bibfield  {title} {\enquote {\bibinfo {title} {Molecular
  individualism},}\ }\href@noop {} {\bibfield  {journal} {\bibinfo  {journal}
  {Science}\ }\textbf {\bibinfo {volume} {276}},\ \bibinfo {pages} {5321}
  (\bibinfo {year} {1997})}\BibitemShut {NoStop}%
\bibitem [{\citenamefont {Mai}\ \emph {et~al.}(2012)\citenamefont {Mai},
  \citenamefont {Brockman},\ and\ \citenamefont {Schroeder}}]{Mai2012}%
  \BibitemOpen
  \bibfield  {author} {\bibinfo {author} {\bibfnamefont {D.~J.}\ \bibnamefont
  {Mai}}, \bibinfo {author} {\bibfnamefont {C.~A.}\ \bibnamefont {Brockman}}, \
  and\ \bibinfo {author} {\bibfnamefont {C.~M.}\ \bibnamefont {Schroeder}},\
  }\bibfield  {title} {\enquote {\bibinfo {title} {Microfluidic systems for
  single dna dynamics},}\ }\href@noop {} {\bibfield  {journal} {\bibinfo
  {journal} {Soft Matter}\ }\textbf {\bibinfo {volume} {8}},\ \bibinfo {pages}
  {10560--10572} (\bibinfo {year} {2012})}\BibitemShut {NoStop}%
\bibitem [{\citenamefont {Rems}\ \emph {et~al.}(2016)\citenamefont {Rems},
  \citenamefont {Kawale}, \citenamefont {Lee},\ and\ \citenamefont
  {Boukany}}]{Rems2016}%
  \BibitemOpen
  \bibfield  {author} {\bibinfo {author} {\bibfnamefont {L.}~\bibnamefont
  {Rems}}, \bibinfo {author} {\bibfnamefont {D.}~\bibnamefont {Kawale}},
  \bibinfo {author} {\bibfnamefont {L.~J.}\ \bibnamefont {Lee}}, \ and\
  \bibinfo {author} {\bibfnamefont {P.~E.}\ \bibnamefont {Boukany}},\
  }\bibfield  {title} {\enquote {\bibinfo {title} {Flow of dna in
  micro/nanofluidics: From fundamentals to applications},}\ }\href@noop {}
  {\bibfield  {journal} {\bibinfo  {journal} {Biomicrofluidics}\ }\textbf
  {\bibinfo {volume} {10}},\ \bibinfo {pages} {043403} (\bibinfo {year}
  {2016})}\BibitemShut {NoStop}%
\bibitem [{\citenamefont {Dai}\ \emph {et~al.}(2016)\citenamefont {Dai},
  \citenamefont {Renner},\ and\ \citenamefont {Doyle}}]{Dai2016}%
  \BibitemOpen
  \bibfield  {author} {\bibinfo {author} {\bibfnamefont {L.}~\bibnamefont
  {Dai}}, \bibinfo {author} {\bibfnamefont {C.~B.}\ \bibnamefont {Renner}}, \
  and\ \bibinfo {author} {\bibfnamefont {P.~S.}\ \bibnamefont {Doyle}},\
  }\bibfield  {title} {\enquote {\bibinfo {title} {The polymer physics of
  single dna confined in nanochannels},}\ }\href@noop {} {\bibfield  {journal}
  {\bibinfo  {journal} {Advances in Colloid and Interface Science}\ }\textbf
  {\bibinfo {volume} {232}},\ \bibinfo {pages} {80--100} (\bibinfo {year}
  {2016})}\BibitemShut {NoStop}%
\bibitem [{\citenamefont {Dorfman}(2010)}]{Dorfman2010}%
  \BibitemOpen
  \bibfield  {author} {\bibinfo {author} {\bibfnamefont {K.~D.}\ \bibnamefont
  {Dorfman}},\ }\bibfield  {title} {\enquote {\bibinfo {title} {Dna
  electrophoresis in microfabricated devices},}\ }\href@noop {} {\bibfield
  {journal} {\bibinfo  {journal} {Rev. Mod. Phys.}\ }\textbf {\bibinfo {volume}
  {82}},\ \bibinfo {pages} {2903--2947} (\bibinfo {year} {2010})}\BibitemShut
  {NoStop}%
\bibitem [{\citenamefont {Mai}\ \emph {et~al.}(2015)\citenamefont {Mai},
  \citenamefont {Marciel}, \citenamefont {Sing},\ and\ \citenamefont
  {Schroeder}}]{Mai2015}%
  \BibitemOpen
  \bibfield  {author} {\bibinfo {author} {\bibfnamefont {D.~J.}\ \bibnamefont
  {Mai}}, \bibinfo {author} {\bibfnamefont {A.~B.}\ \bibnamefont {Marciel}},
  \bibinfo {author} {\bibfnamefont {C.~E.}\ \bibnamefont {Sing}}, \ and\
  \bibinfo {author} {\bibfnamefont {C.~M.}\ \bibnamefont {Schroeder}},\
  }\bibfield  {title} {\enquote {\bibinfo {title} {Topology controlled
  relaxation dynamics of single branched polymers},}\ }\href@noop {} {\bibfield
   {journal} {\bibinfo  {journal} {ACS Macro Letters}\ }\textbf {\bibinfo
  {volume} {4}},\ \bibinfo {pages} {446--452} (\bibinfo {year}
  {2015})}\BibitemShut {NoStop}%
\bibitem [{\citenamefont {Laib}\ \emph {et~al.}(2006)\citenamefont {Laib},
  \citenamefont {Robertson},\ and\ \citenamefont {Smith}}]{Laib2006}%
  \BibitemOpen
  \bibfield  {author} {\bibinfo {author} {\bibfnamefont {S.}~\bibnamefont
  {Laib}}, \bibinfo {author} {\bibfnamefont {R.~M.}\ \bibnamefont {Robertson}},
  \ and\ \bibinfo {author} {\bibfnamefont {D.~E.}\ \bibnamefont {Smith}},\
  }\bibfield  {title} {\enquote {\bibinfo {title} {Preparation and
  characterization of a set of linear dna molecules for polymer physics and
  rheology studies},}\ }\href@noop {} {\bibfield  {journal} {\bibinfo
  {journal} {Macromolecules}\ }\textbf {\bibinfo {volume} {39}},\ \bibinfo
  {pages} {4115--4119} (\bibinfo {year} {2006})}\BibitemShut {NoStop}%
\bibitem [{\citenamefont {Robertson}\ and\ \citenamefont
  {Smith}(2007{\natexlab{a}})}]{Robertson2007}%
  \BibitemOpen
  \bibfield  {author} {\bibinfo {author} {\bibfnamefont {R.~M.}\ \bibnamefont
  {Robertson}}\ and\ \bibinfo {author} {\bibfnamefont {D.~E.}\ \bibnamefont
  {Smith}},\ }\bibfield  {title} {\enquote {\bibinfo {title} {Self-diffusion of
  entangled linear and circular dna molecules: dependence on length and
  concentration},}\ }\href@noop {} {\bibfield  {journal} {\bibinfo  {journal}
  {Macromolecules}\ }\textbf {\bibinfo {volume} {40}},\ \bibinfo {pages}
  {3373--3377} (\bibinfo {year} {2007}{\natexlab{a}})}\BibitemShut {NoStop}%
\bibitem [{\citenamefont {Kratky}\ and\ \citenamefont
  {Porod}(1949)}]{Kratky1949}%
  \BibitemOpen
  \bibfield  {author} {\bibinfo {author} {\bibfnamefont {O.}~\bibnamefont
  {Kratky}}\ and\ \bibinfo {author} {\bibfnamefont {G.}~\bibnamefont {Porod}},\
  }\bibfield  {title} {\enquote {\bibinfo {title} {R{\"o}ntgenuntersuchung
  gel{\"o}ster fadenmolek{\"u}le},}\ }\href@noop {} {\bibfield  {journal}
  {\bibinfo  {journal} {Rev. Trav. Chim.}\ }\textbf {\bibinfo {volume} {68}},\
  \bibinfo {pages} {1106--1122} (\bibinfo {year} {1949})}\BibitemShut {NoStop}%
\bibitem [{\citenamefont {Hagerman}(1998)}]{Hagerman1998}%
  \BibitemOpen
  \bibfield  {author} {\bibinfo {author} {\bibfnamefont {P.~J.}\ \bibnamefont
  {Hagerman}},\ }\bibfield  {title} {\enquote {\bibinfo {title} {Flexibility of
  dna},}\ }\href@noop {} {\bibfield  {journal} {\bibinfo  {journal} {Ann. Rev.
  Biophys. Biophys. Chem.}\ }\textbf {\bibinfo {volume} {17}},\ \bibinfo
  {pages} {265--286} (\bibinfo {year} {1998})}\BibitemShut {NoStop}%
\bibitem [{\citenamefont {Dobrynin}(2006)}]{Dobrynin2006}%
  \BibitemOpen
  \bibfield  {author} {\bibinfo {author} {\bibfnamefont {A.~V.}\ \bibnamefont
  {Dobrynin}},\ }\bibfield  {title} {\enquote {\bibinfo {title} {Effect of
  counterion condensation on rigidity of semiflexible polyelectrolytes},}\
  }\href@noop {} {\bibfield  {journal} {\bibinfo  {journal} {Macromolecules}\
  }\textbf {\bibinfo {volume} {39}},\ \bibinfo {pages} {9519--9527} (\bibinfo
  {year} {2006})}\BibitemShut {NoStop}%
\bibitem [{\citenamefont {Bustamante}\ \emph {et~al.}(1994)\citenamefont
  {Bustamante}, \citenamefont {Marko}, \citenamefont {Siggia},\ and\
  \citenamefont {Smith}}]{Bustamante1994}%
  \BibitemOpen
  \bibfield  {author} {\bibinfo {author} {\bibfnamefont {C.}~\bibnamefont
  {Bustamante}}, \bibinfo {author} {\bibfnamefont {J.~F.}\ \bibnamefont
  {Marko}}, \bibinfo {author} {\bibfnamefont {E.~D.}\ \bibnamefont {Siggia}}, \
  and\ \bibinfo {author} {\bibfnamefont {S.}~\bibnamefont {Smith}},\ }\bibfield
   {title} {\enquote {\bibinfo {title} {Entropic elasticity of lambda-phage
  dna},}\ }\href@noop {} {\bibfield  {journal} {\bibinfo  {journal} {Science}\
  }\textbf {\bibinfo {volume} {265}},\ \bibinfo {pages} {1599--1600} (\bibinfo
  {year} {1994})}\BibitemShut {NoStop}%
\bibitem [{\citenamefont {Kundukad}\ \emph {et~al.}(2014)\citenamefont
  {Kundukad}, \citenamefont {Yan},\ and\ \citenamefont {Doyle}}]{Kundukad2014}%
  \BibitemOpen
  \bibfield  {author} {\bibinfo {author} {\bibfnamefont {B.}~\bibnamefont
  {Kundukad}}, \bibinfo {author} {\bibfnamefont {J.}~\bibnamefont {Yan}}, \
  and\ \bibinfo {author} {\bibfnamefont {P.~S.}\ \bibnamefont {Doyle}},\
  }\bibfield  {title} {\enquote {\bibinfo {title} {Effect of yoyo-1 on the
  mechanical properties of dna},}\ }\href@noop {} {\bibfield  {journal}
  {\bibinfo  {journal} {Soft Matter}\ }\textbf {\bibinfo {volume} {10}},\
  \bibinfo {pages} {9721--9728} (\bibinfo {year} {2014})}\BibitemShut {NoStop}%
\bibitem [{\citenamefont {Tree}\ \emph {et~al.}(2013)\citenamefont {Tree},
  \citenamefont {Muralidhar}, \citenamefont {Doyle},\ and\ \citenamefont
  {Dorfman}}]{Tree2013}%
  \BibitemOpen
  \bibfield  {author} {\bibinfo {author} {\bibfnamefont {D.~R.}\ \bibnamefont
  {Tree}}, \bibinfo {author} {\bibfnamefont {A.}~\bibnamefont {Muralidhar}},
  \bibinfo {author} {\bibfnamefont {P.~S.}\ \bibnamefont {Doyle}}, \ and\
  \bibinfo {author} {\bibfnamefont {K.~D.}\ \bibnamefont {Dorfman}},\
  }\bibfield  {title} {\enquote {\bibinfo {title} {Is dna a good model
  polymer?}}\ }\href@noop {} {\bibfield  {journal} {\bibinfo  {journal}
  {Macromolecules}\ }\textbf {\bibinfo {volume} {46}},\ \bibinfo {pages}
  {8369--8382} (\bibinfo {year} {2013})}\BibitemShut {NoStop}%
\bibitem [{\citenamefont {Stigter}(1977)}]{Stigter1977}%
  \BibitemOpen
  \bibfield  {author} {\bibinfo {author} {\bibfnamefont {D.}~\bibnamefont
  {Stigter}},\ }\bibfield  {title} {\enquote {\bibinfo {title} {Interactions of
  highly charged colloidal cylinders with applications to double-stranded
  dna},}\ }\href@noop {} {\bibfield  {journal} {\bibinfo  {journal}
  {Biopolymers}\ }\textbf {\bibinfo {volume} {16}},\ \bibinfo {pages}
  {1435--1448} (\bibinfo {year} {1977})}\BibitemShut {NoStop}%
\bibitem [{\citenamefont {Chen}\ \emph {et~al.}(2012)\citenamefont {Chen},
  \citenamefont {Meisburger}, \citenamefont {Pabit}, \citenamefont {Sutton},
  \citenamefont {Webb},\ and\ \citenamefont {Pollack}}]{Chen2012}%
  \BibitemOpen
  \bibfield  {author} {\bibinfo {author} {\bibfnamefont {H.}~\bibnamefont
  {Chen}}, \bibinfo {author} {\bibfnamefont {S.~P.}\ \bibnamefont
  {Meisburger}}, \bibinfo {author} {\bibfnamefont {S.~A.}\ \bibnamefont
  {Pabit}}, \bibinfo {author} {\bibfnamefont {J.~L.}\ \bibnamefont {Sutton}},
  \bibinfo {author} {\bibfnamefont {W.~W.}\ \bibnamefont {Webb}}, \ and\
  \bibinfo {author} {\bibfnamefont {L.}~\bibnamefont {Pollack}},\ }\bibfield
  {title} {\enquote {\bibinfo {title} {Ionic strength-dependent persistence
  lengths of single-stranded rna and dna},}\ }\href@noop {} {\bibfield
  {journal} {\bibinfo  {journal} {Proc. Nat. Acad. Sci.}\ }\textbf {\bibinfo
  {volume} {109}},\ \bibinfo {pages} {799--804} (\bibinfo {year}
  {2012})}\BibitemShut {NoStop}%
\bibitem [{\citenamefont {Saleh}\ \emph {et~al.}(2009)\citenamefont {Saleh},
  \citenamefont {McIntosh}, \citenamefont {Pincus},\ and\ \citenamefont
  {Ribeck}}]{Saleh2009}%
  \BibitemOpen
  \bibfield  {author} {\bibinfo {author} {\bibfnamefont {O.~A.}\ \bibnamefont
  {Saleh}}, \bibinfo {author} {\bibfnamefont {D.~B.}\ \bibnamefont {McIntosh}},
  \bibinfo {author} {\bibfnamefont {P.}~\bibnamefont {Pincus}}, \ and\ \bibinfo
  {author} {\bibfnamefont {N.}~\bibnamefont {Ribeck}},\ }\bibfield  {title}
  {\enquote {\bibinfo {title} {Non-linear low force elasticity of
  single-stranded dna molecules},}\ }\href@noop {} {\bibfield  {journal}
  {\bibinfo  {journal} {Phys. Rev. Lett.}\ }\textbf {\bibinfo {volume} {102}},\
  \bibinfo {pages} {068301} (\bibinfo {year} {2009})}\BibitemShut {NoStop}%
\bibitem [{\citenamefont {McIntosh}\ \emph {et~al.}(2009)\citenamefont
  {McIntosh}, \citenamefont {Ribeck},\ and\ \citenamefont
  {Saleh}}]{McIntosh2009}%
  \BibitemOpen
  \bibfield  {author} {\bibinfo {author} {\bibfnamefont {D.~B.}\ \bibnamefont
  {McIntosh}}, \bibinfo {author} {\bibfnamefont {N.}~\bibnamefont {Ribeck}}, \
  and\ \bibinfo {author} {\bibfnamefont {O.~A.}\ \bibnamefont {Saleh}},\
  }\bibfield  {title} {\enquote {\bibinfo {title} {Detailed scaling analysis of
  low-force polyelectrolyte elasticity},}\ }\href@noop {} {\bibfield  {journal}
  {\bibinfo  {journal} {Phys. Rev. E}\ }\textbf {\bibinfo {volume} {80}},\
  \bibinfo {pages} {041803} (\bibinfo {year} {2009})}\BibitemShut {NoStop}%
\bibitem [{\citenamefont {de~Gennes}(1979)}]{deGennesbook}%
  \BibitemOpen
  \bibfield  {author} {\bibinfo {author} {\bibfnamefont {P.~G.}\ \bibnamefont
  {de~Gennes}},\ }\href@noop {} {\emph {\bibinfo {title} {Scaling Concepts in
  Polymer Physics}}}\ (\bibinfo  {publisher} {Cornell University Press},\
  \bibinfo {year} {1979})\BibitemShut {NoStop}%
\bibitem [{\citenamefont {Rubinstein}\ and\ \citenamefont
  {Colby}(2003)}]{Rubinsteinbook}%
  \BibitemOpen
  \bibfield  {author} {\bibinfo {author} {\bibfnamefont {M.}~\bibnamefont
  {Rubinstein}}\ and\ \bibinfo {author} {\bibfnamefont {R.~H.}\ \bibnamefont
  {Colby}},\ }\href@noop {} {\emph {\bibinfo {title} {Polymer Physics}}}\
  (\bibinfo  {publisher} {Oxford University Press},\ \bibinfo {year}
  {2003})\BibitemShut {NoStop}%
\bibitem [{\citenamefont {Jain}\ \emph
  {et~al.}(2012{\natexlab{a}})\citenamefont {Jain}, \citenamefont {Dunweg},\
  and\ \citenamefont {Prakash}}]{Jain2012}%
  \BibitemOpen
  \bibfield  {author} {\bibinfo {author} {\bibfnamefont {A.}~\bibnamefont
  {Jain}}, \bibinfo {author} {\bibfnamefont {B.}~\bibnamefont {Dunweg}}, \ and\
  \bibinfo {author} {\bibfnamefont {J.~R.}\ \bibnamefont {Prakash}},\
  }\bibfield  {title} {\enquote {\bibinfo {title} {Dynamic crossover scaling in
  polymer solutions},}\ }\href@noop {} {\bibfield  {journal} {\bibinfo
  {journal} {Phys. Rev. Lett.}\ }\textbf {\bibinfo {volume} {109}},\ \bibinfo
  {pages} {088302} (\bibinfo {year} {2012}{\natexlab{a}})}\BibitemShut
  {NoStop}%
\bibitem [{\citenamefont {Pan}\ \emph {et~al.}(2014{\natexlab{a}})\citenamefont
  {Pan}, \citenamefont {At~Nguyen}, \citenamefont {Sridhar}, \citenamefont
  {Sunthar},\ and\ \citenamefont {Prakash}}]{Pan2014}%
  \BibitemOpen
  \bibfield  {author} {\bibinfo {author} {\bibfnamefont {S.}~\bibnamefont
  {Pan}}, \bibinfo {author} {\bibfnamefont {D.}~\bibnamefont {At~Nguyen}},
  \bibinfo {author} {\bibfnamefont {T.}~\bibnamefont {Sridhar}}, \bibinfo
  {author} {\bibfnamefont {P.}~\bibnamefont {Sunthar}}, \ and\ \bibinfo
  {author} {\bibfnamefont {R.~J.}\ \bibnamefont {Prakash}},\ }\bibfield
  {title} {\enquote {\bibinfo {title} {Universal solvent quality crossover of
  the zero shear rate viscosity of semidilute {DNA} solutions},}\ }\href@noop
  {} {\bibfield  {journal} {\bibinfo  {journal} {J. Rheo.}\ }\textbf {\bibinfo
  {volume} {58}},\ \bibinfo {pages} {339} (\bibinfo {year}
  {2014}{\natexlab{a}})}\BibitemShut {NoStop}%
\bibitem [{\citenamefont {Tominaga}\ \emph {et~al.}(2002)\citenamefont
  {Tominaga}, \citenamefont {Suda}, \citenamefont {Osa}, \citenamefont
  {Yoshizaki},\ and\ \citenamefont {Yamakawa}}]{Tominaga2002}%
  \BibitemOpen
  \bibfield  {author} {\bibinfo {author} {\bibfnamefont {Y.}~\bibnamefont
  {Tominaga}}, \bibinfo {author} {\bibfnamefont {I.~I.}\ \bibnamefont {Suda}},
  \bibinfo {author} {\bibfnamefont {T.}~\bibnamefont {Osa}}, \bibinfo {author}
  {\bibfnamefont {T.}~\bibnamefont {Yoshizaki}}, \ and\ \bibinfo {author}
  {\bibfnamefont {H.}~\bibnamefont {Yamakawa}},\ }\bibfield  {title} {\enquote
  {\bibinfo {title} {Viscosity and hydrodynamic-radius expansion factors of
  oligo- and poly(a-methylstyrene) in dilute solution},}\ }\href@noop {}
  {\bibfield  {journal} {\bibinfo  {journal} {Macromolecules}\ }\textbf
  {\bibinfo {volume} {35}},\ \bibinfo {pages} {1381--1388} (\bibinfo {year}
  {2002})}\BibitemShut {NoStop}%
\bibitem [{\citenamefont {Kumar}\ and\ \citenamefont
  {Prakash}(2003)}]{Kumar2003}%
  \BibitemOpen
  \bibfield  {author} {\bibinfo {author} {\bibfnamefont {K.~S.}\ \bibnamefont
  {Kumar}}\ and\ \bibinfo {author} {\bibfnamefont {J.~R.}\ \bibnamefont
  {Prakash}},\ }\bibfield  {title} {\enquote {\bibinfo {title} {Equilibrium
  swelling and universal ratios in dilute polymer solutions: Exact brownian
  dynamics simulations for a delta function excluded volume potential},}\
  }\href@noop {} {\bibfield  {journal} {\bibinfo  {journal} {Macromolecules}\
  }\textbf {\bibinfo {volume} {36}},\ \bibinfo {pages} {7842--7856} (\bibinfo
  {year} {2003})}\BibitemShut {NoStop}%
\bibitem [{\citenamefont {Chen}\ and\ \citenamefont
  {Noolandi}(1992)}]{Chen1992}%
  \BibitemOpen
  \bibfield  {author} {\bibinfo {author} {\bibfnamefont {Z.~Y.}\ \bibnamefont
  {Chen}}\ and\ \bibinfo {author} {\bibfnamefont {J.}~\bibnamefont
  {Noolandi}},\ }\bibfield  {title} {\enquote {\bibinfo {title}
  {Renormalization‐group scaling theory for flexible and wormlike polymer
  chains},}\ }\href@noop {} {\bibfield  {journal} {\bibinfo  {journal} {J.
  Chem. Phys.}\ }\textbf {\bibinfo {volume} {96}},\ \bibinfo {pages}
  {1540--1548} (\bibinfo {year} {1992})}\BibitemShut {NoStop}%
\bibitem [{\citenamefont {Sunthar}\ and\ \citenamefont
  {Prakash}(2006)}]{Sunthar2006}%
  \BibitemOpen
  \bibfield  {author} {\bibinfo {author} {\bibfnamefont {P.}~\bibnamefont
  {Sunthar}}\ and\ \bibinfo {author} {\bibfnamefont {J.~R.}\ \bibnamefont
  {Prakash}},\ }\bibfield  {title} {\enquote {\bibinfo {title} {Dynamic scaling
  in dilute polymer solutions: The importance of dynamic correlations},}\
  }\href@noop {} {\bibfield  {journal} {\bibinfo  {journal} {Europhysics
  Letters}\ }\textbf {\bibinfo {volume} {75}},\ \bibinfo {pages} {77--83}
  (\bibinfo {year} {2006})}\BibitemShut {NoStop}%
\bibitem [{\citenamefont {Clisby}(2010)}]{Clisby2010}%
  \BibitemOpen
  \bibfield  {author} {\bibinfo {author} {\bibfnamefont {N.}~\bibnamefont
  {Clisby}},\ }\bibfield  {title} {\enquote {\bibinfo {title} {Accurate
  estimate of the critical exponent for self-avoiding walks via a fast
  implementation of the pivot algorithm},}\ }\href@noop {} {\bibfield
  {journal} {\bibinfo  {journal} {Phys. Rev. Lett.}\ }\textbf {\bibinfo
  {volume} {104}},\ \bibinfo {pages} {055702} (\bibinfo {year}
  {2010})}\BibitemShut {NoStop}%
\bibitem [{\citenamefont {Douglas}\ and\ \citenamefont
  {Freed}(1985)}]{Douglas1985}%
  \BibitemOpen
  \bibfield  {author} {\bibinfo {author} {\bibfnamefont {J.~F.}\ \bibnamefont
  {Douglas}}\ and\ \bibinfo {author} {\bibfnamefont {K.~F.}\ \bibnamefont
  {Freed}},\ }\bibfield  {title} {\enquote {\bibinfo {title} {Renormalization
  and the two-parameter theory. 2. comparison with experiment and other
  two-parameter theories},}\ }\href@noop {} {\bibfield  {journal} {\bibinfo
  {journal} {Macromolecules}\ }\textbf {\bibinfo {volume} {18}},\ \bibinfo
  {pages} {201--211} (\bibinfo {year} {1985})}\BibitemShut {NoStop}%
\bibitem [{\citenamefont {Clisby}\ and\ \citenamefont
  {Dunweg}(2016)}]{Clisby2016}%
  \BibitemOpen
  \bibfield  {author} {\bibinfo {author} {\bibfnamefont {N.}~\bibnamefont
  {Clisby}}\ and\ \bibinfo {author} {\bibfnamefont {B.}~\bibnamefont
  {Dunweg}},\ }\bibfield  {title} {\enquote {\bibinfo {title} {High-precision
  estimate of the hydrodynamic radius for self-avoiding walks},}\ }\href@noop
  {} {\bibfield  {journal} {\bibinfo  {journal} {Phys. Rev. E}\ }\textbf
  {\bibinfo {volume} {94}},\ \bibinfo {pages} {052102} (\bibinfo {year}
  {2016})}\BibitemShut {NoStop}%
\bibitem [{\citenamefont {Edwards}(1965)}]{Edwards1965}%
  \BibitemOpen
  \bibfield  {author} {\bibinfo {author} {\bibfnamefont {S.~F.}\ \bibnamefont
  {Edwards}},\ }\bibfield  {title} {\enquote {\bibinfo {title} {The statistical
  mechanics of polymers with excluded volume},}\ }\href@noop {} {\bibfield
  {journal} {\bibinfo  {journal} {Proc. Phys. Soc.}\ }\textbf {\bibinfo
  {volume} {85}},\ \bibinfo {pages} {613--624} (\bibinfo {year}
  {1965})}\BibitemShut {NoStop}%
\bibitem [{\citenamefont {Smith}\ \emph {et~al.}(1996)\citenamefont {Smith},
  \citenamefont {Perkins},\ and\ \citenamefont {Chu}}]{Smith1996}%
  \BibitemOpen
  \bibfield  {author} {\bibinfo {author} {\bibfnamefont {D.~E.}\ \bibnamefont
  {Smith}}, \bibinfo {author} {\bibfnamefont {T.~T.}\ \bibnamefont {Perkins}},
  \ and\ \bibinfo {author} {\bibfnamefont {S.}~\bibnamefont {Chu}},\ }\bibfield
   {title} {\enquote {\bibinfo {title} {Dynamical scaling of dna diffusion
  coefficients},}\ }\href@noop {} {\bibfield  {journal} {\bibinfo  {journal}
  {Macromolecules}\ }\textbf {\bibinfo {volume} {29}},\ \bibinfo {pages}
  {1372--1373} (\bibinfo {year} {1996})}\BibitemShut {NoStop}%
\bibitem [{\citenamefont {Robertson-Anderson}\ \emph
  {et~al.}(2006)\citenamefont {Robertson-Anderson}, \citenamefont {Laib},\ and\
  \citenamefont {Smith}}]{Robertson2006}%
  \BibitemOpen
  \bibfield  {author} {\bibinfo {author} {\bibfnamefont {R.~M.}\ \bibnamefont
  {Robertson-Anderson}}, \bibinfo {author} {\bibfnamefont {S.}~\bibnamefont
  {Laib}}, \ and\ \bibinfo {author} {\bibfnamefont {D.~E.}\ \bibnamefont
  {Smith}},\ }\bibfield  {title} {\enquote {\bibinfo {title} {Diffusion of
  isolated dna molecules: Dependence on length and topology},}\ }\href@noop {}
  {\bibfield  {journal} {\bibinfo  {journal} {Proc. Nat. Acad. Sci.}\ }\textbf
  {\bibinfo {volume} {103}},\ \bibinfo {pages} {7310--7314} (\bibinfo {year}
  {2006})}\BibitemShut {NoStop}%
\bibitem [{\citenamefont {Hsiao}\ \emph
  {et~al.}(2017{\natexlab{a}})\citenamefont {Hsiao}, \citenamefont {Sasmal},
  \citenamefont {Prakash},\ and\ \citenamefont {Schroeder}}]{Hsiao2017}%
  \BibitemOpen
  \bibfield  {author} {\bibinfo {author} {\bibfnamefont {K.}~\bibnamefont
  {Hsiao}}, \bibinfo {author} {\bibfnamefont {C.}~\bibnamefont {Sasmal}},
  \bibinfo {author} {\bibfnamefont {J.~R.}\ \bibnamefont {Prakash}}, \ and\
  \bibinfo {author} {\bibfnamefont {C.~M.}\ \bibnamefont {Schroeder}},\
  }\bibfield  {title} {\enquote {\bibinfo {title} {Direct observation of {DNA}
  dynamics in semidilute solutions in extensional flow},}\ }\href@noop {}
  {\bibfield  {journal} {\bibinfo  {journal} {J. Rheo.}\ }\textbf {\bibinfo
  {volume} {61}},\ \bibinfo {pages} {151} (\bibinfo {year}
  {2017}{\natexlab{a}})}\BibitemShut {NoStop}%
\bibitem [{\citenamefont {Tsortos}\ \emph {et~al.}(2011)\citenamefont
  {Tsortos}, \citenamefont {Papadakis},\ and\ \citenamefont
  {Gizeli}}]{Tsortos2011}%
  \BibitemOpen
  \bibfield  {author} {\bibinfo {author} {\bibfnamefont {A.}~\bibnamefont
  {Tsortos}}, \bibinfo {author} {\bibfnamefont {G.}~\bibnamefont {Papadakis}},
  \ and\ \bibinfo {author} {\bibfnamefont {E.}~\bibnamefont {Gizeli}},\
  }\bibfield  {title} {\enquote {\bibinfo {title} {The intrinsic viscosity of
  linear dna},}\ }\href@noop {} {\bibfield  {journal} {\bibinfo  {journal}
  {Biopolymers}\ }\textbf {\bibinfo {volume} {95}},\ \bibinfo {pages}
  {824--832} (\bibinfo {year} {2011})}\BibitemShut {NoStop}%
\bibitem [{\citenamefont {Latinwo}\ and\ \citenamefont
  {Schroeder}(2011)}]{Latinwo2011}%
  \BibitemOpen
  \bibfield  {author} {\bibinfo {author} {\bibfnamefont {F.}~\bibnamefont
  {Latinwo}}\ and\ \bibinfo {author} {\bibfnamefont {C.~M.}\ \bibnamefont
  {Schroeder}},\ }\bibfield  {title} {\enquote {\bibinfo {title} {Model systems
  for single molecule polymer dynamics},}\ }\href@noop {} {\bibfield  {journal}
  {\bibinfo  {journal} {Soft Matter}\ }\textbf {\bibinfo {volume} {7}},\
  \bibinfo {pages} {7907--7913} (\bibinfo {year} {2011})}\BibitemShut {NoStop}%
\bibitem [{\citenamefont {Schaefer}\ \emph {et~al.}(1980)\citenamefont
  {Schaefer}, \citenamefont {Joanny},\ and\ \citenamefont
  {Pincus}}]{Schaefer1980}%
  \BibitemOpen
  \bibfield  {author} {\bibinfo {author} {\bibfnamefont {D.~W.}\ \bibnamefont
  {Schaefer}}, \bibinfo {author} {\bibfnamefont {J.~F.}\ \bibnamefont
  {Joanny}}, \ and\ \bibinfo {author} {\bibfnamefont {P.}~\bibnamefont
  {Pincus}},\ }\bibfield  {title} {\enquote {\bibinfo {title} {Dynamics of
  semiflexible polymers in solution},}\ }\href@noop {} {\bibfield  {journal}
  {\bibinfo  {journal} {Macromolecules}\ }\textbf {\bibinfo {volume} {13}},\
  \bibinfo {pages} {1280--1289} (\bibinfo {year} {1980})}\BibitemShut {NoStop}%
\bibitem [{\citenamefont {Smith}\ \emph {et~al.}(1992)\citenamefont {Smith},
  \citenamefont {Finzi},\ and\ \citenamefont {Bustamante}}]{Smith1992}%
  \BibitemOpen
  \bibfield  {author} {\bibinfo {author} {\bibfnamefont {S.~B.}\ \bibnamefont
  {Smith}}, \bibinfo {author} {\bibfnamefont {L.}~\bibnamefont {Finzi}}, \ and\
  \bibinfo {author} {\bibfnamefont {C.}~\bibnamefont {Bustamante}},\ }\bibfield
   {title} {\enquote {\bibinfo {title} {Direct measurements of the elasticity
  of single dna molecules by using magnetic beads},}\ }\href@noop {} {\bibfield
   {journal} {\bibinfo  {journal} {Science}\ }\textbf {\bibinfo {volume}
  {258}},\ \bibinfo {pages} {1122--1126} (\bibinfo {year} {1992})}\BibitemShut
  {NoStop}%
\bibitem [{\citenamefont {Larson}\ \emph {et~al.}(1997)\citenamefont {Larson},
  \citenamefont {Perkins}, \citenamefont {Smith},\ and\ \citenamefont
  {Chu}}]{Larson1997}%
  \BibitemOpen
  \bibfield  {author} {\bibinfo {author} {\bibfnamefont {R.~G.}\ \bibnamefont
  {Larson}}, \bibinfo {author} {\bibfnamefont {T.~T.}\ \bibnamefont {Perkins}},
  \bibinfo {author} {\bibfnamefont {D.~E.}\ \bibnamefont {Smith}}, \ and\
  \bibinfo {author} {\bibfnamefont {S.}~\bibnamefont {Chu}},\ }\bibfield
  {title} {\enquote {\bibinfo {title} {Hydrodynamics of a dna molecule in a
  flow field},}\ }\href@noop {} {\bibfield  {journal} {\bibinfo  {journal}
  {Phys. Rev. E}\ }\textbf {\bibinfo {volume} {55}},\ \bibinfo {pages}
  {1794--1797} (\bibinfo {year} {1997})}\BibitemShut {NoStop}%
\bibitem [{\citenamefont {Hur}\ \emph {et~al.}(2000)\citenamefont {Hur},
  \citenamefont {Shaqfeh},\ and\ \citenamefont {Larson}}]{Hur2000a}%
  \BibitemOpen
  \bibfield  {author} {\bibinfo {author} {\bibfnamefont {J.~S.}\ \bibnamefont
  {Hur}}, \bibinfo {author} {\bibfnamefont {E.~S.~G.}\ \bibnamefont {Shaqfeh}},
  \ and\ \bibinfo {author} {\bibfnamefont {R.~G.}\ \bibnamefont {Larson}},\
  }\bibfield  {title} {\enquote {\bibinfo {title} {Brownian dynamics
  simulations of single dna molecules in shear flow},}\ }\href@noop {}
  {\bibfield  {journal} {\bibinfo  {journal} {Journal of Rheology}\ }\textbf
  {\bibinfo {volume} {44}},\ \bibinfo {pages} {713--742} (\bibinfo {year}
  {2000})}\BibitemShut {NoStop}%
\bibitem [{\citenamefont {Pincus}(1976)}]{Pincus1976}%
  \BibitemOpen
  \bibfield  {author} {\bibinfo {author} {\bibfnamefont {P.}~\bibnamefont
  {Pincus}},\ }\bibfield  {title} {\enquote {\bibinfo {title} {Excluded volume
  effects and stretched polymer chains},}\ }\href@noop {} {\bibfield  {journal}
  {\bibinfo  {journal} {Macromolecules}\ }\textbf {\bibinfo {volume} {9}},\
  \bibinfo {pages} {386--388} (\bibinfo {year} {1976})}\BibitemShut {NoStop}%
\bibitem [{\citenamefont {Dittmore}\ \emph {et~al.}(2011)\citenamefont
  {Dittmore}, \citenamefont {McIntosh}, \citenamefont {Halliday},\ and\
  \citenamefont {Saleh}}]{Dittmore2011}%
  \BibitemOpen
  \bibfield  {author} {\bibinfo {author} {\bibfnamefont {A.}~\bibnamefont
  {Dittmore}}, \bibinfo {author} {\bibfnamefont {D.~B.}\ \bibnamefont
  {McIntosh}}, \bibinfo {author} {\bibfnamefont {S.}~\bibnamefont {Halliday}},
  \ and\ \bibinfo {author} {\bibfnamefont {O.~A.}\ \bibnamefont {Saleh}},\
  }\bibfield  {title} {\enquote {\bibinfo {title} {Single-molecule elasticity
  measurements of the onset of excluded volume in poly(ethylene glycol)},}\
  }\href@noop {} {\bibfield  {journal} {\bibinfo  {journal} {Phys. Rev. Lett.}\
  }\textbf {\bibinfo {volume} {107}},\ \bibinfo {pages} {148301} (\bibinfo
  {year} {2011})}\BibitemShut {NoStop}%
\bibitem [{\citenamefont {Radhakrishnan}\ and\ \citenamefont
  {Underhill}(2012{\natexlab{a}})}]{Radhakrishnan2012}%
  \BibitemOpen
  \bibfield  {author} {\bibinfo {author} {\bibfnamefont {R.}~\bibnamefont
  {Radhakrishnan}}\ and\ \bibinfo {author} {\bibfnamefont {P.~T.}\ \bibnamefont
  {Underhill}},\ }\bibfield  {title} {\enquote {\bibinfo {title} {Models of
  flexible polymers in good solvents: Relaxation and coil-stretch
  transition},}\ }\href@noop {} {\bibfield  {journal} {\bibinfo  {journal}
  {Soft Matter}\ }\textbf {\bibinfo {volume} {8}},\ \bibinfo {pages}
  {6991--7003} (\bibinfo {year} {2012}{\natexlab{a}})}\BibitemShut {NoStop}%
\bibitem [{\citenamefont {Jacobson}\ \emph {et~al.}(2017)\citenamefont
  {Jacobson}, \citenamefont {McIntosh}, \citenamefont {Stevens}, \citenamefont
  {Rubinstein},\ and\ \citenamefont {Saleh}}]{Jacobson2017}%
  \BibitemOpen
  \bibfield  {author} {\bibinfo {author} {\bibfnamefont {D.~R.}\ \bibnamefont
  {Jacobson}}, \bibinfo {author} {\bibfnamefont {D.~B.}\ \bibnamefont
  {McIntosh}}, \bibinfo {author} {\bibfnamefont {M.~J.}\ \bibnamefont
  {Stevens}}, \bibinfo {author} {\bibfnamefont {M.}~\bibnamefont {Rubinstein}},
  \ and\ \bibinfo {author} {\bibfnamefont {O.~A.}\ \bibnamefont {Saleh}},\
  }\bibfield  {title} {\enquote {\bibinfo {title} {Single-stranded nucleic acid
  elasticity arises from internal electrostatic tension},}\ }\href@noop {}
  {\bibfield  {journal} {\bibinfo  {journal} {Proc. Nat. Acad. Sci.}\ }\textbf
  {\bibinfo {volume} {114}},\ \bibinfo {pages} {5095--5100} (\bibinfo {year}
  {2017})}\BibitemShut {NoStop}%
\bibitem [{\citenamefont {Cohen}(1991)}]{Cohen1991}%
  \BibitemOpen
  \bibfield  {author} {\bibinfo {author} {\bibfnamefont {A.}~\bibnamefont
  {Cohen}},\ }\bibfield  {title} {\enquote {\bibinfo {title} {A pade
  approximant to the inverse langevin function},}\ }\href@noop {} {\bibfield
  {journal} {\bibinfo  {journal} {Rheo. Acta.}\ }\textbf {\bibinfo {volume}
  {30}},\ \bibinfo {pages} {270--273} (\bibinfo {year} {1991})}\BibitemShut
  {NoStop}%
\bibitem [{\citenamefont {Radhakrishnan}\ and\ \citenamefont
  {Underhill}(2012{\natexlab{b}})}]{Radhakrishnan2012b}%
  \BibitemOpen
  \bibfield  {author} {\bibinfo {author} {\bibfnamefont {R.}~\bibnamefont
  {Radhakrishnan}}\ and\ \bibinfo {author} {\bibfnamefont {P.~T.}\ \bibnamefont
  {Underhill}},\ }\bibfield  {title} {\enquote {\bibinfo {title} {Impact of
  solvent quality on the hystersis in the coil-stretch transition of flexible
  polymers in a good solvent},}\ }\href@noop {} {\bibfield  {journal} {\bibinfo
   {journal} {Macromolecules}\ }\textbf {\bibinfo {volume} {46}},\ \bibinfo
  {pages} {548--554} (\bibinfo {year} {2012}{\natexlab{b}})}\BibitemShut
  {NoStop}%
\bibitem [{\citenamefont {Li}\ \emph {et~al.}(2015{\natexlab{a}})\citenamefont
  {Li}, \citenamefont {Schroeder},\ and\ \citenamefont {Dorfman}}]{Li2015}%
  \BibitemOpen
  \bibfield  {author} {\bibinfo {author} {\bibfnamefont {X.}~\bibnamefont
  {Li}}, \bibinfo {author} {\bibfnamefont {C.~M.}\ \bibnamefont {Schroeder}}, \
  and\ \bibinfo {author} {\bibfnamefont {K.~D.}\ \bibnamefont {Dorfman}},\
  }\bibfield  {title} {\enquote {\bibinfo {title} {Modeling the stretching of
  wormlike chains in the presence of excluded volume},}\ }\href@noop {}
  {\bibfield  {journal} {\bibinfo  {journal} {Soft Matter}\ }\textbf {\bibinfo
  {volume} {11}},\ \bibinfo {pages} {5947--5954} (\bibinfo {year}
  {2015}{\natexlab{a}})}\BibitemShut {NoStop}%
\bibitem [{\citenamefont {Saadat}\ and\ \citenamefont
  {Khomami}(2016)}]{Saadat2016}%
  \BibitemOpen
  \bibfield  {author} {\bibinfo {author} {\bibfnamefont {A.}~\bibnamefont
  {Saadat}}\ and\ \bibinfo {author} {\bibfnamefont {B.}~\bibnamefont
  {Khomami}},\ }\bibfield  {title} {\enquote {\bibinfo {title} {A new
  bead-spring model for simulation of semi-flexible macromolecules},}\
  }\href@noop {} {\bibfield  {journal} {\bibinfo  {journal} {J. Chem. Phys.}\
  }\textbf {\bibinfo {volume} {145}},\ \bibinfo {pages} {204902} (\bibinfo
  {year} {2016})}\BibitemShut {NoStop}%
\bibitem [{\citenamefont {Shaqfeh}\ \emph {et~al.}(2004)\citenamefont
  {Shaqfeh}, \citenamefont {McKinley}, \citenamefont {Woo}, \citenamefont
  {Nguyen},\ and\ \citenamefont {Sridhar}}]{Shaqfeh2004}%
  \BibitemOpen
  \bibfield  {author} {\bibinfo {author} {\bibfnamefont {E.~S.~G.}\
  \bibnamefont {Shaqfeh}}, \bibinfo {author} {\bibfnamefont {G.~H.}\
  \bibnamefont {McKinley}}, \bibinfo {author} {\bibfnamefont {N.}~\bibnamefont
  {Woo}}, \bibinfo {author} {\bibfnamefont {D.~A.}\ \bibnamefont {Nguyen}}, \
  and\ \bibinfo {author} {\bibfnamefont {T.}~\bibnamefont {Sridhar}},\
  }\bibfield  {title} {\enquote {\bibinfo {title} {On the polymer entropic
  force singularity and its relation to extensional stress relaxation and
  filament recoil},}\ }\href@noop {} {\bibfield  {journal} {\bibinfo  {journal}
  {Journal of Rheology}\ }\textbf {\bibinfo {volume} {48}},\ \bibinfo {pages}
  {209--221} (\bibinfo {year} {2004})}\BibitemShut {NoStop}%
\bibitem [{\citenamefont {Dobrynin}\ \emph {et~al.}(2010)\citenamefont
  {Dobrynin}, \citenamefont {Carrillo},\ and\ \citenamefont
  {Rubinstein}}]{Dobrynin2010}%
  \BibitemOpen
  \bibfield  {author} {\bibinfo {author} {\bibfnamefont {A.~V.}\ \bibnamefont
  {Dobrynin}}, \bibinfo {author} {\bibfnamefont {J.-M.~Y.}\ \bibnamefont
  {Carrillo}}, \ and\ \bibinfo {author} {\bibfnamefont {M.}~\bibnamefont
  {Rubinstein}},\ }\bibfield  {title} {\enquote {\bibinfo {title} {Chains are
  more flexible under tension},}\ }\href@noop {} {\bibfield  {journal}
  {\bibinfo  {journal} {Macromolecules}\ }\textbf {\bibinfo {volume} {43}},\
  \bibinfo {pages} {9181--9190} (\bibinfo {year} {2010})}\BibitemShut {NoStop}%
\bibitem [{\citenamefont {Toan}\ and\ \citenamefont
  {Thirumalai}(2012)}]{Toan2012}%
  \BibitemOpen
  \bibfield  {author} {\bibinfo {author} {\bibfnamefont {N.~M.}\ \bibnamefont
  {Toan}}\ and\ \bibinfo {author} {\bibfnamefont {D.}~\bibnamefont
  {Thirumalai}},\ }\bibfield  {title} {\enquote {\bibinfo {title} {On the
  origin of the unusual behavior in the stretching of single-stranded dna},}\
  }\href@noop {} {\bibfield  {journal} {\bibinfo  {journal} {Journal of
  Chemical Physics}\ }\textbf {\bibinfo {volume} {136}},\ \bibinfo {pages}
  {235103} (\bibinfo {year} {2012})}\BibitemShut {NoStop}%
\bibitem [{\citenamefont {Stevens}\ \emph {et~al.}(2012)\citenamefont
  {Stevens}, \citenamefont {McIntosh},\ and\ \citenamefont
  {Saleh}}]{Stevens2012}%
  \BibitemOpen
  \bibfield  {author} {\bibinfo {author} {\bibfnamefont {M.~J.}\ \bibnamefont
  {Stevens}}, \bibinfo {author} {\bibfnamefont {D.~B.}\ \bibnamefont
  {McIntosh}}, \ and\ \bibinfo {author} {\bibfnamefont {O.~A.}\ \bibnamefont
  {Saleh}},\ }\bibfield  {title} {\enquote {\bibinfo {title} {Simulations of
  stretching a strong, flexible polyelectrolyte},}\ }\href@noop {} {\bibfield
  {journal} {\bibinfo  {journal} {Macromolecules}\ }\textbf {\bibinfo {volume}
  {45}},\ \bibinfo {pages} {5757--5765} (\bibinfo {year} {2012})}\BibitemShut
  {NoStop}%
\bibitem [{\citenamefont {Saleh}(2015)}]{Saleh2015}%
  \BibitemOpen
  \bibfield  {author} {\bibinfo {author} {\bibfnamefont {O.~A.}\ \bibnamefont
  {Saleh}},\ }\bibfield  {title} {\enquote {\bibinfo {title} {Single polymer
  mechanics across the force regimes},}\ }\href@noop {} {\bibfield  {journal}
  {\bibinfo  {journal} {Journal of Chemical Physics}\ }\textbf {\bibinfo
  {volume} {142}},\ \bibinfo {pages} {194902} (\bibinfo {year}
  {2015})}\BibitemShut {NoStop}%
\bibitem [{\citenamefont {Crothers}\ and\ \citenamefont
  {Zimm}(1965)}]{Zimm1965}%
  \BibitemOpen
  \bibfield  {author} {\bibinfo {author} {\bibfnamefont {D.~M.}\ \bibnamefont
  {Crothers}}\ and\ \bibinfo {author} {\bibfnamefont {B.~H.}\ \bibnamefont
  {Zimm}},\ }\bibfield  {title} {\enquote {\bibinfo {title} {Viscosity and
  sedimentation of the dna from bacteriophages t2 and t7 and the relation to
  molecular weight},}\ }\href@noop {} {\bibfield  {journal} {\bibinfo
  {journal} {Journal of Molecular Biology}\ }\textbf {\bibinfo {volume} {89}},\
  \bibinfo {pages} {525--536} (\bibinfo {year} {1965})}\BibitemShut {NoStop}%
\bibitem [{\citenamefont {Sorlie}\ and\ \citenamefont
  {Pecora}(1990)}]{Pecora1990}%
  \BibitemOpen
  \bibfield  {author} {\bibinfo {author} {\bibfnamefont {S.~S.}\ \bibnamefont
  {Sorlie}}\ and\ \bibinfo {author} {\bibfnamefont {R.}~\bibnamefont
  {Pecora}},\ }\bibfield  {title} {\enquote {\bibinfo {title} {A dynamic light
  scattering study of four dna restriction fragments},}\ }\href@noop {}
  {\bibfield  {journal} {\bibinfo  {journal} {Macromolecules}\ }\textbf
  {\bibinfo {volume} {23}},\ \bibinfo {pages} {487--497} (\bibinfo {year}
  {1990})}\BibitemShut {NoStop}%
\bibitem [{\citenamefont {Pecora}(1991)}]{Pecora1991}%
  \BibitemOpen
  \bibfield  {author} {\bibinfo {author} {\bibfnamefont {R.}~\bibnamefont
  {Pecora}},\ }\bibfield  {title} {\enquote {\bibinfo {title} {Dna: a model
  compound for solution studies of macromolecules},}\ }\href@noop {} {\bibfield
   {journal} {\bibinfo  {journal} {Science}\ }\textbf {\bibinfo {volume}
  {251}},\ \bibinfo {pages} {893--898} (\bibinfo {year} {1991})}\BibitemShut
  {NoStop}%
\bibitem [{\citenamefont {Perkins}\ \emph {et~al.}(1995)\citenamefont
  {Perkins}, \citenamefont {Smith}, \citenamefont {Larson},\ and\ \citenamefont
  {Chu}}]{Perkins1995}%
  \BibitemOpen
  \bibfield  {author} {\bibinfo {author} {\bibfnamefont {T.~T.}\ \bibnamefont
  {Perkins}}, \bibinfo {author} {\bibfnamefont {D.~E.}\ \bibnamefont {Smith}},
  \bibinfo {author} {\bibfnamefont {R.~G.}\ \bibnamefont {Larson}}, \ and\
  \bibinfo {author} {\bibfnamefont {S.}~\bibnamefont {Chu}},\ }\bibfield
  {title} {\enquote {\bibinfo {title} {Stretching of a single tethered polymer
  in a uniform flow},}\ }\href@noop {} {\bibfield  {journal} {\bibinfo
  {journal} {Science}\ }\textbf {\bibinfo {volume} {268}},\ \bibinfo {pages}
  {83--87} (\bibinfo {year} {1995})}\BibitemShut {NoStop}%
\bibitem [{\citenamefont {Smith}\ \emph {et~al.}(1999)\citenamefont {Smith},
  \citenamefont {Babcock},\ and\ \citenamefont {Chu}}]{Smith1999}%
  \BibitemOpen
  \bibfield  {author} {\bibinfo {author} {\bibfnamefont {D.~E.}\ \bibnamefont
  {Smith}}, \bibinfo {author} {\bibfnamefont {H.~P.}\ \bibnamefont {Babcock}},
  \ and\ \bibinfo {author} {\bibfnamefont {S.}~\bibnamefont {Chu}},\ }\bibfield
   {title} {\enquote {\bibinfo {title} {Single-polymer dynamics in steady shear
  flow},}\ }\href@noop {} {\bibfield  {journal} {\bibinfo  {journal} {Science}\
  }\textbf {\bibinfo {volume} {283}},\ \bibinfo {pages} {1724--1727} (\bibinfo
  {year} {1999})}\BibitemShut {NoStop}%
\bibitem [{\citenamefont {Perkins}\ \emph {et~al.}(1997)\citenamefont
  {Perkins}, \citenamefont {Smith},\ and\ \citenamefont {Chu}}]{Perkins1997}%
  \BibitemOpen
  \bibfield  {author} {\bibinfo {author} {\bibfnamefont {T.~T.}\ \bibnamefont
  {Perkins}}, \bibinfo {author} {\bibfnamefont {D.~E.}\ \bibnamefont {Smith}},
  \ and\ \bibinfo {author} {\bibfnamefont {S.}~\bibnamefont {Chu}},\ }\bibfield
   {title} {\enquote {\bibinfo {title} {Single polymer dynamics in an
  elongational flow},}\ }\href@noop {} {\bibfield  {journal} {\bibinfo
  {journal} {Science}\ }\textbf {\bibinfo {volume} {276}},\ \bibinfo {pages}
  {2016--2021} (\bibinfo {year} {1997})}\BibitemShut {NoStop}%
\bibitem [{\citenamefont {Smith}\ and\ \citenamefont {Chu}(1998)}]{Smith1998}%
  \BibitemOpen
  \bibfield  {author} {\bibinfo {author} {\bibfnamefont {D.~E.}\ \bibnamefont
  {Smith}}\ and\ \bibinfo {author} {\bibfnamefont {S.}~\bibnamefont {Chu}},\
  }\bibfield  {title} {\enquote {\bibinfo {title} {Response of flexible
  polymers to a sudden elongational flow},}\ }\href@noop {} {\bibfield
  {journal} {\bibinfo  {journal} {Science}\ }\textbf {\bibinfo {volume}
  {281}},\ \bibinfo {pages} {1335--1340} (\bibinfo {year} {1998})}\BibitemShut
  {NoStop}%
\bibitem [{\citenamefont {Larson}\ \emph {et~al.}(1999)\citenamefont {Larson},
  \citenamefont {Hu}, \citenamefont {Smith},\ and\ \citenamefont
  {Chu}}]{Larson1999}%
  \BibitemOpen
  \bibfield  {author} {\bibinfo {author} {\bibfnamefont {R.~G.}\ \bibnamefont
  {Larson}}, \bibinfo {author} {\bibfnamefont {H.}~\bibnamefont {Hu}}, \bibinfo
  {author} {\bibfnamefont {D.~E.}\ \bibnamefont {Smith}}, \ and\ \bibinfo
  {author} {\bibfnamefont {S.}~\bibnamefont {Chu}},\ }\bibfield  {title}
  {\enquote {\bibinfo {title} {Brownian dynamics simulations of a dna molecule
  in an extensional flow field},}\ }\href@noop {} {\bibfield  {journal}
  {\bibinfo  {journal} {Journal of Rheology}\ }\textbf {\bibinfo {volume}
  {43}},\ \bibinfo {pages} {267--304} (\bibinfo {year} {1999})}\BibitemShut
  {NoStop}%
\bibitem [{\citenamefont {Teixeira}\ \emph {et~al.}(2005)\citenamefont
  {Teixeira}, \citenamefont {Babcock}, \citenamefont {Shaqfeh},\ and\
  \citenamefont {Chu}}]{Teixeira2005}%
  \BibitemOpen
  \bibfield  {author} {\bibinfo {author} {\bibfnamefont {R.~E.}\ \bibnamefont
  {Teixeira}}, \bibinfo {author} {\bibfnamefont {H.~P.}\ \bibnamefont
  {Babcock}}, \bibinfo {author} {\bibfnamefont {E.~S.~G.}\ \bibnamefont
  {Shaqfeh}}, \ and\ \bibinfo {author} {\bibfnamefont {S.}~\bibnamefont
  {Chu}},\ }\bibfield  {title} {\enquote {\bibinfo {title} {Shear thinning and
  tumbling dynamics of single polymers in the flow-gradient plane},}\
  }\href@noop {} {\bibfield  {journal} {\bibinfo  {journal} {Macromolecules}\
  }\textbf {\bibinfo {volume} {38}},\ \bibinfo {pages} {581--592} (\bibinfo
  {year} {2005})}\BibitemShut {NoStop}%
\bibitem [{\citenamefont {Schroeder}\ \emph
  {et~al.}(2005{\natexlab{a}})\citenamefont {Schroeder}, \citenamefont
  {Teixeira}, \citenamefont {Shaqfeh},\ and\ \citenamefont
  {Chu}}]{Schroeder2005b}%
  \BibitemOpen
  \bibfield  {author} {\bibinfo {author} {\bibfnamefont {C.~M.}\ \bibnamefont
  {Schroeder}}, \bibinfo {author} {\bibfnamefont {R.~E.}\ \bibnamefont
  {Teixeira}}, \bibinfo {author} {\bibfnamefont {E.~S.~G.}\ \bibnamefont
  {Shaqfeh}}, \ and\ \bibinfo {author} {\bibfnamefont {S.}~\bibnamefont
  {Chu}},\ }\bibfield  {title} {\enquote {\bibinfo {title} {Dynamics of dna in
  the flow-gradient plane of steady shear flow: Observations and
  simulations},}\ }\href@noop {} {\bibfield  {journal} {\bibinfo  {journal}
  {Macromolecules}\ }\textbf {\bibinfo {volume} {38}},\ \bibinfo {pages}
  {1967--1978} (\bibinfo {year} {2005}{\natexlab{a}})}\BibitemShut {NoStop}%
\bibitem [{\citenamefont {Doyle}\ \emph {et~al.}(2000)\citenamefont {Doyle},
  \citenamefont {Ladoux},\ and\ \citenamefont {Viovy}}]{Doyle2000}%
  \BibitemOpen
  \bibfield  {author} {\bibinfo {author} {\bibfnamefont {P.~S.}\ \bibnamefont
  {Doyle}}, \bibinfo {author} {\bibfnamefont {B.}~\bibnamefont {Ladoux}}, \
  and\ \bibinfo {author} {\bibfnamefont {J.-L.}\ \bibnamefont {Viovy}},\
  }\bibfield  {title} {\enquote {\bibinfo {title} {Dynamics of a tethered
  polymer in shear flow},}\ }\href@noop {} {\bibfield  {journal} {\bibinfo
  {journal} {Phys. Rev. Lett.}\ }\textbf {\bibinfo {volume} {84}},\ \bibinfo
  {pages} {4769} (\bibinfo {year} {2000})}\BibitemShut {NoStop}%
\bibitem [{\citenamefont {Schroeder}\ \emph
  {et~al.}(2005{\natexlab{b}})\citenamefont {Schroeder}, \citenamefont
  {Teixeira}, \citenamefont {Shaqfeh},\ and\ \citenamefont
  {Chu}}]{Schroeder2005}%
  \BibitemOpen
  \bibfield  {author} {\bibinfo {author} {\bibfnamefont {C.~M.}\ \bibnamefont
  {Schroeder}}, \bibinfo {author} {\bibfnamefont {R.~E.}\ \bibnamefont
  {Teixeira}}, \bibinfo {author} {\bibfnamefont {E.~S.~G.}\ \bibnamefont
  {Shaqfeh}}, \ and\ \bibinfo {author} {\bibfnamefont {S.}~\bibnamefont
  {Chu}},\ }\bibfield  {title} {\enquote {\bibinfo {title} {Characteristic
  periodic motion of polymers in shear flow},}\ }\href@noop {} {\bibfield
  {journal} {\bibinfo  {journal} {Phys. Rev. Lett.}\ }\textbf {\bibinfo
  {volume} {95}},\ \bibinfo {pages} {018301} (\bibinfo {year}
  {2005}{\natexlab{b}})}\BibitemShut {NoStop}%
\bibitem [{\citenamefont {Babcock}\ \emph {et~al.}(2000)\citenamefont
  {Babcock}, \citenamefont {Smith}, \citenamefont {Hur}, \citenamefont
  {Shaqfeh},\ and\ \citenamefont {Chu}}]{Babcock2000}%
  \BibitemOpen
  \bibfield  {author} {\bibinfo {author} {\bibfnamefont {H.~P.}\ \bibnamefont
  {Babcock}}, \bibinfo {author} {\bibfnamefont {D.~E.}\ \bibnamefont {Smith}},
  \bibinfo {author} {\bibfnamefont {J.~S.}\ \bibnamefont {Hur}}, \bibinfo
  {author} {\bibfnamefont {E.~S.~G.}\ \bibnamefont {Shaqfeh}}, \ and\ \bibinfo
  {author} {\bibfnamefont {S.}~\bibnamefont {Chu}},\ }\bibfield  {title}
  {\enquote {\bibinfo {title} {Relating the microscopic and macroscopic
  response of a polymeric fluid in a shearing flow},}\ }\href@noop {}
  {\bibfield  {journal} {\bibinfo  {journal} {Phys. Rev. Lett.}\ }\textbf
  {\bibinfo {volume} {85}},\ \bibinfo {pages} {2018--2021} (\bibinfo {year}
  {2000})}\BibitemShut {NoStop}%
\bibitem [{\citenamefont {Hur}\ \emph {et~al.}(2001)\citenamefont {Hur},
  \citenamefont {Shaqfeh}, \citenamefont {Babcock}, \citenamefont {Smith},\
  and\ \citenamefont {Chu}}]{Hur2001}%
  \BibitemOpen
  \bibfield  {author} {\bibinfo {author} {\bibfnamefont {J.~S.}\ \bibnamefont
  {Hur}}, \bibinfo {author} {\bibfnamefont {E.~S.~G.}\ \bibnamefont {Shaqfeh}},
  \bibinfo {author} {\bibfnamefont {H.~P.}\ \bibnamefont {Babcock}}, \bibinfo
  {author} {\bibfnamefont {D.~E.}\ \bibnamefont {Smith}}, \ and\ \bibinfo
  {author} {\bibfnamefont {S.}~\bibnamefont {Chu}},\ }\bibfield  {title}
  {\enquote {\bibinfo {title} {Dynamics of dilute and semidilute dna solutions
  in the start-up of shear flow},}\ }\href@noop {} {\bibfield  {journal}
  {\bibinfo  {journal} {Journal of Rheology}\ }\textbf {\bibinfo {volume}
  {45}},\ \bibinfo {pages} {421--450} (\bibinfo {year} {2001})}\BibitemShut
  {NoStop}%
\bibitem [{\citenamefont {Sasmal}\ \emph {et~al.}(2017)\citenamefont {Sasmal},
  \citenamefont {Hsiao}, \citenamefont {Schroeder},\ and\ \citenamefont
  {Prakash}}]{Sasmal2017}%
  \BibitemOpen
  \bibfield  {author} {\bibinfo {author} {\bibfnamefont {C.}~\bibnamefont
  {Sasmal}}, \bibinfo {author} {\bibfnamefont {K.}~\bibnamefont {Hsiao}},
  \bibinfo {author} {\bibfnamefont {C.~M.}\ \bibnamefont {Schroeder}}, \ and\
  \bibinfo {author} {\bibfnamefont {J.~R.}\ \bibnamefont {Prakash}},\
  }\bibfield  {title} {\enquote {\bibinfo {title} {Parameter-free prediction of
  dna dynamics in planar extensional flow of semi-dilute solutions},}\
  }\href@noop {} {\bibfield  {journal} {\bibinfo  {journal} {Journal of
  Rheology}\ }\textbf {\bibinfo {volume} {61}},\ \bibinfo {pages} {169--186}
  (\bibinfo {year} {2017})}\BibitemShut {NoStop}%
\bibitem [{\citenamefont {Babcock}\ \emph {et~al.}(2003)\citenamefont
  {Babcock}, \citenamefont {Teixeira}, \citenamefont {Hur}, \citenamefont
  {Shaqfeh},\ and\ \citenamefont {Chu}}]{Babcock2003}%
  \BibitemOpen
  \bibfield  {author} {\bibinfo {author} {\bibfnamefont {H.~P.}\ \bibnamefont
  {Babcock}}, \bibinfo {author} {\bibfnamefont {R.~E.}\ \bibnamefont
  {Teixeira}}, \bibinfo {author} {\bibfnamefont {J.~S.}\ \bibnamefont {Hur}},
  \bibinfo {author} {\bibfnamefont {E.~S.~G.}\ \bibnamefont {Shaqfeh}}, \ and\
  \bibinfo {author} {\bibfnamefont {S.}~\bibnamefont {Chu}},\ }\bibfield
  {title} {\enquote {\bibinfo {title} {Visualization of molecular fluctuations
  near the critical point of the coil-stretch transition in polymer
  elongation},}\ }\href@noop {} {\bibfield  {journal} {\bibinfo  {journal}
  {Macromolecules}\ }\textbf {\bibinfo {volume} {36}},\ \bibinfo {pages}
  {4544--4548} (\bibinfo {year} {2003})}\BibitemShut {NoStop}%
\bibitem [{\citenamefont {Hur}\ \emph {et~al.}(2002)\citenamefont {Hur},
  \citenamefont {Shaqfeh}, \citenamefont {Babcock},\ and\ \citenamefont
  {Chu}}]{Hur2002}%
  \BibitemOpen
  \bibfield  {author} {\bibinfo {author} {\bibfnamefont {J.~S.}\ \bibnamefont
  {Hur}}, \bibinfo {author} {\bibfnamefont {E.~S.~G.}\ \bibnamefont {Shaqfeh}},
  \bibinfo {author} {\bibfnamefont {H.~P.}\ \bibnamefont {Babcock}}, \ and\
  \bibinfo {author} {\bibfnamefont {S.}~\bibnamefont {Chu}},\ }\bibfield
  {title} {\enquote {\bibinfo {title} {Dynamics and configurational
  fluctuations of single dna molecules in linear mixed flows},}\ }\href@noop {}
  {\bibfield  {journal} {\bibinfo  {journal} {Phys. Rev. E}\ }\textbf {\bibinfo
  {volume} {66}},\ \bibinfo {pages} {011915} (\bibinfo {year}
  {2002})}\BibitemShut {NoStop}%
\bibitem [{\citenamefont {Fuller}\ and\ \citenamefont
  {Leal}(1980)}]{Fuller1980}%
  \BibitemOpen
  \bibfield  {author} {\bibinfo {author} {\bibfnamefont {G.~G.}\ \bibnamefont
  {Fuller}}\ and\ \bibinfo {author} {\bibfnamefont {L.~G.}\ \bibnamefont
  {Leal}},\ }\bibfield  {title} {\enquote {\bibinfo {title} {Flow birefringence
  of dilute polymer solutions in two-dimensional flows},}\ }\href@noop {}
  {\bibfield  {journal} {\bibinfo  {journal} {Rheo. Acta.}\ }\textbf {\bibinfo
  {volume} {19}},\ \bibinfo {pages} {580--600} (\bibinfo {year}
  {1980})}\BibitemShut {NoStop}%
\bibitem [{\citenamefont {Dua}\ and\ \citenamefont {Cherayil}(2003)}]{Dua2003}%
  \BibitemOpen
  \bibfield  {author} {\bibinfo {author} {\bibfnamefont {A.}~\bibnamefont
  {Dua}}\ and\ \bibinfo {author} {\bibfnamefont {B.~J.}\ \bibnamefont
  {Cherayil}},\ }\bibfield  {title} {\enquote {\bibinfo {title} {Polymer
  dynamics in linear mixed flows},}\ }\href@noop {} {\bibfield  {journal}
  {\bibinfo  {journal} {Journal of Chemical Physics}\ }\textbf {\bibinfo
  {volume} {119}},\ \bibinfo {pages} {5696--5700} (\bibinfo {year}
  {2003})}\BibitemShut {NoStop}%
\bibitem [{\citenamefont {Lee}\ \emph {et~al.}(2007{\natexlab{a}})\citenamefont
  {Lee}, \citenamefont {Shaqfeh},\ and\ \citenamefont {Muller}}]{Lee2007}%
  \BibitemOpen
  \bibfield  {author} {\bibinfo {author} {\bibfnamefont {J.~S.}\ \bibnamefont
  {Lee}}, \bibinfo {author} {\bibfnamefont {E.~S.~G.}\ \bibnamefont {Shaqfeh}},
  \ and\ \bibinfo {author} {\bibfnamefont {S.~J.}\ \bibnamefont {Muller}},\
  }\bibfield  {title} {\enquote {\bibinfo {title} {Dynamics of dna tumbling in
  shear to rotational mixed flows: Pathways and periods},}\ }\href@noop {}
  {\bibfield  {journal} {\bibinfo  {journal} {Phys. Rev. E}\ }\textbf {\bibinfo
  {volume} {75}},\ \bibinfo {pages} {040802} (\bibinfo {year}
  {2007}{\natexlab{a}})}\BibitemShut {NoStop}%
\bibitem [{\citenamefont {Lee}\ \emph {et~al.}(2007{\natexlab{b}})\citenamefont
  {Lee}, \citenamefont {Dylla-Spears}, \citenamefont {Teclemariam},\ and\
  \citenamefont {Muller}}]{Lee2007b}%
  \BibitemOpen
  \bibfield  {author} {\bibinfo {author} {\bibfnamefont {J.~S.}\ \bibnamefont
  {Lee}}, \bibinfo {author} {\bibfnamefont {R.}~\bibnamefont {Dylla-Spears}},
  \bibinfo {author} {\bibfnamefont {N.~P.}\ \bibnamefont {Teclemariam}}, \ and\
  \bibinfo {author} {\bibfnamefont {S.~J.}\ \bibnamefont {Muller}},\ }\bibfield
   {title} {\enquote {\bibinfo {title} {Microfluidic four-roll mill for all
  flow types},}\ }\href@noop {} {\bibfield  {journal} {\bibinfo  {journal}
  {Applied Physics Letters}\ }\textbf {\bibinfo {volume} {90}},\ \bibinfo
  {pages} {074103} (\bibinfo {year} {2007}{\natexlab{b}})}\BibitemShut
  {NoStop}%
\bibitem [{\citenamefont {Kirkwood}\ and\ \citenamefont
  {Riseman}(1948)}]{Kirkwood1948}%
  \BibitemOpen
  \bibfield  {author} {\bibinfo {author} {\bibfnamefont {J.~G.}\ \bibnamefont
  {Kirkwood}}\ and\ \bibinfo {author} {\bibfnamefont {J.}~\bibnamefont
  {Riseman}},\ }\bibfield  {title} {\enquote {\bibinfo {title} {The intrinsic
  viscosities and diffusion constants of flexible macromolecules in
  solution},}\ }\href@noop {} {\bibfield  {journal} {\bibinfo  {journal} {J.
  Chem. Phys.}\ }\textbf {\bibinfo {volume} {16}},\ \bibinfo {pages} {565--573}
  (\bibinfo {year} {1948})}\BibitemShut {NoStop}%
\bibitem [{\citenamefont {Kirkwood}(1949)}]{Kirkwood1949}%
  \BibitemOpen
  \bibfield  {author} {\bibinfo {author} {\bibfnamefont {J.~G.}\ \bibnamefont
  {Kirkwood}},\ }\bibfield  {title} {\enquote {\bibinfo {title} {The
  statistical mechanical theory of irreversible processes in solutions of
  flexible macromolecules: Viscoelastic behavior},}\ }\href@noop {} {\bibfield
  {journal} {\bibinfo  {journal} {Rec. Trav. Chim.}\ }\textbf {\bibinfo
  {volume} {68}},\ \bibinfo {pages} {649--660} (\bibinfo {year}
  {1949})}\BibitemShut {NoStop}%
\bibitem [{\citenamefont {Zimm}(1956)}]{Zimm1956}%
  \BibitemOpen
  \bibfield  {author} {\bibinfo {author} {\bibfnamefont {B.~H.}\ \bibnamefont
  {Zimm}},\ }\bibfield  {title} {\enquote {\bibinfo {title} {Dynamics of
  polymer molecule in dilute solutions: Viscoelasticity, flow birefringence,
  and dielectric loss},}\ }\href@noop {} {\bibfield  {journal} {\bibinfo
  {journal} {J. Chem. Phys.}\ }\textbf {\bibinfo {volume} {24}},\ \bibinfo
  {pages} {269--274} (\bibinfo {year} {1956})}\BibitemShut {NoStop}%
\bibitem [{\citenamefont {Rotne}\ and\ \citenamefont
  {Prager}(1969)}]{Rotne1969}%
  \BibitemOpen
  \bibfield  {author} {\bibinfo {author} {\bibfnamefont {J.}~\bibnamefont
  {Rotne}}\ and\ \bibinfo {author} {\bibfnamefont {S.}~\bibnamefont {Prager}},\
  }\bibfield  {title} {\enquote {\bibinfo {title} {Variational treatment of
  hydrodynamic interaction in polymers},}\ }\href@noop {} {\bibfield  {journal}
  {\bibinfo  {journal} {Journal of Chemical Physics}\ }\textbf {\bibinfo
  {volume} {50}},\ \bibinfo {pages} {4831--4837} (\bibinfo {year}
  {1969})}\BibitemShut {NoStop}%
\bibitem [{\citenamefont {Fixman}(1986)}]{Fixman1985}%
  \BibitemOpen
  \bibfield  {author} {\bibinfo {author} {\bibfnamefont {M.}~\bibnamefont
  {Fixman}},\ }\bibfield  {title} {\enquote {\bibinfo {title} {Construction of
  langevin forces in the simulation of hydrodynamic interaction},}\ }\href@noop
  {} {\bibfield  {journal} {\bibinfo  {journal} {Macromolecules}\ }\textbf
  {\bibinfo {volume} {19}},\ \bibinfo {pages} {1204--1207} (\bibinfo {year}
  {1986})}\BibitemShut {NoStop}%
\bibitem [{\citenamefont {Jendrejack}\ \emph {et~al.}(2000)\citenamefont
  {Jendrejack}, \citenamefont {Graham},\ and\ \citenamefont
  {de~Pablo}}]{Jendrejack2000}%
  \BibitemOpen
  \bibfield  {author} {\bibinfo {author} {\bibfnamefont {R.~M.}\ \bibnamefont
  {Jendrejack}}, \bibinfo {author} {\bibfnamefont {M.~D.}\ \bibnamefont
  {Graham}}, \ and\ \bibinfo {author} {\bibfnamefont {J.~J.}\ \bibnamefont
  {de~Pablo}},\ }\bibfield  {title} {\enquote {\bibinfo {title} {Hydrodynamic
  interactions in long chain polymers: Application of the chebyshev polynomial
  approximation in stochastic simulations},}\ }\href@noop {} {\bibfield
  {journal} {\bibinfo  {journal} {J. Chem. Phys.}\ }\textbf {\bibinfo {volume}
  {113}},\ \bibinfo {pages} {2894--2900} (\bibinfo {year} {2000})}\BibitemShut
  {NoStop}%
\bibitem [{\citenamefont {Ottinger}(1985)}]{Ottinger1985}%
  \BibitemOpen
  \bibfield  {author} {\bibinfo {author} {\bibfnamefont {H.~C.}\ \bibnamefont
  {Ottinger}},\ }\bibfield  {title} {\enquote {\bibinfo {title} {Consistently
  averaged hydrodynamic interaction for rouse dumbbells in steady-shear
  flow},}\ }\href@noop {} {\bibfield  {journal} {\bibinfo  {journal} {J. Chem.
  Phys.}\ }\textbf {\bibinfo {volume} {83}},\ \bibinfo {pages} {6535--6536}
  (\bibinfo {year} {1985})}\BibitemShut {NoStop}%
\bibitem [{\citenamefont {Ottinger}(1987)}]{Ottinger1987}%
  \BibitemOpen
  \bibfield  {author} {\bibinfo {author} {\bibfnamefont {H.~C.}\ \bibnamefont
  {Ottinger}},\ }\bibfield  {title} {\enquote {\bibinfo {title} {A model of
  dilute polymer solutions with hydrodynamic interaction and finite
  extensibility: I. basic equations and series expansions},}\ }\href@noop {}
  {\bibfield  {journal} {\bibinfo  {journal} {J. Non-Newtonian Fluid Mech.}\
  }\textbf {\bibinfo {volume} {26}},\ \bibinfo {pages} {207--246} (\bibinfo
  {year} {1987})}\BibitemShut {NoStop}%
\bibitem [{\citenamefont {Ottinger}(1989)}]{Ottinger1989}%
  \BibitemOpen
  \bibfield  {author} {\bibinfo {author} {\bibfnamefont {H.~C.}\ \bibnamefont
  {Ottinger}},\ }\bibfield  {title} {\enquote {\bibinfo {title} {Gaussian
  approximation for rouse chains with hydrodynamic interaction},}\ }\href@noop
  {} {\bibfield  {journal} {\bibinfo  {journal} {J. Chem. Phys.}\ }\textbf
  {\bibinfo {volume} {90}},\ \bibinfo {pages} {463--473} (\bibinfo {year}
  {1989})}\BibitemShut {NoStop}%
\bibitem [{\citenamefont {Zylka}\ and\ \citenamefont
  {Ottinger}(1989)}]{Zylka1989}%
  \BibitemOpen
  \bibfield  {author} {\bibinfo {author} {\bibfnamefont {W.}~\bibnamefont
  {Zylka}}\ and\ \bibinfo {author} {\bibfnamefont {H.~C.}\ \bibnamefont
  {Ottinger}},\ }\bibfield  {title} {\enquote {\bibinfo {title} {A comparison
  between simulations and various approximations for hookean dumbbells with
  hydrodynamic interaction},}\ }\href@noop {} {\bibfield  {journal} {\bibinfo
  {journal} {J. Chem. Phys.}\ }\textbf {\bibinfo {volume} {90}},\ \bibinfo
  {pages} {474--480} (\bibinfo {year} {1989})}\BibitemShut {NoStop}%
\bibitem [{\citenamefont {Ermak}\ and\ \citenamefont
  {McCammon}(1978)}]{Ermak1978}%
  \BibitemOpen
  \bibfield  {author} {\bibinfo {author} {\bibfnamefont {D.~L.}\ \bibnamefont
  {Ermak}}\ and\ \bibinfo {author} {\bibfnamefont {J.~A.}\ \bibnamefont
  {McCammon}},\ }\bibfield  {title} {\enquote {\bibinfo {title} {Brownian
  dynamics with hydrodynamic interactions},}\ }\href@noop {} {\bibfield
  {journal} {\bibinfo  {journal} {J. Chem. Phys.}\ }\textbf {\bibinfo {volume}
  {69}},\ \bibinfo {pages} {1352--1360} (\bibinfo {year} {1978})}\BibitemShut
  {NoStop}%
\bibitem [{\citenamefont {Lopez~Calsales}\ \emph {et~al.}(1992)\citenamefont
  {Lopez~Calsales}, \citenamefont {Navarro},\ and\ \citenamefont {Garcia de~la
  Torre}}]{Cascales1992}%
  \BibitemOpen
  \bibfield  {author} {\bibinfo {author} {\bibfnamefont {J.~J.}\ \bibnamefont
  {Lopez~Calsales}}, \bibinfo {author} {\bibfnamefont {S.}~\bibnamefont
  {Navarro}}, \ and\ \bibinfo {author} {\bibfnamefont {J.}~\bibnamefont {Garcia
  de~la Torre}},\ }\bibfield  {title} {\enquote {\bibinfo {title} {Deformation,
  orientation, and scattering from polymer chains in shear flow: A brownian
  dynamics simulation study},}\ }\href@noop {} {\bibfield  {journal} {\bibinfo
  {journal} {Macromolecules}\ }\textbf {\bibinfo {volume} {25}},\ \bibinfo
  {pages} {3574--3580} (\bibinfo {year} {1992})}\BibitemShut {NoStop}%
\bibitem [{\citenamefont {Lyulin}\ \emph {et~al.}(1999)\citenamefont {Lyulin},
  \citenamefont {Adolf},\ and\ \citenamefont {Davies}}]{Lyulin1999}%
  \BibitemOpen
  \bibfield  {author} {\bibinfo {author} {\bibfnamefont {A.~V.}\ \bibnamefont
  {Lyulin}}, \bibinfo {author} {\bibfnamefont {D.~B.}\ \bibnamefont {Adolf}}, \
  and\ \bibinfo {author} {\bibfnamefont {G.~R.}\ \bibnamefont {Davies}},\
  }\bibfield  {title} {\enquote {\bibinfo {title} {Brownian dynamics
  simulations of linear polymers under shear flow},}\ }\href@noop {} {\bibfield
   {journal} {\bibinfo  {journal} {J. Chem. Phys.}\ }\textbf {\bibinfo {volume}
  {111}},\ \bibinfo {pages} {758--771} (\bibinfo {year} {1999})}\BibitemShut
  {NoStop}%
\bibitem [{\citenamefont {Kroger}\ \emph {et~al.}(2000)\citenamefont {Kroger},
  \citenamefont {Alba-Perez}, \citenamefont {Laso},\ and\ \citenamefont
  {Ottinger}}]{Kroger2000}%
  \BibitemOpen
  \bibfield  {author} {\bibinfo {author} {\bibfnamefont {M.}~\bibnamefont
  {Kroger}}, \bibinfo {author} {\bibfnamefont {A.}~\bibnamefont {Alba-Perez}},
  \bibinfo {author} {\bibfnamefont {M.}~\bibnamefont {Laso}}, \ and\ \bibinfo
  {author} {\bibfnamefont {H.~C.}\ \bibnamefont {Ottinger}},\ }\bibfield
  {title} {\enquote {\bibinfo {title} {Variance reduced brownian dynamics
  simulation of a bead-spring chain under steady shear flow considering
  hydrodynamic interaction effects},}\ }\href@noop {} {\bibfield  {journal}
  {\bibinfo  {journal} {J. Chem. Phys.}\ }\textbf {\bibinfo {volume} {113}},\
  \bibinfo {pages} {4767--4773} (\bibinfo {year} {2000})}\BibitemShut {NoStop}%
\bibitem [{\citenamefont {Agarwal}\ \emph {et~al.}(1998)\citenamefont
  {Agarwal}, \citenamefont {Bhargava},\ and\ \citenamefont
  {Mashelkar}}]{Agarwal1998}%
  \BibitemOpen
  \bibfield  {author} {\bibinfo {author} {\bibfnamefont {U.~S.}\ \bibnamefont
  {Agarwal}}, \bibinfo {author} {\bibfnamefont {R.}~\bibnamefont {Bhargava}}, \
  and\ \bibinfo {author} {\bibfnamefont {R.~A.}\ \bibnamefont {Mashelkar}},\
  }\bibfield  {title} {\enquote {\bibinfo {title} {Brownian dynamics
  simulations of a polymer molecule in solution under elongational flow},}\
  }\href@noop {} {\bibfield  {journal} {\bibinfo  {journal} {J. Chem. Phys.}\
  }\textbf {\bibinfo {volume} {108}},\ \bibinfo {pages} {1610--1617} (\bibinfo
  {year} {1998})}\BibitemShut {NoStop}%
\bibitem [{\citenamefont {Hernandez~Cifre}\ and\ \citenamefont {Garcia de~la
  Torre}(1999)}]{Cifre1999}%
  \BibitemOpen
  \bibfield  {author} {\bibinfo {author} {\bibfnamefont {J.~G.}\ \bibnamefont
  {Hernandez~Cifre}}\ and\ \bibinfo {author} {\bibfnamefont {J.}~\bibnamefont
  {Garcia de~la Torre}},\ }\bibfield  {title} {\enquote {\bibinfo {title}
  {Steady-state behavior of dilute polymers in elongational flow. dependence of
  the critical elongational rate on chain length, hydrodynamic interaction, and
  excluded volume},}\ }\href@noop {} {\bibfield  {journal} {\bibinfo  {journal}
  {Journal of Rheology}\ }\textbf {\bibinfo {volume} {43}},\ \bibinfo {pages}
  {339--358} (\bibinfo {year} {1999})}\BibitemShut {NoStop}%
\bibitem [{\citenamefont {Agarwal}(2000)}]{Agarwal2000}%
  \BibitemOpen
  \bibfield  {author} {\bibinfo {author} {\bibfnamefont {U.~S.}\ \bibnamefont
  {Agarwal}},\ }\bibfield  {title} {\enquote {\bibinfo {title} {Effect of
  initial conformation, flow strength, and hydrodynamic interaction on polymer
  molecules in extensional flows},}\ }\href@noop {} {\bibfield  {journal}
  {\bibinfo  {journal} {J. Chem. Phys.}\ }\textbf {\bibinfo {volume} {113}},\
  \bibinfo {pages} {3397--3403} (\bibinfo {year} {2000})}\BibitemShut {NoStop}%
\bibitem [{\citenamefont {Hernandez~Cifre}\ and\ \citenamefont {Garcia de~la
  Torre}(2001)}]{Cifre2001}%
  \BibitemOpen
  \bibfield  {author} {\bibinfo {author} {\bibfnamefont {J.~G.}\ \bibnamefont
  {Hernandez~Cifre}}\ and\ \bibinfo {author} {\bibfnamefont {J.}~\bibnamefont
  {Garcia de~la Torre}},\ }\bibfield  {title} {\enquote {\bibinfo {title}
  {Kinetic aspects of the coil-stretch transition of polymer chains in dilute
  solution under extensional flow},}\ }\href@noop {} {\bibfield  {journal}
  {\bibinfo  {journal} {J. Chem. Phys.}\ }\textbf {\bibinfo {volume} {115}},\
  \bibinfo {pages} {9578--9584} (\bibinfo {year} {2001})}\BibitemShut {NoStop}%
\bibitem [{\citenamefont {Neelov}\ \emph {et~al.}(2002)\citenamefont {Neelov},
  \citenamefont {Adolf}, \citenamefont {Lyulin},\ and\ \citenamefont
  {Davies}}]{Neelov2002}%
  \BibitemOpen
  \bibfield  {author} {\bibinfo {author} {\bibfnamefont {I.~M.}\ \bibnamefont
  {Neelov}}, \bibinfo {author} {\bibfnamefont {D.~B.}\ \bibnamefont {Adolf}},
  \bibinfo {author} {\bibfnamefont {A.~V.}\ \bibnamefont {Lyulin}}, \ and\
  \bibinfo {author} {\bibfnamefont {G.~R.}\ \bibnamefont {Davies}},\ }\bibfield
   {title} {\enquote {\bibinfo {title} {Brownian dynamics simulation of linear
  polymers under elongational flow: Bead-rod model with hydrodynamic
  interactions},}\ }\href@noop {} {\bibfield  {journal} {\bibinfo  {journal}
  {J. Chem. Phys.}\ }\textbf {\bibinfo {volume} {117}},\ \bibinfo {pages}
  {4030--4041} (\bibinfo {year} {2002})}\BibitemShut {NoStop}%
\bibitem [{\citenamefont {Jendrejack}\ \emph {et~al.}(2002)\citenamefont
  {Jendrejack}, \citenamefont {Graham},\ and\ \citenamefont
  {de~Pablo}}]{Jendrejack2002}%
  \BibitemOpen
  \bibfield  {author} {\bibinfo {author} {\bibfnamefont {R.~M.}\ \bibnamefont
  {Jendrejack}}, \bibinfo {author} {\bibfnamefont {M.~D.}\ \bibnamefont
  {Graham}}, \ and\ \bibinfo {author} {\bibfnamefont {J.~J.}\ \bibnamefont
  {de~Pablo}},\ }\bibfield  {title} {\enquote {\bibinfo {title} {Stochastic
  simulations of dna in flow: Dynamics and the effects of hydrodynamic
  interactions},}\ }\href@noop {} {\bibfield  {journal} {\bibinfo  {journal}
  {J. Chem. Phys.}\ }\textbf {\bibinfo {volume} {116}},\ \bibinfo {pages}
  {7752--7759} (\bibinfo {year} {2002})}\BibitemShut {NoStop}%
\bibitem [{\citenamefont {Hsieh}\ and\ \citenamefont
  {Larson}(2003)}]{Hsieh2003}%
  \BibitemOpen
  \bibfield  {author} {\bibinfo {author} {\bibfnamefont {C.-C.}\ \bibnamefont
  {Hsieh}}\ and\ \bibinfo {author} {\bibfnamefont {R.~G.}\ \bibnamefont
  {Larson}},\ }\bibfield  {title} {\enquote {\bibinfo {title} {Modeling
  hydrodynamic interaction in brownian dynamics: Simulations of extensional
  flows of dilute solutions of dna and polystyrene},}\ }\href@noop {}
  {\bibfield  {journal} {\bibinfo  {journal} {Journal of Non-Newtonian Fluid
  Mechanics}\ }\textbf {\bibinfo {volume} {113}},\ \bibinfo {pages} {147--191}
  (\bibinfo {year} {2003})}\BibitemShut {NoStop}%
\bibitem [{\citenamefont {Somasi}\ \emph {et~al.}(2002)\citenamefont {Somasi},
  \citenamefont {Khomami}, \citenamefont {Woo}, \citenamefont {Hur},\ and\
  \citenamefont {Shaqfeh}}]{Somasi2002}%
  \BibitemOpen
  \bibfield  {author} {\bibinfo {author} {\bibfnamefont {M.}~\bibnamefont
  {Somasi}}, \bibinfo {author} {\bibfnamefont {B.}~\bibnamefont {Khomami}},
  \bibinfo {author} {\bibfnamefont {N.~J.}\ \bibnamefont {Woo}}, \bibinfo
  {author} {\bibfnamefont {J.~S.}\ \bibnamefont {Hur}}, \ and\ \bibinfo
  {author} {\bibfnamefont {E.~S.~G.}\ \bibnamefont {Shaqfeh}},\ }\bibfield
  {title} {\enquote {\bibinfo {title} {Brownian dynamics simulations of
  bead-rod and bead-spring chains: Numerical algorithms and coarse graining
  issues},}\ }\href@noop {} {\bibfield  {journal} {\bibinfo  {journal} {Journal
  of Non-Newtonian Fluid Mechanics}\ }\textbf {\bibinfo {volume} {108}},\
  \bibinfo {pages} {227--255} (\bibinfo {year} {2002})}\BibitemShut {NoStop}%
\bibitem [{\citenamefont {Batchelor}(1970)}]{Batchelor1970}%
  \BibitemOpen
  \bibfield  {author} {\bibinfo {author} {\bibfnamefont {G.~K.}\ \bibnamefont
  {Batchelor}},\ }\bibfield  {title} {\enquote {\bibinfo {title} {Slender-body
  theory for particles of arbitrary cross-section in stokes flow},}\
  }\href@noop {} {\bibfield  {journal} {\bibinfo  {journal} {J. Fluid. Mech.}\
  }\textbf {\bibinfo {volume} {44}},\ \bibinfo {pages} {419--440} (\bibinfo
  {year} {1970})}\BibitemShut {NoStop}%
\bibitem [{\citenamefont {de~Gennes}(1974)}]{deGennes1974}%
  \BibitemOpen
  \bibfield  {author} {\bibinfo {author} {\bibfnamefont {P.~G.}\ \bibnamefont
  {de~Gennes}},\ }\bibfield  {title} {\enquote {\bibinfo {title} {Coil-stretch
  transition of dilute flexible polymers under ultra-high velocity
  gradients},}\ }\href@noop {} {\bibfield  {journal} {\bibinfo  {journal} {J.
  Chem. Phys.}\ }\textbf {\bibinfo {volume} {60}},\ \bibinfo {pages}
  {5030--5042} (\bibinfo {year} {1974})}\BibitemShut {NoStop}%
\bibitem [{\citenamefont {Tanner}(1975)}]{Tanner1975}%
  \BibitemOpen
  \bibfield  {author} {\bibinfo {author} {\bibfnamefont {R.~I.}\ \bibnamefont
  {Tanner}},\ }\bibfield  {title} {\enquote {\bibinfo {title} {Stresses in
  dilute solutions of bead-nonlinear-spring macromolecules. iii. friction
  coefficient varying with dumbbell extension},}\ }\href@noop {} {\bibfield
  {journal} {\bibinfo  {journal} {Trans. Soc. Rheology}\ }\textbf {\bibinfo
  {volume} {19}},\ \bibinfo {pages} {557--582} (\bibinfo {year}
  {1975})}\BibitemShut {NoStop}%
\bibitem [{\citenamefont {Hinch}(1977)}]{Hinch1977}%
  \BibitemOpen
  \bibfield  {author} {\bibinfo {author} {\bibfnamefont {E.~J.}\ \bibnamefont
  {Hinch}},\ }\bibfield  {title} {\enquote {\bibinfo {title} {Mechanical models
  of dilute polymer solutions in strong flows},}\ }\href@noop {} {\bibfield
  {journal} {\bibinfo  {journal} {Phys. Fluids}\ }\textbf {\bibinfo {volume}
  {20}},\ \bibinfo {pages} {S22--S30} (\bibinfo {year} {1977})}\BibitemShut
  {NoStop}%
\bibitem [{\citenamefont {Phan-Thien}\ \emph {et~al.}(1984)\citenamefont
  {Phan-Thien}, \citenamefont {Manero},\ and\ \citenamefont {Leal}}]{Leal1984}%
  \BibitemOpen
  \bibfield  {author} {\bibinfo {author} {\bibfnamefont {N.}~\bibnamefont
  {Phan-Thien}}, \bibinfo {author} {\bibfnamefont {O.}~\bibnamefont {Manero}},
  \ and\ \bibinfo {author} {\bibfnamefont {L.~G.}\ \bibnamefont {Leal}},\
  }\bibfield  {title} {\enquote {\bibinfo {title} {A study of
  conformation-dependent friction in a dumbbell model for dilute solutions},}\
  }\href@noop {} {\bibfield  {journal} {\bibinfo  {journal} {Rheo. Acta.}\
  }\textbf {\bibinfo {volume} {23}},\ \bibinfo {pages} {151--162} (\bibinfo
  {year} {1984})}\BibitemShut {NoStop}%
\bibitem [{\citenamefont {Fan}\ \emph {et~al.}(1985)\citenamefont {Fan},
  \citenamefont {Bird},\ and\ \citenamefont {Renardy}}]{Fan1985}%
  \BibitemOpen
  \bibfield  {author} {\bibinfo {author} {\bibfnamefont {X.-J.}\ \bibnamefont
  {Fan}}, \bibinfo {author} {\bibfnamefont {R.~B.}\ \bibnamefont {Bird}}, \
  and\ \bibinfo {author} {\bibfnamefont {M.}~\bibnamefont {Renardy}},\
  }\bibfield  {title} {\enquote {\bibinfo {title} {Configuration-dependent
  friction coefficients and elastic dumbbell rheology},}\ }\href@noop {}
  {\bibfield  {journal} {\bibinfo  {journal} {Journal of Non-Newtonian Fluid
  Mechanics}\ }\textbf {\bibinfo {volume} {18}},\ \bibinfo {pages} {255--272}
  (\bibinfo {year} {1985})}\BibitemShut {NoStop}%
\bibitem [{\citenamefont {Magda}\ \emph {et~al.}(1988)\citenamefont {Magda},
  \citenamefont {Larson},\ and\ \citenamefont {Mackay}}]{Magda1988}%
  \BibitemOpen
  \bibfield  {author} {\bibinfo {author} {\bibfnamefont {J.~J.}\ \bibnamefont
  {Magda}}, \bibinfo {author} {\bibfnamefont {R.~G.}\ \bibnamefont {Larson}}, \
  and\ \bibinfo {author} {\bibfnamefont {M.~E.}\ \bibnamefont {Mackay}},\
  }\bibfield  {title} {\enquote {\bibinfo {title} {Deformation-dependent
  hydrodynamic interaction in flows of dilute polymer solutions},}\ }\href@noop
  {} {\bibfield  {journal} {\bibinfo  {journal} {J. Chem. Phys.}\ }\textbf
  {\bibinfo {volume} {89}},\ \bibinfo {pages} {2504--2513} (\bibinfo {year}
  {1988})}\BibitemShut {NoStop}%
\bibitem [{\citenamefont {Wiest}\ \emph {et~al.}(1989)\citenamefont {Wiest},
  \citenamefont {Wedgewood},\ and\ \citenamefont {Bird}}]{Wiest1989}%
  \BibitemOpen
  \bibfield  {author} {\bibinfo {author} {\bibfnamefont {J.~M.}\ \bibnamefont
  {Wiest}}, \bibinfo {author} {\bibfnamefont {L.~E.}\ \bibnamefont
  {Wedgewood}}, \ and\ \bibinfo {author} {\bibfnamefont {R.~B.}\ \bibnamefont
  {Bird}},\ }\bibfield  {title} {\enquote {\bibinfo {title} {On coil-stretch
  transitions in dilute polymer solutions},}\ }\href@noop {} {\bibfield
  {journal} {\bibinfo  {journal} {J. Chem. Phys.}\ }\textbf {\bibinfo {volume}
  {90}},\ \bibinfo {pages} {587--594} (\bibinfo {year} {1989})}\BibitemShut
  {NoStop}%
\bibitem [{\citenamefont {Carrington}\ and\ \citenamefont
  {Odell}(1996)}]{Carrington1996}%
  \BibitemOpen
  \bibfield  {author} {\bibinfo {author} {\bibfnamefont {S.~P.}\ \bibnamefont
  {Carrington}}\ and\ \bibinfo {author} {\bibfnamefont {J.~A.}\ \bibnamefont
  {Odell}},\ }\bibfield  {title} {\enquote {\bibinfo {title} {How do polymers
  stretch in stagnation point extensional flow fields?}}\ }\href@noop {}
  {\bibfield  {journal} {\bibinfo  {journal} {Journal of Non-Newtonian Fluid
  Mechanics}\ }\textbf {\bibinfo {volume} {67}},\ \bibinfo {pages} {269--283}
  (\bibinfo {year} {1996})}\BibitemShut {NoStop}%
\bibitem [{\citenamefont {Schroeder}\ \emph {et~al.}(2003)\citenamefont
  {Schroeder}, \citenamefont {Babcock}, \citenamefont {Shaqfeh},\ and\
  \citenamefont {Chu}}]{Schroeder2003}%
  \BibitemOpen
  \bibfield  {author} {\bibinfo {author} {\bibfnamefont {C.~M.}\ \bibnamefont
  {Schroeder}}, \bibinfo {author} {\bibfnamefont {H.~P.}\ \bibnamefont
  {Babcock}}, \bibinfo {author} {\bibfnamefont {E.~S.~G.}\ \bibnamefont
  {Shaqfeh}}, \ and\ \bibinfo {author} {\bibfnamefont {S.}~\bibnamefont
  {Chu}},\ }\bibfield  {title} {\enquote {\bibinfo {title} {Observation of
  polymer conformation hysteresis in extensional flow},}\ }\href@noop {}
  {\bibfield  {journal} {\bibinfo  {journal} {Science}\ }\textbf {\bibinfo
  {volume} {301}},\ \bibinfo {pages} {1515--1519} (\bibinfo {year}
  {2003})}\BibitemShut {NoStop}%
\bibitem [{\citenamefont {Schroeder}\ \emph {et~al.}(2004)\citenamefont
  {Schroeder}, \citenamefont {Shaqfeh},\ and\ \citenamefont
  {Chu}}]{Schroeder2004}%
  \BibitemOpen
  \bibfield  {author} {\bibinfo {author} {\bibfnamefont {C.~M.}\ \bibnamefont
  {Schroeder}}, \bibinfo {author} {\bibfnamefont {E.~S.~G.}\ \bibnamefont
  {Shaqfeh}}, \ and\ \bibinfo {author} {\bibfnamefont {S.}~\bibnamefont
  {Chu}},\ }\bibfield  {title} {\enquote {\bibinfo {title} {Effect of
  hydrodynamic interactions on dna dynamics in extensional flow: Simulation and
  single molecule experiment},}\ }\href@noop {} {\bibfield  {journal} {\bibinfo
   {journal} {Macromolecules}\ }\textbf {\bibinfo {volume} {37}},\ \bibinfo
  {pages} {9242--9256} (\bibinfo {year} {2004})}\BibitemShut {NoStop}%
\bibitem [{\citenamefont {Shenoy}\ \emph {et~al.}(2016)\citenamefont {Shenoy},
  \citenamefont {Rao},\ and\ \citenamefont {Schroeder}}]{Shenoy2016}%
  \BibitemOpen
  \bibfield  {author} {\bibinfo {author} {\bibfnamefont {A.}~\bibnamefont
  {Shenoy}}, \bibinfo {author} {\bibfnamefont {C.~V.}\ \bibnamefont {Rao}}, \
  and\ \bibinfo {author} {\bibfnamefont {C.~M}\ \bibnamefont {Schroeder}},\
  }\bibfield  {title} {\enquote {\bibinfo {title} {Stokes trap for multiplexed
  particle manipulation and assembly using fluidics},}\ }\href@noop {}
  {\bibfield  {journal} {\bibinfo  {journal} {Proc. Nat. Acad. Sci}\ }\textbf
  {\bibinfo {volume} {113}},\ \bibinfo {pages} {3976} (\bibinfo {year}
  {2016})}\BibitemShut {NoStop}%
\bibitem [{\citenamefont {Hsieh}\ and\ \citenamefont
  {Larson}(2005)}]{Hsieh2005}%
  \BibitemOpen
  \bibfield  {author} {\bibinfo {author} {\bibfnamefont {C.-C.}\ \bibnamefont
  {Hsieh}}\ and\ \bibinfo {author} {\bibfnamefont {R.~G.}\ \bibnamefont
  {Larson}},\ }\bibfield  {title} {\enquote {\bibinfo {title} {Prediction of
  coil-stretch hysteresis for dilute polystyrene molecules in extensional
  flow},}\ }\href@noop {} {\bibfield  {journal} {\bibinfo  {journal} {Journal
  of Rheology}\ }\textbf {\bibinfo {volume} {49}},\ \bibinfo {pages}
  {1081--1089} (\bibinfo {year} {2005})}\BibitemShut {NoStop}%
\bibitem [{\citenamefont {Beck}\ and\ \citenamefont
  {Shaqfeh}(2007)}]{Beck2007}%
  \BibitemOpen
  \bibfield  {author} {\bibinfo {author} {\bibfnamefont {V.~A.}\ \bibnamefont
  {Beck}}\ and\ \bibinfo {author} {\bibfnamefont {E.~S.~G.}\ \bibnamefont
  {Shaqfeh}},\ }\bibfield  {title} {\enquote {\bibinfo {title}
  {Ergodicity-breaking and the unraveling dynamics of a polymer in linear and
  nonlinear extensional flows},}\ }\href@noop {} {\bibfield  {journal}
  {\bibinfo  {journal} {Journal of Rheology}\ }\textbf {\bibinfo {volume}
  {51}},\ \bibinfo {pages} {561--574} (\bibinfo {year} {2007})}\BibitemShut
  {NoStop}%
\bibitem [{\citenamefont {Somani}\ \emph {et~al.}(2010)\citenamefont {Somani},
  \citenamefont {Shaqfeh},\ and\ \citenamefont {Prakash}}]{Somani2010}%
  \BibitemOpen
  \bibfield  {author} {\bibinfo {author} {\bibfnamefont {S.}~\bibnamefont
  {Somani}}, \bibinfo {author} {\bibfnamefont {E.~S.~G.}\ \bibnamefont
  {Shaqfeh}}, \ and\ \bibinfo {author} {\bibfnamefont {J.~R.}\ \bibnamefont
  {Prakash}},\ }\bibfield  {title} {\enquote {\bibinfo {title} {Effect of
  solvent quality on the coil-stretch transition},}\ }\href@noop {} {\bibfield
  {journal} {\bibinfo  {journal} {Macromolecules}\ }\textbf {\bibinfo {volume}
  {43}},\ \bibinfo {pages} {10679--10691} (\bibinfo {year} {2010})}\BibitemShut
  {NoStop}%
\bibitem [{\citenamefont {Sridhar}\ \emph {et~al.}(2007)\citenamefont
  {Sridhar}, \citenamefont {Nguyen}, \citenamefont {Prabhakar},\ and\
  \citenamefont {Prakash}}]{Sridhar2007}%
  \BibitemOpen
  \bibfield  {author} {\bibinfo {author} {\bibfnamefont {T.}~\bibnamefont
  {Sridhar}}, \bibinfo {author} {\bibfnamefont {D.~A.}\ \bibnamefont {Nguyen}},
  \bibinfo {author} {\bibfnamefont {R.}~\bibnamefont {Prabhakar}}, \ and\
  \bibinfo {author} {\bibfnamefont {J.~R.}\ \bibnamefont {Prakash}},\
  }\bibfield  {title} {\enquote {\bibinfo {title} {Rheological observation of
  glassy dynamics of dilute polymer solutions near the coil-stretch transition
  in elongational flows},}\ }\href@noop {} {\bibfield  {journal} {\bibinfo
  {journal} {Phys. Rev. Lett.}\ }\textbf {\bibinfo {volume} {98}},\ \bibinfo
  {pages} {167801} (\bibinfo {year} {2007})}\BibitemShut {NoStop}%
\bibitem [{\citenamefont {Prabhaker}\ \emph {et~al.}(2017)\citenamefont
  {Prabhaker}, \citenamefont {Sasmal}, \citenamefont {Nguyen}, \citenamefont
  {Sridhar},\ and\ \citenamefont {Prakash}}]{Prabhakar2017}%
  \BibitemOpen
  \bibfield  {author} {\bibinfo {author} {\bibfnamefont {R.}~\bibnamefont
  {Prabhaker}}, \bibinfo {author} {\bibfnamefont {C.}~\bibnamefont {Sasmal}},
  \bibinfo {author} {\bibfnamefont {D.~A.}\ \bibnamefont {Nguyen}}, \bibinfo
  {author} {\bibfnamefont {T.}~\bibnamefont {Sridhar}}, \ and\ \bibinfo
  {author} {\bibfnamefont {J.~R.}\ \bibnamefont {Prakash}},\ }\bibfield
  {title} {\enquote {\bibinfo {title} {Effect of stretching-induced changes in
  hydrodynamic screening on coil-stretch hysteresis of unentangled polymer
  solutions},}\ }\href@noop {} {\bibfield  {journal} {\bibinfo  {journal}
  {Physical Review Fluids}\ }\textbf {\bibinfo {volume} {2}},\ \bibinfo {pages}
  {011301(R)} (\bibinfo {year} {2017})}\BibitemShut {NoStop}%
\bibitem [{\citenamefont {Prabhaker}\ \emph {et~al.}(2016)\citenamefont
  {Prabhaker}, \citenamefont {Gadkari}, \citenamefont {Gopesh},\ and\
  \citenamefont {Shaw}}]{Prabhakar2016}%
  \BibitemOpen
  \bibfield  {author} {\bibinfo {author} {\bibfnamefont {R.}~\bibnamefont
  {Prabhaker}}, \bibinfo {author} {\bibfnamefont {S.}~\bibnamefont {Gadkari}},
  \bibinfo {author} {\bibfnamefont {T.}~\bibnamefont {Gopesh}}, \ and\ \bibinfo
  {author} {\bibfnamefont {M.~J.}\ \bibnamefont {Shaw}},\ }\bibfield  {title}
  {\enquote {\bibinfo {title} {Influence of stretching induced
  self-concentration and self-dilution on coil-stretch hysteresis and capillary
  thinning of unentangled polymer solutions},}\ }\href@noop {} {\bibfield
  {journal} {\bibinfo  {journal} {J. Rheol.}\ }\textbf {\bibinfo {volume}
  {60}},\ \bibinfo {pages} {345--366} (\bibinfo {year} {2016})}\BibitemShut
  {NoStop}%
\bibitem [{\citenamefont {Celani}\ \emph {et~al.}(2006)\citenamefont {Celani},
  \citenamefont {Puliafito},\ and\ \citenamefont {Vincenzi}}]{Celani2006}%
  \BibitemOpen
  \bibfield  {author} {\bibinfo {author} {\bibfnamefont {A.}~\bibnamefont
  {Celani}}, \bibinfo {author} {\bibfnamefont {A.}~\bibnamefont {Puliafito}}, \
  and\ \bibinfo {author} {\bibfnamefont {D.}~\bibnamefont {Vincenzi}},\
  }\bibfield  {title} {\enquote {\bibinfo {title} {Dynamical slowdown of
  polymers in laminar and random flows},}\ }\href@noop {} {\bibfield  {journal}
  {\bibinfo  {journal} {Phys. Rev. Lett.}\ }\textbf {\bibinfo {volume} {97}},\
  \bibinfo {pages} {118301} (\bibinfo {year} {2006})}\BibitemShut {NoStop}%
\bibitem [{\citenamefont {Gerashchenko}\ and\ \citenamefont
  {Steinberg}(2008)}]{Steinberg2008}%
  \BibitemOpen
  \bibfield  {author} {\bibinfo {author} {\bibfnamefont {S.}~\bibnamefont
  {Gerashchenko}}\ and\ \bibinfo {author} {\bibfnamefont {V.}~\bibnamefont
  {Steinberg}},\ }\bibfield  {title} {\enquote {\bibinfo {title} {Critical
  slowing down in polymer dynamics near the coil-stretch transition in
  elongational flow},}\ }\href@noop {} {\bibfield  {journal} {\bibinfo
  {journal} {Phys. Rev. E}\ }\textbf {\bibinfo {volume} {78}},\ \bibinfo
  {pages} {040801} (\bibinfo {year} {2008})}\BibitemShut {NoStop}%
\bibitem [{\citenamefont {Reif}(1965)}]{Reifbook}%
  \BibitemOpen
  \bibfield  {author} {\bibinfo {author} {\bibfnamefont {F.}~\bibnamefont
  {Reif}},\ }\href@noop {} {\emph {\bibinfo {title} {Fundamentals of
  Statistical and Thermal Physics}}}\ (\bibinfo  {publisher} {McGraw-Hill},\
  \bibinfo {year} {1965})\BibitemShut {NoStop}%
\bibitem [{\citenamefont {Volkmuth}\ and\ \citenamefont
  {Austin}(1992)}]{Volkmuth1992}%
  \BibitemOpen
  \bibfield  {author} {\bibinfo {author} {\bibfnamefont {W.~D.}\ \bibnamefont
  {Volkmuth}}\ and\ \bibinfo {author} {\bibfnamefont {R.~H.}\ \bibnamefont
  {Austin}},\ }\bibfield  {title} {\enquote {\bibinfo {title} {Dna
  electrophoresis in microlithographic arrays},}\ }\href@noop {} {\bibfield
  {journal} {\bibinfo  {journal} {Nature}\ }\textbf {\bibinfo {volume} {358}},\
  \bibinfo {pages} {600--602} (\bibinfo {year} {1992})}\BibitemShut {NoStop}%
\bibitem [{\citenamefont {Randall}\ and\ \citenamefont
  {Doyle}(2004)}]{Randall2004}%
  \BibitemOpen
  \bibfield  {author} {\bibinfo {author} {\bibfnamefont {G.~C.}\ \bibnamefont
  {Randall}}\ and\ \bibinfo {author} {\bibfnamefont {P.~S.}\ \bibnamefont
  {Doyle}},\ }\bibfield  {title} {\enquote {\bibinfo {title} {Electrophoretic
  collision of a dna molecule with an insulating post},}\ }\href@noop {}
  {\bibfield  {journal} {\bibinfo  {journal} {Phys. Rev. Lett.}\ }\textbf
  {\bibinfo {volume} {93}},\ \bibinfo {pages} {058102} (\bibinfo {year}
  {2004})}\BibitemShut {NoStop}%
\bibitem [{\citenamefont {Randall}\ and\ \citenamefont
  {Doyle}(2005)}]{Randall2005}%
  \BibitemOpen
  \bibfield  {author} {\bibinfo {author} {\bibfnamefont {G.~C.}\ \bibnamefont
  {Randall}}\ and\ \bibinfo {author} {\bibfnamefont {P.~S.}\ \bibnamefont
  {Doyle}},\ }\bibfield  {title} {\enquote {\bibinfo {title} {Dna deformation
  in electric fields:  dna driven past a cylindrical obstruction},}\
  }\href@noop {} {\bibfield  {journal} {\bibinfo  {journal} {Macromolecules}\
  }\textbf {\bibinfo {volume} {38}},\ \bibinfo {pages} {2410--2418} (\bibinfo
  {year} {2005})}\BibitemShut {NoStop}%
\bibitem [{\citenamefont {Randall}\ and\ \citenamefont
  {Doyle}(2006)}]{Randall2006}%
  \BibitemOpen
  \bibfield  {author} {\bibinfo {author} {\bibfnamefont {G.~C.}\ \bibnamefont
  {Randall}}\ and\ \bibinfo {author} {\bibfnamefont {P.~S.}\ \bibnamefont
  {Doyle}},\ }\bibfield  {title} {\enquote {\bibinfo {title} {Collision of a
  dna polymer with a small obstacle},}\ }\href@noop {} {\bibfield  {journal}
  {\bibinfo  {journal} {Macromolecules}\ }\textbf {\bibinfo {volume} {39}},\
  \bibinfo {pages} {7734--7745} (\bibinfo {year} {2006})}\BibitemShut {NoStop}%
\bibitem [{\citenamefont {Volkmuth}\ \emph {et~al.}(1994)\citenamefont
  {Volkmuth}, \citenamefont {Duke}, \citenamefont {Wu}, \citenamefont
  {Austin},\ and\ \citenamefont {Szabo}}]{Volkmuth1994}%
  \BibitemOpen
  \bibfield  {author} {\bibinfo {author} {\bibfnamefont {W.~D.}\ \bibnamefont
  {Volkmuth}}, \bibinfo {author} {\bibfnamefont {T.}~\bibnamefont {Duke}},
  \bibinfo {author} {\bibfnamefont {M.~C.}\ \bibnamefont {Wu}}, \bibinfo
  {author} {\bibfnamefont {R.~H.}\ \bibnamefont {Austin}}, \ and\ \bibinfo
  {author} {\bibfnamefont {A.}~\bibnamefont {Szabo}},\ }\bibfield  {title}
  {\enquote {\bibinfo {title} {Dna electrodiffusion in a 2d array of posts},}\
  }\href@noop {} {\bibfield  {journal} {\bibinfo  {journal} {Phys. Rev. Lett.}\
  }\textbf {\bibinfo {volume} {72}},\ \bibinfo {pages} {2117--2120} (\bibinfo
  {year} {1994})}\BibitemShut {NoStop}%
\bibitem [{\citenamefont {Teclemariam}\ \emph {et~al.}(2007)\citenamefont
  {Teclemariam}, \citenamefont {Beck}, \citenamefont {Shaqfeh},\ and\
  \citenamefont {Muller}}]{Teclemariam2007}%
  \BibitemOpen
  \bibfield  {author} {\bibinfo {author} {\bibfnamefont {N.~P.}\ \bibnamefont
  {Teclemariam}}, \bibinfo {author} {\bibfnamefont {V.~A.}\ \bibnamefont
  {Beck}}, \bibinfo {author} {\bibfnamefont {E.~S.~G.}\ \bibnamefont
  {Shaqfeh}}, \ and\ \bibinfo {author} {\bibfnamefont {S.~J.}\ \bibnamefont
  {Muller}},\ }\bibfield  {title} {\enquote {\bibinfo {title} {Dynamics of dna
  polymers in post arrays: Comparison of single molecule experiments and
  simulations},}\ }\href@noop {} {\bibfield  {journal} {\bibinfo  {journal}
  {Macromolecules}\ }\textbf {\bibinfo {volume} {40}},\ \bibinfo {pages}
  {3848--3859} (\bibinfo {year} {2007})}\BibitemShut {NoStop}%
\bibitem [{\citenamefont {Zhou}\ \emph {et~al.}(2017)\citenamefont {Zhou},
  \citenamefont {Wang}, \citenamefont {Menard}, \citenamefont {Panyukov},
  \citenamefont {Rubinstein},\ and\ \citenamefont {Ramsey}}]{Zhou2017}%
  \BibitemOpen
  \bibfield  {author} {\bibinfo {author} {\bibfnamefont {J.}~\bibnamefont
  {Zhou}}, \bibinfo {author} {\bibfnamefont {Y.}~\bibnamefont {Wang}}, \bibinfo
  {author} {\bibfnamefont {L.~D.}\ \bibnamefont {Menard}}, \bibinfo {author}
  {\bibfnamefont {S.}~\bibnamefont {Panyukov}}, \bibinfo {author}
  {\bibfnamefont {M.}~\bibnamefont {Rubinstein}}, \ and\ \bibinfo {author}
  {\bibfnamefont {J.~M.}\ \bibnamefont {Ramsey}},\ }\bibfield  {title}
  {\enquote {\bibinfo {title} {Enhanced nanochannel translocation and
  localization of genomic dna molecules using three-dimensional nanofunnels},}\
  }\href@noop {} {\bibfield  {journal} {\bibinfo  {journal} {Nature
  Communications}\ }\textbf {\bibinfo {volume} {8}} (\bibinfo {year}
  {2017})}\BibitemShut {NoStop}%
\bibitem [{\citenamefont {Brockman}\ \emph {et~al.}(2011)\citenamefont
  {Brockman}, \citenamefont {Kim},\ and\ \citenamefont
  {Schroeder}}]{Brockman2011}%
  \BibitemOpen
  \bibfield  {author} {\bibinfo {author} {\bibfnamefont {C.~A.}\ \bibnamefont
  {Brockman}}, \bibinfo {author} {\bibfnamefont {S.~J.}\ \bibnamefont {Kim}}, \
  and\ \bibinfo {author} {\bibfnamefont {C.~M.}\ \bibnamefont {Schroeder}},\
  }\bibfield  {title} {\enquote {\bibinfo {title} {Direct observation of single
  flexible polymers using single stranded dna},}\ }\href@noop {} {\bibfield
  {journal} {\bibinfo  {journal} {Soft Matter}\ }\textbf {\bibinfo {volume}
  {7}},\ \bibinfo {pages} {8005--8012} (\bibinfo {year} {2011})}\BibitemShut
  {NoStop}%
\bibitem [{\citenamefont {McIntosh}\ \emph {et~al.}(2014)\citenamefont
  {McIntosh}, \citenamefont {Duggan}, \citenamefont {Gouil},\ and\
  \citenamefont {Saleh}}]{McIntosh2014}%
  \BibitemOpen
  \bibfield  {author} {\bibinfo {author} {\bibfnamefont {D.~B.}\ \bibnamefont
  {McIntosh}}, \bibinfo {author} {\bibfnamefont {G.}~\bibnamefont {Duggan}},
  \bibinfo {author} {\bibfnamefont {Q.}~\bibnamefont {Gouil}}, \ and\ \bibinfo
  {author} {\bibfnamefont {O.~A.}\ \bibnamefont {Saleh}},\ }\bibfield  {title}
  {\enquote {\bibinfo {title} {Sequence-dependent elasticity and electrostatics
  of single-stranded dna: Signatures of base-stacking},}\ }\href@noop {}
  {\bibfield  {journal} {\bibinfo  {journal} {Biophysical Journal}\ }\textbf
  {\bibinfo {volume} {106}},\ \bibinfo {pages} {659--666} (\bibinfo {year}
  {2014})}\BibitemShut {NoStop}%
\bibitem [{\citenamefont {Marciel}\ and\ \citenamefont
  {Schroeder}(2013)}]{Marciel2013}%
  \BibitemOpen
  \bibfield  {author} {\bibinfo {author} {\bibfnamefont {A.~B.}\ \bibnamefont
  {Marciel}}\ and\ \bibinfo {author} {\bibfnamefont {C.~M.}\ \bibnamefont
  {Schroeder}},\ }\bibfield  {title} {\enquote {\bibinfo {title} {New
  directions in single polymer dynamics},}\ }\href@noop {} {\bibfield
  {journal} {\bibinfo  {journal} {Journal of Polymer Science B: Polymer
  Physics}\ }\textbf {\bibinfo {volume} {51}},\ \bibinfo {pages} {556--566}
  (\bibinfo {year} {2013})}\BibitemShut {NoStop}%
\bibitem [{\citenamefont {Ewoldt}\ \emph {et~al.}(2010)\citenamefont {Ewoldt},
  \citenamefont {Winter}, \citenamefont {Maxey},\ and\ \citenamefont
  {McKinley}}]{Ewoldt2010}%
  \BibitemOpen
  \bibfield  {author} {\bibinfo {author} {\bibfnamefont {R.~H.}\ \bibnamefont
  {Ewoldt}}, \bibinfo {author} {\bibfnamefont {H.~P.}\ \bibnamefont {Winter}},
  \bibinfo {author} {\bibfnamefont {J.}~\bibnamefont {Maxey}}, \ and\ \bibinfo
  {author} {\bibfnamefont {G.~H.}\ \bibnamefont {McKinley}},\ }\bibfield
  {title} {\enquote {\bibinfo {title} {Large amplitude oscillatory shear of
  pseudoplastic and elastoviscoplastic materials},}\ }\href@noop {} {\bibfield
  {journal} {\bibinfo  {journal} {Rheo. Acta.}\ }\textbf {\bibinfo {volume}
  {49}},\ \bibinfo {pages} {191--212} (\bibinfo {year} {2010})}\BibitemShut
  {NoStop}%
\bibitem [{\citenamefont {Dimitriou}\ \emph {et~al.}(2013)\citenamefont
  {Dimitriou}, \citenamefont {Ewoldt},\ and\ \citenamefont
  {McKinley}}]{Ewoldt2013}%
  \BibitemOpen
  \bibfield  {author} {\bibinfo {author} {\bibfnamefont {C.~J.}\ \bibnamefont
  {Dimitriou}}, \bibinfo {author} {\bibfnamefont {R.~H.}\ \bibnamefont
  {Ewoldt}}, \ and\ \bibinfo {author} {\bibfnamefont {G.~H.}\ \bibnamefont
  {McKinley}},\ }\bibfield  {title} {\enquote {\bibinfo {title} {Describing and
  prescribing the constitutive response of yield stress fluids using large
  amplitude oscillatory shear stress (laostress)},}\ }\href@noop {} {\bibfield
  {journal} {\bibinfo  {journal} {Journal of Rheology}\ }\textbf {\bibinfo
  {volume} {57}},\ \bibinfo {pages} {27--70} (\bibinfo {year}
  {2013})}\BibitemShut {NoStop}%
\bibitem [{\citenamefont {Rogers}(2012)}]{Rogers2012a}%
  \BibitemOpen
  \bibfield  {author} {\bibinfo {author} {\bibfnamefont {S.~A.}\ \bibnamefont
  {Rogers}},\ }\bibfield  {title} {\enquote {\bibinfo {title} {A sequence of
  physical processes determined and quantified in laos: An instantaneous local
  2d/3d approach},}\ }\href@noop {} {\bibfield  {journal} {\bibinfo  {journal}
  {Journal of Rheology}\ }\textbf {\bibinfo {volume} {56}},\ \bibinfo {pages}
  {1129--1151} (\bibinfo {year} {2012})}\BibitemShut {NoStop}%
\bibitem [{\citenamefont {Rogers}\ and\ \citenamefont
  {Lettinga}(2012)}]{Rogers2012b}%
  \BibitemOpen
  \bibfield  {author} {\bibinfo {author} {\bibfnamefont {S.~A.}\ \bibnamefont
  {Rogers}}\ and\ \bibinfo {author} {\bibfnamefont {P.}~\bibnamefont
  {Lettinga}},\ }\bibfield  {title} {\enquote {\bibinfo {title} {A sequence of
  physical processes determined and quantified in laos: Application to
  theoretical nonlinear models},}\ }\href@noop {} {\bibfield  {journal}
  {\bibinfo  {journal} {Journal of Rheology}\ }\textbf {\bibinfo {volume}
  {56}},\ \bibinfo {pages} {1--25} (\bibinfo {year} {2012})}\BibitemShut
  {NoStop}%
\bibitem [{\citenamefont {Zhou}\ and\ \citenamefont
  {Schroeder}(2016{\natexlab{a}})}]{Zhou2016a}%
  \BibitemOpen
  \bibfield  {author} {\bibinfo {author} {\bibfnamefont {Y.}~\bibnamefont
  {Zhou}}\ and\ \bibinfo {author} {\bibfnamefont {C.~M.}\ \bibnamefont
  {Schroeder}},\ }\bibfield  {title} {\enquote {\bibinfo {title} {Single
  polymer dynamics under large amplitude oscillatory extension},}\ }\href@noop
  {} {\bibfield  {journal} {\bibinfo  {journal} {Phys. Rev. Fluids}\ }\textbf
  {\bibinfo {volume} {1}},\ \bibinfo {pages} {053301} (\bibinfo {year}
  {2016}{\natexlab{a}})}\BibitemShut {NoStop}%
\bibitem [{\citenamefont {Zhou}\ and\ \citenamefont
  {Schroeder}(2016{\natexlab{b}})}]{Zhou2016b}%
  \BibitemOpen
  \bibfield  {author} {\bibinfo {author} {\bibfnamefont {Y.}~\bibnamefont
  {Zhou}}\ and\ \bibinfo {author} {\bibfnamefont {C.~M.}\ \bibnamefont
  {Schroeder}},\ }\bibfield  {title} {\enquote {\bibinfo {title} {Transient and
  average unsteady dynamics of single polymers in large-amplitude oscillatory
  extension},}\ }\href@noop {} {\bibfield  {journal} {\bibinfo  {journal}
  {Macromolecules}\ }\textbf {\bibinfo {volume} {49}},\ \bibinfo {pages}
  {8018--8030} (\bibinfo {year} {2016}{\natexlab{b}})}\BibitemShut {NoStop}%
\bibitem [{\citenamefont {Thomas}\ \emph {et~al.}(2009)\citenamefont {Thomas},
  \citenamefont {DePuit},\ and\ \citenamefont {Khomami}}]{Thomas2009}%
  \BibitemOpen
  \bibfield  {author} {\bibinfo {author} {\bibfnamefont {D.~G.}\ \bibnamefont
  {Thomas}}, \bibinfo {author} {\bibfnamefont {R.~J.}\ \bibnamefont {DePuit}},
  \ and\ \bibinfo {author} {\bibfnamefont {B.}~\bibnamefont {Khomami}},\
  }\bibfield  {title} {\enquote {\bibinfo {title} {Dynamic simulation of
  individual macromolecules in oscillatory shear flow},}\ }\href@noop {}
  {\bibfield  {journal} {\bibinfo  {journal} {Journal of Rheology}\ }\textbf
  {\bibinfo {volume} {53}},\ \bibinfo {pages} {275--291} (\bibinfo {year}
  {2009})}\BibitemShut {NoStop}%
\bibitem [{\citenamefont {Latinwo}\ and\ \citenamefont
  {Schroeder}(8345)}]{Latinwo2013}%
  \BibitemOpen
  \bibfield  {author} {\bibinfo {author} {\bibfnamefont {F.}~\bibnamefont
  {Latinwo}}\ and\ \bibinfo {author} {\bibfnamefont {C.~M.}\ \bibnamefont
  {Schroeder}},\ }\bibfield  {title} {\enquote {\bibinfo {title}
  {Nonequilibrium work relations for polymer dynamics in dilute solutions},}\
  }\href@noop {} {\bibfield  {journal} {\bibinfo  {journal} {Macromolecules}\
  }\textbf {\bibinfo {volume} {46}},\ \bibinfo {pages} {8345--8355} (\bibinfo
  {year} {8345})}\BibitemShut {NoStop}%
\bibitem [{\citenamefont {Jarzynski}(1997)}]{Jarzynski1997}%
  \BibitemOpen
  \bibfield  {author} {\bibinfo {author} {\bibfnamefont {C.}~\bibnamefont
  {Jarzynski}},\ }\bibfield  {title} {\enquote {\bibinfo {title}
  {Nonequilibrium equality for free energy differences},}\ }\href@noop {}
  {\bibfield  {journal} {\bibinfo  {journal} {Phys. Rev. Lett.}\ }\textbf
  {\bibinfo {volume} {78}},\ \bibinfo {pages} {2690--2693} (\bibinfo {year}
  {1997})}\BibitemShut {NoStop}%
\bibitem [{\citenamefont {Liphardt}\ \emph {et~al.}(2002)\citenamefont
  {Liphardt}, \citenamefont {Dumont}, \citenamefont {Smith}, \citenamefont
  {Tinoco},\ and\ \citenamefont {Bustamante}}]{Liphardt2002}%
  \BibitemOpen
  \bibfield  {author} {\bibinfo {author} {\bibfnamefont {J.}~\bibnamefont
  {Liphardt}}, \bibinfo {author} {\bibfnamefont {S.}~\bibnamefont {Dumont}},
  \bibinfo {author} {\bibfnamefont {S.~B.}\ \bibnamefont {Smith}}, \bibinfo
  {author} {\bibfnamefont {L.}~\bibnamefont {Tinoco}}, \ and\ \bibinfo {author}
  {\bibfnamefont {C.}~\bibnamefont {Bustamante}},\ }\bibfield  {title}
  {\enquote {\bibinfo {title} {Equilibrium information from nonequilibrium
  measurements in an experimental test of jarzynski's equality},}\ }\href@noop
  {} {\bibfield  {journal} {\bibinfo  {journal} {Science}\ }\textbf {\bibinfo
  {volume} {296}},\ \bibinfo {pages} {1832--1835} (\bibinfo {year}
  {2002})}\BibitemShut {NoStop}%
\bibitem [{\citenamefont {Speck}\ \emph {et~al.}(2008)\citenamefont {Speck},
  \citenamefont {Mehl},\ and\ \citenamefont {Seifert}}]{Speck2008}%
  \BibitemOpen
  \bibfield  {author} {\bibinfo {author} {\bibfnamefont {T.}~\bibnamefont
  {Speck}}, \bibinfo {author} {\bibfnamefont {J.}~\bibnamefont {Mehl}}, \ and\
  \bibinfo {author} {\bibfnamefont {U.}~\bibnamefont {Seifert}},\ }\bibfield
  {title} {\enquote {\bibinfo {title} {Role of external flow and frame
  invariance in stochastic thermodynamics},}\ }\href@noop {} {\bibfield
  {journal} {\bibinfo  {journal} {Phys. Rev. Lett.}\ }\textbf {\bibinfo
  {volume} {100}},\ \bibinfo {pages} {178302} (\bibinfo {year}
  {2008})}\BibitemShut {NoStop}%
\bibitem [{\citenamefont {Latinwo}\ \emph {et~al.}(2014)\citenamefont
  {Latinwo}, \citenamefont {Hsiao},\ and\ \citenamefont
  {Schroeder}}]{Latinwo2014b}%
  \BibitemOpen
  \bibfield  {author} {\bibinfo {author} {\bibfnamefont {F.}~\bibnamefont
  {Latinwo}}, \bibinfo {author} {\bibfnamefont {K.}~\bibnamefont {Hsiao}}, \
  and\ \bibinfo {author} {\bibfnamefont {C.~M.}\ \bibnamefont {Schroeder}},\
  }\bibfield  {title} {\enquote {\bibinfo {title} {Nonequilibrium
  thermodynamics of dilute polymer solutions in flow},}\ }\href@noop {}
  {\bibfield  {journal} {\bibinfo  {journal} {J. Chem. Phys.}\ }\textbf
  {\bibinfo {volume} {141}},\ \bibinfo {pages} {174903} (\bibinfo {year}
  {2014})}\BibitemShut {NoStop}%
\bibitem [{\citenamefont {Latinwo}\ and\ \citenamefont
  {Schroeder}(2014)}]{Latinwo2014a}%
  \BibitemOpen
  \bibfield  {author} {\bibinfo {author} {\bibfnamefont {F.}~\bibnamefont
  {Latinwo}}\ and\ \bibinfo {author} {\bibfnamefont {C.~M.}\ \bibnamefont
  {Schroeder}},\ }\bibfield  {title} {\enquote {\bibinfo {title} {Determining
  elasticity from single polymer dynamics},}\ }\href@noop {} {\bibfield
  {journal} {\bibinfo  {journal} {Soft Matter}\ }\textbf {\bibinfo {volume}
  {10}},\ \bibinfo {pages} {2178--2187} (\bibinfo {year} {2014})}\BibitemShut
  {NoStop}%
\bibitem [{\citenamefont {Ghosal}\ and\ \citenamefont
  {Cherayil}(2016{\natexlab{a}})}]{Ghosal2016a}%
  \BibitemOpen
  \bibfield  {author} {\bibinfo {author} {\bibfnamefont {A.}~\bibnamefont
  {Ghosal}}\ and\ \bibinfo {author} {\bibfnamefont {B.~J.}\ \bibnamefont
  {Cherayil}},\ }\bibfield  {title} {\enquote {\bibinfo {title} {Polymer
  extension under flow: A path integral evaluation of the free energy change
  using the jarzynski equality},}\ }\href@noop {} {\bibfield  {journal}
  {\bibinfo  {journal} {J. Chem. Phys.}\ }\textbf {\bibinfo {volume} {144}},\
  \bibinfo {pages} {214902} (\bibinfo {year} {2016}{\natexlab{a}})}\BibitemShut
  {NoStop}%
\bibitem [{\citenamefont {Ghosal}\ and\ \citenamefont
  {Cherayil}(2016{\natexlab{b}})}]{Ghosal2016b}%
  \BibitemOpen
  \bibfield  {author} {\bibinfo {author} {\bibfnamefont {A.}~\bibnamefont
  {Ghosal}}\ and\ \bibinfo {author} {\bibfnamefont {B.~J.}\ \bibnamefont
  {Cherayil}},\ }\bibfield  {title} {\enquote {\bibinfo {title} {Polymer
  extension under flow: Some statistical properties of the work distribution
  function},}\ }\href@noop {} {\bibfield  {journal} {\bibinfo  {journal} {J.
  Chem. Phys.}\ }\textbf {\bibinfo {volume} {145}},\ \bibinfo {pages} {204901}
  (\bibinfo {year} {2016}{\natexlab{b}})}\BibitemShut {NoStop}%
\bibitem [{\citenamefont {Schneider}\ \emph {et~al.}(2007)\citenamefont
  {Schneider}, \citenamefont {Nuschele}, \citenamefont {Wixforth},
  \citenamefont {Gorzelanny}, \citenamefont {Alexander-Katz}, \citenamefont
  {Netz},\ and\ \citenamefont {Schneider}}]{Schneider2007}%
  \BibitemOpen
  \bibfield  {author} {\bibinfo {author} {\bibfnamefont {S.~W.}\ \bibnamefont
  {Schneider}}, \bibinfo {author} {\bibfnamefont {S.}~\bibnamefont {Nuschele}},
  \bibinfo {author} {\bibfnamefont {A.}~\bibnamefont {Wixforth}}, \bibinfo
  {author} {\bibfnamefont {C.}~\bibnamefont {Gorzelanny}}, \bibinfo {author}
  {\bibfnamefont {A.}~\bibnamefont {Alexander-Katz}}, \bibinfo {author}
  {\bibfnamefont {R.~R.}\ \bibnamefont {Netz}}, \ and\ \bibinfo {author}
  {\bibfnamefont {M.~F.}\ \bibnamefont {Schneider}},\ }\bibfield  {title}
  {\enquote {\bibinfo {title} {Shear-induced unfolding triggers adhesion of von
  willebrand factor fibers},}\ }\href@noop {} {\bibfield  {journal} {\bibinfo
  {journal} {Proc. Nat. Acad. Sci.}\ }\textbf {\bibinfo {volume} {104}},\
  \bibinfo {pages} {7899--7903} (\bibinfo {year} {2007})}\BibitemShut {NoStop}%
\bibitem [{\citenamefont {Alexander-Katz}\ \emph {et~al.}(2006)\citenamefont
  {Alexander-Katz}, \citenamefont {Schneider}, \citenamefont {Schneider},
  \citenamefont {Wixforth},\ and\ \citenamefont {Netz}}]{Katz2006}%
  \BibitemOpen
  \bibfield  {author} {\bibinfo {author} {\bibfnamefont {A.}~\bibnamefont
  {Alexander-Katz}}, \bibinfo {author} {\bibfnamefont {M.~F.}\ \bibnamefont
  {Schneider}}, \bibinfo {author} {\bibfnamefont {S.~W.}\ \bibnamefont
  {Schneider}}, \bibinfo {author} {\bibfnamefont {A.}~\bibnamefont {Wixforth}},
  \ and\ \bibinfo {author} {\bibfnamefont {R.~R.}\ \bibnamefont {Netz}},\
  }\bibfield  {title} {\enquote {\bibinfo {title} {Shear-flow-induced unfolding
  of polymeric globules},}\ }\href@noop {} {\bibfield  {journal} {\bibinfo
  {journal} {Phys. Rev. Lett.}\ }\textbf {\bibinfo {volume} {97}},\ \bibinfo
  {pages} {138101} (\bibinfo {year} {2006})}\BibitemShut {NoStop}%
\bibitem [{\citenamefont {Alexander-Katz}\ and\ \citenamefont
  {Netz}(2008)}]{Katz2008}%
  \BibitemOpen
  \bibfield  {author} {\bibinfo {author} {\bibfnamefont {A.}~\bibnamefont
  {Alexander-Katz}}\ and\ \bibinfo {author} {\bibfnamefont {R.~R.}\
  \bibnamefont {Netz}},\ }\bibfield  {title} {\enquote {\bibinfo {title}
  {Dynamics and instabilities of collapsed polymers in shear flow},}\
  }\href@noop {} {\bibfield  {journal} {\bibinfo  {journal} {Macromolecules}\
  }\textbf {\bibinfo {volume} {41}},\ \bibinfo {pages} {3363--3374} (\bibinfo
  {year} {2008})}\BibitemShut {NoStop}%
\bibitem [{\citenamefont {Sing}\ and\ \citenamefont
  {Alexander-Katz}(2010{\natexlab{a}})}]{Sing2010}%
  \BibitemOpen
  \bibfield  {author} {\bibinfo {author} {\bibfnamefont {C.~E.}\ \bibnamefont
  {Sing}}\ and\ \bibinfo {author} {\bibfnamefont {A.}~\bibnamefont
  {Alexander-Katz}},\ }\bibfield  {title} {\enquote {\bibinfo {title}
  {Globule-stretch transitions of collapsed polymers in elongational flow
  fields},}\ }\href@noop {} {\bibfield  {journal} {\bibinfo  {journal}
  {Macromolecules}\ }\textbf {\bibinfo {volume} {43}},\ \bibinfo {pages}
  {3532--3541} (\bibinfo {year} {2010}{\natexlab{a}})}\BibitemShut {NoStop}%
\bibitem [{\citenamefont {Sing}\ and\ \citenamefont
  {Alexander-Katz}(2010{\natexlab{b}})}]{Sing2010b}%
  \BibitemOpen
  \bibfield  {author} {\bibinfo {author} {\bibfnamefont {C.~E.}\ \bibnamefont
  {Sing}}\ and\ \bibinfo {author} {\bibfnamefont {A.}~\bibnamefont
  {Alexander-Katz}},\ }\bibfield  {title} {\enquote {\bibinfo {title}
  {Elongational flow induces the unfolding of von willebrand factor at
  physiological flow rates},}\ }\href@noop {} {\bibfield  {journal} {\bibinfo
  {journal} {Biophysical Journal}\ }\textbf {\bibinfo {volume} {98}},\ \bibinfo
  {pages} {L35--L37} (\bibinfo {year} {2010}{\natexlab{b}})}\BibitemShut
  {NoStop}%
\bibitem [{\citenamefont {Sing}\ and\ \citenamefont
  {Alexander-Katz}(2011{\natexlab{a}})}]{Sing2011}%
  \BibitemOpen
  \bibfield  {author} {\bibinfo {author} {\bibfnamefont {C.~E.}\ \bibnamefont
  {Sing}}\ and\ \bibinfo {author} {\bibfnamefont {A.}~\bibnamefont
  {Alexander-Katz}},\ }\bibfield  {title} {\enquote {\bibinfo {title} {Dynamics
  of collapsed polymers under the simultaneous influence of elongational and
  shear flows},}\ }\href@noop {} {\bibfield  {journal} {\bibinfo  {journal} {J.
  Chem. Phys.}\ }\textbf {\bibinfo {volume} {135}},\ \bibinfo {pages} {014902}
  (\bibinfo {year} {2011}{\natexlab{a}})}\BibitemShut {NoStop}%
\bibitem [{\citenamefont {Einert}\ \emph {et~al.}(2011)\citenamefont {Einert},
  \citenamefont {Sing}, \citenamefont {Alexander-Katz},\ and\ \citenamefont
  {Netz}}]{Einert2011}%
  \BibitemOpen
  \bibfield  {author} {\bibinfo {author} {\bibfnamefont {T.~R.}\ \bibnamefont
  {Einert}}, \bibinfo {author} {\bibfnamefont {C.~E.}\ \bibnamefont {Sing}},
  \bibinfo {author} {\bibfnamefont {A.}~\bibnamefont {Alexander-Katz}}, \ and\
  \bibinfo {author} {\bibfnamefont {R.~R.}\ \bibnamefont {Netz}},\ }\bibfield
  {title} {\enquote {\bibinfo {title} {Conformational dynamics and internal
  friction in homopolymer globules: equilibrium vs. non-equilibrium
  simulations},}\ }\href@noop {} {\bibfield  {journal} {\bibinfo  {journal}
  {European Physical Journal E}\ }\textbf {\bibinfo {volume} {34}},\ \bibinfo
  {pages} {130} (\bibinfo {year} {2011})}\BibitemShut {NoStop}%
\bibitem [{\citenamefont {Sing}\ and\ \citenamefont
  {Alexander-Katz}(2011{\natexlab{b}})}]{Sing2011b}%
  \BibitemOpen
  \bibfield  {author} {\bibinfo {author} {\bibfnamefont {C.~E.}\ \bibnamefont
  {Sing}}\ and\ \bibinfo {author} {\bibfnamefont {A.}~\bibnamefont
  {Alexander-Katz}},\ }\bibfield  {title} {\enquote {\bibinfo {title}
  {Non-monotonic hydrodynamic lift force on highly extended polymers near
  surfaces},}\ }\href@noop {} {\bibfield  {journal} {\bibinfo  {journal}
  {Europhysics Letters}\ }\textbf {\bibinfo {volume} {95}},\ \bibinfo {pages}
  {48001} (\bibinfo {year} {2011}{\natexlab{b}})}\BibitemShut {NoStop}%
\bibitem [{\citenamefont {Sing}\ and\ \citenamefont
  {Alexander-Katz}(2011{\natexlab{c}})}]{Sing2011c}%
  \BibitemOpen
  \bibfield  {author} {\bibinfo {author} {\bibfnamefont {C.~E.}\ \bibnamefont
  {Sing}}\ and\ \bibinfo {author} {\bibfnamefont {A.}~\bibnamefont
  {Alexander-Katz}},\ }\bibfield  {title} {\enquote {\bibinfo {title} {Theory
  of tethered polymers in shear flow: The strong stretching limit},}\
  }\href@noop {} {\bibfield  {journal} {\bibinfo  {journal} {Macromolecules}\
  }\textbf {\bibinfo {volume} {44}},\ \bibinfo {pages} {9020--9028} (\bibinfo
  {year} {2011}{\natexlab{c}})}\BibitemShut {NoStop}%
\bibitem [{\citenamefont {Sing}\ and\ \citenamefont
  {Alexander-Katz}(2011{\natexlab{d}})}]{Sing2011d}%
  \BibitemOpen
  \bibfield  {author} {\bibinfo {author} {\bibfnamefont {C.~E.}\ \bibnamefont
  {Sing}}\ and\ \bibinfo {author} {\bibfnamefont {A.}~\bibnamefont
  {Alexander-Katz}},\ }\bibfield  {title} {\enquote {\bibinfo {title}
  {Equilibrium structure and dynamics of self-associating single polymers},}\
  }\href@noop {} {\bibfield  {journal} {\bibinfo  {journal} {Macromolecules}\
  }\textbf {\bibinfo {volume} {44}},\ \bibinfo {pages} {6962--6971} (\bibinfo
  {year} {2011}{\natexlab{d}})}\BibitemShut {NoStop}%
\bibitem [{\citenamefont {Sing}\ and\ \citenamefont
  {Alexander-Katz}(2011{\natexlab{e}})}]{Sing2011e}%
  \BibitemOpen
  \bibfield  {author} {\bibinfo {author} {\bibfnamefont {C.~E.}\ \bibnamefont
  {Sing}}\ and\ \bibinfo {author} {\bibfnamefont {A.}~\bibnamefont
  {Alexander-Katz}},\ }\bibfield  {title} {\enquote {\bibinfo {title} {Giant
  nonmonotonic stretching response of a self-associating polymer in shear
  flow},}\ }\href@noop {} {\bibfield  {journal} {\bibinfo  {journal} {Phys.
  Rev. Lett.}\ }\textbf {\bibinfo {volume} {107}},\ \bibinfo {pages} {198302}
  (\bibinfo {year} {2011}{\natexlab{e}})}\BibitemShut {NoStop}%
\bibitem [{\citenamefont {Kong}\ \emph {et~al.}(2014)\citenamefont {Kong},
  \citenamefont {Dalal}, \citenamefont {Li},\ and\ \citenamefont
  {Larson}}]{Kong2014}%
  \BibitemOpen
  \bibfield  {author} {\bibinfo {author} {\bibfnamefont {M.}~\bibnamefont
  {Kong}}, \bibinfo {author} {\bibfnamefont {I.~S.}\ \bibnamefont {Dalal}},
  \bibinfo {author} {\bibfnamefont {G.}~\bibnamefont {Li}}, \ and\ \bibinfo
  {author} {\bibfnamefont {R.~G.}\ \bibnamefont {Larson}},\ }\bibfield  {title}
  {\enquote {\bibinfo {title} {Systematic coarse-graining of the dynamics of
  self-attractive semiflexible polymers},}\ }\href@noop {} {\bibfield
  {journal} {\bibinfo  {journal} {Macromolecules}\ }\textbf {\bibinfo {volume}
  {47}},\ \bibinfo {pages} {1494--1502} (\bibinfo {year} {2014})}\BibitemShut
  {NoStop}%
\bibitem [{\citenamefont {Radhakrishnan}\ and\ \citenamefont
  {Underhill}(2015)}]{Radhakrishnan2015}%
  \BibitemOpen
  \bibfield  {author} {\bibinfo {author} {\bibfnamefont {R.}~\bibnamefont
  {Radhakrishnan}}\ and\ \bibinfo {author} {\bibfnamefont {P.~T.}\ \bibnamefont
  {Underhill}},\ }\bibfield  {title} {\enquote {\bibinfo {title} {Influence of
  shear on globule formation in dilute solutions of flexible polymers},}\
  }\href@noop {} {\bibfield  {journal} {\bibinfo  {journal} {J. Chem. Phys.}\
  }\textbf {\bibinfo {volume} {142}},\ \bibinfo {pages} {144901} (\bibinfo
  {year} {2015})}\BibitemShut {NoStop}%
\bibitem [{\citenamefont {Prakash}\ and\ \citenamefont
  {Ottinger}(1999)}]{Prakash1999}%
  \BibitemOpen
  \bibfield  {author} {\bibinfo {author} {\bibfnamefont {J.~R.}\ \bibnamefont
  {Prakash}}\ and\ \bibinfo {author} {\bibfnamefont {H.~C.}\ \bibnamefont
  {Ottinger}},\ }\bibfield  {title} {\enquote {\bibinfo {title} {Viscometric
  functions for a dilute solution of polymers in a good solvent},}\ }\href@noop
  {} {\bibfield  {journal} {\bibinfo  {journal} {Macromolecules}\ }\textbf
  {\bibinfo {volume} {32}},\ \bibinfo {pages} {2028--2043} (\bibinfo {year}
  {1999})}\BibitemShut {NoStop}%
\bibitem [{\citenamefont {Prakash}(2001)}]{Prakash2001}%
  \BibitemOpen
  \bibfield  {author} {\bibinfo {author} {\bibfnamefont {J.~R.}\ \bibnamefont
  {Prakash}},\ }\bibfield  {title} {\enquote {\bibinfo {title} {Rouse chains
  with excluded volume interactions: Linear viscoelasticity},}\ }\href@noop {}
  {\bibfield  {journal} {\bibinfo  {journal} {Macromolecules}\ }\textbf
  {\bibinfo {volume} {34}},\ \bibinfo {pages} {3396--3411} (\bibinfo {year}
  {2001})}\BibitemShut {NoStop}%
\bibitem [{\citenamefont {Prakash}(2002)}]{Prakash2002a}%
  \BibitemOpen
  \bibfield  {author} {\bibinfo {author} {\bibfnamefont {J.~R.}\ \bibnamefont
  {Prakash}},\ }\bibfield  {title} {\enquote {\bibinfo {title} {Rouse chains
  with excluded volume interactions in steady simple shear flow},}\ }\href@noop
  {} {\bibfield  {journal} {\bibinfo  {journal} {Journal of Rheology}\ }\textbf
  {\bibinfo {volume} {46}},\ \bibinfo {pages} {1353--1380} (\bibinfo {year}
  {2002})}\BibitemShut {NoStop}%
\bibitem [{\citenamefont {Prabhaker}\ and\ \citenamefont
  {Prakash}(2002)}]{Prakash2002b}%
  \BibitemOpen
  \bibfield  {author} {\bibinfo {author} {\bibfnamefont {R.}~\bibnamefont
  {Prabhaker}}\ and\ \bibinfo {author} {\bibfnamefont {J.~R.}\ \bibnamefont
  {Prakash}},\ }\bibfield  {title} {\enquote {\bibinfo {title} {Viscometric
  functions for hookean dumbbells with excluded volume and hydrodynamic
  interactions},}\ }\href@noop {} {\bibfield  {journal} {\bibinfo  {journal}
  {Journal of Rheology}\ }\textbf {\bibinfo {volume} {46}},\ \bibinfo {pages}
  {1191--1220} (\bibinfo {year} {2002})}\BibitemShut {NoStop}%
\bibitem [{\citenamefont {Prabhaker}\ \emph
  {et~al.}(2004{\natexlab{a}})\citenamefont {Prabhaker}, \citenamefont
  {Prakash},\ and\ \citenamefont {Sridhar}}]{Prakash2004}%
  \BibitemOpen
  \bibfield  {author} {\bibinfo {author} {\bibfnamefont {R.}~\bibnamefont
  {Prabhaker}}, \bibinfo {author} {\bibfnamefont {J.~R.}\ \bibnamefont
  {Prakash}}, \ and\ \bibinfo {author} {\bibfnamefont {T.}~\bibnamefont
  {Sridhar}},\ }\bibfield  {title} {\enquote {\bibinfo {title} {A successive
  fine-graining scheme for predicting the rheological properties of dilute
  polymer solutions},}\ }\href@noop {} {\bibfield  {journal} {\bibinfo
  {journal} {Journal of Rheology}\ }\textbf {\bibinfo {volume} {48}},\ \bibinfo
  {pages} {1251--1278} (\bibinfo {year} {2004}{\natexlab{a}})}\BibitemShut
  {NoStop}%
\bibitem [{\citenamefont {Sunthar}\ and\ \citenamefont
  {Prakash}(2005)}]{Prakash2005}%
  \BibitemOpen
  \bibfield  {author} {\bibinfo {author} {\bibfnamefont {P.}~\bibnamefont
  {Sunthar}}\ and\ \bibinfo {author} {\bibfnamefont {J.~R.}\ \bibnamefont
  {Prakash}},\ }\bibfield  {title} {\enquote {\bibinfo {title} {Parameter-free
  prediction of dna conformations in elongational flow by successive fine
  graining},}\ }\href@noop {} {\bibfield  {journal} {\bibinfo  {journal}
  {Macromolecules}\ }\textbf {\bibinfo {volume} {38}},\ \bibinfo {pages}
  {617--640} (\bibinfo {year} {2005})}\BibitemShut {NoStop}%
\bibitem [{\citenamefont {Prabhaker}\ \emph
  {et~al.}(2004{\natexlab{b}})\citenamefont {Prabhaker}, \citenamefont
  {Sunthar},\ and\ \citenamefont {Prakash}}]{Prakash2004b}%
  \BibitemOpen
  \bibfield  {author} {\bibinfo {author} {\bibfnamefont {R.}~\bibnamefont
  {Prabhaker}}, \bibinfo {author} {\bibfnamefont {P.}~\bibnamefont {Sunthar}},
  \ and\ \bibinfo {author} {\bibfnamefont {J.~R.}\ \bibnamefont {Prakash}},\
  }\bibfield  {title} {\enquote {\bibinfo {title} {Exploring the universal
  dynamics of dilute polymer solutions in extensional flows},}\ }\href@noop {}
  {\bibfield  {journal} {\bibinfo  {journal} {Physics A}\ }\textbf {\bibinfo
  {volume} {339}},\ \bibinfo {pages} {34--39} (\bibinfo {year}
  {2004}{\natexlab{b}})}\BibitemShut {NoStop}%
\bibitem [{\citenamefont {Saadat}\ and\ \citenamefont
  {Khomami}(2014)}]{Saadat2014}%
  \BibitemOpen
  \bibfield  {author} {\bibinfo {author} {\bibfnamefont {A.}~\bibnamefont
  {Saadat}}\ and\ \bibinfo {author} {\bibfnamefont {B.}~\bibnamefont
  {Khomami}},\ }\bibfield  {title} {\enquote {\bibinfo {title} {Computationally
  efficient algorithms for incorporation of hydrodynamic and excluded volume
  interactions in brownian dynamics simulations: A comparative study of the
  krylov subspace and chebyshev based techniques},}\ }\href@noop {} {\bibfield
  {journal} {\bibinfo  {journal} {J. Chem. Phys.}\ }\textbf {\bibinfo {volume}
  {140}},\ \bibinfo {pages} {184903} (\bibinfo {year} {2014})}\BibitemShut
  {NoStop}%
\bibitem [{\citenamefont {Saadat}\ and\ \citenamefont
  {Khomami}(2015{\natexlab{a}})}]{Saadat2015b}%
  \BibitemOpen
  \bibfield  {author} {\bibinfo {author} {\bibfnamefont {A.}~\bibnamefont
  {Saadat}}\ and\ \bibinfo {author} {\bibfnamefont {B.}~\bibnamefont
  {Khomami}},\ }\bibfield  {title} {\enquote {\bibinfo {title} {Molecular based
  prediction of the extensional rheology of high molecular weight polystyrene
  dilute solutions: A hi-fidelity brownian dynamics approach},}\ }\href@noop {}
  {\bibfield  {journal} {\bibinfo  {journal} {Journal of Rheology}\ }\textbf
  {\bibinfo {volume} {59}},\ \bibinfo {pages} {1507--1525} (\bibinfo {year}
  {2015}{\natexlab{a}})}\BibitemShut {NoStop}%
\bibitem [{\citenamefont {Moghani}\ and\ \citenamefont
  {Khomami}(2017)}]{Moghani2017}%
  \BibitemOpen
  \bibfield  {author} {\bibinfo {author} {\bibfnamefont {M.~M.}\ \bibnamefont
  {Moghani}}\ and\ \bibinfo {author} {\bibfnamefont {B.}~\bibnamefont
  {Khomami}},\ }\bibfield  {title} {\enquote {\bibinfo {title} {Computationally
  efficient algorithms for brownian dynamics simulation of long flexible
  macromolecules modeled as bead-rod chains},}\ }\href@noop {} {\bibfield
  {journal} {\bibinfo  {journal} {Phys. Rev. Fluids}\ }\textbf {\bibinfo
  {volume} {2}},\ \bibinfo {pages} {023303} (\bibinfo {year}
  {2017})}\BibitemShut {NoStop}%
\bibitem [{\citenamefont {Moghani}\ and\ \citenamefont
  {Khomami}(2016)}]{Moghani2016}%
  \BibitemOpen
  \bibfield  {author} {\bibinfo {author} {\bibfnamefont {M.~M.}\ \bibnamefont
  {Moghani}}\ and\ \bibinfo {author} {\bibfnamefont {B.}~\bibnamefont
  {Khomami}},\ }\bibfield  {title} {\enquote {\bibinfo {title} {Flexible
  polyelectrolyte chain in a strong electrolyte solution: Insight into
  equilibrium properties and force-extension behavior from mesoscale
  simulation},}\ }\href@noop {} {\bibfield  {journal} {\bibinfo  {journal} {J.
  Chem. Phys.}\ }\textbf {\bibinfo {volume} {144}},\ \bibinfo {pages} {024903}
  (\bibinfo {year} {2016})}\BibitemShut {NoStop}%
\bibitem [{\citenamefont {Miao}\ \emph {et~al.}(2017)\citenamefont {Miao},
  \citenamefont {Young},\ and\ \citenamefont {Sing}}]{Sing2017}%
  \BibitemOpen
  \bibfield  {author} {\bibinfo {author} {\bibfnamefont {L.}~\bibnamefont
  {Miao}}, \bibinfo {author} {\bibfnamefont {C.~D.}\ \bibnamefont {Young}}, \
  and\ \bibinfo {author} {\bibfnamefont {C.~E.}\ \bibnamefont {Sing}},\
  }\bibfield  {title} {\enquote {\bibinfo {title} {An iterative method for
  hydrodynamic interactions in brownian dynamics simulations of polymer
  dynamics},}\ }\href@noop {} {\bibfield  {journal} {\bibinfo  {journal} {J.
  Chem. Phys.}\ }\textbf {\bibinfo {volume} {147}},\ \bibinfo {pages} {024904}
  (\bibinfo {year} {2017})}\BibitemShut {NoStop}%
\bibitem [{\citenamefont {Pan}\ \emph {et~al.}(2014{\natexlab{b}})\citenamefont
  {Pan}, \citenamefont {Ahirwal}, \citenamefont {Nguyen}, \citenamefont
  {Sridhar}, \citenamefont {Sunthar},\ and\ \citenamefont
  {Prakash}}]{Pan2014b}%
  \BibitemOpen
  \bibfield  {author} {\bibinfo {author} {\bibfnamefont {S.}~\bibnamefont
  {Pan}}, \bibinfo {author} {\bibfnamefont {D.}~\bibnamefont {Ahirwal}},
  \bibinfo {author} {\bibfnamefont {D.~A.}\ \bibnamefont {Nguyen}}, \bibinfo
  {author} {\bibfnamefont {T.}~\bibnamefont {Sridhar}}, \bibinfo {author}
  {\bibfnamefont {P.}~\bibnamefont {Sunthar}}, \ and\ \bibinfo {author}
  {\bibfnamefont {J.~R.}\ \bibnamefont {Prakash}},\ }\bibfield  {title}
  {\enquote {\bibinfo {title} {Viscosity radius of polymers in dilute
  solutions: Universal behavior from dna rheology and brownian dynamics
  simulations},}\ }\href@noop {} {\bibfield  {journal} {\bibinfo  {journal}
  {Macromolecules}\ }\textbf {\bibinfo {volume} {47}},\ \bibinfo {pages}
  {7548--7560} (\bibinfo {year} {2014}{\natexlab{b}})}\BibitemShut {NoStop}%
\bibitem [{\citenamefont {Liu}\ \emph {et~al.}(2009)\citenamefont {Liu},
  \citenamefont {Jun},\ and\ \citenamefont {Steinberg}}]{Liu2009}%
  \BibitemOpen
  \bibfield  {author} {\bibinfo {author} {\bibfnamefont {Y.}~\bibnamefont
  {Liu}}, \bibinfo {author} {\bibfnamefont {Y.}~\bibnamefont {Jun}}, \ and\
  \bibinfo {author} {\bibfnamefont {V.}~\bibnamefont {Steinberg}},\ }\bibfield
  {title} {\enquote {\bibinfo {title} {Concentration dependence of the longest
  relaxation times of dilute and semi-dilute polymer solutions},}\ }\href@noop
  {} {\bibfield  {journal} {\bibinfo  {journal} {Journal of Rheology}\ }\textbf
  {\bibinfo {volume} {53}},\ \bibinfo {pages} {1069--1085} (\bibinfo {year}
  {2009})}\BibitemShut {NoStop}%
\bibitem [{\citenamefont {Stoltz}\ \emph {et~al.}(2006)\citenamefont {Stoltz},
  \citenamefont {de~Pablo},\ and\ \citenamefont {Graham}}]{Stoltz2006}%
  \BibitemOpen
  \bibfield  {author} {\bibinfo {author} {\bibfnamefont {C.}~\bibnamefont
  {Stoltz}}, \bibinfo {author} {\bibfnamefont {J.~J.}\ \bibnamefont
  {de~Pablo}}, \ and\ \bibinfo {author} {\bibfnamefont {M.~D.}\ \bibnamefont
  {Graham}},\ }\bibfield  {title} {\enquote {\bibinfo {title} {Concentration
  dependence of shear and extensional rheology of polymer solutions: Brownian
  dynamics simulations},}\ }\href@noop {} {\bibfield  {journal} {\bibinfo
  {journal} {Journal of Rheology}\ }\textbf {\bibinfo {volume} {50}},\ \bibinfo
  {pages} {137--167} (\bibinfo {year} {2006})}\BibitemShut {NoStop}%
\bibitem [{\citenamefont {Huang}\ \emph {et~al.}(2010)\citenamefont {Huang},
  \citenamefont {Winkler}, \citenamefont {Sutman},\ and\ \citenamefont
  {Gompper}}]{Huang2010}%
  \BibitemOpen
  \bibfield  {author} {\bibinfo {author} {\bibfnamefont {C.~C.}\ \bibnamefont
  {Huang}}, \bibinfo {author} {\bibfnamefont {R.~G.}\ \bibnamefont {Winkler}},
  \bibinfo {author} {\bibfnamefont {G.}~\bibnamefont {Sutman}}, \ and\ \bibinfo
  {author} {\bibfnamefont {G.}~\bibnamefont {Gompper}},\ }\bibfield  {title}
  {\enquote {\bibinfo {title} {Semidilute polymer solutions at equilibrium and
  under shear flow},}\ }\href@noop {} {\bibfield  {journal} {\bibinfo
  {journal} {Macromolecules}\ }\textbf {\bibinfo {volume} {43}},\ \bibinfo
  {pages} {10107--10116} (\bibinfo {year} {2010})}\BibitemShut {NoStop}%
\bibitem [{\citenamefont {Jain}\ \emph
  {et~al.}(2012{\natexlab{b}})\citenamefont {Jain}, \citenamefont {Sunthar},
  \citenamefont {Dunweg},\ and\ \citenamefont {Prakash}}]{Jain2012b}%
  \BibitemOpen
  \bibfield  {author} {\bibinfo {author} {\bibfnamefont {A.}~\bibnamefont
  {Jain}}, \bibinfo {author} {\bibfnamefont {P.}~\bibnamefont {Sunthar}},
  \bibinfo {author} {\bibfnamefont {B.}~\bibnamefont {Dunweg}}, \ and\ \bibinfo
  {author} {\bibfnamefont {J.~R.}\ \bibnamefont {Prakash}},\ }\bibfield
  {title} {\enquote {\bibinfo {title} {Optimization of a brownian dynamics
  algorithm for semi-dilute polymer solutions},}\ }\href@noop {} {\bibfield
  {journal} {\bibinfo  {journal} {Phys. Rev. E}\ }\textbf {\bibinfo {volume}
  {85}},\ \bibinfo {pages} {066703} (\bibinfo {year}
  {2012}{\natexlab{b}})}\BibitemShut {NoStop}%
\bibitem [{\citenamefont {Saadat}\ and\ \citenamefont
  {Khomami}(2015{\natexlab{b}})}]{Saadat2015}%
  \BibitemOpen
  \bibfield  {author} {\bibinfo {author} {\bibfnamefont {A.}~\bibnamefont
  {Saadat}}\ and\ \bibinfo {author} {\bibfnamefont {B.}~\bibnamefont
  {Khomami}},\ }\bibfield  {title} {\enquote {\bibinfo {title} {Matrix-free
  brownian dynamics simulation technique for semidilute polymeric solutions},}\
  }\href@noop {} {\bibfield  {journal} {\bibinfo  {journal} {Phys. Rev. E}\
  }\textbf {\bibinfo {volume} {95}},\ \bibinfo {pages} {033307} (\bibinfo
  {year} {2015}{\natexlab{b}})}\BibitemShut {NoStop}%
\bibitem [{\citenamefont {Smith}\ \emph {et~al.}(1995)\citenamefont {Smith},
  \citenamefont {Perkins},\ and\ \citenamefont {Chu}}]{Smith1995}%
  \BibitemOpen
  \bibfield  {author} {\bibinfo {author} {\bibfnamefont {D.~E.}\ \bibnamefont
  {Smith}}, \bibinfo {author} {\bibfnamefont {T.~T.}\ \bibnamefont {Perkins}},
  \ and\ \bibinfo {author} {\bibfnamefont {S.}~\bibnamefont {Chu}},\ }\bibfield
   {title} {\enquote {\bibinfo {title} {Self-diffusion of an entangled dna
  molecule by reptation},}\ }\href@noop {} {\bibfield  {journal} {\bibinfo
  {journal} {Phys. Rev. Lett.}\ }\textbf {\bibinfo {volume} {75}},\ \bibinfo
  {pages} {4146--4149} (\bibinfo {year} {1995})}\BibitemShut {NoStop}%
\bibitem [{\citenamefont {Lodge}(1999)}]{Lodge1999}%
  \BibitemOpen
  \bibfield  {author} {\bibinfo {author} {\bibfnamefont {T.~P.}\ \bibnamefont
  {Lodge}},\ }\bibfield  {title} {\enquote {\bibinfo {title} {Reconciliation of
  the molecular weight dependence of diffusion and viscosity in entangled
  polymers},}\ }\href@noop {} {\bibfield  {journal} {\bibinfo  {journal} {Phys.
  Rev. Lett.}\ }\textbf {\bibinfo {volume} {83}},\ \bibinfo {pages} {3218}
  (\bibinfo {year} {1999})}\BibitemShut {NoStop}%
\bibitem [{\citenamefont {Robertson}\ and\ \citenamefont
  {Smith}(2007{\natexlab{b}})}]{Robertson2007b}%
  \BibitemOpen
  \bibfield  {author} {\bibinfo {author} {\bibfnamefont {R.~M.}\ \bibnamefont
  {Robertson}}\ and\ \bibinfo {author} {\bibfnamefont {D.~E.}\ \bibnamefont
  {Smith}},\ }\bibfield  {title} {\enquote {\bibinfo {title} {Direct
  measurement of the intermolecular forces confining a single molecule in an
  entangled polymer solution},}\ }\href@noop {} {\bibfield  {journal} {\bibinfo
   {journal} {Phys. Rev. Lett.}\ }\textbf {\bibinfo {volume} {99}},\ \bibinfo
  {pages} {126001} (\bibinfo {year} {2007}{\natexlab{b}})}\BibitemShut
  {NoStop}%
\bibitem [{\citenamefont {Mason}\ \emph {et~al.}(1998)\citenamefont {Mason},
  \citenamefont {Dhople},\ and\ \citenamefont {Wirtz}}]{Mason1998}%
  \BibitemOpen
  \bibfield  {author} {\bibinfo {author} {\bibfnamefont {T.~G.}\ \bibnamefont
  {Mason}}, \bibinfo {author} {\bibfnamefont {A.}~\bibnamefont {Dhople}}, \
  and\ \bibinfo {author} {\bibfnamefont {D.}~\bibnamefont {Wirtz}},\ }\bibfield
   {title} {\enquote {\bibinfo {title} {Linear viscoelastic moduli of
  concentrated dna solutions},}\ }\href@noop {} {\bibfield  {journal} {\bibinfo
   {journal} {Macromolecules}\ }\textbf {\bibinfo {volume} {31}},\ \bibinfo
  {pages} {3600--3603} (\bibinfo {year} {1998})}\BibitemShut {NoStop}%
\bibitem [{\citenamefont {Jary}\ \emph {et~al.}(1999)\citenamefont {Jary},
  \citenamefont {Sikorav},\ and\ \citenamefont {Lairez}}]{Jary1999}%
  \BibitemOpen
  \bibfield  {author} {\bibinfo {author} {\bibfnamefont {D.}~\bibnamefont
  {Jary}}, \bibinfo {author} {\bibfnamefont {J.-L.}\ \bibnamefont {Sikorav}}, \
  and\ \bibinfo {author} {\bibfnamefont {D.}~\bibnamefont {Lairez}},\
  }\bibfield  {title} {\enquote {\bibinfo {title} {Nonlinear viscoelasticity of
  entangled dna molecules},}\ }\href@noop {} {\bibfield  {journal} {\bibinfo
  {journal} {Europhysics Letters}\ }\textbf {\bibinfo {volume} {46}},\ \bibinfo
  {pages} {251--255} (\bibinfo {year} {1999})}\BibitemShut {NoStop}%
\bibitem [{\citenamefont {Zhou}\ and\ \citenamefont {Larson}(2006)}]{Zhou2006}%
  \BibitemOpen
  \bibfield  {author} {\bibinfo {author} {\bibfnamefont {Q.}~\bibnamefont
  {Zhou}}\ and\ \bibinfo {author} {\bibfnamefont {R.~G.}\ \bibnamefont
  {Larson}},\ }\bibfield  {title} {\enquote {\bibinfo {title} {Direct
  calculation of the tube potential confining entangled polymers},}\
  }\href@noop {} {\bibfield  {journal} {\bibinfo  {journal} {Macromolecules}\
  }\textbf {\bibinfo {volume} {39}},\ \bibinfo {pages} {6737} (\bibinfo {year}
  {2006})}\BibitemShut {NoStop}%
\bibitem [{\citenamefont {Teixeira}\ \emph {et~al.}(2007)\citenamefont
  {Teixeira}, \citenamefont {Dambal}, \citenamefont {Richter}, \citenamefont
  {Shaqfeh},\ and\ \citenamefont {Chu}}]{Teixeira2007}%
  \BibitemOpen
  \bibfield  {author} {\bibinfo {author} {\bibfnamefont {R.~E.}\ \bibnamefont
  {Teixeira}}, \bibinfo {author} {\bibfnamefont {A.~K.}\ \bibnamefont
  {Dambal}}, \bibinfo {author} {\bibfnamefont {D.~H.}\ \bibnamefont {Richter}},
  \bibinfo {author} {\bibfnamefont {E.~S.~G.}\ \bibnamefont {Shaqfeh}}, \ and\
  \bibinfo {author} {\bibfnamefont {S.}~\bibnamefont {Chu}},\ }\bibfield
  {title} {\enquote {\bibinfo {title} {The individualistic dynamics of
  entangled dna in solution},}\ }\href@noop {} {\bibfield  {journal} {\bibinfo
  {journal} {Macromolecules}\ }\textbf {\bibinfo {volume} {40}},\ \bibinfo
  {pages} {2461--2476} (\bibinfo {year} {2007})}\BibitemShut {NoStop}%
\bibitem [{\citenamefont {Pakdel}\ and\ \citenamefont
  {McKinley}(1996)}]{Pakdel1996}%
  \BibitemOpen
  \bibfield  {author} {\bibinfo {author} {\bibfnamefont {P.}~\bibnamefont
  {Pakdel}}\ and\ \bibinfo {author} {\bibfnamefont {G.~H.}\ \bibnamefont
  {McKinley}},\ }\bibfield  {title} {\enquote {\bibinfo {title} {Elastic
  instability and curved streamlines},}\ }\href@noop {} {\bibfield  {journal}
  {\bibinfo  {journal} {Phys. Rev. Lett.}\ }\textbf {\bibinfo {volume} {77}},\
  \bibinfo {pages} {2459--2462} (\bibinfo {year} {1996})}\BibitemShut {NoStop}%
\bibitem [{\citenamefont {Shaqfeh}(1996)}]{Shaqfeh1996}%
  \BibitemOpen
  \bibfield  {author} {\bibinfo {author} {\bibfnamefont {E.~S.~G.}\
  \bibnamefont {Shaqfeh}},\ }\bibfield  {title} {\enquote {\bibinfo {title}
  {Purely elastic instabilities in viscometric flows},}\ }\href@noop {}
  {\bibfield  {journal} {\bibinfo  {journal} {Annu. Rev. Fluid. Mech.}\
  }\textbf {\bibinfo {volume} {28}},\ \bibinfo {pages} {129--185} (\bibinfo
  {year} {1996})}\BibitemShut {NoStop}%
\bibitem [{\citenamefont {Rodd}\ \emph {et~al.}(2005)\citenamefont {Rodd},
  \citenamefont {Scott}, \citenamefont {Boger}, \citenamefont {Cooper-White},\
  and\ \citenamefont {McKinley}}]{Rodd2005b}%
  \BibitemOpen
  \bibfield  {author} {\bibinfo {author} {\bibfnamefont {L.~E.}\ \bibnamefont
  {Rodd}}, \bibinfo {author} {\bibfnamefont {T.~P.}\ \bibnamefont {Scott}},
  \bibinfo {author} {\bibfnamefont {D.~V.}\ \bibnamefont {Boger}}, \bibinfo
  {author} {\bibfnamefont {J.~J.}\ \bibnamefont {Cooper-White}}, \ and\
  \bibinfo {author} {\bibfnamefont {G.~H.}\ \bibnamefont {McKinley}},\
  }\bibfield  {title} {\enquote {\bibinfo {title} {The inertio-elastic planar
  entry flow of low-viscosity elastic fluids in micro-fabricated geometries},}\
  }\href@noop {} {\bibfield  {journal} {\bibinfo  {journal} {Journal of
  Non-Newtonian Fluid Mechanics}\ }\textbf {\bibinfo {volume} {129}},\ \bibinfo
  {pages} {1--22} (\bibinfo {year} {2005})}\BibitemShut {NoStop}%
\bibitem [{\citenamefont {Gulati}\ \emph {et~al.}(2008)\citenamefont {Gulati},
  \citenamefont {Liepmann},\ and\ \citenamefont {Muller}}]{Gulati2008}%
  \BibitemOpen
  \bibfield  {author} {\bibinfo {author} {\bibfnamefont {S.}~\bibnamefont
  {Gulati}}, \bibinfo {author} {\bibfnamefont {D.}~\bibnamefont {Liepmann}}, \
  and\ \bibinfo {author} {\bibfnamefont {S.~J.}\ \bibnamefont {Muller}},\
  }\bibfield  {title} {\enquote {\bibinfo {title} {Elastic secondary flows of
  semidilute dna solutions in abrupt 90 degree microbends},}\ }\href@noop {}
  {\bibfield  {journal} {\bibinfo  {journal} {Phys. Rev. E}\ }\textbf {\bibinfo
  {volume} {78}},\ \bibinfo {pages} {036314} (\bibinfo {year}
  {2008})}\BibitemShut {NoStop}%
\bibitem [{\citenamefont {Gulati}\ \emph {et~al.}(2015)\citenamefont {Gulati},
  \citenamefont {Muller},\ and\ \citenamefont {Liepmann}}]{Gulati2015}%
  \BibitemOpen
  \bibfield  {author} {\bibinfo {author} {\bibfnamefont {S.}~\bibnamefont
  {Gulati}}, \bibinfo {author} {\bibfnamefont {S.~J.}\ \bibnamefont {Muller}},
  \ and\ \bibinfo {author} {\bibfnamefont {D.}~\bibnamefont {Liepmann}},\
  }\bibfield  {title} {\enquote {\bibinfo {title} {Flow of dna solutions in a
  microfluidic gradual contraction},}\ }\href@noop {} {\bibfield  {journal}
  {\bibinfo  {journal} {Biomicrofluidics}\ }\textbf {\bibinfo {volume} {9}},\
  \bibinfo {pages} {054102} (\bibinfo {year} {2015})}\BibitemShut {NoStop}%
\bibitem [{\citenamefont {Hemminger}\ \emph {et~al.}(2010)\citenamefont
  {Hemminger}, \citenamefont {Boukany}, \citenamefont {Wang},\ and\
  \citenamefont {Lee}}]{Hemminger2010}%
  \BibitemOpen
  \bibfield  {author} {\bibinfo {author} {\bibfnamefont {O.}~\bibnamefont
  {Hemminger}}, \bibinfo {author} {\bibfnamefont {P.~E.}\ \bibnamefont
  {Boukany}}, \bibinfo {author} {\bibfnamefont {S.-Q.}\ \bibnamefont {Wang}}, \
  and\ \bibinfo {author} {\bibfnamefont {L.~J.}\ \bibnamefont {Lee}},\
  }\bibfield  {title} {\enquote {\bibinfo {title} {Flow pattern and molecular
  visualization of dna solutions through a 4:1 planarflow pattern and molecular
  visualization of dna solutions through a 4:1 planar micro-contraction},}\
  }\href@noop {} {\bibfield  {journal} {\bibinfo  {journal} {Journal of
  Non-Newtonian Fluid Mechanics}\ }\textbf {\bibinfo {volume} {165}},\ \bibinfo
  {pages} {1613--1624} (\bibinfo {year} {2010})}\BibitemShut {NoStop}%
\bibitem [{\citenamefont {Sachdev}\ \emph {et~al.}(2016)\citenamefont
  {Sachdev}, \citenamefont {Muralidharan},\ and\ \citenamefont
  {Boukany}}]{Sachdev2016}%
  \BibitemOpen
  \bibfield  {author} {\bibinfo {author} {\bibfnamefont {S.}~\bibnamefont
  {Sachdev}}, \bibinfo {author} {\bibfnamefont {A.}~\bibnamefont
  {Muralidharan}}, \ and\ \bibinfo {author} {\bibfnamefont {P.~E.}\
  \bibnamefont {Boukany}},\ }\bibfield  {title} {\enquote {\bibinfo {title}
  {Molecular processes leading to necking in extensional flow},}\ }\href@noop
  {} {\bibfield  {journal} {\bibinfo  {journal} {Macromolecules}\ }\textbf
  {\bibinfo {volume} {49}},\ \bibinfo {pages} {9578--9585} (\bibinfo {year}
  {2016})}\BibitemShut {NoStop}%
\bibitem [{\citenamefont {Kawale}\ \emph {et~al.}(2017)\citenamefont {Kawale},
  \citenamefont {Bouwman}, \citenamefont {Sachdev}, \citenamefont {Zitha},
  \citenamefont {Kreutzer}, \citenamefont {Rossen},\ and\ \citenamefont
  {Boukany}}]{Kawale2017}%
  \BibitemOpen
  \bibfield  {author} {\bibinfo {author} {\bibfnamefont {D.}~\bibnamefont
  {Kawale}}, \bibinfo {author} {\bibfnamefont {G.}~\bibnamefont {Bouwman}},
  \bibinfo {author} {\bibfnamefont {S.}~\bibnamefont {Sachdev}}, \bibinfo
  {author} {\bibfnamefont {P.~L.~J.}\ \bibnamefont {Zitha}}, \bibinfo {author}
  {\bibfnamefont {M.~T.}\ \bibnamefont {Kreutzer}}, \bibinfo {author}
  {\bibfnamefont {W.~R.}\ \bibnamefont {Rossen}}, \ and\ \bibinfo {author}
  {\bibfnamefont {P.~E.}\ \bibnamefont {Boukany}},\ }\bibfield  {title}
  {\enquote {\bibinfo {title} {Polymer conformation during flow in porous
  media},}\ }\href@noop {} {\bibfield  {journal} {\bibinfo  {journal} {Soft
  Matter}\ ,\ \bibinfo {pages} {DOI: 10.1039/C7SM00817A}} (\bibinfo {year}
  {2017})}\BibitemShut {NoStop}%
\bibitem [{\citenamefont {Boukany}\ \emph {et~al.}(2008)\citenamefont
  {Boukany}, \citenamefont {Hu},\ and\ \citenamefont {Wang}}]{Boukany2008}%
  \BibitemOpen
  \bibfield  {author} {\bibinfo {author} {\bibfnamefont {P.~E.}\ \bibnamefont
  {Boukany}}, \bibinfo {author} {\bibfnamefont {Y.~T.}\ \bibnamefont {Hu}}, \
  and\ \bibinfo {author} {\bibfnamefont {S.-Q.}\ \bibnamefont {Wang}},\
  }\bibfield  {title} {\enquote {\bibinfo {title} {Observations of wall slip
  and shear banding in an entangled dna solution},}\ }\href@noop {} {\bibfield
  {journal} {\bibinfo  {journal} {Macromolecules}\ }\textbf {\bibinfo {volume}
  {41}},\ \bibinfo {pages} {2644--2650} (\bibinfo {year} {2008})}\BibitemShut
  {NoStop}%
\bibitem [{\citenamefont {Boukany}\ and\ \citenamefont
  {Wang}(2009{\natexlab{a}})}]{Boukany2009}%
  \BibitemOpen
  \bibfield  {author} {\bibinfo {author} {\bibfnamefont {P.~E.}\ \bibnamefont
  {Boukany}}\ and\ \bibinfo {author} {\bibfnamefont {S.-Q.}\ \bibnamefont
  {Wang}},\ }\bibfield  {title} {\enquote {\bibinfo {title} {Shear banding or
  not in entangled dna solutions depending on the level of entanglement},}\
  }\href@noop {} {\bibfield  {journal} {\bibinfo  {journal} {Journal of
  Rheology}\ }\textbf {\bibinfo {volume} {53}},\ \bibinfo {pages} {73--83}
  (\bibinfo {year} {2009}{\natexlab{a}})}\BibitemShut {NoStop}%
\bibitem [{\citenamefont {Boukany}\ and\ \citenamefont
  {Wang}(2009{\natexlab{b}})}]{Boukany2009b}%
  \BibitemOpen
  \bibfield  {author} {\bibinfo {author} {\bibfnamefont {P.~E.}\ \bibnamefont
  {Boukany}}\ and\ \bibinfo {author} {\bibfnamefont {S.-Q.}\ \bibnamefont
  {Wang}},\ }\bibfield  {title} {\enquote {\bibinfo {title} {Exploring the
  transition from wall slip to bulk shearing banding in well-entangled dna
  solutions},}\ }\href@noop {} {\bibfield  {journal} {\bibinfo  {journal} {Soft
  Matter}\ }\textbf {\bibinfo {volume} {5}},\ \bibinfo {pages} {780--789}
  (\bibinfo {year} {2009}{\natexlab{b}})}\BibitemShut {NoStop}%
\bibitem [{\citenamefont {Boukany}\ and\ \citenamefont
  {Wang}(2010)}]{Boukany2010}%
  \BibitemOpen
  \bibfield  {author} {\bibinfo {author} {\bibfnamefont {P.~E.}\ \bibnamefont
  {Boukany}}\ and\ \bibinfo {author} {\bibfnamefont {S.-Q.}\ \bibnamefont
  {Wang}},\ }\bibfield  {title} {\enquote {\bibinfo {title} {Shear banding or
  not in entangled dna solutions},}\ }\href@noop {} {\bibfield  {journal}
  {\bibinfo  {journal} {Macromolecules}\ }\textbf {\bibinfo {volume} {43}},\
  \bibinfo {pages} {6950--6952} (\bibinfo {year} {2010})}\BibitemShut {NoStop}%
\bibitem [{\citenamefont {Boukany}\ \emph {et~al.}(2010)\citenamefont
  {Boukany}, \citenamefont {Hemminger}, \citenamefont {Wang},\ and\
  \citenamefont {Lee}}]{Boukany2010b}%
  \BibitemOpen
  \bibfield  {author} {\bibinfo {author} {\bibfnamefont {P.~E.}\ \bibnamefont
  {Boukany}}, \bibinfo {author} {\bibfnamefont {O.}~\bibnamefont {Hemminger}},
  \bibinfo {author} {\bibfnamefont {S.-Q.}\ \bibnamefont {Wang}}, \ and\
  \bibinfo {author} {\bibfnamefont {L.~J.}\ \bibnamefont {Lee}},\ }\bibfield
  {title} {\enquote {\bibinfo {title} {Molecular imaging of slip in entangled
  dna solutions},}\ }\href@noop {} {\bibfield  {journal} {\bibinfo  {journal}
  {Phys. Rev. Lett.}\ }\textbf {\bibinfo {volume} {105}},\ \bibinfo {pages}
  {027802} (\bibinfo {year} {2010})}\BibitemShut {NoStop}%
\bibitem [{\citenamefont {Hemminger}\ and\ \citenamefont
  {Boukany}(2017)}]{Hemminger2017}%
  \BibitemOpen
  \bibfield  {author} {\bibinfo {author} {\bibfnamefont {O.}~\bibnamefont
  {Hemminger}}\ and\ \bibinfo {author} {\bibfnamefont {P.~E.}\ \bibnamefont
  {Boukany}},\ }\bibfield  {title} {\enquote {\bibinfo {title} {Microscopic
  origin of wall slip during flow of an entangled dna solution in
  microfluidics: Flow induced chain stretching versus chain desorption},}\
  }\href@noop {} {\bibfield  {journal} {\bibinfo  {journal} {Biomicrofluidics}\
  }\textbf {\bibinfo {volume} {11}},\ \bibinfo {pages} {044118} (\bibinfo
  {year} {2017})}\BibitemShut {NoStop}%
\bibitem [{\citenamefont {Malm}\ and\ \citenamefont {Waigh}(2017)}]{Malm2017}%
  \BibitemOpen
  \bibfield  {author} {\bibinfo {author} {\bibfnamefont {A.~V.}\ \bibnamefont
  {Malm}}\ and\ \bibinfo {author} {\bibfnamefont {T.~A.}\ \bibnamefont
  {Waigh}},\ }\bibfield  {title} {\enquote {\bibinfo {title} {Elastic
  turbulence in entangled semi-dilute dna solutions measured with optical
  coherence tomography velocimetry},}\ }\href@noop {} {\bibfield  {journal}
  {\bibinfo  {journal} {Scientific Reports}\ }\textbf {\bibinfo {volume} {7}},\
  \bibinfo {pages} {1186} (\bibinfo {year} {2017})}\BibitemShut {NoStop}%
\bibitem [{\citenamefont {Mohagheghi}\ and\ \citenamefont
  {Khomami}(2015)}]{Mohagheghi2015}%
  \BibitemOpen
  \bibfield  {author} {\bibinfo {author} {\bibfnamefont {M.}~\bibnamefont
  {Mohagheghi}}\ and\ \bibinfo {author} {\bibfnamefont {B.}~\bibnamefont
  {Khomami}},\ }\bibfield  {title} {\enquote {\bibinfo {title} {Molecular
  processes leading to shear banding in well entangled polymeric melts},}\
  }\href@noop {} {\bibfield  {journal} {\bibinfo  {journal} {ACS Macro Lett.}\
  }\textbf {\bibinfo {volume} {4}},\ \bibinfo {pages} {684--688} (\bibinfo
  {year} {2015})}\BibitemShut {NoStop}%
\bibitem [{\citenamefont {Mohagheghi}\ and\ \citenamefont
  {Khomami}(2016{\natexlab{a}})}]{Mohagheghi2016a}%
  \BibitemOpen
  \bibfield  {author} {\bibinfo {author} {\bibfnamefont {M.}~\bibnamefont
  {Mohagheghi}}\ and\ \bibinfo {author} {\bibfnamefont {B.}~\bibnamefont
  {Khomami}},\ }\bibfield  {title} {\enquote {\bibinfo {title} {Elucidating the
  flow-microstructure coupling in the entangled polymer melts. part i: Single
  chain dynamics in shear flow},}\ }\href@noop {} {\bibfield  {journal}
  {\bibinfo  {journal} {Journal of Rheology}\ }\textbf {\bibinfo {volume}
  {60}},\ \bibinfo {pages} {849--859} (\bibinfo {year}
  {2016}{\natexlab{a}})}\BibitemShut {NoStop}%
\bibitem [{\citenamefont {Mohagheghi}\ and\ \citenamefont
  {Khomami}(2016{\natexlab{b}})}]{Mohagheghi2016b}%
  \BibitemOpen
  \bibfield  {author} {\bibinfo {author} {\bibfnamefont {M.}~\bibnamefont
  {Mohagheghi}}\ and\ \bibinfo {author} {\bibfnamefont {B.}~\bibnamefont
  {Khomami}},\ }\bibfield  {title} {\enquote {\bibinfo {title} {Elucidating the
  flow-microstructure coupling in entangled polymer melts. part ii: Molecular
  mechanism of shear banding},}\ }\href@noop {} {\bibfield  {journal} {\bibinfo
   {journal} {Journal of Rheology}\ }\textbf {\bibinfo {volume} {60}},\
  \bibinfo {pages} {861--872} (\bibinfo {year}
  {2016}{\natexlab{b}})}\BibitemShut {NoStop}%
\bibitem [{\citenamefont {Mohagheghi}\ and\ \citenamefont
  {Khomami}(2016{\natexlab{c}})}]{Mohagheghi2016c}%
  \BibitemOpen
  \bibfield  {author} {\bibinfo {author} {\bibfnamefont {M.}~\bibnamefont
  {Mohagheghi}}\ and\ \bibinfo {author} {\bibfnamefont {B.}~\bibnamefont
  {Khomami}},\ }\bibfield  {title} {\enquote {\bibinfo {title} {Molecularly
  based criteria for shear banding in transient flow of entangled polymeric
  fluids},}\ }\href@noop {} {\bibfield  {journal} {\bibinfo  {journal} {Phys.
  Rev. Fluids}\ }\textbf {\bibinfo {volume} {93}},\ \bibinfo {pages} {062606}
  (\bibinfo {year} {2016}{\natexlab{c}})}\BibitemShut {NoStop}%
\bibitem [{\citenamefont {van Ruymbeke}\ \emph {et~al.}(2014)\citenamefont {van
  Ruymbeke}, \citenamefont {Lee}, \citenamefont {T.}, \citenamefont
  {Nikopoulou}, \citenamefont {Hadjichristidis}, \citenamefont {Snijkers},\
  and\ \citenamefont {Vlassopoulos}}]{Ruymbeke2014}%
  \BibitemOpen
  \bibfield  {author} {\bibinfo {author} {\bibfnamefont {E.}~\bibnamefont {van
  Ruymbeke}}, \bibinfo {author} {\bibfnamefont {H.}~\bibnamefont {Lee}},
  \bibinfo {author} {\bibfnamefont {Chang}\ \bibnamefont {T.}}, \bibinfo
  {author} {\bibfnamefont {A.}~\bibnamefont {Nikopoulou}}, \bibinfo {author}
  {\bibfnamefont {N.}~\bibnamefont {Hadjichristidis}}, \bibinfo {author}
  {\bibfnamefont {F.}~\bibnamefont {Snijkers}}, \ and\ \bibinfo {author}
  {\bibfnamefont {D.}~\bibnamefont {Vlassopoulos}},\ }\bibfield  {title}
  {\enquote {\bibinfo {title} {Molecular rheology of branched polymers:
  decoding and exploring the role of architectural dispersity through a synergy
  of anionic synthesis, interaction chromatography, rheometry and modeling},}\
  }\href@noop {} {\bibfield  {journal} {\bibinfo  {journal} {Soft Matter}\
  }\textbf {\bibinfo {volume} {10}},\ \bibinfo {pages} {4762--4777} (\bibinfo
  {year} {2014})}\BibitemShut {NoStop}%
\bibitem [{\citenamefont {Daniels}\ \emph {et~al.}(2001)\citenamefont
  {Daniels}, \citenamefont {McLeish}, \citenamefont {Crosby}, \citenamefont
  {Young},\ and\ \citenamefont {Fernyhough}}]{Daniels2001}%
  \BibitemOpen
  \bibfield  {author} {\bibinfo {author} {\bibfnamefont {D.~R.}\ \bibnamefont
  {Daniels}}, \bibinfo {author} {\bibfnamefont {T.~C.~B.}\ \bibnamefont
  {McLeish}}, \bibinfo {author} {\bibfnamefont {B.~J.}\ \bibnamefont {Crosby}},
  \bibinfo {author} {\bibfnamefont {R.~N.}\ \bibnamefont {Young}}, \ and\
  \bibinfo {author} {\bibfnamefont {C.~M.}\ \bibnamefont {Fernyhough}},\
  }\bibfield  {title} {\enquote {\bibinfo {title} {Molecular rheology of comb
  polymer melts: 1. linear viscoelastic response},}\ }\href@noop {} {\bibfield
  {journal} {\bibinfo  {journal} {Macromolecules}\ }\textbf {\bibinfo {volume}
  {34}},\ \bibinfo {pages} {7025--7033} (\bibinfo {year} {2001})}\BibitemShut
  {NoStop}%
\bibitem [{\citenamefont {Heuer}\ \emph {et~al.}(2003)\citenamefont {Heuer},
  \citenamefont {Saha},\ and\ \citenamefont {Archer}}]{Heuer2003}%
  \BibitemOpen
  \bibfield  {author} {\bibinfo {author} {\bibfnamefont {D.~M.}\ \bibnamefont
  {Heuer}}, \bibinfo {author} {\bibfnamefont {S.}~\bibnamefont {Saha}}, \ and\
  \bibinfo {author} {\bibfnamefont {L.~A.}\ \bibnamefont {Archer}},\ }\bibfield
   {title} {\enquote {\bibinfo {title} {Electrophoretic dynamics of large dna
  stars in polymer solutions and gels},}\ }\href@noop {} {\bibfield  {journal}
  {\bibinfo  {journal} {Electrophoresis}\ }\textbf {\bibinfo {volume} {24}},\
  \bibinfo {pages} {3314--3322} (\bibinfo {year} {2003})}\BibitemShut {NoStop}%
\bibitem [{\citenamefont {Saha}\ \emph {et~al.}(2006)\citenamefont {Saha},
  \citenamefont {Heuer},\ and\ \citenamefont {Archer}}]{Saha2006}%
  \BibitemOpen
  \bibfield  {author} {\bibinfo {author} {\bibfnamefont {S.}~\bibnamefont
  {Saha}}, \bibinfo {author} {\bibfnamefont {D.~M.}\ \bibnamefont {Heuer}}, \
  and\ \bibinfo {author} {\bibfnamefont {L.~A.}\ \bibnamefont {Archer}},\
  }\bibfield  {title} {\enquote {\bibinfo {title} {Electrophoretic mobility of
  of linear and star-branched dna in semidilute polymer solutions},}\
  }\href@noop {} {\bibfield  {journal} {\bibinfo  {journal} {Electrophoresis}\
  }\textbf {\bibinfo {volume} {27}},\ \bibinfo {pages} {3181--3194} (\bibinfo
  {year} {2006})}\BibitemShut {NoStop}%
\bibitem [{\citenamefont {Saha}\ \emph {et~al.}(2004)\citenamefont {Saha},
  \citenamefont {Heuer},\ and\ \citenamefont {Archer}}]{Saha2004}%
  \BibitemOpen
  \bibfield  {author} {\bibinfo {author} {\bibfnamefont {S.}~\bibnamefont
  {Saha}}, \bibinfo {author} {\bibfnamefont {D.~M.}\ \bibnamefont {Heuer}}, \
  and\ \bibinfo {author} {\bibfnamefont {L.~A.}\ \bibnamefont {Archer}},\
  }\bibfield  {title} {\enquote {\bibinfo {title} {Effect of matrix chain
  length on the electrophoretic mobility of large linear and branched dna in
  polymer solutions},}\ }\href@noop {} {\bibfield  {journal} {\bibinfo
  {journal} {Electrophoresis}\ }\textbf {\bibinfo {volume} {25}},\ \bibinfo
  {pages} {396--404} (\bibinfo {year} {2004})}\BibitemShut {NoStop}%
\bibitem [{\citenamefont {Grosberg}\ and\ \citenamefont
  {Rabin}(2007)}]{Grosberg2007}%
  \BibitemOpen
  \bibfield  {author} {\bibinfo {author} {\bibfnamefont {A.~Y.}\ \bibnamefont
  {Grosberg}}\ and\ \bibinfo {author} {\bibfnamefont {Y.}~\bibnamefont
  {Rabin}},\ }\bibfield  {title} {\enquote {\bibinfo {title} {Metastable tight
  knots in a wormlike polymer},}\ }\href@noop {} {\bibfield  {journal}
  {\bibinfo  {journal} {Phys. Rev. Lett.}\ }\textbf {\bibinfo {volume} {99}},\
  \bibinfo {pages} {217801} (\bibinfo {year} {2007})}\BibitemShut {NoStop}%
\bibitem [{\citenamefont {Tang}\ \emph {et~al.}(2011)\citenamefont {Tang},
  \citenamefont {Du},\ and\ \citenamefont {Doyle}}]{Tang2011}%
  \BibitemOpen
  \bibfield  {author} {\bibinfo {author} {\bibfnamefont {J.}~\bibnamefont
  {Tang}}, \bibinfo {author} {\bibfnamefont {N.}~\bibnamefont {Du}}, \ and\
  \bibinfo {author} {\bibfnamefont {P.~S.}\ \bibnamefont {Doyle}},\ }\bibfield
  {title} {\enquote {\bibinfo {title} {Compression and self-entanglement of
  single dna molecules under uniform electric fields},}\ }\href@noop {}
  {\bibfield  {journal} {\bibinfo  {journal} {Proc. Nat. Acad. Sci.}\ }\textbf
  {\bibinfo {volume} {108}},\ \bibinfo {pages} {16153--16158} (\bibinfo {year}
  {2011})}\BibitemShut {NoStop}%
\bibitem [{\citenamefont {Renner}\ and\ \citenamefont
  {Doyle}(2014)}]{Renner2014}%
  \BibitemOpen
  \bibfield  {author} {\bibinfo {author} {\bibfnamefont {C.~B.}\ \bibnamefont
  {Renner}}\ and\ \bibinfo {author} {\bibfnamefont {P.~S.}\ \bibnamefont
  {Doyle}},\ }\bibfield  {title} {\enquote {\bibinfo {title} {Untying knotted
  dna with elongational flows},}\ }\href@noop {} {\bibfield  {journal}
  {\bibinfo  {journal} {ACS Macro Lett.}\ }\textbf {\bibinfo {volume} {3}},\
  \bibinfo {pages} {963--967} (\bibinfo {year} {2014})}\BibitemShut {NoStop}%
\bibitem [{\citenamefont {Renner}\ and\ \citenamefont
  {Doyle}(2015)}]{Renner2015}%
  \BibitemOpen
  \bibfield  {author} {\bibinfo {author} {\bibfnamefont {C.~B.}\ \bibnamefont
  {Renner}}\ and\ \bibinfo {author} {\bibfnamefont {P.~S.}\ \bibnamefont
  {Doyle}},\ }\bibfield  {title} {\enquote {\bibinfo {title} {Stretching
  self-entangled dna molecules in elongational fields},}\ }\href@noop {}
  {\bibfield  {journal} {\bibinfo  {journal} {Soft Matter}\ }\textbf {\bibinfo
  {volume} {11}},\ \bibinfo {pages} {3105--3114} (\bibinfo {year}
  {2015})}\BibitemShut {NoStop}%
\bibitem [{\citenamefont {Reifenberger}\ \emph {et~al.}(2015)\citenamefont
  {Reifenberger}, \citenamefont {Dorfman},\ and\ \citenamefont
  {Cao}}]{Reifenberger2015}%
  \BibitemOpen
  \bibfield  {author} {\bibinfo {author} {\bibfnamefont {J.~G.}\ \bibnamefont
  {Reifenberger}}, \bibinfo {author} {\bibfnamefont {K.~D.}\ \bibnamefont
  {Dorfman}}, \ and\ \bibinfo {author} {\bibfnamefont {H.}~\bibnamefont
  {Cao}},\ }\bibfield  {title} {\enquote {\bibinfo {title} {Topological events
  in single molecules of e. coli dna confined in nanochannels},}\ }\href@noop
  {} {\bibfield  {journal} {\bibinfo  {journal} {Analyst}\ }\textbf {\bibinfo
  {volume} {140}},\ \bibinfo {pages} {4887--4894} (\bibinfo {year}
  {2015})}\BibitemShut {NoStop}%
\bibitem [{\citenamefont {Marciel}\ \emph {et~al.}(2015)\citenamefont
  {Marciel}, \citenamefont {Mai},\ and\ \citenamefont
  {Schroeder}}]{Marciel2015}%
  \BibitemOpen
  \bibfield  {author} {\bibinfo {author} {\bibfnamefont {A.~B.}\ \bibnamefont
  {Marciel}}, \bibinfo {author} {\bibfnamefont {D.~J.}\ \bibnamefont {Mai}}, \
  and\ \bibinfo {author} {\bibfnamefont {C.~M.}\ \bibnamefont {Schroeder}},\
  }\bibfield  {title} {\enquote {\bibinfo {title} {Template-directed synthesis
  of structurally defined branched polymers},}\ }\href@noop {} {\bibfield
  {journal} {\bibinfo  {journal} {Macromolecules}\ }\textbf {\bibinfo {volume}
  {48}},\ \bibinfo {pages} {1296--1303} (\bibinfo {year} {2015})}\BibitemShut
  {NoStop}%
\bibitem [{\citenamefont {Berezney}\ \emph {et~al.}(2017)\citenamefont
  {Berezney}, \citenamefont {Marciel}, \citenamefont {Schroeder},\ and\
  \citenamefont {Saleh}}]{Berezney2017}%
  \BibitemOpen
  \bibfield  {author} {\bibinfo {author} {\bibfnamefont {J.~P.}\ \bibnamefont
  {Berezney}}, \bibinfo {author} {\bibfnamefont {A.~B.}\ \bibnamefont
  {Marciel}}, \bibinfo {author} {\bibfnamefont {C.~M.}\ \bibnamefont
  {Schroeder}}, \ and\ \bibinfo {author} {\bibfnamefont {O.~A.}\ \bibnamefont
  {Saleh}},\ }\bibfield  {title} {\enquote {\bibinfo {title} {Scale-dependent
  stiffness and internal tension of a model brush polymer},}\ }\href@noop {}
  {\bibfield  {journal} {\bibinfo  {journal} {Phys. Rev. Lett.}\ }\textbf
  {\bibinfo {volume} {116}},\ \bibinfo {pages} {127801} (\bibinfo {year}
  {2017})}\BibitemShut {NoStop}%
\bibitem [{\citenamefont {Kapnistos}\ \emph {et~al.}(2008)\citenamefont
  {Kapnistos}, \citenamefont {Lang}, \citenamefont {Vlassopoulos},
  \citenamefont {Pyckhout-Hintzen}, \citenamefont {Richter},\ and\
  \citenamefont {et~al.}}]{Kapnistos2008}%
  \BibitemOpen
  \bibfield  {author} {\bibinfo {author} {\bibfnamefont {M.}~\bibnamefont
  {Kapnistos}}, \bibinfo {author} {\bibfnamefont {M.}~\bibnamefont {Lang}},
  \bibinfo {author} {\bibfnamefont {D.}~\bibnamefont {Vlassopoulos}}, \bibinfo
  {author} {\bibfnamefont {W.}~\bibnamefont {Pyckhout-Hintzen}}, \bibinfo
  {author} {\bibfnamefont {D.~H.}\ \bibnamefont {Richter}}, \ and\ \bibinfo
  {author} {\bibnamefont {et~al.}},\ }\bibfield  {title} {\enquote {\bibinfo
  {title} {Unexpected power-law stress relaxation of entangled ring
  polymers},}\ }\href@noop {} {\bibfield  {journal} {\bibinfo  {journal}
  {Nature Materials}\ }\textbf {\bibinfo {volume} {7}},\ \bibinfo {pages}
  {997--1002} (\bibinfo {year} {2008})}\BibitemShut {NoStop}%
\bibitem [{\citenamefont {Pasquino}\ \emph {et~al.}(2013)\citenamefont
  {Pasquino}, \citenamefont {Vasilakopoulos}, \citenamefont {Jeong},
  \citenamefont {Lee}, \citenamefont {Rogers},\ and\ \citenamefont
  {et~al.}}]{Pasquino2013}%
  \BibitemOpen
  \bibfield  {author} {\bibinfo {author} {\bibfnamefont {R.}~\bibnamefont
  {Pasquino}}, \bibinfo {author} {\bibfnamefont {T.~C.}\ \bibnamefont
  {Vasilakopoulos}}, \bibinfo {author} {\bibfnamefont {Y.~C.}\ \bibnamefont
  {Jeong}}, \bibinfo {author} {\bibfnamefont {H.}~\bibnamefont {Lee}}, \bibinfo
  {author} {\bibfnamefont {S.~A.}\ \bibnamefont {Rogers}}, \ and\ \bibinfo
  {author} {\bibnamefont {et~al.}},\ }\bibfield  {title} {\enquote {\bibinfo
  {title} {Viscosity of ring polymer melts},}\ }\href@noop {} {\bibfield
  {journal} {\bibinfo  {journal} {ACS Macro Lett.}\ }\textbf {\bibinfo {volume}
  {2}},\ \bibinfo {pages} {874--878} (\bibinfo {year} {2013})}\BibitemShut
  {NoStop}%
\bibitem [{\citenamefont {Yan}\ \emph {et~al.}(2016)\citenamefont {Yan},
  \citenamefont {Costanzo}, \citenamefont {Jeong}, \citenamefont {T.},\ and\
  \citenamefont {Vlassopoulos}}]{Yan2016}%
  \BibitemOpen
  \bibfield  {author} {\bibinfo {author} {\bibfnamefont {Z.-C.}\ \bibnamefont
  {Yan}}, \bibinfo {author} {\bibfnamefont {S.}~\bibnamefont {Costanzo}},
  \bibinfo {author} {\bibfnamefont {Y.~C.}\ \bibnamefont {Jeong}}, \bibinfo
  {author} {\bibfnamefont {Chang}\ \bibnamefont {T.}}, \ and\ \bibinfo {author}
  {\bibfnamefont {D.}~\bibnamefont {Vlassopoulos}},\ }\bibfield  {title}
  {\enquote {\bibinfo {title} {Linear and nonlinear shear rheology of a
  marginally entangled ring polymer},}\ }\href@noop {} {\bibfield  {journal}
  {\bibinfo  {journal} {Macromolecules}\ }\textbf {\bibinfo {volume} {49}},\
  \bibinfo {pages} {1444--1453} (\bibinfo {year} {2016})}\BibitemShut {NoStop}%
\bibitem [{\citenamefont {Robertson}\ and\ \citenamefont
  {Smith}(2007{\natexlab{c}})}]{Robertson2007c}%
  \BibitemOpen
  \bibfield  {author} {\bibinfo {author} {\bibfnamefont {R.~M.}\ \bibnamefont
  {Robertson}}\ and\ \bibinfo {author} {\bibfnamefont {D.~E.}\ \bibnamefont
  {Smith}},\ }\bibfield  {title} {\enquote {\bibinfo {title} {Strong effects of
  molecular topology on diffusion of entangled dna molecules},}\ }\href@noop {}
  {\bibfield  {journal} {\bibinfo  {journal} {Proc. Nat. Acad. Sci.}\ }\textbf
  {\bibinfo {volume} {104}},\ \bibinfo {pages} {4824--4827} (\bibinfo {year}
  {2007}{\natexlab{c}})}\BibitemShut {NoStop}%
\bibitem [{\citenamefont {Chapman}\ \emph {et~al.}(2012)\citenamefont
  {Chapman}, \citenamefont {Shanbhag}, \citenamefont {Smith},\ and\
  \citenamefont {Robertson-Anderson}}]{Chapman2012}%
  \BibitemOpen
  \bibfield  {author} {\bibinfo {author} {\bibfnamefont {C.~D.}\ \bibnamefont
  {Chapman}}, \bibinfo {author} {\bibfnamefont {S.}~\bibnamefont {Shanbhag}},
  \bibinfo {author} {\bibfnamefont {D.~E.}\ \bibnamefont {Smith}}, \ and\
  \bibinfo {author} {\bibfnamefont {R.~M.}\ \bibnamefont
  {Robertson-Anderson}},\ }\bibfield  {title} {\enquote {\bibinfo {title}
  {Complex effects of molecular topology on diffusion in entangled biopolymer
  blends},}\ }\href@noop {} {\bibfield  {journal} {\bibinfo  {journal} {Soft
  Matter}\ }\textbf {\bibinfo {volume} {8}},\ \bibinfo {pages} {9177--9182}
  (\bibinfo {year} {2012})}\BibitemShut {NoStop}%
\bibitem [{\citenamefont {Serag}\ \emph {et~al.}(2014)\citenamefont {Serag},
  \citenamefont {Abadi},\ and\ \citenamefont {Habuchi}}]{Serag2014}%
  \BibitemOpen
  \bibfield  {author} {\bibinfo {author} {\bibfnamefont {M.~F.}\ \bibnamefont
  {Serag}}, \bibinfo {author} {\bibfnamefont {M.}~\bibnamefont {Abadi}}, \ and\
  \bibinfo {author} {\bibfnamefont {S.}~\bibnamefont {Habuchi}},\ }\bibfield
  {title} {\enquote {\bibinfo {title} {Single-molecule diffusion and
  conformational dynamics by spatial integration of temporal fluctuations},}\
  }\href@noop {} {\bibfield  {journal} {\bibinfo  {journal} {Nature
  Communications}\ }\textbf {\bibinfo {volume} {5}} (\bibinfo {year}
  {2014})}\BibitemShut {NoStop}%
\bibitem [{\citenamefont {Abadi}\ \emph {et~al.}(2015)\citenamefont {Abadi},
  \citenamefont {Serag},\ and\ \citenamefont {Habuchi}}]{Abadi2015}%
  \BibitemOpen
  \bibfield  {author} {\bibinfo {author} {\bibfnamefont {M.}~\bibnamefont
  {Abadi}}, \bibinfo {author} {\bibfnamefont {M.~F.}\ \bibnamefont {Serag}}, \
  and\ \bibinfo {author} {\bibfnamefont {S.}~\bibnamefont {Habuchi}},\
  }\bibfield  {title} {\enquote {\bibinfo {title} {Single-molecule imaging
  reveals topology dependent mutual relaxation of polymer chains},}\
  }\href@noop {} {\bibfield  {journal} {\bibinfo  {journal} {Macromolecules}\
  }\textbf {\bibinfo {volume} {48}},\ \bibinfo {pages} {6263--6371} (\bibinfo
  {year} {2015})}\BibitemShut {NoStop}%
\bibitem [{\citenamefont {Li}\ \emph {et~al.}(2015{\natexlab{b}})\citenamefont
  {Li}, \citenamefont {Hsiao}, \citenamefont {Brockman}, \citenamefont {Yates},
  \citenamefont {Robertson-Anderson}, \citenamefont {Kornfield}, \citenamefont
  {Schroeder},\ and\ \citenamefont {McKenna}}]{Li2015b}%
  \BibitemOpen
  \bibfield  {author} {\bibinfo {author} {\bibfnamefont {Y.}~\bibnamefont
  {Li}}, \bibinfo {author} {\bibfnamefont {K.}~\bibnamefont {Hsiao}}, \bibinfo
  {author} {\bibfnamefont {C.~A.}\ \bibnamefont {Brockman}}, \bibinfo {author}
  {\bibfnamefont {D.~Y.}\ \bibnamefont {Yates}}, \bibinfo {author}
  {\bibfnamefont {R.~M.}\ \bibnamefont {Robertson-Anderson}}, \bibinfo {author}
  {\bibfnamefont {J.~A.}\ \bibnamefont {Kornfield}}, \bibinfo {author}
  {\bibfnamefont {C.~M.}\ \bibnamefont {Schroeder}}, \ and\ \bibinfo {author}
  {\bibfnamefont {G.~B.}\ \bibnamefont {McKenna}},\ }\bibfield  {title}
  {\enquote {\bibinfo {title} {When ends meet: circular dna stretches
  differently in elongational flows},}\ }\href@noop {} {\bibfield  {journal}
  {\bibinfo  {journal} {Macromolecules}\ }\textbf {\bibinfo {volume} {48}},\
  \bibinfo {pages} {5997--6001} (\bibinfo {year}
  {2015}{\natexlab{b}})}\BibitemShut {NoStop}%
\bibitem [{\citenamefont {Hsiao}\ \emph {et~al.}(2016)\citenamefont {Hsiao},
  \citenamefont {Schroeder},\ and\ \citenamefont {Sing}}]{Hsiao2016}%
  \BibitemOpen
  \bibfield  {author} {\bibinfo {author} {\bibfnamefont {K.}~\bibnamefont
  {Hsiao}}, \bibinfo {author} {\bibfnamefont {C.~M.}\ \bibnamefont
  {Schroeder}}, \ and\ \bibinfo {author} {\bibfnamefont {C.~E.}\ \bibnamefont
  {Sing}},\ }\bibfield  {title} {\enquote {\bibinfo {title} {Ring polymer
  dynamics are governed by a coupling between architecture and hydrodynamic
  interactions},}\ }\href@noop {} {\bibfield  {journal} {\bibinfo  {journal}
  {Macromolecules}\ }\textbf {\bibinfo {volume} {49}},\ \bibinfo {pages}
  {1961--1971} (\bibinfo {year} {2016})}\BibitemShut {NoStop}%
\bibitem [{\citenamefont {Hegde}\ \emph {et~al.}(2011)\citenamefont {Hegde},
  \citenamefont {Chang}, \citenamefont {Chen},\ and\ \citenamefont
  {Khare}}]{Hegde2011}%
  \BibitemOpen
  \bibfield  {author} {\bibinfo {author} {\bibfnamefont {G.~A.}\ \bibnamefont
  {Hegde}}, \bibinfo {author} {\bibfnamefont {J.}~\bibnamefont {Chang}},
  \bibinfo {author} {\bibfnamefont {Y.}~\bibnamefont {Chen}}, \ and\ \bibinfo
  {author} {\bibfnamefont {R.}~\bibnamefont {Khare}},\ }\bibfield  {title}
  {\enquote {\bibinfo {title} {Conformation and diffusion behavior of ring
  polymers in solution: A comparison between molecular dynamics, multiparticle
  collision dynamics, and lattice boltzmann simulations},}\ }\href@noop {}
  {\bibfield  {journal} {\bibinfo  {journal} {J. Chem. Phys.}\ }\textbf
  {\bibinfo {volume} {135}},\ \bibinfo {pages} {184901} (\bibinfo {year}
  {2011})}\BibitemShut {NoStop}%
\bibitem [{\citenamefont {Clasen}\ \emph {et~al.}(2006)\citenamefont {Clasen},
  \citenamefont {Plog}, \citenamefont {Kulicke}, \citenamefont {Owens},
  \citenamefont {Macosko}, \citenamefont {Scriven}, \citenamefont {Verani},\
  and\ \citenamefont {McKinley}}]{Clasen2006}%
  \BibitemOpen
  \bibfield  {author} {\bibinfo {author} {\bibfnamefont {C.}~\bibnamefont
  {Clasen}}, \bibinfo {author} {\bibfnamefont {J.~P.}\ \bibnamefont {Plog}},
  \bibinfo {author} {\bibfnamefont {W.-M.}\ \bibnamefont {Kulicke}}, \bibinfo
  {author} {\bibfnamefont {M.}~\bibnamefont {Owens}}, \bibinfo {author}
  {\bibfnamefont {C.}~\bibnamefont {Macosko}}, \bibinfo {author} {\bibfnamefont
  {L.~E.}\ \bibnamefont {Scriven}}, \bibinfo {author} {\bibfnamefont
  {M.}~\bibnamefont {Verani}}, \ and\ \bibinfo {author} {\bibfnamefont {G.~H.}\
  \bibnamefont {McKinley}},\ }\bibfield  {title} {\enquote {\bibinfo {title}
  {How dilute are dilute solutions in extensional flows?}}\ }\href@noop {}
  {\bibfield  {journal} {\bibinfo  {journal} {J. Rheo.}\ }\textbf {\bibinfo
  {volume} {50}},\ \bibinfo {pages} {849} (\bibinfo {year} {2006})}\BibitemShut
  {NoStop}%
\bibitem [{\citenamefont {Lentzakis}\ \emph {et~al.}(2014)\citenamefont
  {Lentzakis}, \citenamefont {Das}, \citenamefont {Vlassopoulos},\ and\
  \citenamefont {Read}}]{Lentzakis2014}%
  \BibitemOpen
  \bibfield  {author} {\bibinfo {author} {\bibfnamefont {H.}~\bibnamefont
  {Lentzakis}}, \bibinfo {author} {\bibfnamefont {C.}~\bibnamefont {Das}},
  \bibinfo {author} {\bibfnamefont {D.}~\bibnamefont {Vlassopoulos}}, \ and\
  \bibinfo {author} {\bibfnamefont {D.~J.}\ \bibnamefont {Read}},\ }\bibfield
  {title} {\enquote {\bibinfo {title} {Pom-pom-like constitutive equations for
  comb polymers},}\ }\href@noop {} {\bibfield  {journal} {\bibinfo  {journal}
  {Journal of Rheology}\ }\textbf {\bibinfo {volume} {58}},\ \bibinfo {pages}
  {1855--1875} (\bibinfo {year} {2014})}\BibitemShut {NoStop}%
\bibitem [{\citenamefont {Quake}\ \emph {et~al.}(1997)\citenamefont {Quake},
  \citenamefont {Babcock},\ and\ \citenamefont {Chu}}]{Quake1997}%
  \BibitemOpen
  \bibfield  {author} {\bibinfo {author} {\bibfnamefont {S.~R.}\ \bibnamefont
  {Quake}}, \bibinfo {author} {\bibfnamefont {H.~P.}\ \bibnamefont {Babcock}},
  \ and\ \bibinfo {author} {\bibfnamefont {S.}~\bibnamefont {Chu}},\ }\bibfield
   {title} {\enquote {\bibinfo {title} {The dynamics of partially extended
  single molecules of dna},}\ }\href@noop {} {\bibfield  {journal} {\bibinfo
  {journal} {Nature}\ }\textbf {\bibinfo {volume} {388}},\ \bibinfo {pages}
  {151--154} (\bibinfo {year} {1997})}\BibitemShut {NoStop}%
\bibitem [{\citenamefont {Shendruk}\ \emph {et~al.}(2017)\citenamefont
  {Shendruk}, \citenamefont {Sean}, \citenamefont {Berard}, \citenamefont
  {Wolf}, \citenamefont {Dragoman}, \citenamefont {Battat}, \citenamefont
  {Slater},\ and\ \citenamefont {Leslie}}]{Shendruk2017}%
  \BibitemOpen
  \bibfield  {author} {\bibinfo {author} {\bibfnamefont {T.~N.}\ \bibnamefont
  {Shendruk}}, \bibinfo {author} {\bibfnamefont {D.}~\bibnamefont {Sean}},
  \bibinfo {author} {\bibfnamefont {D.~J.}\ \bibnamefont {Berard}}, \bibinfo
  {author} {\bibfnamefont {J.}~\bibnamefont {Wolf}}, \bibinfo {author}
  {\bibfnamefont {J.}~\bibnamefont {Dragoman}}, \bibinfo {author}
  {\bibfnamefont {S.}~\bibnamefont {Battat}}, \bibinfo {author} {\bibfnamefont
  {G.~W.}\ \bibnamefont {Slater}}, \ and\ \bibinfo {author} {\bibfnamefont
  {S.~R.}\ \bibnamefont {Leslie}},\ }\bibfield  {title} {\enquote {\bibinfo
  {title} {Rotation-induced macromolecular spooling of dna},}\ }\href@noop {}
  {\bibfield  {journal} {\bibinfo  {journal} {Physical Review X}\ }\textbf
  {\bibinfo {volume} {7}},\ \bibinfo {pages} {031005} (\bibinfo {year}
  {2017})}\BibitemShut {NoStop}%
\bibitem [{\citenamefont {Hsiao}\ \emph
  {et~al.}(2017{\natexlab{b}})\citenamefont {Hsiao}, \citenamefont {Dinic},
  \citenamefont {Ren}, \citenamefont {Sharma},\ and\ \citenamefont
  {Schroeder}}]{Hsiao2017b}%
  \BibitemOpen
  \bibfield  {author} {\bibinfo {author} {\bibfnamefont {K.}~\bibnamefont
  {Hsiao}}, \bibinfo {author} {\bibfnamefont {J.}~\bibnamefont {Dinic}},
  \bibinfo {author} {\bibfnamefont {Y.}~\bibnamefont {Ren}}, \bibinfo {author}
  {\bibfnamefont {V.}~\bibnamefont {Sharma}}, \ and\ \bibinfo {author}
  {\bibfnamefont {C.~M.}\ \bibnamefont {Schroeder}},\ }\bibfield  {title}
  {\enquote {\bibinfo {title} {Passive non-linear microrheology for determining
  extensional viscosity},}\ }\href@noop {} {\bibfield  {journal} {\bibinfo
  {journal} {Phys. Fluids}\ }\textbf {\bibinfo {volume} {29}},\ \bibinfo
  {pages} {121603} (\bibinfo {year} {2017}{\natexlab{b}})}\BibitemShut
  {NoStop}%
\end{thebibliography}%

\end{document}